\documentclass{article}

\usepackage{PRIMEarxiv}

\usepackage[T1]{fontenc}    
\usepackage{hyperref}       
\usepackage{url}            
\usepackage{booktabs}       
\usepackage{amssymb,amsmath,eurosym,geometry,ulem,graphicx,caption,color,setspace,sectsty,comment,footmisc,pdflscape,subfigure,array}

\usepackage{amsfonts}       
\usepackage{nicefrac}       
\usepackage{microtype}      
\usepackage{lipsum}
\usepackage{fancyhdr}       
\usepackage{graphicx}       
\usepackage{doi}
\usepackage{rotating}
\usepackage[scientific-notation=true]{siunitx}
\usepackage[section]{placeins}
\usepackage{gensymb} 

\title{Social Cost of Greenhouse Gases---OPTiMEM and the Heat Conjecture ($s$)}

\author{ \href{https://orcid.org/0000-0003-1202-2795}{\includegraphics[scale=0.06]{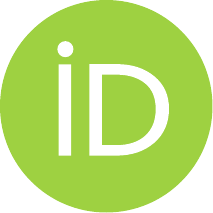}\hspace{1mm}Brian P. Hanley} \\
	Butterfly Sciences, Post Falls, Idaho, USA \\
	\texttt{brian.hanley@bf-sci.com} \\
	\And
	\href{https://orcid.org/0000-0002-9883-0787}{\includegraphics[scale=0.06]{orcid.pdf}\hspace{1mm}Pieter Tans} \\
	Institute of Arctic and Alpine Research, University of Colorado Boulder, Boulder, Colorado, USA \\	
    Global Monitoring Laboratory, National Oceanic and Atmospheric Administration, Boulder, Colorado, USA \\	
    \And
	\href{https://orcid.org/0000-0002-1096-2436}{\includegraphics[scale=0.06]{orcid.pdf}\hspace{1mm}Edward A.G. Schuur} \\
	Center for Ecosystem Science and Society, Northern Arizona University, Flagstaff, Arizona, USA \\	
    \And
	Geoffrey Gardiner \\
	Chartered Governance Institute Fellow, Associate---London Institute of Banking and Finance, UK \\
	\And
    Adam Smith \\
    Climate Central, Princeton, New Jersey, USA \\
	Formerly: NOAA/NESDIS/NCEI, Asheville, North Carolina, USA \\
}

\hypersetup{
pdftitle={Supplement: Social Cost of Greenhouse Gases---OPTiMEM and the Heat Conjecture},
pdfsubject={climate economics},
pdfauthor={Brian P. Hanley, Pieter Tans, Edward A.G. Schuur, Geoffrey Gardiner, Adam  Smith},
pdfkeywords={SCC, SC-GHG, SC-CO2, SC-N2O, social cost of carbon, social cost of CH4, social cost of N2O, social cost of GHG, social cost of F-gases, climate economics, climate model, heat conjecture, simplified climate model, ocean heat model}
}

\begin{document}

\maketitle

\begin{quote}
    "But what would you do differently?" - Anonymous climate economist, 2021
\end{quote}

\section{Genesis of this work}
Many iterations ago, this work started with the above question. At the time, various flavours of integrated assessment models (IAMs) were virtually the only way to get at the cost of emitting carbon dioxide. As of submission of this manuscript, this is still true. The gulf between natural scientists and economists \cite{lenton2013tipping} seems to be unbridgeable. Can a bridge be built? We attempt to do that here. Is it perfect? No. Is it helpful for estimating social cost of carbon, nitrogen, and halogenated gases? Yes. Is it firmly connected to physics? Yes. 

\begin{figure}[hbt!]
\centering \includegraphics[width=3.5in]{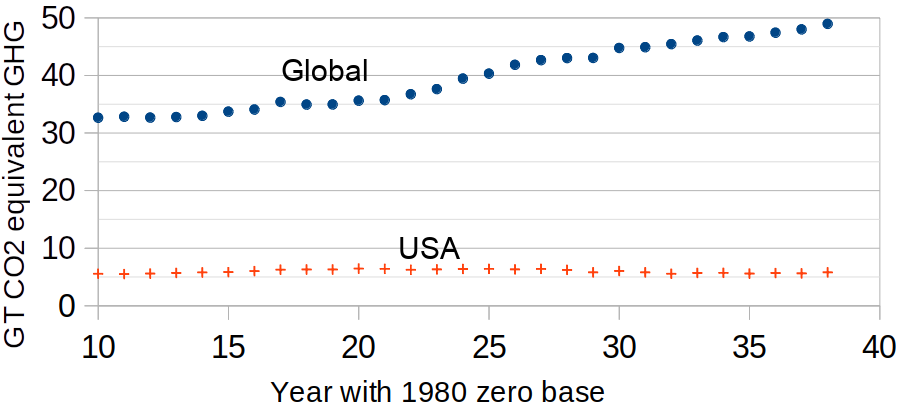}
\caption{\textbf{Gigatons of CO$_2$ equivalent greenhouse gases emitted by year} \cite{ClimateWatch2021DataExplorerGHG}. Upper dots global emissions, lower crosses, USA emissions. USA emissions as percentage of global emissions declined from high of 18\% of global emissions to 12\% due to global growth by remaining fairly flat. However, cost per tonne increases with global CO$_2$, not national CO$_2$. Given the longevity of the climate problem, the CO$_2$ equivalent metrics have understated the climate impact of CO$_2$ relative to other GHGs, leading to unwarranted emphasis on GHGs like CH$_4$ \cite{Tans_2022UseAbuseOfC-14andC-13InAtmosphericCO2}. }
\label{fig_MT_CO2_by_year}
\end{figure}

Despite well-meaning scenarios that propose global CO$_2$ emissions will decline presented in every IPCC report since 1988, the trend of global CO$_2$ increase continues without significant change (fig. \ref{fig_MT_CO2_by_year}). Even if any individual nation manages to flatten its emissions, what matters is the trajectory of the globe. Together this gulf plus the urgent need for alternative methods of estimation provided the incentives for this project. In response, our Ocean-Heat-Content (OHC) Physics and Time Macro Economic Model (OPTiMEM) system was developed. 

\subsection{Heuristic model---not a viable path forward for us.} 
\begin{figure}[!hbt]
\centering\includegraphics[width=6.25in]{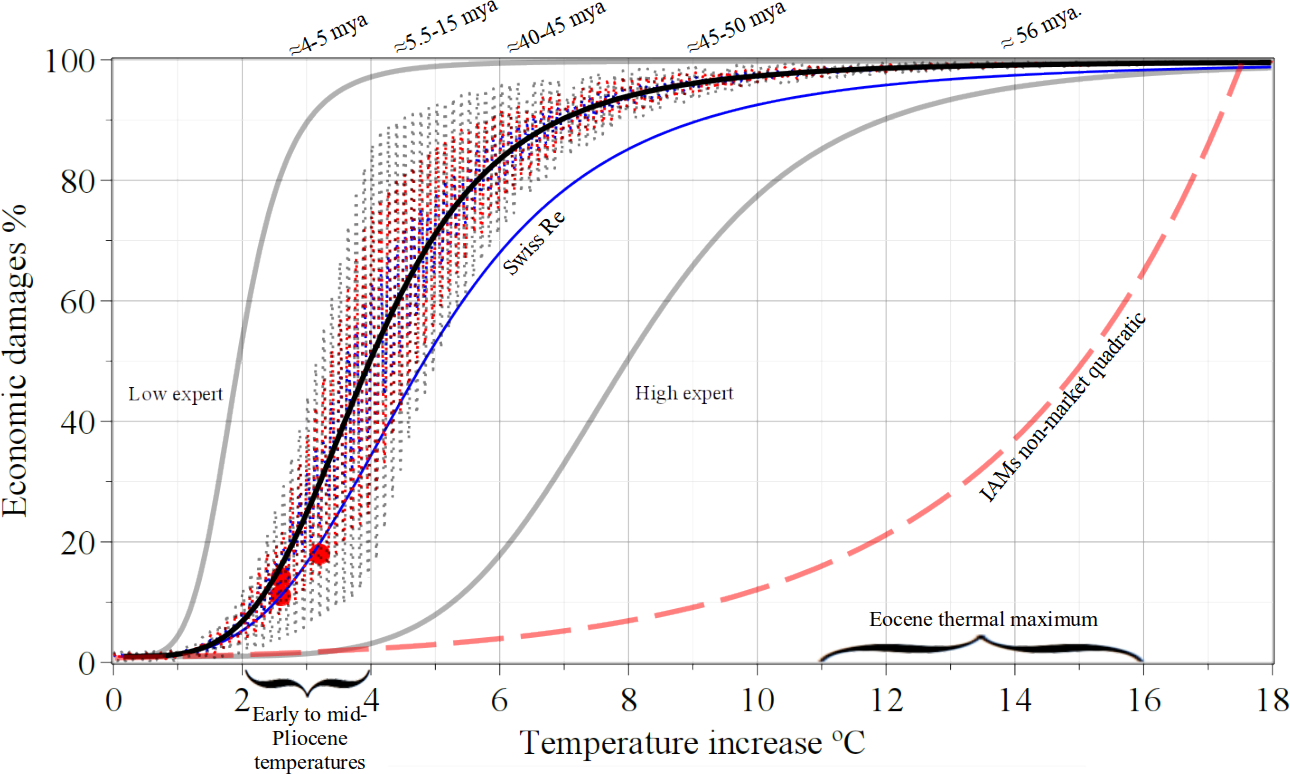}
\caption{\textbf{Expert heuristic estimate of economic damages---interesting, but not a viable path forward.} \emph{This graph presents a relation between temperature ($\Delta$T) and damages that simply does not exist.} We present this graph to show the evolution of our thinking over time. Here we have responded to the criticisms regarding quadratic (exponential) growth, smooth curves, catastrophic events and discontinuities \cite{STANTON2009Inside30ClimateModels,Pindyck2017UseandMisuse, Pindyck2013PolicyDilemma, Pindyck2013Whatmodelstell, WEITZMAN2010, WEITZMAN2012, Keen2020appallingly, keen2021erroneous, Keen2022EstimatesTippingCannotBeReconciled}, providing 3 layers of "cloud", where the actual results could be likely to land. The heavy smooth black curve is the expert's mean.  IAMs non-market damages quadratic function (long dash light red) \cite{nordhaussztorc2016Rdice}.
The small blue/red/gray heat map points suggest probability potential for deviation from the expert's mean curve. Blue dots is highest probability damages years up to 1 bad year in 10, red dots is intermediate 1 year in 100, gray dots is lowest probability, up to 1 year in 1000. (These are, in concept, like 10 year, 100 year and 1000 year storms.) Three large red dots are Swiss Re published estimates \cite{SwissReInstClimate}. The smooth curves should not occur in the real world. Instead, a random walk centred on a curve, bounded by the blue/red/gray probability regions should occur. However, again, the relationship shown here does not actually exist.} 
\label{Fig-Expert-Sigmoid}
\end{figure}

Before any iterations of our heat conjecture social cost of greenhouse gases (SC-GHG) model began, the first author created a logistic type of curve for a heuristic (fig. \ref{Fig-Expert-Sigmoid}) that was first used in a PNAS letter pointing out a problem with willingness to pay (WTP) and curve calibration \cite{Keen2022EstimatesTippingCannotBeReconciled}.  This heuristic conforms to Pindyck's direction to make use of simple models with expert opinion \cite{Pindyck2017UseandMisuse}. What is good about this heuristic model is simplicity, and that it is founded on the concept of risk. Any real world damages that occur will display small and large excursions from the primary curve similar to what the Swiss Re estimates do in figure \ref{Fig-Expert-Sigmoid}. The more time sampled data, the larger the variation. We present this heuristic as a snapshot of expert opinion that we could not carry forward further.

However, this heuristic, presented in the style of IAMs---IAMs climate economics models show damages versus temperature as a shortcut (viz. DICE, etc.)---is problematic. \emph{There is just not any such simple relationship between climate economic damages and temperature. It is impossible to remove time from the equation, as heat buildup can take centuries.} Still, climate scientists communicating with the public, climate economists, and the IPCC report, use temperature increase as a shorthand, and it has some value though misleading. 

The reason for being unable to remove time on the scale of centuries from modelling of future damages is primarily the huge "flywheel" of the ocean that absorbs massive amounts heat. Second to this is land and ice (fig. \ref{Supp_Von_Schuckmann_Full}). It takes a long time to fill the ocean's heat-sink. The average age of the deep ocean is on the order of 1,000 years with the north Atlantic being on the order of 600 years, and the north Pacific 2,000 years old \cite[Fig 1, p. 3]{Matsumoto2007RadiocarbonCirculationAgeOceans},\cite{Cimoli2023CFCsSF6_GlobalOcean8Decades}. Benthic temperatures correlate with polar ocean temperature over time \cite[fig. 6]{Valdes2021DeepOceanTempsThruTime} and remain lower due to polar overturning circulation. The north Atlantic circulation is slowing \cite{Ditlevsen2023AMOCcollapse}, which should result in the north Atlantic benthic water's age rising. 

Climate economics may continue to speak to the public in terms of temperature, but internally, the field should be clear that it does not work that way. Temperature is the gradient that heat energy flows across. Heat is the energy that gets stored. Heat provides the power capacity of the global weather system, and the ocean is the "big flywheel". 

\subsection{NOAA weather damages---an accessible entry point for correlative approach.} The NOAA weather damages data seemed to be an accessible starting point, though there were other possible choices. For instance, the EdGCM \cite{EDGCM2023} is a straightforward global climate model (GCM) to operate. However, it was not available to use or modify, and high resolution global climate models seemed too complex to tackle. All GCMs have severe limitations on weather damages they can provide. Plus, GCMs provide varying results \cite{Archer2009AtmosphericLifetimeOfCO2}. Even to the extent that GCMs are capable of providing GIS damages data, the amount of work required to subdivide regions on the earth and estimate what model output would mean for economies year-on-year is not tractable for a small team---quite challenging for a large team. Thus the amount of work to obtain and modify multiple climate models was prohibitive. So, the NOAA weather damages dataset appeared to be a good direction. 

\subsubsection{Simple NOAA dataset analysis first model} The first iteration of our modelling effort attempted to show Social Cost of Carbon (SCC) by simply analysing the NOAA weather damages dataset using standard regression methods to find fitted curves. Our  regression curves were exponential equations of the form $f(y)= C \cdot e^{m \cdot y}$, where $C$ and $m$ are constants. However, at the end of that work, only the first 50 years of projections seemed reasonable to believe, and in retrospect, it looks more like that limit of believability is 25 years. In the physical world, continued exponential increase does not happen very long, not over centuries. We were aware that the IAMs family of model's implicit assumption was that carbon fuel energy $could$ continue to be the basis for an economy over the next 300 years. \emph{Our analysis is that this idea that carbon fuel energy could continue is quite impossible based on models that used the DICE assumptions. (See: \S \ref{DICE_Model_scenario}, Fig. \ref{Fig_Supp_DICE_Unhinged_Scenario_18.1})} It was in this phase of model development that we identified the need for a NOAA dataset containing totals of all available weather damages.   

\subsection{The OPTiMEM system for economist use.} 
To link the NOAA dataset with projections required creation of a simplified climate model that produces results based on climate science, with outputs usable by climate economists. Gall's Law was the design philosophy for development of this climate model and its connected economic/financial model. Thus, it was developed in stages with modules that chain together through external files, so that results of one stage are used by the next \cite{Gall1986Systemantics} and can be audited. This avoided the "ball of yarn" problem.  

\subsubsection{Carbon consumption model and CO$_2$ data correlation.} To link NOAA damages to climate required creating a carbon consumption model to drive a physics model of climate. How fast could carbon be burned and how much coal, oil and natural gas was reasonably available? A carbon model driving climate meant burning the carbon, and modelling how the earth heated up. Obviously, this also meant finding datasets for each step of the simplified climate model to calibrate to, and finding or deriving equations. 

For carbon-consumption, we made primary use of two papers, one by Tans \cite{Tans2009_OceanicandAtmosCO2} who used fossil fuel reserves as a limit on human ability to emit CO$_2$. The Richards equations used to model carbon production come from Malanichev \cite{Malanichev2018LimitsShaleOil} who studied the productivity of mines and oil wells. For CO$_2$ there are quite good records going back many years.  

\subsubsection{Earth energy imbalance (EEI)} The EEI is the forcing energy flux for the earth in watts per square meter ($m^{-2}$). The way this works is that earth emits heat (infrared, IR), radiation to space. GHGs prevent a portion of the IR from radiating into space. The part of the IR spectrum that is not absorbed, sometimes called "window regions", will radiate out into space uninterrupted.  The EEI is the step where there are few records. After a lot of work evaluating methods, the EEI was solved by using equations from the current best paper published on radiative forcing by GHG, from Byrne, et al \cite{Byrne2014RadiativeForcingEquations}. Other gases, including methane, N$_2$O \cite{Tian2020N2OSources_sinks}, and halogenated gases (F-gases), are handled by Byrne's equations \cite{Byrne2014RadiativeForcingEquations}. NOx \cite{Lammel1995GreenhouseOFNoX} is not, but may have some unclear GHG effects which we could not treat separately from the Tian 2020 reference that was used for nitrogen \cite{Tian2020N2OSources_sinks}. 

Please note that in our paper the acronym GWP has been taken for Gross World Product. We do \emph{not} use it to mean Global Warming Potential.

 \subsubsection{Conversion of EEI to temperature forcing} For the next step, conversion of EEI to an EEI-caused surface temperature (EEI$_T$), there was plenty of global temperature data for the industrial period starting 1880 CE to use to create a linear function with EEI. EEI implements the power law for warming, so a linear temperature translation is acceptable, and simplifying. With the temperature curve, we could drive heating of the ocean as if it were a surface in contact. 

\subsubsection{Ocean heat content (OHC)} OHC data started to be gathered later than other weather and climate data and it is a more technological matter. We chose a paper by Von Shuckmann, et al \cite{vonSchuckmann2020HeatstoredEarthSystem}, who also published an excellent dataset \cite{vonSchuckmann2020GCOSEHIEXPv2} (fig. \ref{Supp_Von_Schuckmann_Full}). To simulate heating of the ocean as a whole we used the basic heat equation. It was a pleasant surprise that the result of a straightforward heat equation algorithm landed almost exactly on the Von Shuckmann, et al dataset (fig. \ref{Fig_Supp_OHCMatchToQHeat}) on the first try. Fitting the ocean heat data to near perfection required small adjustments to two parameters: the ocean heat we start with, and thickness of the layers that interact with each other in the ocean. This latter is a standard parameter of the simple heat equation. This indicated that the physics was good. 

Computing the overall heating of the entire ocean was a much easier task than what global climate models do. GCMs represent ocean circulation, wind, etcetera, at local scale, part by part. In principle, what global climate models do is a familiar concept to economists, many of whom work on micro to macro derivations. Global climate models use a grid of 3 dimensional cells in layers covering the globe. Cell sizes have been 4.5$\degree$x7.5$\degree$ ($\approx$320,000 km$^{2}$ at 40$\degree$ latitude $\approx$565 km on a side). This has since been improved to $\approx$100 km on a side ($\approx$10,000 km$^{2}$), and then again slow improvement toward $\approx$5 km \cite{Warscher2019A5kmClimateSimCentralEurope} and $\approx$1 km on a side \cite{Schar2020KilometerScaleClimateModelsProspectsandChallenges} resolution.  In the future, when GCM cell sizes are refined enough to be capable of modelling weather reasonably well, economists may be able to use a set to do Bayesian modelling using GCMs for prediction of damages. This would be a large and complex project, requiring immense computing resources. 

\subsubsection{A single weather cell for the whole planet} Our simplified climate model can be thought of as having a single cell for the whole planet. We work strictly with the averages. Not until the conversion of heat curves to financial damages do we estimate deviation from averages using straightforward statistical methods. 

\subsection{Heat conjecture introduction} Our linkage to NOAA weather damages is a conjecture that we think is reasonable, and better than anything available (\S\ref{lb:HeatConjecture}). Thus, OPTiMEM focuses on ocean heat (fig. \ref{Supp_Von_Schuckmann_Full} \& \ref{Fig_Supp_OHC_ZJ-v1}) and like many economics models, we ignore what we cannot yet practically model, which include the single digit global heat fractions of ice, land, and atmosphere. 

We would prefer to include land, permafrost, cryosphere (ice caps, glaciers), and atmosphere fully. We tried to find existing models and equations, but this aspect is too difficult for several reasons. First, water conducts heat well with lots of data for it. Land has nothing comparable, and ice models have been shown to be incorrect. Land and ice models must be GIS based and interactive with ocean and atmosphere, just as ocean circulation models are. These models must take into account global and local interactions \cite{Valdivia2023WarmingRossGyre}. We also found papers on cryosphere melting and discussion of how simple heat diffusion used in climate models is incorrect \cite{Batchelor2023RapidIcdRetreatHundredsMetres, Milillo2022RapidGlacierRetreat, Pettit2021ThwaitesEasternCollapse}. Thus, the foundation for us to use in our simplified model is present for the ocean, but not yet for land and cryosphere. 

The biggest caveat with our climate model is that our method mostly assumes that the fundamentals of the EEI going forward will be the same as they are now. We acknowledge that this is so, and do our best with what is currently available regarding sulfate aerosols changing the EEI. This gives us a set of models, each of which has an EEI variant. However, there could be other surprises and changes to EEI.

\subsection{What is new in our work?} 

\subsubsection{Our ocean heat conjecture for damages.} The first thing that is new is that we make our heat conjecture---that the shape and size of the weather damages curves can be modelled using scaled versions of the ocean heat curve (\S\ref{lb:HeatConjecture}). In other words, our conjecture is that weather damages should be proportional to heat energy increase in the ocean. This allows us to project forward based on empirical modelling with a reasonable basis for believing the results. It is a quite simple conjecture, and we think it makes sense. It has basic physics going for it, because the ocean is the greatest heat buffer for our planet. The atmosphere and land can get warmer than  the ocean, but the ocean acts as a drag on all of it. And the 88\% of heat energy that goes into the ocean is the largest proportion of energy for global weather. 

\subsubsection{We prove that the sign of risk for the exponent on future losses must be negative.} The second thing that is new is that we examine discount rates (\S\ref{lb:DiscountingClimateDamages}), and there is a clever proof worked out regarding the sign of risk for a future stream of losses. Economists will be familiar with present value (PV) calculation using discount rates. In the usual case, to model future risk to a future income stream, the discount rate must rise because for an income stream risk means income is lower. For damages though, we have stream of future losses, and it is rare that finance needs to model a stream of future losses. When this is done, instead of increasing the discount rate, one must decrease it---because the risk with losses is that they get larger. This means that when using an exponential equation, the sign on that risk exponent is negative. And, of course, when doing it arithmetically, the rate must be divided by some factor, not multiplied by it.

\subsubsection{We propose a carbon bond to implement real-world discounting.} The third thing that is new is that, because this is a real world model, and to bypass the discounting controversy, we propose that the only way to have a meaningful discount rate is to use a real-world mechanism (\S\ref{Sect:Carbon_Bond}). Otherwise we are depending on the Pure Rate of Time Preference (PRTP) for the discount rate \cite{Kelleher_2017PRTP(RPTP)}. In our examination of the PRTP below, we show that such a discount should be quite small, and that it must decline rapidly (\S\ref{sect:PRTP}). HM Treasury's implementation in 2018 maintains an arbitrary floor on PRTP, but that is not supported by any academic work we could find. Professor Duffie at Stanford was kind enough to suggest the use of a bond for implementing discounting. Gardiner made the 300 year term of such a bond reasonable. US T-bill data generated a mean of 1.57\% (all real dollars, adjusted for inflation.)

\subsubsection{Our novel carbon bond term is 300 years.} This carbon bond needs to have a term equal to the minimum term that we use for calculation of the social cost of greenhouse gases (SC-GHG). That term is 300 years going forward. That very long term government debt aspect is something that Geoffrey Gardiner studied in his banking career \cite[p. 18]{Gardiner2006EvoCreidtaryControls}. This bit of esoterica about the success of the British Empire is not well known. Thus, stable nations should not shy away from bonds with 300 year maturities, as long as they are used properly. The proven term on the British war debt from the American Revolutionary war is not quite 300 years, but it is close enough to show the principle of long term bonds, and this American revolutionary war debt is still open, still paying interest. 

\subsubsection{Willingness to pay (WTP) as used in climate economics} We give willingness to pay (WTP) a good long rethink. This leads us to a new version of WTP that we think is heartening and helpful (\S\ref{WTP_StadiumVGlobe}, eq.'s \ref{eq:WTP_for_stadium},\ref{eq:WTP_for_globe}).  

\subsubsection{SC-GHG cannot be a single cost, it rises} Our SC-GHG computation does not support a single cost of carbon, nor a single cost of CH$_4$, N$_2$O, nor F-gases. The SC-CO$_2$ rises each year, and so does every other SC-GHG, because each year's SC-GHG encompasses a future period of impacts going forward where each year has greater energy for damages. Rising SC-GHG cannot be modelled as a straightforward relationship to CO$_2$ or temperature, because each forcing, GHG $\rightarrow \Delta\degree$C $\rightarrow$ $\Delta$Q (heat content), has total system lags by as much as one or two centuries or more (fig. \ref{Fig_Supp_OHC_ZJ-v1}). There are immediate effects and long-term effects.  This is also a system that will \emph{not} reach an equilibrium state.  Remember that on a geological scale, warming can be treated as nearly instantaneous, but those are time scales in which the human species just appeared.

\subsubsection{Energy is the foundation of monetary valuation} Energy is a primary basis of monetary valuation, and thus critical for policy. Of course, inflation arises from multiple sources, from oil price \cite{HA2023UnderstandingGlobalDriversOfInflation_HowImportantAreOilPrices} which supports the energy thesis, to politics \cite{WHITEHEAD1979THEPOLITICALCAUSESOFINFLATION}. Jarvis documents a strong relation between production and energy consumption \cite{JARVIS2018EnergyReturns_TheLong-runGrowthOfGlobalIndustrialSociety}. This has resulted in Keen, et al updating the Cobb-Douglas production function to an Energy Based Cobb-Douglas Production Function (EBCDPF) \cite{KEEN2019NoteOnEnergyInProduction}. 

The implication of Jarvis' finding is that the idea that we can cut energy use by 10\%, 50\% or 90\% (all figures mentioned by advocates) and simplify our civilisation while maintaining a viable economy is not going to work. This finding means that energy poverty = monetary/economic poverty. There is little reason not to pursue efficiency and a degree of frugality---however, cue the paradox of thrift \cite{Corden2012GlobalImbalancesAndTheParadoxOfThrift,Vermann2012ParadoxThrift,Yiannis2011DeficitReductionTheAgeOfAusterityAndTheParadoxOfInsolvency,Das2019ParadoxOfAusterityMulti-CountryEvidence}.

Overall, the world is going to become a more difficult place to live in. We think, however, that if we do the right things, civilisation can survive and thrive. 

\section{Discounting of climate damages}
\label{lb:DiscountingClimateDamages}
 Kelleher notes that IPCC authors appear to interpret value in a Broomean manner, emphasising justice and rights.  (Broome provided an example in which there is greater value in killing one person to save the lives of 5 others with transplants---and yet this is unjust.) In the context of climate change, Broomean morality is the argument for a pure rate of time preference (PRTP),  (symbolized by the greek letter $rho$ ($\rho$)) of 0\%---because current generations do not have the right to subject later generations to the damaging effects of our actions \cite[pp. 448-451]{Kelleher_2017PRTP(RPTP),Dietz2008EconEthicsClimate}.  

We examined the discounting discussion of the Nordhaus 2017 supplement \cite{Nordhaus2017RevisitingSCC_Supplement} and tried to find justification for it in the Stern report referenced \cite{Stern2007EconomicsofClimateChangeSternReview}.  Stern states in this reference that the kind of discounting used is only valid for marginal difference cases. Stern also clearly supports that climate damages must be handled as a prescriptive (ethical) matter, not a descriptive one, as the below quote defines. 

\begin{quote}
    He [Arrow 1995], like the authors described in Chapter 2 on this issue, is very clear that this should be seen as a prescriptive or ethical issue rather than one which depends on the revealed preference of individuals in allocating their own consumption and wealth (the descriptive approach). The allocation an individual makes in her own lifetime may well reflect the possibility of her death and the probability that she will survive a hundred years may indeed be very small. But this intertemporal allocation by the individual has only limited relevance for the long-run ethical question associated with climate change. -- Stern 2017 \cite[p. 47]{Stern2007EconomicsofClimateChangeSternReview}
\end{quote}

We agree that climate change damages are not properly handled as a marginal preference case. Stern clarifies in his 2007 review that the only sound ethical basis for placing less value on the utility of future generations was uncertainty over whether or not those generations will  be present due to \emph{exogenous} cause of extinction such as nuclear war. In this context, $e^{-\delta} t$ should be the probability that the generations exist at time $t$ (table \ref{Table_2A1_fromSternReview}). This probability of future generations existing reduces at rate $\delta$. 

Climate change is a "project" that is the \emph{cause} of the probability that future generations may not exist. The approach discussed by Stern \cite[Table 2A1, p. 47]{Stern2007EconomicsofClimateChangeSternReview} that allows use of discounting, requires the probability future generations are extinct be \emph{exogenous}. However, the effects of climate change threatening human survival are \emph{endogenous} to the climate change "project". 

We note that the results of IAMs provide very low estimates of damages \cite{Howard2021GaugingConsensusClimateChange,Howard2017Meta-analysisOFClimateDamageEstimates}, even 300 years from now, in a scenarioo that assumes a continuously growing economy, which would require continuously growing energy consumption, could exist based on burning carbon. If we could assume that this underlying assumption were true, then there might be some path to justification. However, this assumption is divorced from physics---temperature increase should hit roughly $+110\degree C$ by 2300 (See: \S \ref{DICE_Model_scenario}, fig. \ref{Fig_Supp_DICE_Unhinged_Scenario_18.1}). This means the upper ocean would boil. Clearly, global civilisation would have grave problems long before that. 

What appears to be present in the DICE system of estimation is a closed loop. Because the assumption of damages is relatively small, even over 300 years of exponentially growing fossil fuel based economy, the high discount rates may be ethically usable. However, the major reason why damages are computed to be so low is that high discount rates are used.  

Nordhaus summarizes, "\emph{Under these assumptions, the SCC is proportional to $1/(\rho + \delta)$,}" but does not specify what the values of $\rho$ and $\delta$ might be. The figure 3 of Nordhaus 2017 \cite{Nordhaus2017RevisitingTheSCC} shows values of discount rate for Stern and DICE. Stern's discount rate appears to be $\approx 0.9$\% and DICE base $\approx 2.5$\%. It is unclear where this value attributed to Stern came from, and the DICE base of $\approx 2.5$\% is roughly half what is implemented in DICE \cite{nordhaussztorc2016Rdice,Nordhaus2018ChangesinDICE_Model}.

Circling back to $1/(\rho + \delta)$, let us examine \emph{Table 2A.1} \cite[p. 47]{Stern2007EconomicsofClimateChangeSternReview} reproduced below (Table \ref{Table_2A1_fromSternReview}). Here, Stern provides a set of computed values based on choices of $\delta$. There is only one value of $\delta$ that appears at all reasonable to use aside from zero, $\delta = 0.1\%$. Look in the last column "Not surviving 100 years," and the probability is nearly 90\%, as Stern points out. Leaving aside the question of whether nuclear war in a warming world that causes shortages could be considered endogenous to climate change, a 90.5\% chance of a nuclear war is a defensible assumption we could accept. 

\begin{table}[!hbt] 
\caption{Stern review Table 2A.1} 
\label{Table_2A1_fromSternReview}
\begin{tabular}{l|llll}
       & Probability of &               & Probability of &               \\
       & human race     &               & human race     &               \\
       & surviving      & Not surviving & surviving      & Not surviving \\
$\delta$ (\%) & 10 years       & 10 years      & 100 years      & 100 years     \\ \hline
0.1    & 0.99           & 0.01          & 0.095          & 0.905         \\
0.5    & 0.951          & 0.049         & 0.607          & 0.393         \\
1      & 0.905          & 0.095         & 0.368          & 0.632         \\
1.5    & 0.861          & 0.139         & 0.223          & 0.777        
\end{tabular}
\end{table}

Consequently, using the Nordhaus 2017 formula, $1/(\rho + \delta)$, we conclude that this DICE base discount rate of $\approx$ 2.5\% has a maximum $\delta$ of 0.1\%. That means that $\rho$ is 2.4\%, ignoring that this 2.5\% value is roughly half of what is found in the DICE model itself \cite{nordhaussztorc2016Rdice}. Consequently, we needed to understand $\rho$, which is discussed below at some length, and the conclusion is in \S\ref{sect:PRTP_Conclusion}.  

\subsection{Pure rate of time preference (PRTP) $\rho$}
\label{sect:PRTP}
The pure rate of time preference (PRTP) (also rate of pure time preference (RPTP), or utility discount rate) was created to \emph{discount utility/welfare benefits} in the future versus the present.  PRTP was created to account for the decrease in present value something has because it is in the future rather than the present. This is in line with the concept of discounting of future income in present value (PV) computations, because it was created to account for future benefit. 

Two practical examples of PRTP applied to a social benefit are construction of an electrical generation plant or a hospital. At the end of construction, there is social welfare benefit in the form of electricity for the electricity generation plant, and health care services for the hospital. For these two examples, the PRTP would be applied for the time it takes to build an operating plant or hospital, thus discounting the benefits provided in the future. In the practical world of accounting, PRTP is part of the discounting that is applied to proposals for public projects by HM Treasury's Greenbook in the UK \cite{HMTreasury2022GreenBook}. 

\emph{To think about the case of a future loss, let us say that a gangster has promised to harm us.} With a bow to Blatt's "Utility of being hanged at the gallows," \cite{Blatt1979UtilitiyOfHangedGallows}, let us imagine the gangster will cut off one of our legs in the future. The farther in the future this loss of a leg is, the less worried we will be today about getting our leg cut off, although how much we will be worried may be hard to pin down. From this thought experiment we see that PRTP allows discounting of a future loss as well as a future benefit. Thus, it could, perhaps, be a primary discount rate factor. However, we will see below that in some implementations, risk is part of defining PRTP, which is problematic because we have proven that risk must make a discount rate for losses smaller, never larger \cite{Hanley2022SignOfRiskForFutureLosses}.   

Kelleher argues against a single common definition of PRTP and discounting in social welfare, outlining a variety of views. He agrees with the Broomean view that justice and rights can, in some circumstances, trump value. Broome provided an example in which there is greater value in killing one person to save the lives of 5 others with transplants---and yet this is unjust. Kelleher notes that IPCC authors appear to interpret value in a Broomean manner, emphasising justice and rights. In the context of climate change, this kind of morality is the argument for a PRTP of 0\%---because current generations do not have the right to subject later generations to the damaging effects of our actions \cite[p. 448-451]{Kelleher_2017PRTP(RPTP)}. 

\subsubsection{Views on PRTP discounting and use of formula parameters}
The broader idea of PRTP discounting is mathematically described in equations \ref{eq:RamsayFormula} and \ref{eq:RamsayFormulaprecedent} below. Relative to climate change, prescriptivists take philosophical positions based on ethics to future generations. Descriptivists look at human preference choices and interest rates for guidance \cite[p. 442]{Kelleher_2017PRTP(RPTP)}. 

Examining how prescriptivist and descriptivist views affect equations \ref{eq:RamsayFormula} and \ref{eq:RamsayFormulaprecedent}, parameters do multiple duty to implement different concepts. A single value may be chosen for PRTP such as 0.1\% by Stern, or 0\% by Cline \cite[p. 10]{Dietz2008EconEthicsClimate}. By contrast, Nordhaus and Weitzman prefer a 2\%-3\% value for PRTP. However, as we will see, the PRTP equations should not result in a single PRTP value, but rather a power law declining function, because according to V (eq. \ref{eq:RamsayFormulaprecedent}) the value of $\rho$ (rho) in the Ramsay formula (eq. \ref{eq:RamsayFormula}) cannot be a single, stable value, nor can $\rho$ be a rising value. We find the elision of this feature by the Ramsay formula (eq. \ref{eq:RamsayFormula}) questionable. 

Within the formulas (eq's. \ref{eq:RamsayFormula}, \ref{eq:RamsayFormulaprecedent} ) the value of $\eta$ (eta) has been used to capture attitude towards risk as well as inequality within and between generations, with choice of a higher value for $\eta$ indicating greater risk and social inequality aversion \cite[p. 11]{Dietz2008EconEthicsClimate}. In this case, high $\eta$ means greater aversion (lower discount) and low $\eta$ means lower aversion (higher discount). We note there are other formulas presented for computing these \cite[p. 11]{Dietz2008EconEthicsClimate}.  

Alternatively, a term like per capita income ($C_t$) may need to be used, as Dagupta does, to account for both inequality of consumption and well-being, in addition to rising $C_t$ \cite[pp. 151-2]{Dasgupta2008DiscountngClimate}. The message here is that economists discount value of future consumption, and may similarly discount the future well-being that results from future consumption. Those methods may combine multiple things that sometimes may be in opposition, into one number. 

\begin{flalign}
\begin{aligned}
\label{eq:RamsayFormula}
\text{Ramsay formula} & \text{ for discount rate of future consumption. Derived from eq. \ref{eq:RamsayFormulaprecedent}}\\
\delta =& \, \eta \cdot g + \rho \\
\text{Where: } \delta =& \text{ (delta) discount rate}\\ 
\eta =& \text{ (eta) marginal utility of consumption elasticity, but in climate practice,} \\
& \text{ a coefficient of aversion to social inequality \cite[p. 10]{Dietz2008EconEthicsClimate}\cite[pp. 151-2]{Dasgupta2008DiscountngClimate}} \\
g =& \text{ annual consumption growth rate } \\
\rho =& \text{ (rho) annual rate of pure time preference, or utility discount rate} 
\end{aligned}
\end{flalign}

\begin{flalign}
\begin{aligned}
\label{eq:RamsayFormulaprecedent}
V =& \sum_{t=0}^{\infty} N_t \cdot (\frac{C_t^{1-\eta}}{1-\eta}) \cdot (\frac{1}{(1+\rho)^t}) \\
\text{Where: } V =& \text{ V, which  is the name of this equation's result}\\ 
N_t =& \text{ number of people alive at time }t \\
t =& \text{ time} \\
C_t =& \text{ per capita consumption at time }t \\
(\frac{C_t^{1-\eta}}{1-\eta}) =& \, U(C_t) \text{ Utility function. Here we do not see any terms for} \\ &\text{ "well-being" which is hard to measure.}\\
(\frac{1}{(1+\rho)^t}) =& \text{ PRTP} 
\end{aligned}
\end{flalign}
In eq. \ref{eq:RamsayFormulaprecedent} the three terms may  be revised. $N_t$ may be set to 1 to simplify it with a constant population. Likewise, $C_t$ may be assumed constant. These sorts of operations are how the Ramsay function (eq. \ref{eq:RamsayFormula}) was derived.

The utility function may have $\eta$ be set to a value that is supposed to account for inequality, and for well-being, and for the formally present per capita consumption \cite[p. 445]{Kelleher_2017PRTP(RPTP)}. Thus, one can see that by choice of parameters for these equations, discount rate can give a wide range of results. However, given the difficulty of representing things like morality, this is understandable.

Years ago, after examining discounting and PRTP, Dasgupta came to the troubling evaluation that none of the choices when applied to climate change were good. "\emph{...if future uncertainties are large, the formulation of intergenerational well-being we economists have grown used to could lead to ethical paradoxes even if the uncertainties are thin-tailed. Various modelling avenues that offer a way out of the dilemma are discussed. None is entirely satisfactory}" \cite{Dasgupta2008DiscountngClimate}. 

\subsubsection{Reconciling PRTP theory to practical usage and risk of greater damages}
A rare practical implementation (the only one found) of PRTP discount rates may be found in the U.K.'s HM Treasury Green Book, 2018 version (see table \ref{Table_A10_fromDasgupta} taken from Dasgupta). The Green Book has clear instructions on the use of discounting \cite[pp. 45, 116-119]{HMTreasury2022GreenBook}. These discussions in the Green Book assume that projects should accrue benefits. Thus, in these HM Treasury guidelines, a greater risk (for instance, catastrophic risk) requires a larger discount rate, because risk relative to a positive benefit is that the benefit be lower than  expected. This is conceptually identical to what is done when evaluating a business proposition using present value (PV). However, because HM Treasury increases discount rate to account for risk, this risk element is incorrect for evaluation of risk relative to future losses, as we have shown (see \S2(d) of main paper).

\subsubsection{Risk and future streams of losses applied to DICE model}
\label{Risk_and_future_losses_streams}
Using calculation methods made for evaluating public benefits or streams of future income to evaluate future streams of \emph{losses} cannot provide correct results, because the sign of risk is opposite for losses. We treat this problem in the main paper in \S 2(d). The damages from climate change when evaluated for risk need to have the discount rate get smaller, not larger---because risk means greater damage, and raising the discount rate lowers PV of damages. Thus, for our discussion we assume that the maximum DICE discount rate of 5.1\% \cite{Nordhaus2018ChangesinDICE_Model} is a risk adjusted multiple of a baseline discount rate. Thus, the baseline is important to determine.

The highest baseline discount rate that we can justify with empirical methods is $d_{M{T30yr}}$, the 1.57\% rate discussed in \S \ref{Sect:Carbon_Bond} \emph{Carbon bond to implement climate damages discounting}. So, to find the multiple of that baseline the 5.1\% discount is divided by the 1.57\% discount, yielding the risk adjustment multiplier $r_{_{DF}}$ .
$\\ \hbox{} \qquad \qquad \qquad \qquad r_{_{DF}}={\frac{d_{DICE}}{d_{M{T30yr}}}} = 3.423 $. \\ 

We will assume then, that to obtain the correct maximum DICE discount rate, we need to divide $d_{M{T30yr}}$ by the $r_{_{DF}}$ multiple. We divide here because the sign of risk for losses ($d_{rar}$) is a negative exponent of \emph{e}. 

$ \hbox{} \qquad \qquad \qquad \qquad 
E_{rar}(t) = I \cdot e^{(g_{_E} - d_{tvm} - \mathbf{d_{rar}} )t}  \qquad \qquad \qquad \qquad $eq.(2.5) from main paper.

Applying this $r_{_{DF}}$ factor to the 1.57\% discount rate gives a revised DICE maximum rate ($d_{r_{DICE}}$), \\
$\hbox{} \qquad \qquad \qquad \qquad d_{r_{DICE}}= \frac{d_{M{T30yr}}}{r_{DF}} = 0.435\%$. \\ 
However, note that this possible stand-in for a PRTP is a fixed value, and it seems to bear little relationship to HM Treasury's Greenbook, or to PRTP theory discussed above. The Greenbook value for 31-75 years (table \ref{Table_A10_fromDasgupta}) happens to be quite close, but by what rationale should this value become a set value for PRTP? There is none. 

\subsubsection{U.K.'s HM Treasury Green Book implementation of PRTP}
\begin{figure}[hbt!]
\centering\includegraphics[width=5.25in]{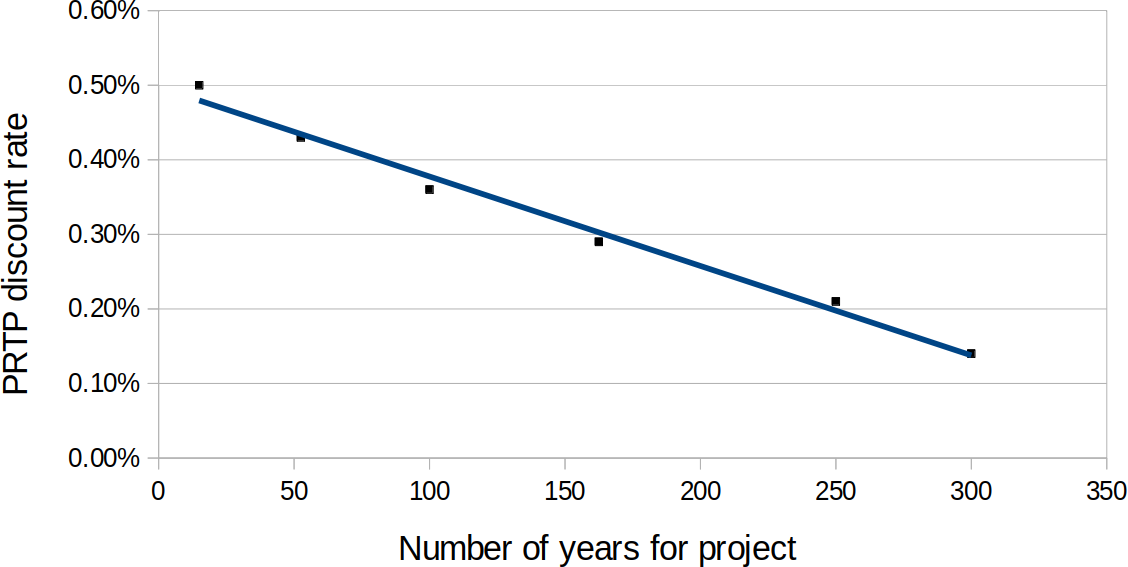}
\caption{\textbf{Declining Discount Rates In the Green Book} \\ $f(x) =  -1.1997 \cdot 10^{-5} \cdot x + 0.004976 $ (eq. \ref{eq:HMTreasuryGreenBookFittedCurve}) $R^2 = 0.988 \qquad$  0 intercept = 414.77 years  } Area under eq. \ref{eq:HMTreasuryGreenBookFittedCurve}, 0-300 $x_{F1} = 0.953 \qquad$       Area under the table \ref{Table_A10_fromDasgupta} values 0-300, $x_T = 0.951$ \\
Area under the table \ref{Table_A10_fromDasgupta} values 0-1500, $x_T = 2.631 \qquad$ Area units are square PRTP years.
\label{Fig_Supp_PRTPGreenbook}
\end{figure}

Table \ref{Table_A10_fromDasgupta} and figure \ref{Fig_Supp_PRTPGreenbook} show the U.K.'s HM Treasury Green Book values for pure rate of time preference. Note that these values are not constant. We established at the beginning of this subsection that a rapidly declining PRTP is correct to use for discounting of future losses. Here we see an interpretation of PRTP that is linear, not power law, which is hard to explain except as the decision of a committee faced with the difficult task of making an implementation of PRTP theory. However, because the Greenbook is the only known practical implementation of PRTP, it will be used below in various ways to search for a reasonable equation for implementation.  

\begin{table}[!hbt] 
\caption{Green Book pure rate of time preference} 
\label{Table_A10_fromDasgupta}
\begin{tabular}{l|rrrrrr}
Table A10.2.1 from Dasgupta \cite[p. 282]{Dasgupta2021Discounting}              & \multicolumn{6}{l}{Declining Discount Rates In the 2018 Green Book} \\
Years duration               & 0-30     & 31-75    & 76-125   & 126-200  & 201-300  & 301    \\ \hline 
Green Book discount rate     & 3.50\%   & 3.00\%   & 2.50\%   & 2.00\%   & 1.50\%   & 1.00\%  \\
Reduced discount rate PRTP=0 & 3.00\%   & 2.57\%   & 2.14\%   & 1.71\%   & 1.29\%   & 0.86\%  \\
Calculated PRTP              & \textbf{0.50\%}   & \textbf{0.43\%}   & \textbf{0.36\%}   & \textbf{0.29\%}   & \textbf{0.21\%}   & \textbf{0.14\%} 
\end{tabular}
\end{table}

\begin{flalign}
\begin{aligned}
\label{eq:HMTreasuryGreenBookFittedCurve}
\textbf{HM Treasury Green Book} & \text{ curve fit, figure \ref{Fig_Supp_PRTPGreenbook}}\\
f(x) =&  -1.1997 \cdot 10^{-5} \cdot x + 0.004976 \\
\text{Where: } x =& \text{ is something related to PRTP basis in equation \ref{eq:RamsayFormulaAmended}}\\ 
\end{aligned}
\end{flalign}

Figure \ref{Fig_Supp_PRTPGreenbook} shows a plot of computed PRTP ($\rho$) from table \ref{Table_A10_fromDasgupta} that is a linear fit to data. This linear fit would produce a negative $\rho$ at $\approx$415 years. 

\subsubsection{Reconciling HM Treasury Green Book with equations \ref{eq:RamsayFormula} and \ref{eq:RamsayFormulaprecedent}}
Per equations \ref{eq:RamsayFormulaprecedent} \& \ref{eq:RamsayFormulaAmended}, the value of equation \ref{eq:RamsayFormula} $\rho$ should vary  as $\frac{1}{(1+\rho)^t}$. Simply comparing this with the fitted equation \ref{eq:HMTreasuryGreenBookFittedCurve} (fig. \ref{Fig_Supp_PRTPGreenbook}) there is an obvious conflict, as eq. \ref{eq:HMTreasuryGreenBookFittedCurve} is linear, and $\frac{1}{(1+\rho)^t}$ is a power law. However, because HM Treasury Green Book is the primary source of practical use, we will attempt to see to what degree we can reconcile them. 

Unfortunately, Ramsay named his $\rho$ used in equation \ref{eq:RamsayFormula} the same as the $\rho$ in eq. \ref{eq:RamsayFormulaprecedent}. Rather than write a head exploding equation, we will do the sensible thing. We will substitute $P$ for the right hand side $\rho$ and call $P$ the PRTP basis value. \\ 
$\hbox{} \qquad \qquad \qquad \qquad \rho = \frac{1}{(1+{P})^t} \qquad \qquad \qquad \qquad \qquad $ (Formula $P$)\\
$\hbox{} \qquad \qquad \qquad $ where: $P$ is the PRTP basis rate and $t$ is years. 

Formula $P$ indicates that we should amend Ramsay's formula (eq. \ref{eq:RamsayFormula}) as well, which we do in eq. \ref{eq:RamsayFormulaAmended}. 

\begin{flalign}
\begin{aligned}
\label{eq:RamsayFormulaAmended}
\textbf{Amended Ramsay formula} & \text{ for discount rate of future consumption. Derived from eq. \ref{eq:RamsayFormulaprecedent}}\\
\delta =& \eta \cdot g + \frac{1}{(1+{P})^t} \\
\text{Where: } \delta =& \text{ discount rate}\\ 
\eta =& \text{ marginal utility of consumption elasticity, but in climate practice,} \\
& \text{ a coefficient of aversion to social inequality \cite[p. 10]{Dietz2008EconEthicsClimate}\cite[pp. 151-2]{Dasgupta2008DiscountngClimate}} \\
g =& \text{ annual consumption growth rate } \\
\rho = \frac{1}{(1+{P})^t} =& \text{ annual rate of pure time preference, or utility discount rate} \\
P =& \text{ PRTP basis value}
\end{aligned}
\end{flalign}

The area under the curves for HM Treasury Green Book will be evaluated relative to the integral of $\rho$.  (eq. \ref{eq:IntegralsF1-3}).  
\begin{flalign}
\begin{aligned}
\label{eq:IntegralsF1-3}
\int \rho\,dt = P_S =&\int_0^{t}\frac{1}{(1+P)^t} dt\\
\text{ Where: } t =& \text{time in years}  \\ P =& \text{PRTP basis value}\\ 
\end{aligned}
\end{flalign}

\subsubsection{First estimate of $\rho$ by year: find $P$ for $\overline{\rho} = 0.50\%$ bounded by $t=$0-30}
For this first variant estimate, we assume that table 1's value for $\rho$ is 0.50\% for the period $t = 0-30$. We will find our basis rate $P$ assuming that 0.50\% is the average $\rho$ for the set from 0-30 using  Formula $\rho \, 1$.\\\\ 
$\hbox{} \qquad \qquad \qquad \qquad \rho = \frac{\sum_{t=0}^{30}\frac{1}{(1+P)^t}}{30} \qquad \qquad \qquad \qquad$ (Formula $\rho \, 1$)\\
$\hbox{} \qquad \qquad \qquad \qquad$ such that $\rho$ = 0.0050. \\ \\

Using Formula $\rho \, 1$, the basis rate, $P$ $\approx$6.667. 
The area under the curve for $t$ = 0-300 using equation \ref{eq:IntegralsF1-3} is $P_S$ = 0.491. This $P_S$ area is $\approx$0.5 of the $x_{F1}$ value of 0.953 from figure \ref{Fig_Supp_PRTPGreenbook}).   \\

\begin{figure}[hbt!]
\centering\includegraphics[width=4.0in]{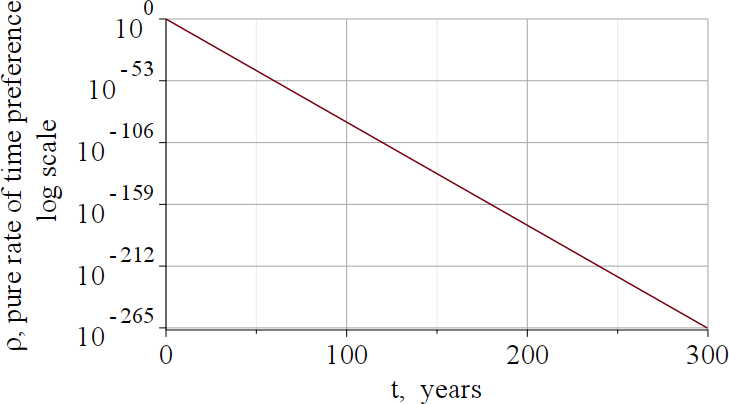}
\caption{\textbf{Declining Discount Rates using Formula $\rho \, 1$  such that $\overline{\rho}$ =0.50\% for $t$=0-30. $P_S \approx6.667$} determined from equation \ref{eq:IntegralsF1-3}} 
\label{Fig_Supp_rhoRealBehavior_667}
\end{figure}

Figure \ref{Fig_Supp_rhoRealBehavior_667} shows in log scale how $\rho$ behaves with this $P \approx6.667$ basis value. At 15 years, if computed using Formula $\rho \, 1$, $\rho$ = 5.378E-14. The decline of $rho$ with $t$ is very rapid. 

\subsubsection{Second estimate of $\rho$ by year: Find $P$ for $\rho = 0.50\%$ when $t=15$}
For our second variant to estimate basis $P$ assume that the value provided for the 0-30 year PRTP in table \ref{Table_A10_fromDasgupta} is the correct value for 15 years. This formula would be: \\
$\hbox{} \qquad \qquad \qquad \qquad \rho = \frac{1}{(1+P)^{15}} \qquad \qquad \qquad \qquad \qquad $ (Formula $\rho \,2$) \\
$\hbox{} \qquad \qquad \qquad \qquad $ such that $\rho = 0.0050$ 

\begin{figure}[hbt!]
\centering\includegraphics[width=4.0in]{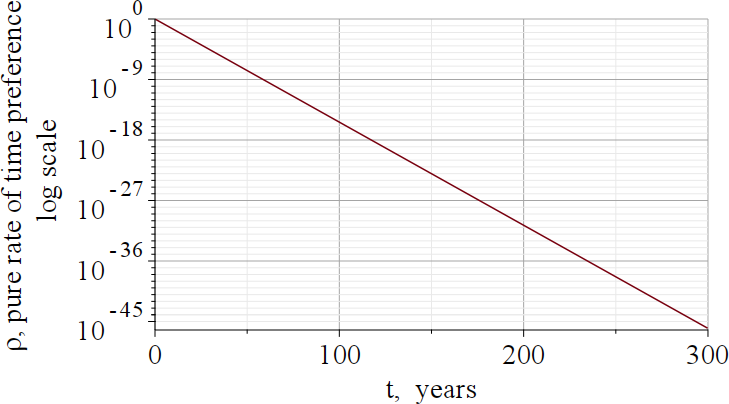}
\caption{\textbf{Declining Discount Rates using Formula $\rho \, 2$ such that $\rho = 0.50\%$.} $P_S \approx 0.424$ determined from Formula $\rho \,2$ }
\label{Fig_Supp_rhoRealBehavior}
\end{figure}

Using Formula $\rho \,2$, the basis rate, $P$ $\approx$0.424. The area under the curve for $t =$ 0-300, using equation \ref{eq:IntegralsF1-3} yields $P_S$ = 2.831 (fig \ref{Fig_Supp_rhoRealBehavior}). This area under the curve value of $P_S$ is $\approx$3 times the $x_{F1}$ value from figure \ref{Fig_Supp_PRTPGreenbook}. 

\subsubsection{Third estimate of $\rho$ by year: Find $P$ for $x{_F1}$ > $P_S$ > $x{T}$}

\begin{figure}[hbt!]
\centering\includegraphics[width=4.0in]{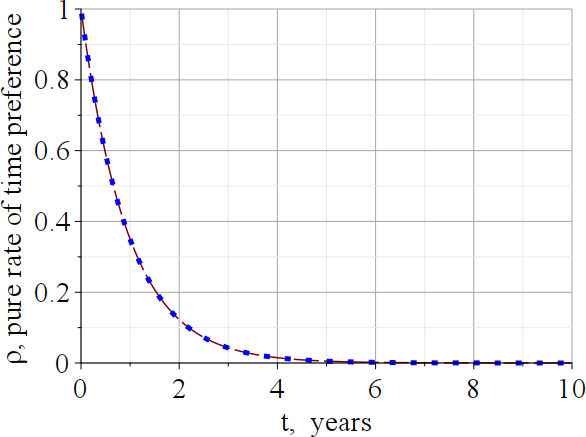}
\caption{\textbf{Declining Discount Rates using equation \ref{eq:IntegralsF1-3} to approximate range of $P$}. For $P$ such that $x_T<P_S <x_{F1}$ where: $P$ is PRTP basis value. At $t$ = 10 years, $P = 0.0000271$ to $0.0000277$.  Area under the curves: High range = 1.856. Low range = 1.862. The two range curves are for most practical purposes identical. }
\label{Fig_Supp_rhoRealBehavior_1856}
\end{figure}

Equation \ref{eq:IntegralsF1-3} can also be used to find a value of $P$, such that $P_S$ for $P$ is between $x_{F1} = 0.953$ and $x_T = 0.951$. This yields a range of $P = 1.856$ when $P_S = 0.953$, and $P= 1.862 $ when $P_S = 0.951$. This is shown in figure \ref{Fig_Supp_rhoRealBehavior_1856} which does not use a log scale in order to show how rapidly PRTP would approach zero. 

\subsubsection{An attempt to reconcile Greenbook estimate of $\rho$ by year: Set 0.14\% minimum and $t=$0-1500}
The Green Book solves CO$_2$ equivalence of negative PRTP caused by their linear projection of PRTP by setting all PRTP values for $t > 300$ to 0.14\%. This indicates that the creators of the Green Book table knew that their system should display some kind of asymptotic-like behaviour. This suggests to us one more option for our equation constructions. This option would take a long view, 1500 years, and set our minimum at 0.14\% as the Green Book does. 

The area under the table's values when extended to 1500 years is 2.631. We modify equation \ref{eq:IntegralsF1-3} by adding the 0.14\% Green Book minimum yielding equation \ref{eq:Integral0014} to estimate the $P$ PRTP basis. 
\begin{flalign}
\begin{aligned}
\label{eq:Integral0014}
P_{S14} =&\int_0^{1500}\frac{1}{(1+x)^t}+0.0014 \, dt\\
\text{ Where: } t =& \text{ time in years } \\ P =& \text{ PRTP basis value}\\ 
\end{aligned}
\end{flalign}
 
\begin{figure}[hbt!]
\centering\includegraphics[width=4.0in]{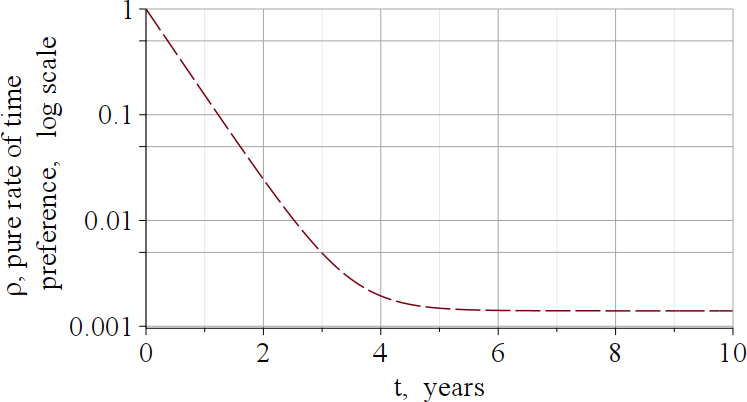}
\caption{\textbf{Declining Discount Rates using equation \ref{eq:Integral0014} to approximate $P$} such that $P_S \approx2.631$ where: $P$ is PRTP basis in revised  Ramsay equation \ref{eq:RamsayFormulaAmended}. As with figure \ref{Fig_Supp_rhoRealBehavior_1856}, most of the PRTP basis is lost in the first 4 years with this method. It is not possible to have a fixed PRTP discount rate.}
\label{Fig_Supp_rhoRealBehavior_0014}
\end{figure}

This equation \ref{eq:Integral0014} yields $P = 5.57$ for $x_{S14} = 2.631$. This is plotted in
figure \ref{Fig_Supp_rhoRealBehavior_0014}.  

\subsection{Summary conclusion of estimates of $\rho$ by year}
\label{sect:PRTP_Conclusion}
None of these set of estimates seen in figures, \ref{Fig_Supp_rhoRealBehavior_667},  \ref{Fig_Supp_rhoRealBehavior}, or \ref{Fig_Supp_rhoRealBehavior_1856} approximate what is seen in figure \ref{Fig_Supp_PRTPGreenbook} very well. The behaviour of these equations with exponential decline must approximate zero, or some other arbitrarily chosen minimum asymptote value, in the first 5 years. We exclude unsupported arbitrary minimums as not justifiable.

Whether the Green Book's values for PRTP applied to future benefits are "correct" is probably moot. In a basic sense, because PRTP appears to have such a degree of subjectivity, and HM Treasury has used them as official values, the Green Book's table represents a significant empiricism, the closest thing to a practical guide regarding use of PRTP that is available. What this means is that HM Treasury's Green Book mavens may not believe Ramsay's formula, or found it as trivial as we did when they used it, and did something else. Thus we see a compromise in the Greenbook that seems to be saying that the  present value of a hospital 20 years in the future is not the same as the value of that same hospital 5 years in the future, but stabilises for the next couple of decades. This seems reasonable on its face, however, it is difficult to support the compromise chosen based on the literature in the area---we cannot find any such justification.  

Thus, the most correct version seems to be something along the lines of Figure \ref{Fig_Supp_rhoRealBehavior_1856}. It is clear is that an implementation of PRTP per equation \ref{eq:RamsayFormulaAmended} is not highly sensitive to the $P$ basis value used. 

What we can say is that using equation \ref{eq:RamsayFormulaAmended}, it is not possible to get a constant, rising, or slowly declining value. We can further say that the impact of implementing PRTP would be very small; so small it is not worth bothering with at this time---certainly not for a quantity that has so many questions. Consequently, our implementation of PRTP is 0\%, and we did not use it. Instead, we turned to a bond system that is more difficult to argue with. 

\section{Carbon bond to implement climate damages discounting} 
\label{Sect:Carbon_Bond}
We defined a carbon bond to implement real-world discounting.  The concept of the carbon bond is the provision of an interest bearing bond that will pay the cost of emitting 1 tonne of CO$_2$ for 300 years into the future. For these purposes, a carbon bond is also a greenhouse gases bond. We assume that the proceeds of the bonds will be optimally invested in effective remediation or minimization of future emissions.

\subsection{Carbon bond term}
Based on our minimum SC-CO$_2$ term, carbon bonds should have a 300 year or longer maturity. Such a long maturity could prove interesting in the world of financial instruments, though most current day financial instruments top out in the range of 30 years with rare exceptions like the 100 year Japanese ancestral real estate mortgage \cite{CHANG1995_100yrJapanMortgage}. This 100 year mortgage is different in that the payer(s) are private individuals with mean life spans and capacity to pay well below 100 years, whereas our long-term carbon bonds would be paid by national governments, which is feasible. Gardiner discusses how the Bank of England recognised very long-term debt as a national asset around the time of the American revolution \cite{Gardiner2006EvoCreidtaryControls}. 
 
\begin{quote}
    "The French financial system had not been able to cope with the financing of the wars, but the British system had amazing success. That was despite the fact that the American War of Independence was twice as costly to the British as it was to the French. How had the British financial system managed so well? One reason may have been the development of a market in interest-bearing long-dated government debt. A large part of the principal of the debts incurred in those wars, and indeed of earlier wars, is still outstanding. Far from being a burden on the nation, the government debt became the assets of the wealthy. One of the greatest ironies of economics is that an indebted nation is also ipso facto a wealthy nation. A proviso is that the indebtedness should be to its own citizens."  Gardiner, G.W. 2006  \cite[p. 18]{Gardiner2006EvoCreidtaryControls} 
\end{quote}

The British debt from the American Revolutionary war was preceded by bonds (Gilts) that were issued to finance the War of the Austrian Succession. These bonds were not redeemed until 2015 when interest rates dropped below the 2.5\% those consolidated bonds bore \cite{Gardiner2023AustrianSuccessionGilts1740_2015}. Those bonds were immediately rolled over, so in practice the end of their term was a technicality, and arguably, de facto, this debt remains unredeemed. 

Gardiner based his proviso that indebtedness of a nation should be to its own citizens on history. However, that 300 year period was in a non-unified global banking and trade system with often warring nation-states in Europe. Today, bonds are traded throughout the globe. War has not disappeared, but it has declined. The climate crisis is a global one, and we will presume that there will be enough wealth to go around if nations pursue this path. 

\subsection{A variable rate carbon bond is probably technically necessary}

The value of bonds varies inversely to interest rate adjustments. In a declining real-interest regime, the effective discount rate should rise for a time, and vice versa. National currencies should inflate slowly over time, but their relative rates vary considerably, which means exchange rates will also have an impact. 

Carbon bonds should be variable-rate bonds, that maintain a real-dollar rate of interest, although this is complex as will be discussed. This requires accounting for inflation, and the real-dollar rate of  their nation's central bank. In  principle, the inflation rate should be reasonably constant in real dollars. In practice, it is not. 

For instance, a carbon bond could be defined to carry an interest rate of 1.0442 times the prime rate, reset at some interval trailing average of the prime rate. However, the prime rate may go above the inflation rate. On the flip side, the prime rate sometimes spends time below the inflation rate. So both need to be accounted for. 

A bond that accounts for both inflation and prime could be preferred by banks and other financial institutions for asset liquidity purposes, which may be all to the good in order to sell them and perhaps help preserve them within their issuer nation. This could help stabilise markets under problematic market conditions. 

\subsection{Operation}
 The carbon bond could be purchased from a government (or other issuer) with the immediate payment to be used for mitigation or replacement of the source of a greenhouse gas. The bond could be re-sold into an open market where banks, pension funds, and other financial entities could buy them. It is assumed here that the bond could be purchased by borrowing from the treasury/central bank due to its arbitrage. 

Alternatively, ownership of a bond could be transferred by the government to parties able to administer them and their funds for specific purposes. There are rare private corporations that have existed for 300 years, such as the insurer, Hamburger Feuerkasse. Lloyds is a government administered organisation, and most likely, stable governments could reasonably be expected to administer carbon bond payments over such a long time period. 

Shares in a carbon bond security, perhaps a tranched instrument composed of a basket of carbon bonds from multiple nations, could be allocated for the benefit of projects intended to address the worst of climatic effects---providing a long term income stream. There are multiple ways that such a bond could be used, provided that the use fulfils the bond purpose optimally. 

For the purposes of this study, exactly who is paid and how is not important. Rather, having a credible instrument to implement a real-world discount rate is the point. We assume that the present value received in payment for the bond at sale represents a reasonable estimate of the that bond's PV. 

\subsection{Estimation of carbon bond interest rate range, mean, and $\sigma$}
\label{CarbonBondInterestRate}
\begin{figure}[hbt!]
\centering\includegraphics[width=5.0in]{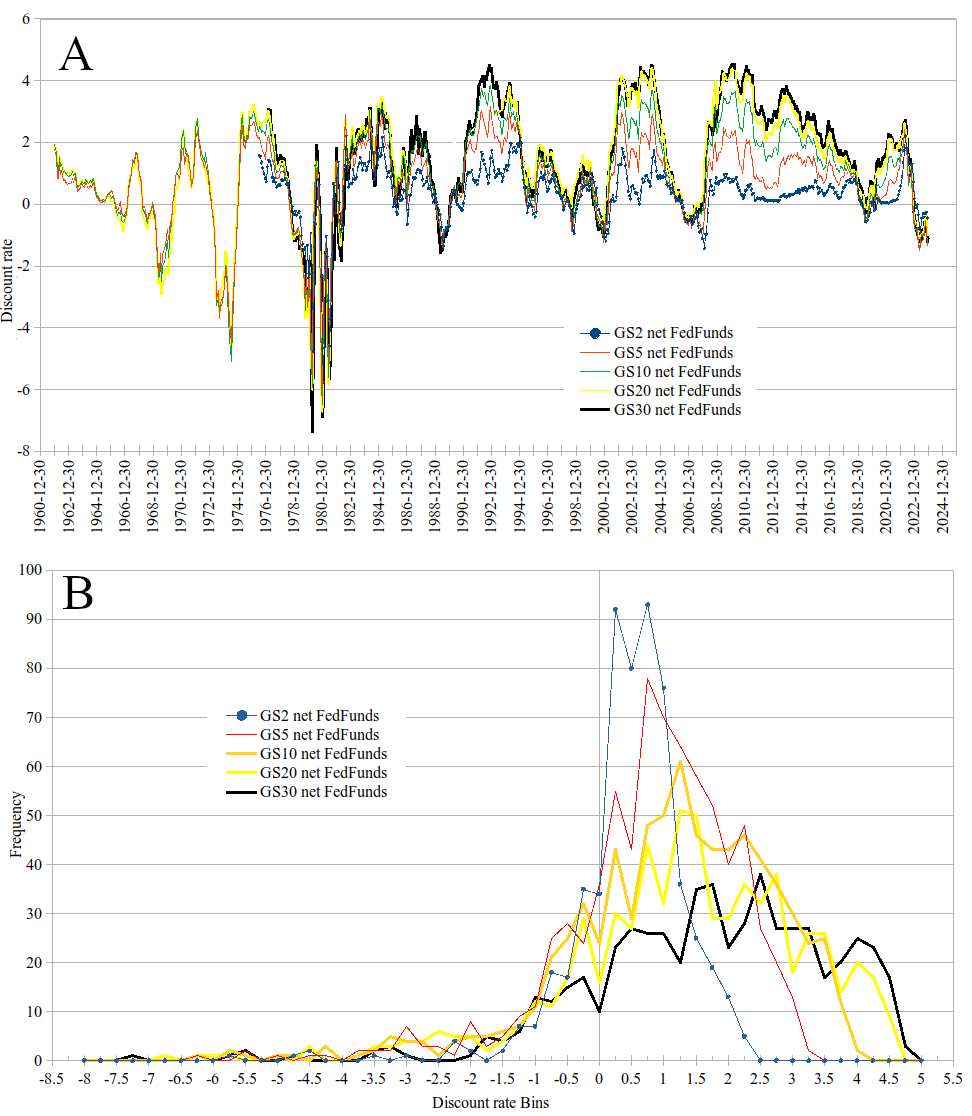}
\caption{\textbf{Monthly Treasury Real dollars net of Effective Federal Funds rate for 2, 5, 10, 20, and 30 year bonds.} Upper figure A shows inflation adjusted treasuries by month. By inspection, rate increase with long term of bond is visible for positive regions. Lower figure B shows frequency distributions for this set of T-bills. This curve is skewed with resemblance to a Poisson distribution. In B, longer term bonds "crawl" to the right, which should be a good direction for investors. The negative outliers extend farther, but negative area under the curve is minimal. The high ratio of positive to negative area under these curves suggests that some investors may prefer a fixed rate long-term bond. This is also visible in table \ref{TableT_Bill-Stats}. The deep negative rate tail below -2\% appears to be the result of the exogenous shock of OPEC. Such negative discount conditions cannot continue very long without breaking the bond system.   
 }
\label{Fig_Treasury_Real_T-bill_returns_1977-2023}
\end{figure}

To estimate long-term carbon bond rates, we examined monthly US Treasury bill data for 2, 5, 10, 20, and 30 year T-Bills with varying amounts of data \cite{FRED2024GS2,FRED2024GS5,FRED2024GS10,FRED2024GS20,FRED2024GS30}. The least amount of data is 47 years for 30 year T-bills from 1977-2023 \cite{FRED2024GS2,FRED2024GS30}. These data were corrected for monthly inflation to obtain real dollars using CPI data \cite{FRED2024CPIAUCSL}. Federal funds effective rate data was used \cite{FRED2024FEDFUNDS}, also corrected for monthly inflation \cite{FRED2024CPIAUCSL}. Figure \ref{Fig_Treasury_Real_T-bill_returns_1977-2023}A shows the resulting 30 year T-bill monthly data. By inspection of figure \ref{Fig_Treasury_Real_T-bill_returns_1977-2023}A, the association of the 1970's energy  crisis years (1973-1980) with a negative bond yield period is suggested. This association empirically provides additional support of Jarvis' energy $\leftrightarrow$ real-dollar relation.

Figure \ref{Fig_Treasury_Real_T-bill_returns_1977-2023}B shows the frequency distribution of these data, which does not appear to be a normal distribution. Corrected to real dollars, these data should display limits determined by capacity of investors to take losses, and capacity of the treasury to pay interest. What exactly those limits are is an open question, but we propose that in a modern economy, -9\% and +6\% are likely to be near those limits. 

It is possible that practical interest rate limits are not as limited as we believe, because deep  history of interest rates shows a long decline over the past 8 centuries \cite{Schmelzing2020_8CenturiesOfGlobalRealInterestRates}. This suggests that should economic times similar to feudal circumstances recur, akin to those of the 14th-19th century, the limits proposed above may be broken. However, that would not be a vibrant economy as exists in the first quarter of the 21st century. 

The distributions  of figure \ref{Fig_Treasury_Real_T-bill_returns_1977-2023}B do not appear to be a normal distribution. These have some similarity to an F distribution, however, for reasons discussed above, it is probably not, because large negative outliers would signal breakdown. Classification is complicated by considering limits imposed upon offerors and investors, as investors will not continue to purchase bonds that cause them losses. This indicates that formally, the Chebyshev theorem (eq. \ref{eq:Chebyshev1}) is appropriate to determine an upper limit of coverage of these distributions, although our discussion above also indicates that Chebyshev should be overly conservative. Thus, we say that the true coverage is somewhere in between as shown in table \ref{TableT_Bill}, which shows how much of the interest/discount rate range should be covered by Chebyshev's theorem, versus a normal distribution curve.

It is possible that some investors might prefer the risk of a fixed rate long-term bond, because history shows a high ratio of positive to negative rate area under the curve of figure \ref{Fig_Treasury_Real_T-bill_returns_1977-2023}. Whether or not fixed-rate long-term bonds could be accepted would also be sensitive to whether mark to market or mark to maturity accounting was required. 

\begin{table}[hbt!]
\caption{Comparison of Chebyshev's theorem (eq. \ref{eq:Chebyshev1}) coverage of the frequency distributions of figure \ref{Fig_Treasury_Real_T-bill_returns_1977-2023}B with a normal distribution, for the 30 year T-bill data of figure \ref{Fig_Treasury_Real_T-bill_returns_1977-2023}. Here, $\sigma$ = 1.86\%.  } 
\begin{tabular}{llllll}
$\sigma$ multiple (\textit{k})                           & \quad2       & \quad3       & \quad4         & \quad5           & \quad6            \\
\textit{k} $ \cdot \, \sigma$ single tail distance of discount rates        & \;\,3.71    & \;\,5.57    & \;\,7.43      & \;\,9.29        & 11.14        \\
Chebyshev theorem \% of data inside $\pm$ \textit{k}$\cdot \, \sigma$           & 75.00\% & 88.89\% & 93.75\%   & 96.00\%     & 97.22\%      \\
Normal distribution \% of data inside $\pm$ k $\cdot \, \sigma$ & 95.44\% & 99.73\% & 99.9937\% & 99.999943\% & 99.9999998\%
\end{tabular}
 \label{TableT_Bill}
\end{table}

\begin{table}[!hbt]
\caption{Skew and kurtosis of the 5 T-Bill  term frequency distributions of figure \ref{Fig_Treasury_Real_T-bill_returns_1977-2023}B. Skew appears to trend towards zero for longer-term bonds, which is desirable. Kurtosis is known to be unreliable with a thin-tail/fat-tail graph.}
\begin{tabular}{llllll}
 T-Bill term        & 2 yr & 5 yr & 10 yr & 20 yr & 30 yr \\ \hline 
Skew     & 2.62             & 1.55             & \;1.05              & 1.06              & \;0.72              \\
Kurtosis & 6.14             & 1.24             & -0.33             & 0.04              & -0.95            
\end{tabular} 
\label{TableT_Bill_Freq_Skew_Kurtosis}
\end{table}

\begin{table}[hbt!]
\caption{Statistics of frequency distributions shown in figure \ref{Fig_Treasury_Real_T-bill_returns_1977-2023}B. There is an increasing trend for the absolute value of all statistics except midrange. } 
\begin{tabular}{llllll}
T-Bill term & 2 yr  & 5 yr  & 10 yr & 20 yr & 30 yr \\ \hline
$\sigma$    & 0.89  & 1.34  & 1.62  & 1.82  & 1.86  \\
Mean        & 0.35  & 0.68  & 1.00  & 1.22  & 1.57  \\
Median      & 0.44  & 0.82  & 1.17  & 1.31  & 1.71  \\
Midrange    & -1.79 & -1.56 & -1.31 & -1.16 & -1.41 \\
Max         & 2.21  & 3.16  & 3.85  & 4.43  & 4.53  \\
Min         & -5.79 & -6.27 & -6.47 & -6.75 & -7.35
\end{tabular}
 \label{TableT_Bill-Stats}
\end{table}

\subsection{Choice of central tendency discount rate}

If we assume a carbon bond with a 300 year term, table \ref{TableT_Bill-Stats} suggests what one should expect the mean or median discount rate should be. If there is no discount rate reset interval in the operational definition of the carbon bond, it would appear that the mean rate should be higher than the 30 year T-bill. However, it was impractical to perform empirical research into the behaviour of very long-term bonds that Bank of England has records of. 

A very long-term bond issued with a fixed rate would need to be priced assuming it would weather every financial condition. So this should mean setting the discount rate at a level it would remain attractive. However, other factors come into play outside of the bond. If a regulator forces long term carbon bonds to immediately mark to market (MTM) instead of allowing mark to maturity valuation, this could force periodic fire-sales and failures. A 300 year carbon bond would seem to generally make sense to be allowed to mark to maturity, and this should help stabilise the bonds. Perhaps an alternative could be fractional mark to market based on remaining term.

With a discount rate reset interval between a month and 20 years, the behaviour of this 300 year carbon bond could be more like a shorter term bond, as with floating rate bonds. Thus, there is a range of potential central rates depending on the rate reset interval. 

\begin{figure}[!ht]
\centering\includegraphics[width=5.0in]{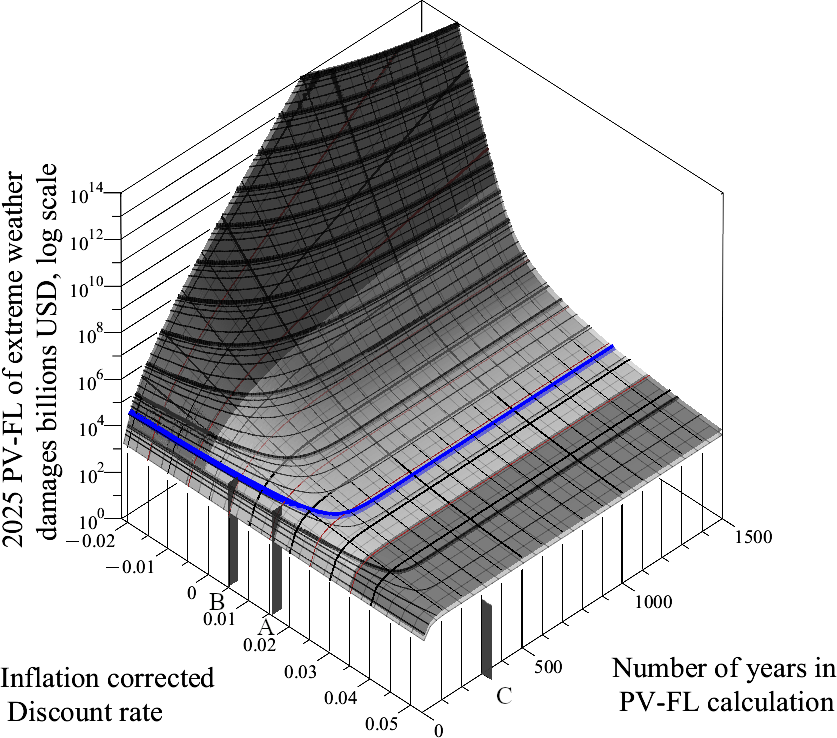} \\ 
\caption{\textbf{Isosurface graph:  $\pm$2$\sigma$ discount rate mapping of \textbf{$PV_{FL}$} for weather damages going forward from 2025. Represents total USA damages for all greenhouse gases together for the year 2025. } 
 The $\mathbf{x}$ axis scale varies the discount rate. The $\mathbf{y}$ axis scale is the number of years included in the $\mathbf{PV_{FL}}$ calculation. Losses are projected by the relevant function (\S11.2) forward from 2025 divided by the discount rate. (See main paper $\S$2.4.3.) 
Vertical \textbf{Tab A} at $\mathbf{d_{M{T30yr}}}$ of 1.57\% is the central rate from which $\sigma$ fluctuation is shown. The \textbf{Tab B} at $\mathbf{d_{r_{DICE}}}$ of 0.483\% discount shows the corrected DICE maximum and may be the most accurate net. The $\sigma$ derives from US 30 year T-bill real rate history \S3.4). Dark grey \textbf{Tab C} is the 300 year term of DICE. Light grey region centred at \textbf{Tab A} 1.57\% discount $\mathbf{d_{M{T30yr}}}$ is the centre of $\pm$$1\sigma$. Dark gray regions extend to $\pm$ $2\sigma$, and encompass the +5.1\% maximum DICE 2016R model rate (main paper $\S$2.4.3.) and the -2.2\% rate. These outer $\pm$ 2$\sigma$ dark grey isosurface segments may be hard to believe, but given the potentially hundreds of years in which energy poverty (main paper \S4.1) or fiscal austerity and the paradox of thrift \cite{Corden2012GlobalImbalancesAndTheParadoxOfThrift,Vermann2012ParadoxThrift,Yiannis2011DeficitReductionTheAgeOfAusterityAndTheParadoxOfInsolvency,Das2019ParadoxOfAusterityMulti-CountryEvidence} could be created, such rates are not out of the question, and if we are to consider it possible that a 5.1\% real discount rate could exist, we should equally accept a -2.2\% discount. This latter value would be expected in an austere or energy poverty period, with a probable root cause of low energy availability. The impact of low energy availability is likely to be societal breakdown as future damages outpace capacity for repair and rebuilding.  
\textbf{Heavy blue contour line:} 2025's estimated GDP of \$24.7 trillion.The \textbf{minor Z log-scale contours} are 2, 4, 6, and 8 times the previous major contour.
}
\label{figPVDrateNumYears}
\end{figure}

There is a similar consideration for deciding what value of $\sigma$ to use. With these caveats, the nominal carbon bond rate used for this study is $d_{MT30yr}$ = 1.57\%, and $\sigma$ = 1.86. 

In practical use, figure \ref{figPVDrateNumYears} (which is the main document's figure 7 isosurface graph reproduced here) shows the the full range of contemplated present value total real cost for the USA in 2025. Similarly, figure 8 of the main document shows this discounting applied to each GHG's social cost. Thus, the higher the real interest rate on the carbon bond, the less the impact should be on GDP, other things being equal. Note that there will be a different 3 dimensional phase space of cost for each year going forward from 2025. 

We decided to keep the name "Carbon bond" even though these bonds would need to cover all the greenhouse gases. Carbon combustion provides the majority of the greenhouse effect and will continue to do so for the next 300-400 years.

\section{Discussion of our special NOAA Total Weather Damages (TWD) version with all costs}

\begin{quote}
 In performing these disaster cost assessments these statistics were taken from a wide variety of sources and represent, to the best of our ability, the estimated total costs of these events -- that is, the costs in terms of dollars that would not have been incurred had the event not taken place. Insured and uninsured losses are included in damage estimates. Sources include the National Weather Service, the Federal Emergency Management Agency, U.S. Department of Agriculture, National Interagency Fire Center, U.S. Army Corps, individual state emergency management agencies, state and regional climate centres, media reports, and insurance industry estimates.” - NCEI  \cite{NOAAbillionDlr} 
\end{quote} 

This project obtained a version of the NOAA Consumer Price Index-adjusted weather disaster damages\cite{NOAAbillionDlr} with \emph{all} available cost adjustment made for each year, the Total Weather Damages (TWD) dataset. This step was taken because potentially, a large number of smaller losses can add up significantly, and the "gap year" of 1987 needed a realistic value for our purposes. (One of the expectations of climate change is that much of the precipitation increase will occur as longer-lived lower wind energy storms delivering more precipitation.) Thus, the TWD dataset provided is unique for use developing future climate damages estimates, and supports production of an empirically based social cost of GHG (SC-GHG). This NOAA data has been through decades of refinement and reanalysis, with the first major review \cite{Smith2013billion-dollarClimateBiases} finding systematic underestimates of 10-15\%, and the second showing good confidence after sensitivity analysis \cite{Smith2015QuantifyingUncertaintyClimateDisaster}. 

We present US-centric OPTiMEM risk curves against estimates probable economic growth,  with a wide range depending on the rate of growth. Visible in figure \ref{Fig_WD_Heat_Conjecture_Risk_graph} is that outlier year events may rise to large fractions of GDP. Properly done this requires risk curves for maximum expected single year damages in any one year. 

\subsection{Coverage of climate effects and risk}
For this total NOAA weather damages dataset to be meaningful, it needs to exist over a time span when relevant climate effects occur: sea level rise, precipitation increase, increased storm activity, and evidence of unusual heat. 
Our NOAA TWD dataset covers 42 years that comprise: \textbf{Increased precipitation} \cite{Ayat2022Subhourlyrainintensity,EPA2022GlobalMeanTempPrecip} (fig. \ref{EPA_Global_temp-precip_exhibit-6}, \ref{Fig_Supp_Rain_Wind_Ratio_PowerLaw}). \textbf{Increased storm activity} \cite{hurdat2-1851-2021-041922} (fig.'s \ref{Fig_Supp_Norm-ZJ-to-ACS}, \ref{Fig_Supp_AllStorms_1ZJ}). \textbf{Sea level rise} 1880-1993, sea level rose $\approx$ 16.1 cm \cite{Church2011SeaLevelRiseLate19th_Early21stCentury}. Sea level rose $\approx$ 9.1 cm ($\approx$ 3.6 inches) 1993-2021 \cite{NASA_GSFC2021GlobalMeanSeaLevelTrend}, for a total of $\approx$ 26.3 cm ($\approx$10.3 inches) 1880-2021 CE.  1961 is a rough inflection point for acceleration of sea level rise. \textbf{Unusual heat and evidence} Fire increase in quantity and destructiveness \cite{Zheng2023RecorodBorealCO2,Jones2024GlobalRiseInForestFire_ClimateChangeInExtratropics,Balch2024Fastest_most_destructive_fires2001to2020,Higuerae2021ForestsBurnGT2000yr}. Flash drought acceleration \cite{Yuan2023FlashDroughtsClimateChange}. Temperature increase (fig. \ref{EPA_Global_temp-precip_exhibit-6}). Thus, these aspects are present within our total NOAA weather damages dataset. See: \S \ref{lb:NOAAvaliditySummation}.

Tipping points are not all baked in, as discussed in the next section \S \ref{sect:TippingPoints}, so our method cannot reliably predict most tipping point effects without special code to account for it. This method shares this problem with all known climate economic models at this time. However, thinking is modifying towards tipping points manifesting more smoothly \cite{Lenton2021TippingClimateSystem}, so it is possible that to the extent tipping point beginnings are present, these could be baked into these data. 

With the caveat that our damages are partial, inclusion of the above drivers of economic damages in the total NOAA weather dataset time span allows us to make believable estimates, and also estimate risk. Risk is derived from the variance of the total NOAA weather damages dataset from the fitted curve over the time period of the total NOAA weather damages model (\S\ref{lb:RiskMethods}).

\subsection{Inputs to NOAA TWD}
\label{InputsToNOAATWD}
More than one dozen public and private sector data sources help capture the total, direct costs (both insured and uninsured) of the weather and climate events. One of the key transformations is scaling up insured loss data to account for uninsured and underinsured losses, which differs by hazard (peril), geography, and asset class.

Market-based losses where property insurance for a hazard exists are either insured or un(der)insured. Several FEMA Public Assistance, Individual Assistance, and other needs assistance programs, in addition to SBA programs, help cover uninsured or underinsured losses.  However, it is often the case that these Federal sources of funding are slow and fractional to the likely need. Therefore, NOAA refers to transforming the residential, commercial and automotive, etc. policies in relation to the specific hazard, geography, and asset class that was damaged based on statistical review.  NOAA carefully tries to capture the total, direct costs (both insured and uninsured) of the weather and climate events, but there are significant non-market losses that cannot be currently captured, and other bases for damages that are likewise not captured.

The primary costs comprise: physical damage to residential, commercial, and municipal buildings; material assets (content) within buildings; time element losses such as business interruption or loss of living quarters; damage to vehicles and boats; public assets including damage to roads, bridges, levees; damage to electrical infrastructure, lost hydropower production, offshore energy platforms, US military bases; agricultural assets including crops, livestock, and commercial timber; and wildfire suppression costs. An exhaustive list is not practical here. 

\subsection{Limitations of total NOAA weather damages---indirect and non-market}
\label{NOAADamagesLimitations}
Because the OPTiMEM model is based on the TWD, which does not bontain indirect and non-market damages, disaster costs do not take into account losses to: natural capital or environmental degradation (ecosystems, fisheries, air quality); costs and impacts due to climate change related disease; mental or physical healthcare related costs; the value of a statistical life (VSL); or supply chain contingent business interruption costs beyond the hazard area.

The cost of climate related disease is a major item. Even cancer is documented as increased from climate change \cite{Nogueira2020ClimateChangeAndCancer}. That contagious disease such as the 2020-2023 COVID-19 pandemic is a cost of climate change driven by animal migration, human incursions, and temperature, is strongly supported, as is increased tropical and vector borne diseases for humans and animals \cite{Carlson2022ClimateChangeIncreasesCross-speciesViralTransmission,Casadevall2020ClimateNewDisease,Chen2021COVID}, \cite{Tidman2021NTDs}, \cite{BOOTH2018NTDs}. To account for the cost of events like the novel SARS-CoV-2 (COVID-19) pandemic, or incursions of tropical diseases like malaria, dengue, or rising babesiosis \cite{Ssentongo2024BeyondHumanBabesiosis_PrevalenceMortality}, is complex for multiple reasons. The first is identifying applicable records going back to the inception of this NOAA dataset that are complete enough. Second, for each such event, developing totals comprising: deaths above expected by demographic, economic losses to GDP during the pandemic, direct costs to respond to the pandemic, and, arguably, the differential between GDP projected without a pandemic, and GDP going forward from the start of the pandemic for some arbitrary period, for instance, 5 or 10 years from the official pandemic start. Typically, a recovery can achieve the previous growth rate, but rarely exceeds it, so the losses relative to a normal economy can be quite high. 

In addition to novel, tropical, and insect dependent disease, there are animal pandemic diseases such as highly pathogenic avian influenza (HPAI) \cite{LIANG2022HPAI_H5N6} in an overlapping time period (2021-2023) that impacted the price of eggs \cite{Knight2022USDA_HPAI_egg_prices,FRED2023APU0000708111} and significantly affected the price of chicken \cite{FRED2023APU0000706111}. This HPAI has infected and killed animals from 17 mammalian species \cite{USDA2022_HPAI_mammal_detection}. Non-avian species deaths detected attributable to this HPAI range from bottlenose dolphin and seals to bears, raccoon's and foxes, signalling some possible non-market ecological damages. Eggs are an important inexpensive source of high quality protein worldwide. This loss must be substituted with other protein sources or face nutritional deficit, which in  turn represents an inflationary pressure. 

Therefore, the NOAA estimates should be considered conservative with respect to what is truly lost, but cannot be completely measured due to a lack of consistently available data.

This exclusion of indirect and non-market damages could potentially be addressed at a future time, and does not impact the usefulness of this alternative approach. 

\subsection{Error margins of the NOAA Total Weather Damages (TWD) dataset}
The margins for error in the NOAA dataset are in relation to the insurance penetration assumptions by hazard and asset, among other time-element uncertainties such as inflation in materials and labor cost after a disaster event. The takeaway is that the NOAA estimate is a solid, and possibly overly conservative estimate with respect to what is truly lost, but cannot (yet) be completely measured.

\subsection{Statistical variance of the NOAA Total Weather Damages (TWD) dataset and risk use} 
The statistical variance methods development are treated in \S\ref{EstimationOfsigmafunction}-\S\ref{Examination_Slope_Change}. The application of this to risk estimation is treated in \S\ref{lb:RiskMethods}.

\subsection{Summation of validity of the Total Weather Damages NOAA dataset}
\label{lb:NOAAvaliditySummation}
Economic data inevitably contains influences from a wide variety of factors over time spans. The method of finding correlates in economic data that is subject to a wide variety of influences has a long history and is common practice \cite{Lee2001PopulationInequalityHomicide,Burke2009WarmingIncreasesWarAfrica,Burke2015GlobalNonlinearEffectOfTemp,Mirza2021GlobalInequalityRemote}. The fundamental physics driving climate is understood, geological climate history has strong evidence, and the interval covered by these total NOAA TWD data comprises significant changes to climate that are generally predicted by climate science.  Thus, taken together, we judge the method of correlation to be robust within the assumptions of our emissions scenarios comprised of CO$_2$, CH$_4$, N$_2$O, Fgas, driven by carbon consumption, GDP and population. 

\section{Tipping points}
\label{sect:TippingPoints}
Table \ref{table_tipping} shows one set of climate tipping points and evaluates whether they are represented in climate models. Most of these are only present in OPTiMEM's results to the degree that they are baked in, although CH$_4$ arctic permafrost melt scenarios are present. These are not an absolute list, and surprises should be expected. For instance, in addition to permafrost thaw \cite{Schuur2022PermafrostFeedbacksWarming} (which is included in OPTiMEM), troubling surprises regarding amounts and mechanism of methane increase have come to light \cite{Qu2022Attribution2020methaneSurgeGOSAT,Shaw2022MethaneFluxesZambia,Feng2022MethaneResponsibleRecords2020_2021}, with 80-85\% of increased emissions due to tropical wetlands and microbial sources \cite{Feng2022TropicalMethaneEmissionsLargeFraction}  \cite{Basu2022EstimatingMethaneConsistentC-13}. That current tipping point predictions will be incomplete is consistent with the inherent conservatism of science \cite{Kemp2022ClimateEndgame}. Thus, it is reasonable that real world damages should surpass our projections. 

\begin{table}[!ht]
\caption{Climate tipping points.} 
\label{table_tipping}
\begin{tabular}{ll} 
\hline
Tipping point &Representation in GCMs  \\
\hline
\textbf{Arctic} summer sea ice melt & Probably represented well for albedo effect adding heat 
    \cr &however interlocked with arctic ocean methane clathrates 
    \cr &which are shallow \cite{Connor2010methancyclereview}. \\
\textbf{Antarctic} ice melt & Not represented well \cite{Pettit2021ThwaitesEasternCollapse,Batchelor2023RapidIcdRetreatHundredsMetres,Milillo2022RapidGlacierRetreat,Schulz2022ImprovedMeltRate,Ohenhen2024DisappearingCitiesOnUScoasts,Abram2025EmergingEvidenceOfAbruptChangesInTheAntarcticEnvironment} \S\ref{lb:CryosphereIceMelt}. \\
\textbf{Ocean} methane clathrates &May not be represented well due to data being entirely 
    \cr &from modelling \cite{BUFFETT2004GlobalOceanClathrateInventory,Frieling2016,Frieling2019,Inglis2020,Inglis2020_Supp}. 
    Methane release from clathrates and 
    \cr &clathrate blocked methane seeps may not be accurately quantified \cite{Schuur2022PermafrostFeedbacksWarming,Isaksen2001ThreeDeepAlpine-permafrostBoreholesInSvalbardandScandinavia,Frieling2016,Frieling2019,Inglis2020,Seabrook2025AntarcticSeepEmergenceAndDiscoveryInTheShallowCoastalEnvironment} \\
\textbf{Greenland} ice sheet melt &Recently revised upward in an RCM \cite{Hofer2020GISsealevel} \& serious ice modelling problems. \cite{Schulz2022ImprovedMeltRate,Milillo2022RapidGlacierRetreat,Batchelor2023RapidIcdRetreatHundredsMetres}.    \\
\textbf{Northern permafrost melt}  &Probably well represented soon \cite{Schuur2022PermafrostFeedbacksWarming}
    \cr & This is represented in our heat conjecture model. \\
\textbf{AMOC} shutdown &Probably fairly well represented 
    \cr & caveat: storm severity not well modelled. \\
\textbf{West} African monsoon shift &Probably fairly well represented 
    \cr & caveat: storm severity not well modelled.\\
\textbf{Indian} monsoon shift &Probably fairly well represented 
    \cr & caveat: storm severity not well modelled.\\
\textbf{Boreal} forest shift &Probably fairly well represented, however, while
    \cr & tree disease emergence is certain\cite{REED2022Beechdisease,Ewing2020BeechDisease,Singh2019Appledisease} 
    \cr &disease details difficult/impossible to predict in advance.
    \cr &Fire may not be fully represented \S\ref{lb:ForestAndWildfire}.\\
\textbf{Amazon} rainforest dieback & Rainforest carbon sink failure. Models may not account 
    \cr &for human exacerbation \cite{NASA2022UptickAmazonFire, Boulton2022Amazonresilience, Gatti2021Amazonia_as_a_carbon_source}.\\
 \\
\hline
\end{tabular}
\vspace*{-4pt}
\end{table}

Tipping points may be triggered at 1$\degree$ to 2$\degree$C warming \cite{lenton2019risky}. There are two tipping points that have evidence of being breached---the Amazon rainforest and permafrost thaw. The tipping point that is formally included in our climate model is permafrost thaw release of carbon as CO$_2$ and CH$_4$ based on Schuur, et al \cite{Schuur2022PermafrostFeedbacksWarming}. Antarctic methane has not been much studied, and concerning seeps have been found \cite{Seabrook2025AntarcticSeepEmergenceAndDiscoveryInTheShallowCoastalEnvironment}.  The Amazon was a carbon source in 2021 \cite{Gatti2021Amazonia_as_a_carbon_source}, and may be transitioning permanently in no small part from human forest burning \cite{Boulton2022Amazonresilience}. These two elements appear late in the 42 year time span of the total NOAA weeather damages, but there is some very early stage coverage baked into the total NOAA weather damages dataset.
CMIP attempts tipping points to some degree \cite{Dingley2023CMIPAnnualRpt2022-2023}. Tipping point thinking from climate science appears to be shifting somewhat toward smoother effects \cite{Lenton2021TippingClimateSystem}.  

The implication of the discussion of this section is that our climate model may project lower rather than higher. This is the inherent conservatism of science \cite{Herrando-Perez2019StatisticalLanguageBacksConservatisminClimateAssessments,BRYSSE2013ClimateChangePredictionErringOnTheSideOfLeastDrama}. 

\subsection{Halogen catalysis destroying ozone over the northern hemisphere}
Future CO$_2$ is virtually guaranteed to pass 500 ppm, which is Eocene CO$_2$ \cite{Burke2018Plio}. The Eocene world was 6$\degree$C to 14$\degree$C warmer and had long term methane releases. However, the Eocene climate's temperatures were \emph{declining} from the PETM, whereas the current era is rising from what will be the low temperature for tens of thousands of years. The 500 ppm CO$_2$ level is also a trigger point for destruction of northern hemisphere ozone, causing loss of ultraviolet shielding from a rise in chlorine and bromine ions in the atmosphere \cite{Anderson2017ChlorineBromineCatalysis}. The same effect is a possible result of volcanic activity \cite{StauntonSykes2021HalogensRadiativeForcing}. The effect of chlorine and bromine catalysis on the $e$fold time of methane is an open question. We made no attempt to model potential impact from chlorine/bromine catalysis of ozone by increased catalysis and convection \cite{Anderson2017ChlorineBromineCatalysis}. That is a task for GCMs to perform. This impact could be significant. 

\subsection{Volcanism}
In geologic time, volcanism played a central role in supplying CO$_2$ for global thermal maximums \cite{Hochuli2016,Li2021a,Li2021b,Tong_Du2022103878,IEA2022CO2Emissions2021}. Geologically, the role of volcanic injection of water into the stratosphere to drive thermal maximums is unknown. A 2022 undersea volcanic eruption in Tonga increased stratospheric water by $\approx$10\% of normal stratospheric load, which is expected to have a warming effect \cite{Millan2022HungaTongaHydrationStratosphere}. Ice sheet loss causes continental unloading and magma production \cite{Sternai2016DeglaciationMagmaUnloading}. North America is still rising after loss of its ice sheet. Greenland and Antarctica are losing ice, and previously unknown heat sources have been identified. Much land volcanism acts to cool, at least in the short term \cite{ZielinskiVolcanism7000BCGISP2}. Thus, the highly uncertain role of possible exacerbation of global warming by natural volcanism remains to be seen. 

Arguably, volcanism could be included as a tipping point because volcanoes played a role in previous thermal maximums \cite{Hochuli2016,Li2021a,Li2021b,Tong_Du2022103878,IEA2022CO2Emissions2021}. Volcanoes have complex effects \cite{StauntonSykes2021HalogensRadiativeForcing}, both decreasing \cite{ZielinskiVolcanism7000BCGISP2} and increasing \cite{Millan2022HungaTongaHydrationStratosphere} warming. The argument for making volcanism a tipping point is that unloading of continents should increase volcanism \cite{Sternai2016DeglaciationMagmaUnloading, VanVliet-lanoe2020VolcanismEnhanced}. Recent work shows that climate models do not represent ice melting very well, so continental unloading could occur faster than expected \cite{Pettit2021ThwaitesEasternCollapse, Fox2022TheComingCollapseSciAm}. Recent estimates of ice loss rates range from 610 m day$^{-1}$ \cite{Batchelor2023RapidIcdRetreatHundredsMetres} to 1.3-1.7 km y$^{-1}$ \cite{Milillo2022RapidGlacierRetreat}.  We only note these potential factors in climate. We make no attempt to model volcanic emissions.  

\subsection{Silicate weathering}
Our understanding of silicate weathering is incomplete: "\emph{We find a CO$_2$ release of megatons of carbon annually from weathering of OC$_{petro}$ [rock organic carbon] in near-surface rocks, rivalling or even exceeding the CO$_2$ drawdown by silicate weathering at the global scale.}" \cite{Zondervan2023RockCarbonCO2releaseOffsetsSilicateWeatheringSink}. The meaning of this is that some amount of CO$_2$ offset in near term models of the next 1500 years or so may not be real, so slightly higher CO$_2$ may be expected. In the very long term, this may slow, or negate projections based on silicate rock absorption such as the UCC \cite{Archer2020UltimateCostofCarbon, Archer2009AtmosphericLifetimeOfCO2}. Thus, the morality of harm to descendants far in the future rears its head \cite[p. 24]{Stern2021TheEconomicsofImmenseRiskUrgentActionandRadicalChange}\cite{Dasgupta2008DiscountngClimate}. We make no attempt to model this concern, using the largest model survey available for CO$_2$ drawdown (fig. \ref{Fig-RemainderFractions}).

\section{Climate model and scenarios}
\label{lb:ClimateModelandScenarios}
\begin{figure}[hbt!]
\centering\includegraphics[width=5.25in]{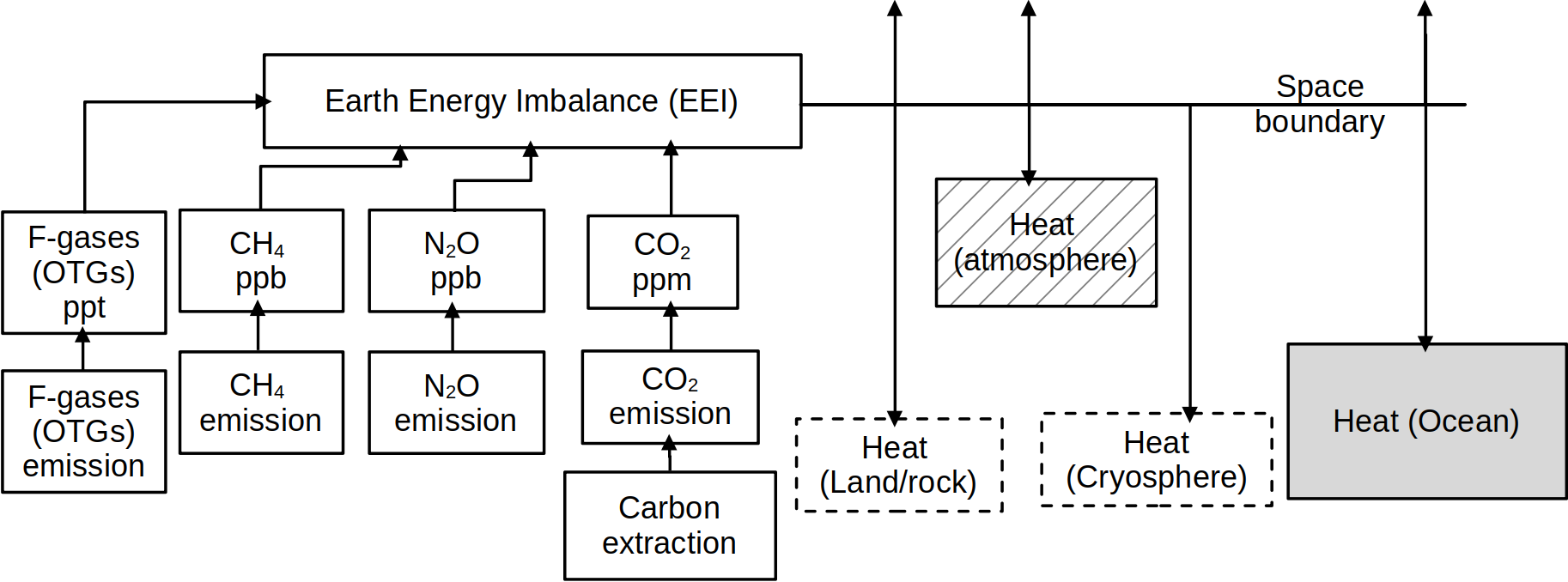}
\caption{\textbf{Climate model atmospheric physics schema.} This diagram describes the climate model used for this work. Shaded ocean heat box is where heat in OPTiMEM is stored. This OHC is validated by von Shuckmann and NASA datasets, representing 88.0\% of planetary heat energy relevant to climate/weather. Cross hatched atmosphere heat box is modelled as a pseudo-surface in contact with the ocean that does not store the 0.9\% that is atmospheric heat. Dashed boxes for land and cryosphere indicate these are not used. Land and cryosphere are much more difficult to model and it was not attempted. (\S\ref{lb:Heat}, fig. \ref{Supp_Von_Schuckmann_Full}) We start with carbon extraction limits driving CO$_2$ emissions which translates to CO$_2$ parts per million. N$_2$O is driven by population and Gross World Product (GWP). Methane emissions are historical fractions of CO$_2$ with the addition of permafrost emissions. Methane converts to CO$_2$ based on its $e$fold chemical lifetime. Fgases also participate in changing the heat flux (EEI), which in turn heats land, ice, atmosphere, and ocean. The ocean is the largest heat reservoir by far, and the only one for which we have a usable dataset for validation. Thus, ocean heat content (OHC) is our chosen heat curve.}
\label{Fig_Climate_Drivers_diagram}
\end{figure}

 Figure \ref{Fig_Climate_Drivers_diagram} presents the atmospheric physics schema of our climate physics model.

The fundamental problem of climate economics is that while the importance of correctly determining the social cost of carbon (SCC), is incontrovertible -- alternative methods of CO$_2$ cost assessment are non-existent for most practical purposes. OPTiMEM makes extensive use of validations of our carbon driven climate forcing. 

 The basis of climate forcing in OPTiMEM is greenhouse gases. Greenhouse gases are composed of CO$_2$, methane (CH$_4$), nitrous oxide (N$_2$O), and Montreal Protocol trace gases (MPTGs) The Montreal Protocol covers halogenated hydrocarbons containing chlorine and fluorine(CFCs) or bromine (BFCs). These are also potent greenhouse gases with significant GHG forcing at parts per trillion.  Other trace gases (OTGs) comprise Paris agreement gases, sulfur hexafluoride (SF$_6$), nitrogen trifluoride (NF$_3$), sulfur dioxide (SO$_2$) and ozone (O$_3$) plus a growing set of hydrofluorocarbons (HFC), perfluorinated compounds (PFC), fluorinated ethers (HFE), perfluoropolyethers (PFPMIE) with long life spans \cite{Hansen2017YoungPeoplesBurden}\cite[table 2.14, p. 212]{Forster2007GHG-changes}. Certain chlorinated fluorocarbons (CFCs) are of exceptional interest due to their potent GHG effects. Fluoroform (CHF$_3$, HFC-23), and trifluoromethyl sulfur pentafluoride (SF$_5$CF$_3$) have long dwell time, estimated at 1000 years in the stratosphere, as do (CHF$_3$) \cite{Oram1998Fluoroform} (SF$_5$CF$_3$) \cite{Sturges2000PotentGHG1000yr}.

 There appear to be problems with  the reporting of carbon emissions as reported by He, et al \cite{He2024MajorGapsPetrochemicalEmissionsReporting}. These problems are not going to be worked out for some time, and we made no attempt to introduce a correction in this model. 

\subsection{Nitrogen model}
\label{Nitrogen_model}
\begin{figure}[!ht]
\centering\includegraphics[width=6.25in]{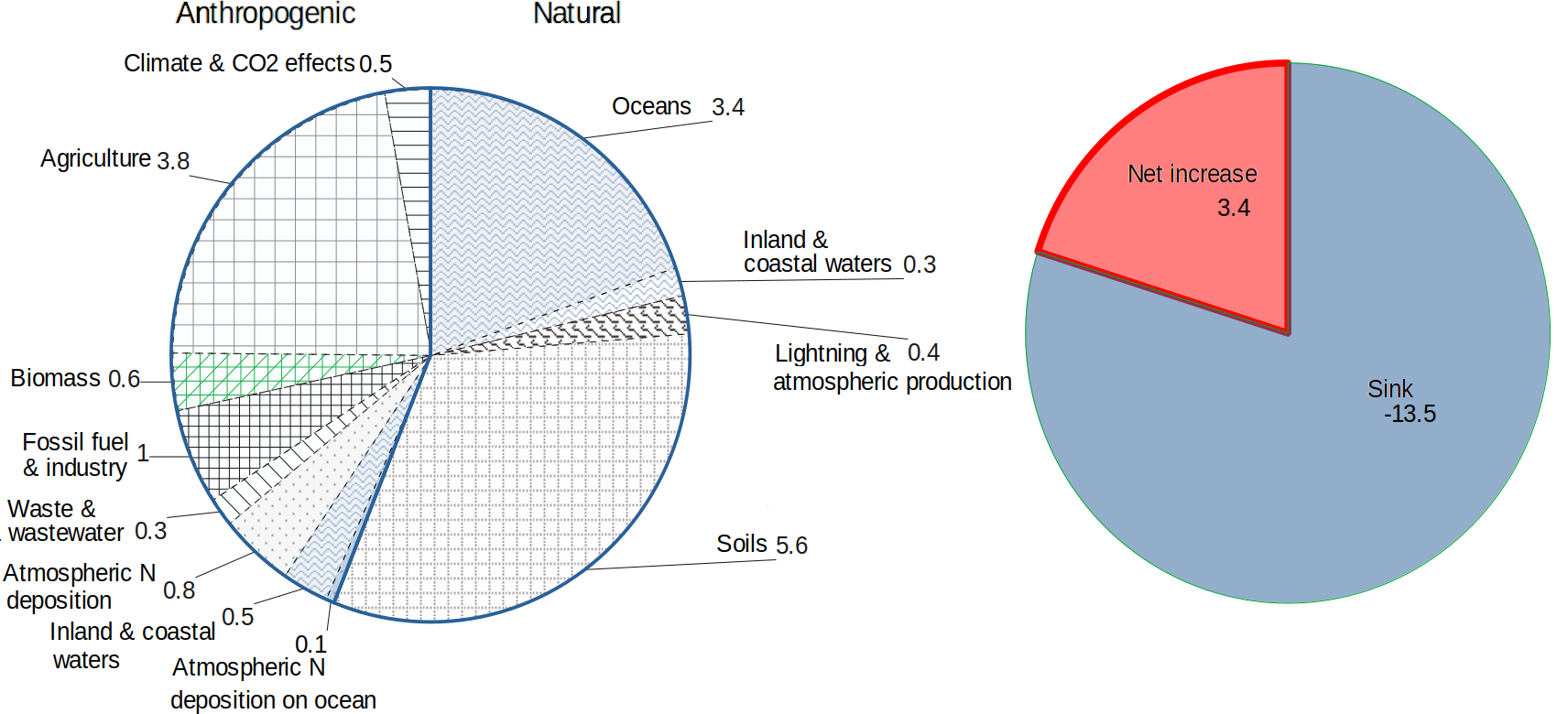}
\centering\includegraphics[width=6.25in]{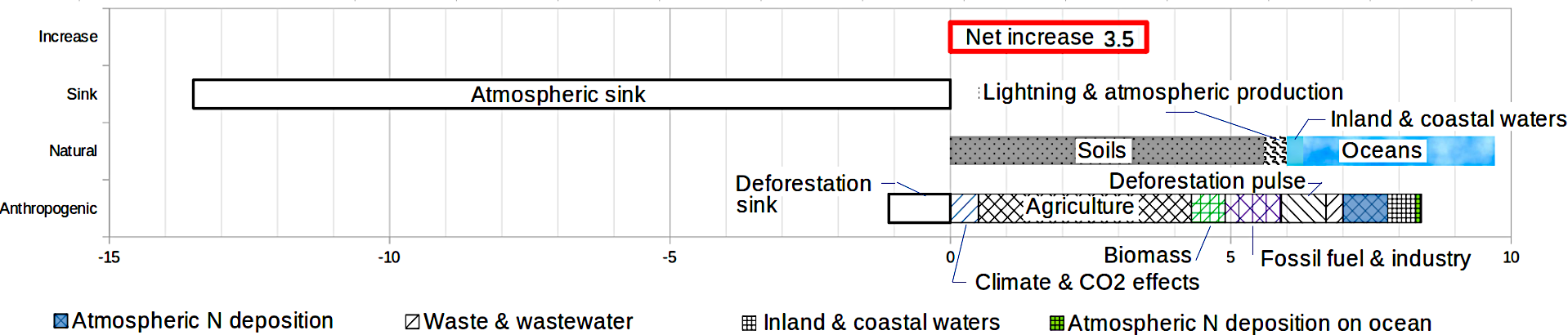}
\caption{\textbf{Global N$_2$O Budget, in Teragrams (Tg) \cite{Tian2020N2OSources_sinks}} Upper Left graph: N$_2$O sources. Blue lines demarcate Anthropogenic sources on left, and Natural sources on right. Upper Right graph: N$_2$O sink with net increases. Lower graph: Sources, sinks and Net increase of N$_2$O. Lower graph includes deforestation effects.}
\label{Fig_Supp_N2OBudget}
\end{figure}

N$_2$O has been modelled in the past as fairly well correlated to CO$_2$ \cite[p. 49, SM2]{Hansen2023WarmingInPipeline}, however, this correlation is because both are related to the economy and population. Economy and population are related to each other. The global N$_2$O budget \cite{Tian2020N2OSources_sinks}, has net 3.4-4.8 Tg anthropogenic increase for the 2007-2016 decade (Fig. \ref{Fig_Supp_N2OBudget}). Approximately 14\% of the anthropogenic N$_2$O production is directly from fossil fuels, with an $\approx$7\% additional indirectly from CO$_2$, for roughly 21\%.

Consequently, for OPTiMEM, we correlate agricultural N$_2$O to human population growth, using UN consensus \cite{UN2022PopGrowth2100} and high growth \cite{UN2004WorldPopTo2300} population scenarios. For practical purposes, the high population scenarios mean that carbon synfuel would have to be used in place of fossil fuel in the absence of electrification.

The primary nitrous oxide reference to guide this model is Tian, et al, 2020, which accounts for it as nitrogen, which we translate to N$_2$O.  Because Tian clarifies proportions from agriculture and industry separately, and we are attempting long-term projections, we separate agriculture and industry. This ties agricultural N$_2$O to population, and industrial N$_2$O to gross world product (GWP) \cite{Tian2020N2OSources_sinks}. The argument for splitting them is that population growth is expected to level off and/or decline, however, economic growth should still increase, and abatement of industrial sources is not yet done. Because we found no usable dataset(s) or references for NOx impact on climate, only N$_2$O was modelled. 

\begin{table}[!hbt]
\caption{N$_2$O emission calibration for 2007-2016.}
\label{Table_N2OEm} \begin{tabular}{rrrr}
N$_2$O sink Tg & N$_2$O soil-ocean em Tg & Anthropogenic N$_2$O em Tg & Net N$_2$O em Tg   \\
-12.4       & 11.707               & 9.194        & 8.501 \\            
\end{tabular}
\end{table}

To fit existing data, values shown in table \ref{Table_N2OEm} were chosen for the 2007-2016 decade that Tian studied, within their range parameters to fit existing N$_2$O data. The anthropogenic N$_2$O was divided 60.81\% for population, and 39.19\% for GWP based on Tian's figures. The relationships of both population and GWP to N$_2$O were verified to be linear, with $R^2$ of 0.53 and 0.57 respectively (not shown). Parts per billion (ppb) were converted to teragrams (Tg) with a factor of 7.7988 Tg per ppb, and each  year's N$_2$O increase divided according to the 60.81\% for population, and 39.19\% for GWP. 

Linear equation fits to these resulting population and GWP N$_2$O datsets yielded two linear equations (eq. \ref{eq:TianEqN2OtoPopGWP} for projecting forward, with standard deviation of 0.908 Tg for population and 0.601 Tg for GWP. 

GWP data used is FRED data from 1960-2021 \cite{FRED2021NYGDPMKTPCDWLD}, which was then corrected for inflation using US Bureau of Labor Statistics inflation calculation. This dataset matched well with Maddison \cite{MaddisonDatabase2010} where they overlap. 
 
\begin{figure}[hbt!]
\centering\includegraphics[width=5.25in]{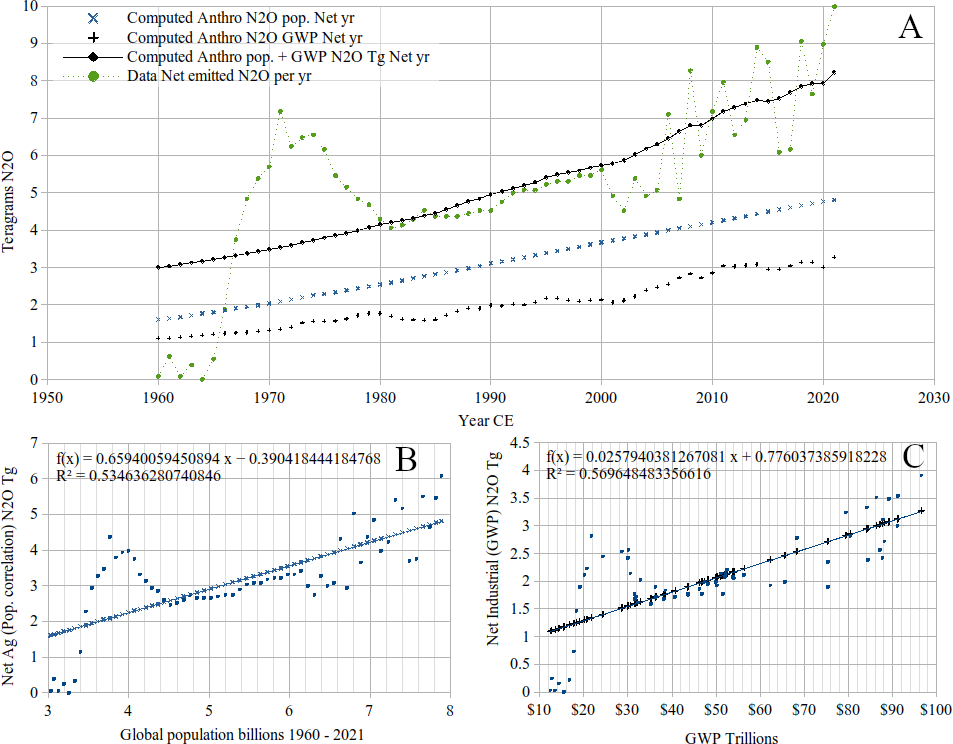}
\caption{\textbf{N$_2$O component equation graphs.} Panel \textbf{A} shows the computed curve of N$_2$O for industry correlated to GWP as black "+" symbols. Computed curve of N$_2$O for agriculture correlated to population is shown as blue "x" symbols. The added curves for full anthropogenic N$_2$O net increase for a year is shown as black diamonds, superimposed on the green circle historical N$_2$O data of net emission by year. Panel \textbf{B} shows fit to data for agriculture. In panel \textbf{C} the fit to data for industry  is shown.   }
\label{Fig_Supp_N2O_Equation_Graphs}
\end{figure} 

\begin{flalign}
  \begin{aligned} 
\label{eq:TianEqN2OtoPopGWP}
 \text{Population N$_2$O Tg } & \text{from Agriculture} \\
AgN_2O(p) &=  -0.3904 + p \cdot 0.659400 \\
 \text{ GWP N$_2$O Tg from} & \text{ industry and transportation}   \\ 
 GWPN_2O(Y) &= 0.776037 + Y \cdot 0.0257940 \\
\end{aligned}
 \end{flalign}
 \qquad\qquad \qquad \qquad \text{Where: } p  =  \text{ population in billions}  \qquad\qquad   \text{Y = GWP real dollars, trillions} \\
 \text{ } \qquad\qquad \qquad \text{Note that as fitted equations both of these equations are inclusive of N$_2$O's $e$fold life span. }

\subsubsection{Population data and projections for N$_2$O}

\begin{figure}[hbt!]
\centering\includegraphics[width=5.25in]{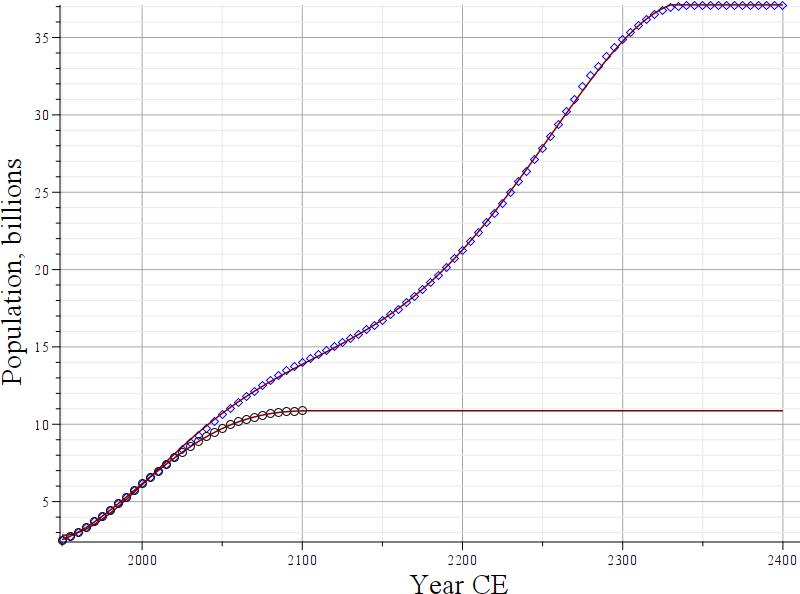}
\caption{\textbf{UN population growth.} Black circles, composite of  UN population projection to 2300 CE \cite{UN2004WorldPopTo2300}, UN projection to 2100 CE \cite{UNDESApopDiv2019}, and FRED data \cite{FRED2021SPPOPTOTLWLD}. Black curve is PopNorm(y) (eq's. \ref{eq:PopProj1} \& \ref{eq:PopProj2}). Blue diamonds, composite of a UN population projection to 2300 CE \cite{UN2004WorldPopTo2300}, and FRED data 1960-2021\cite{FRED2021SPPOPTOTLWLD}. Blue curve is PopHigh(y) (eq. \ref{eq:PopProj2}). }
\label{Fig_Supp_UNPopGrowth}
\end{figure} 
The population datasets and projections are a composite of a high UN population projection to 2300 CE \cite{UN2004WorldPopTo2300}, a normative UN projection to 2100 CE \cite{UNDESApopDiv2019}, Gapminder data \cite{Gapminder2021}, Maddison data \cite{MaddisonDatabase2010} and FRED data \cite{FRED2021SPPOPTOTLWLD}. These are fitted with equations for use in the model and set to stable values for long term projections past their respective limits as shown in equations \ref{eq:PopProj1} and \ref{eq:PopProj2} and figure \ref{Fig_Supp_UNPopGrowth}. 

The long term population values are 10.9 billion \cite{UNDESApopDiv2019} and 37.2 billion \cite{UN2004WorldPopTo2300} people, for the normal and high projections, respectively, after 2100 and 2300. We do not assume population decline for either population projection. Particularly the high  projection of 37.2 billion people seems implausible.  Most ecologists, climate scientists and some schools of economics would not find it likely. Sustainability of industrial society depends on energy and resources to support the  GDP per-capita to create society, with capacity to use energy being primary. To operate a society at such such population levels on this planet would require what are now science fiction types of adaptation, possibly moving population and/or manufacturing to off-planet habitats. However, this 37.2 billion population projection was made and accepted by the UN. 

Accepting modelling of the 37.2 billion population scenario is a bit problematic, because on the one hand, OPTiMEM tries to create grounded projections, thus we avoid what appear to be fantasy declines in carbon emissions that have been popular over the past 40 years. Many would reject the possibility of a 37.2 billion population as fantasy. On the other hand, while there are indications that growth  may be slowing, most motivations visible in history are obviously favouring growth. Further, the field of economics assumes exponential growth as a matter of course, and has been correct overall for quite a long time. World population has passed population levels previously declared impossible. 

\begin{align}
    \begin{split} 
\label{eq:PopProj1}
\text{Population equations are broken into} & \text{ 3 segments: Year 1 to 1699, 1700 to 1949, and 1950 forward (eq. \ref{eq:PopProj2}). }\\
 Pop1to1699(y) &=  0.000245y + 0.118240047961 \quad \text{| y = 1 ..1699 |} \\
 Pop1700to1950(y) &= 0.000011 \cdot e^{0.00634999} \qquad \text{| y = 1700 ..1949 |}\\
\text{Where: } y  &=  \text{ year CE}
\end{split}
\end{align}
\begin{align}
    \begin{split} 
\label{eq:PopProj2}
 \text{Population proje} & \text{ction to 10.9 billion in 2107.  } \\ \text{probabilistic UN} & \text{ Scenario fit.} \\
 PopNorm(y) =& \, 2.64969450476903\text{E-8} \cdot y^4 \\ 
- &\, 2.17800322941767\text{E-4} \cdot y^3 \\ 
+ &\, 0.670800158391551 \cdot y^2 \\
 - \, 917.412621331478 \cdot y &\,
  + \, 470082.225075675 \\ 
\text{Max} &= 10.9 \text{ billion} \\
 \text{| y }&= \text{ 1950 ..2107 |} \\
\end{split}
 \begin{split}
 \text{Population proje} & \text{ction to 37.2 billion in 2330 } \\ \text{probabilistic U} & \text{N Scenario fit.} \\
PopHigh(y) &=  9.12508314 \text{E-14} \cdot y^6  \\
- \, 1.2207291841 \text{E-9} \cdot y^5 &\,
+ \, 0.6785342706\text{E-5} \cdot y^4 \\ 
- \, 0.0200612724127 \cdot y^3 &\,
+ \, 33.277680978726 \cdot y^2 \\ 
- \, 29368.1480547 \cdot y  &\,
+ \, 10773554.975 \\
\text{Max} &= 37.2 \text{ billion} \\
 \text{| y }&= \text{ 1950 ..2330 |} \\
\end{split}
\end{align}
\qquad \qquad \qquad \qquad \qquad \qquad \qquad \qquad \qquad \text{Where: } y  =  \text{ year CE}  

We break the population N$_2$O scenario into two components, agriculture and industry. For agriculture to sustain a global population of 37.2 billion people, based on 7.88 billion people alive in 2021, it would be necessary to raise 2021 agricultural productivity by roughly 4.7 times. This is plausibly possible using three methods: Genetically modified organism (GMO) crops and animals, some amount of factory (greenhouse) farming, and robotic farming. Of these three basic categories, factory farming has gotten significant investment and is supplying crops to our tables today. 

The first author worked on robotic farming for a number of years. The promise of robotic farming is based on the observation that human labour high intensity gardening has productivity in the range of 3 to 20 times that of our current mechanised agriculture. This is due to lower impact on soil, and practical intercropping/multi-cropping. 

GMO crops are controversial, however, gene editing can make the promise of better production and targeting of traits easier and more practical \cite{Labant2020RedesignFoodSupply} . On the scale of 3 centuries that the UN high scenario \cite{UN2004WorldPopTo2300} has to reach its peak of 37.2 billion people, such techniques may be viable. The first author's experience indicates that lack of progress is due to zealous regulation that functionally restricts most commercial applications to things like herbicide resistance that increase sales for a secondary product like an herbicide. Novel plant diseases are quite practical to prevent, which will prove necessary in a future that is warmer \cite{Juroszek202030yrReviewClimateCrop}\cite{Parker2020FoodSecurityClimate}. Similarly, food animals have potential for greater efficiency, however, regulation not based on science has made such animals rare \cite{Murray2016GeneEngLivestock}. A future society with a high population will have little choice. 

Total use of energy today would require roughly double the current 27,295 TwH of electricity produced in 2021 \cite{EIA2021IntNatElectWorld}. This 27,295 TwH was produced with 8004.8 gigawatts of capacity \cite{GlobalEconomy2021lectCapacity}. Assuming 3 times current energy use should be a safe estimate for a future society that might need to rely in part on a synthetic fuel cycle, and applying the 4.7 expansion factor discussed above gives:   $3\cdot 4.7 \cdot 27,295 = 384,860 $ TwH per year; using $3\cdot 4.7 \cdot 8004.8 $ Gw$ = 112.9$ terawatts of capacity. 

Thus, while we cannot suggest that a 37.2 billion population is desirable, we do not think that it can be ruled out. Plausibly supporting it does not require resort to inventions that do not yet exist, just sensible use of inventions we have. Thus, this high population scenario represents the upper limit for OPTiMEM's current scenarios.

\subsubsection{Synfuel notes}

Synfuel is defined here as synthetic production of hydrocarbons from CO$_2$ and water or hydrogen, without use of any fossil fuel. 
Such a closed-cycle 100\%  synthetic fuel system based on recycling CO$_2$ could provide high energy density fuel primarily for transport and heavy industrial processes such as mining. Such a synthetic fuel system is very energy intensive, requiring on the order of 2.5-3 times the energy of direct use of electricity.  However, some applications are quite difficult to electrify, including most diesel heavy machinery and aircraft, and possibly ocean shipping. Synfuel can also be co-generated from waste heat of industrial processes \cite{Nguyen2015SynfuelCurrentStatus,Li2016GHGSynfuel,OZCAN2021SynfuelThermoEconAnalaysisWasteHeat}. 

NOx are not considered GHGs because they have a short chemical lifetime. NOx is the sum of NO and NO2. For a synfuel system to work, NOx abatement may be required, however, a relationship between NOx and N$_2$O is not shown. NOx is directly produced by combustion, and and Lammel comments that NOx is not fully understood \cite{Lammel1995GreenhouseOFNoX}. Therefore use of synfuel may underestimate the effect  on warming. 

There are good NOx abatement catalytic systems \cite{Ali2021NOx_reduction}, as long as they are used.

\subsubsection{Gross World Product (GWP) and projections for N$_2$O}
GWP is projected by obtaining GWP (eq. \ref{eq:GWPProj1}) based on population for the year (eq. \ref{eq:PopProj1} \& \ref{eq:PopProj2} \& fig. \ref{Fig_Supp_UNPopGrowth}). For OPTiMEM, after population stabilises, GWP linear growth with a slope of 1.5\% is assumed. This is not intended to be predictive of population or GWP, but to bound a scenario range.

\begin{align}
    \begin{split} 
\label{eq:GWPProj1}
 \text{GWP proje} & \text{ction for 10.9 billion in 2100  } \\
 GWP2100(p, y) =& 16.64p - 40.7232 \\
 &+ 0.015\cdot (y-2100)_+ \\
\text{Max} =& 10.9 \text{ billion} \\
\end{split}
 \begin{split}
 \text{GWP proje} & \text{ction for 37.2 billion in 2300  } \\
 GWP2300(p, y) =& 16.64p - 40.7232 \\
 &+ 0.015\cdot (y-2300)_+ \\
\text{Max} =& 37.2 \text{ billion} \\
\end{split}
\end{align}
$\qquad \qquad \qquad  \text{Where: } y  =  \text{ year CE   } \qquad p = \text{population} \qquad ()_+ \text{ is Macaulay notation for negative results = 0}$

\subsubsection{N$_2$O projection fit}
\begin{figure}[hbt!]
\centering\includegraphics[width=5.25in]{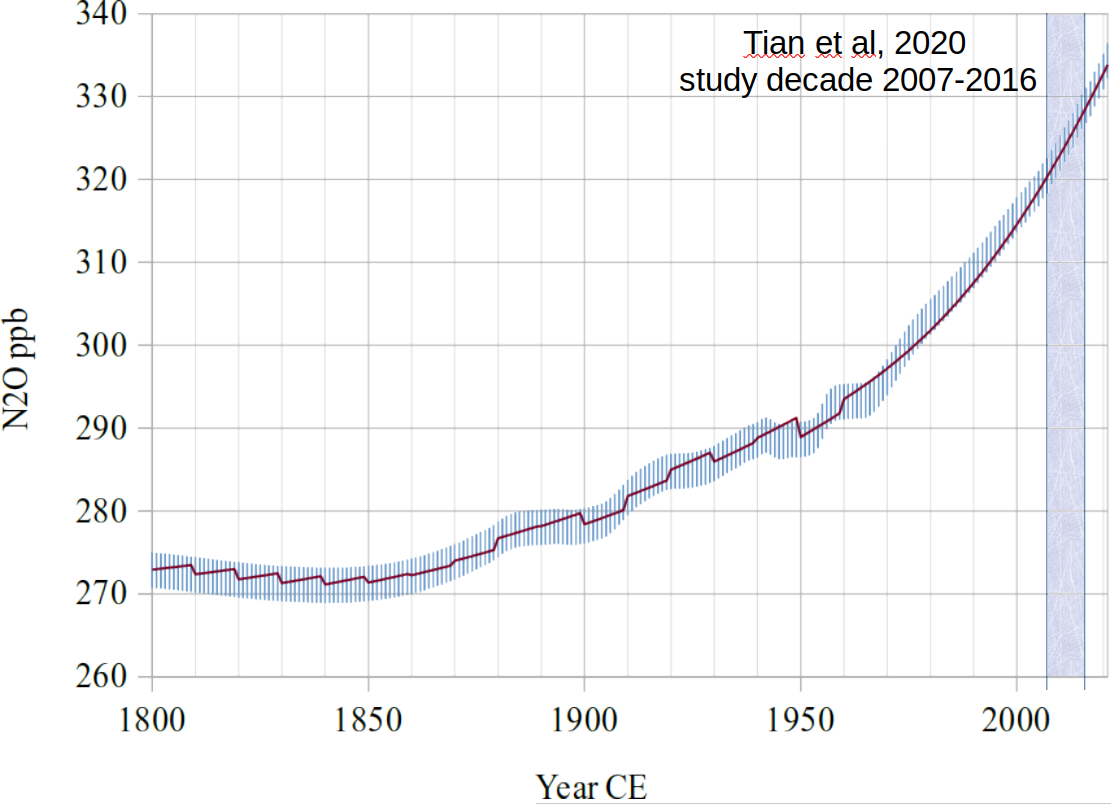}
\caption{\textbf{N$_2$O total.} Black curve is projection from equations \ref{eq:PopProj1} \& \ref{eq:PopProj2}. Blue bars are N$_2$O calibration dataset \cite{NOAA2022AGGIClimateTable}. Sawtooth pattern prior to 1960 is from resetting the curve against the calibration dataset from 1800-1960. This is not meaningful for projection going forward. The Tian reference used to build this model covers the 2007-2016 region, indicated by shaded vertical band above 2000 CE. }
\label{Fig_Supp_N2O_calibration_Tian}
\end{figure} 

This model does not have a chemistry/physics based model of N$_2$O. These equations are correlated fits. To create a chemistry model would require modelling of all the sources and sinks identified in Tian, et al \cite{Tian2020N2OSources_sinks}, shown in figure \ref{Fig_Supp_N2OBudget}, as well as Anderson's work \cite{Anderson2017ChlorineBromineCatalysis, Anderson2018CouplingCatalysis}. The task of creating a complete N$_2$O chemistry behaviour model is suited for large climate models. 

Examining figure \ref{Fig_Supp_N2O_calibration_Tian}, N$_2$O data smooths to the projection after 1960-1970. While imperfect, this projection is the best available, and when population does level out, it will account for possible industrial growth---assuming the level of abatement during Tian's study \cite{Tian2020N2OSources_sinks}. The result of running the system of equations of this model is that the N$_2$O level enters a linear growth phase after 2107 for the normal population scenario and after 2330 for the high population scenario. (Not shown.) The reason for this is seen  in the  GWP projection above, as N$_2$O is correlated to population. So, in this model, when the population stabilises, then N$_2$O growth also flattens. The significance of this is that, that without any remediation of N$2$O emissions for agriculture and for industry, a severe problem should occur---assuming civilisation does not crash.  

\begin{figure}[!ht]
\centering\includegraphics[width=5.5in]{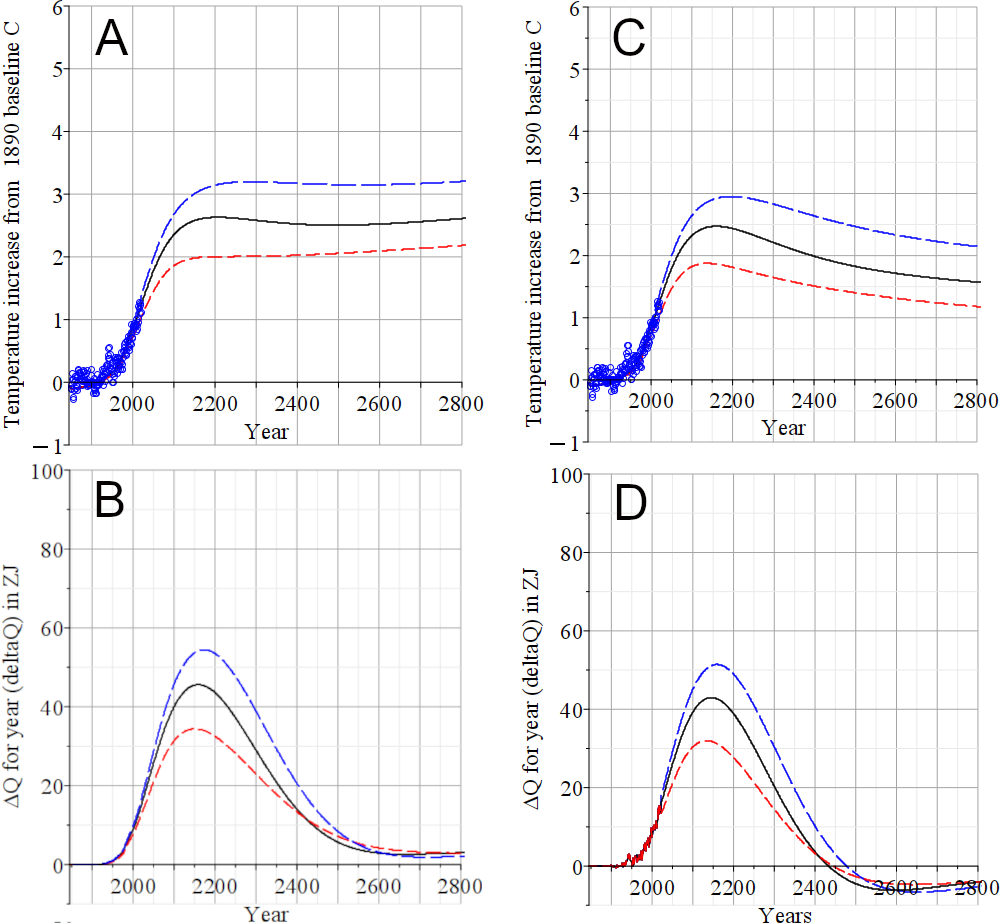}
\caption{\textbf{Effect of nitrogen modelling without Tian 2020's N$_2$O. } Compare this figure's panel D to the D panel of figures \ref{Fig_Supp_Baseline_PopNorm_19.5}-\ref{Fig_Supp_PermaFrost_PopNorm_S-aerosol_19.5}. Here we compare the $\Delta$T $\degree$C, and $\Delta$Q (dQ) in Zettajoules (ZJ), between Tian 2020 N$_2$O (A \& B) vs previous model N$_2$O (C \& D). Both scenarios are Baseline\_PopNorm, no permafrost thaw, no S-aerosol correction. Blue circles, historical temperature anomaly records. Upper blue long dash and lowest red dash curves are highest and lowest remainder CO$_2$ per eqs. \ref{eqs:NetCO2Remainders} and fig. \ref{Fig-RemainderFractions}. Black curve is the result of the Archer density plot harmonic mean, that we call "central value". Note dQ curve 'D' has several centuries of high rate of heat loss from the ocean, which should make superstorms plausible \cite{Hansen2016IceMeltSeaLevelRiseAndSuperstorms}. Such a dQ curve will happen if necessary abatement of N$_2$O and CO$_2$ occurs.}
\label{Fig_Supp_Baseline_PopNorm_n2O-comparison_T}
\end{figure} 

\subsubsection{Serious concerns over nitrous oxide long-term impact on climate}
The impact of this nitrogen model based on Tian' is profound as we see if we compare figure \ref{Fig_Supp_Baseline_PopNorm_n2O-comparison_T}D to the D panel of figures \ref{Fig_Supp_Baseline_PopNorm_19.5}-\ref{Fig_Supp_PermaFrost_PopNorm_S-aerosol_19.5}. After updating our nitrogen model with Tian \cite{Tian2020N2OSources_sinks}, no scenario enters a net yearly OHC release period in the next 800 years. Outside of nitrogen cycling processes, nitrous oxide (N$_2$O) has a long $e$fold time (\S\ref{N2O_efold}). The implication of this profound long-term effect is that N$_2$O should become a priority to abate. Industrial abatement is understood, and should make a significant difference. Likewise, some degree of abatement of agricultural N$_2$O is practicable \cite[p. 255]{Tian2020N2OSources_sinks}. 

We note that Tian, et al say that \emph{"growth in N$_2$O emissions exceeds some of the highest projected emission scenarios"} \cite{Tian2020N2OSources_sinks}. Additionally, nitrogen dynamics are still not fully understood \cite{Chamba2023NitrateBasedNighttimeAtmosNucleation}, so OPTiMEM's estimate may be low.  

N$_2$O abatement is straightforward for industrial installations. Our global industrial society will find N$_2$O very problematic over time. OPTiMEM shows N$_2$O can replace CO$_2$ as a primary greenhouse gas circa 2460 CE (fig. \ref{Fig_T_by_gas_contribution}), and prevent global temperature from dropping.

\subsection{Carbon model Scenarios to model uncertainty (CO$_2$ and CH$_4$)}
\label{lb:carbon_model_scenarios}
\begin{figure}[hbt!]
\centering\includegraphics[width=5.25in]{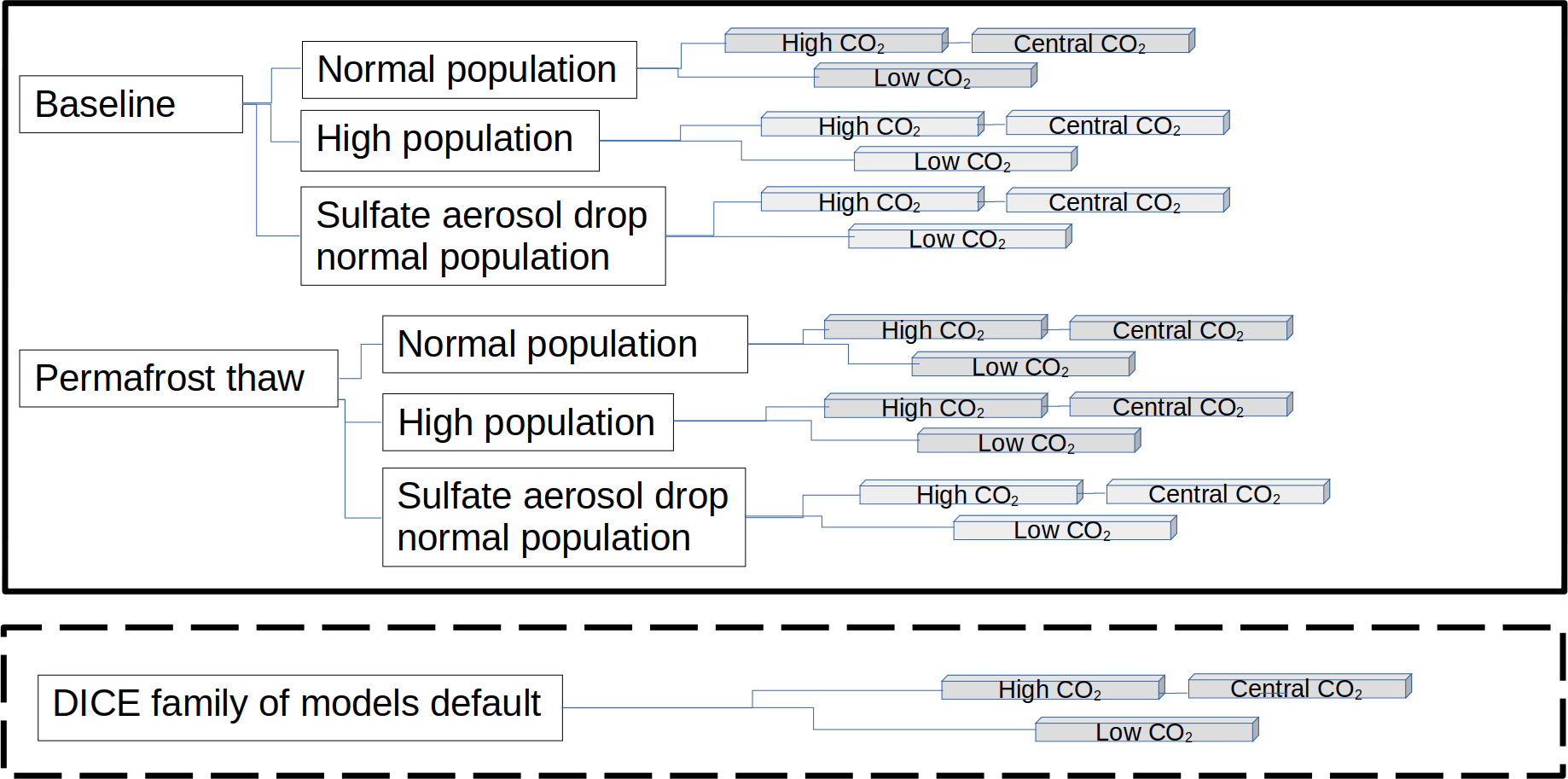}
\caption{\textbf{Scenarios tree.} Upper solid box shows the 18 primary scenarios for this study. These use the normal population curve (Fig. \ref{Fig_Supp_UNPopGrowth}). Lower dashed box shows the 3 DICE family of models scenarios for the DICE implicit assumptions test. }
\label{Fig_Supp_Scenario_tree}
\end{figure} 
We developed two primary scenarios, baseline and permafrost thaw, with three warming variants for each, normal population \cite{UN2022PopGrowth2100}, high  population \cite{UN2004WorldPopTo2300}, and sulfate aerosol drop   \cite{Manshausen2022Invisibleshiptracks,Diamond2023CloudSulfurRegulation,Hansen2023WarmingInPipeline,Gettelman2024HasReducingShipEmissionsBroughtForwardGlobalWarming} with normal population. Opposing views on sulfate aerosols attribute the heating to El Ni\~{n}o \cite{Raghuraman2024WarmingSpikeElNinoSouthernOscillation}, or statistical variation \cite{Watson-Parris2024SO2ConfoundedByInternalVariability,Samset2024_2023TempsReflectSSTVariability}. However, the temperatures were unique \cite{Cattiaux2024HowExtremeWereDailyGlobalTemperaturesin2023andEarly2024}. As these scenarios are intended to account for uncertainty, we retain the sulfate aerosol drop scenario. For each of these scenarios, there are 3 variants, high, central and low CO$_2$ (Fig. \ref{Fig_Supp_Scenario_tree}). Note that the sulfate aerosol drop conditions (S-aerosol) should prevail in a low carbon consumption world, however, here it is assumed that a 37.2 billion population will have performed geoengineering to abate loss of S-aerosol, even if this is just adding sulfur into some synfuels. We did not create high population s-aerosol scenarios.

These scenarios are driven using carbon consumption and methane release equations, with contributions from N$_2$O and fluorinated gases (F-gases). F-gases are also known as other trace gases (OTGs), and are visible in our results as a small, sharp, rise in 1970. Those gases feed into the Earth Energy Imbalance (EEI). The EEI is implemented by the energy forcing equations from Byrne, et al, 2014 \cite[p. 157, Table 2]{Byrne2014RadiativeForcingEquations}. For each of the variants, the three CO$_2$ remainder equations and other GHGs were applied using these EEI equations, for a total of 16 outcomes (fig. \ref{Fig_Supp_Scenario_tree}). EEI is used "as if" it were a heat body in contact with the ocean surface having a temperature (EEI$_T$). Then EEI$_T$ drives heat content via the heat equations.

\subsubsection{Baseline carbon combustion CO$_2$ emissions }
\begin{figure}[hbt!]
\centering\includegraphics[width=4.5in]{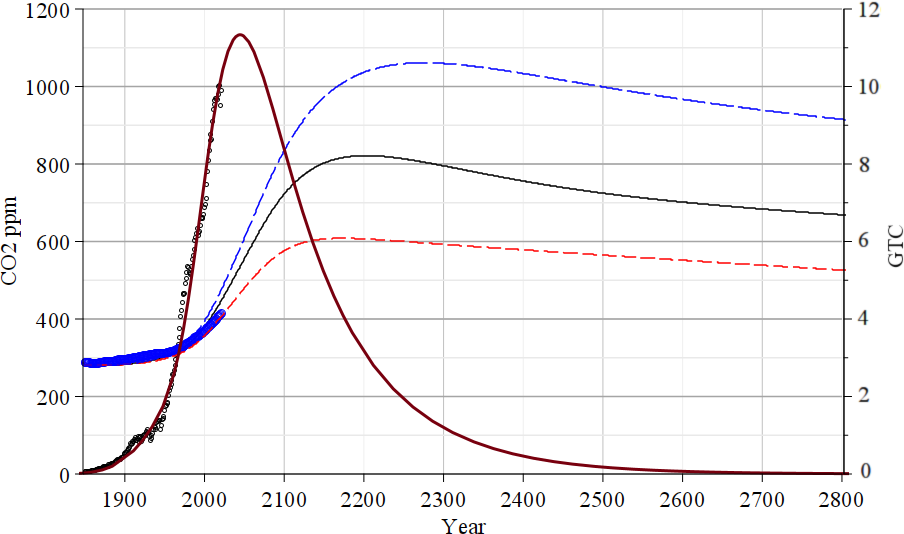}
\caption{\textbf{Results of carbon model for Baseline } starting 1890. Note that the carbon curve gigatonnes of carbon (GtC) is invariant between scenarios, with the exception of the DICE implicit assumptions scenario (fig. \ref{DICE_Model_scenario}).  Solid black (right scale) is GtC from equation \ref{eq:1500GtC} curve, Base GtC+Modern GtC. Black circles are GtC consumption historical data.  Red dash and blue long dash curves are low and high boundaries of CO$_2$ ppm (left scale) based on Archer eq's. \ref{eq:NetCO2RemainderL} \& \ref{eq:NetCO2RemainderH}. Solid black is central value of CO$_2$ (eq. \ref{eq:NetCO2RemainderC}).}
\label{Fig_Supp_C&CO2_BaselineScenario1}
\end{figure}

To construct our baseline carbon release scenario, we settled on using Tans \cite{Tans2009_OceanicandAtmosCO2} and Malanichev \cite{Malanichev2018LimitsShaleOil} as primary guidance. The primary logic here is the conjecture that the rise and fall of the aggregate system of carbon extraction should have a similar curve to that of the individual wells. Thus, our carbon consumption curves (eq. \ref{eq:1500GtC}) are based on the Richards equations \cite{Richards1959FlexibleGrowthFunction} versions published later by Malanichev \cite{Malanichev2018LimitsShaleOil}. This included examination of limits to known reserves, plus significant reserve expansion based on history. In our carbon model, 1500 GtC reserves expanded by 647 GtC to a total of 2147 GtC. To fit historical data required adding two such functions together, with the second equation offset by 95 years. The result was a beautiful fit (fig. \ref{Fig_Supp_C&CO2_BaselineScenario1}). 

\begin{align}
    \begin{aligned} \label{eq:1500GtC}
 \text{OPTiMEM Carbon Consumption equation}   \\ 
 GtC(x, k, g, G_c, E_g, t, s)  &= (t-1849) \cdot \frac{(x \cdot k \cdot G_c \cdot {E_g}^t)\cdot e^{(-k \cdot g \cdot (t-s))}}{(e^{(-k \cdot g \cdot (t-s))}+1)^{\frac{1}{g} +1}} \\
\text{ Called with two sets} & \text{ of parameters: } \\
\text{Parameters (Qdt1000) Base GtC: } x&= 0.08, k=1.031, g=0.010, G_c=1000, E_g=1.0008, t= year, s=0 \qquad \\
\text{Parameters (Qdt500) Modern GtC: } x&= 0.1, k=1.81, g=0.01250, G_c=500, E_g=1.00275, t= year, s=95\\
\text{Base GtC + Modern GtC }&= \text{ see carbon curve fig. \ref{Fig_Supp_C&CO2_BaselineScenario1}. }\\
\text{Where: } t &= \text{ year}  \qquad 
  x, \, k, \, \& \, g  = \text{fit coefficients} \qquad 
  s = \text{ shift, years} \\
 G_c &= \text{ Gigatonnes of carbon } \qquad
 E_g = \text{ Growth rate of proven reserves }  \\
    \end{aligned}   
\end{align}

Eq. \ref{eq:1500GtC} drives the carbon model baseline scenario (fig. \ref{Fig_Supp_C&CO2_BaselineScenario1}). This baseline scenario is our current most probable without any tipping point effects except what may be baked into the NOAA data. We assume that carbon emissions will drop due to reaching peak carbon. For this computation, we assume growth of proven reserves, beginning with an assumed total 1500 GtC \cite{Tans2009_OceanicandAtmosCO2} possible through 2800 CE. Growth of carbon fuel reserves was added during the curve fit phase due to inability to obtain a good fit using the most aggressive total of 1500 GtC as a limit. This baseline curve's integral to 2700 CE reaches 2147 GtC as the integral limit, and hits peak carbon extraction in 2042 CE. 

We note, as a caveat, that Hubbert's first 1956 predictions were 10-15 years in the future for the lower 48 states, and for the globe approximately 50 years \cite{Hubbert1956NuclearAndFossilFuels}. In recent decades peak extraction has appeared to be $\approx$ 10 to 15 years in the future.  Extraction improvement carbon is currently being pumped in the USA and a few other nations (fracking) and it can potentially be applied all over the world. This leads one to scepticism regarding true peak carbon. Countering scepticism of peak carbon are some indications of carbon fuel Energy Return On energy Invested (EROEI) dropping due to higher energy cost of extraction \cite{Tverberg2021HubbertLimit,Tverberg2021FossilFuelProblem}, and the fact that well over one third of all emissions since pre-industrial times have taken place since the year 2000. Carbon reserves are under great pressure.

\subsubsection{Ocean \& biosphere CO$_2$ absorption remainder model }
\begin{figure}[hbt!]
\centering\includegraphics[width=6.25in]{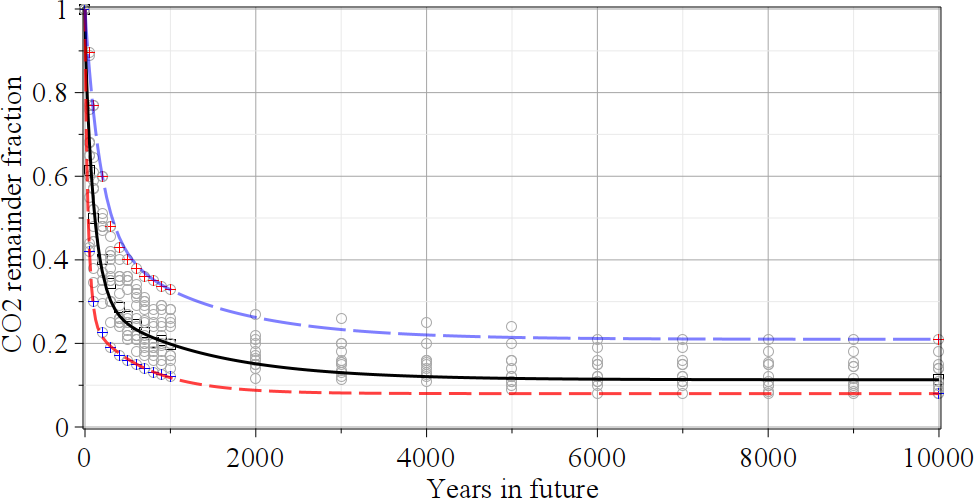}
\caption{\textbf{Remaining CO$_2$ by year fitted to Archer'} \cite[Fig. 1b]{Archer2009AtmosphericLifetimeOfCO2}. Bottom dashed red curve is fitted to lowest remainder points, C$_{RL}$ (eq. \ref{eq:NetCO2RemainderL}). Central solid black curve is harmonic mean of all data C$_{RC}$  (eq. \ref{eq:NetCO2RemainderC}).   Upper long-dash blue curve is fitted to points from highest Archer results C$_{RH}$ (eq.  \ref{eq:NetCO2RemainderH}). Grey circles are density plot of Archer data. Red, green and blue crosses are calibration points calculated for curve fits. These three CO$_2$ remainder equations provide the fraction for future years equivalent to the $e$fold computations. } 
\label{Fig-RemainderFractions}
\end{figure}

A model created from Archer, et al's 1000 petagram (1000 gigatonne) results of eight climate models \cite{Archer2009AtmosphericLifetimeOfCO2} was used to estimate the range of CO$_2$ absorption (primarily oceanic). ( fig. \ref{Fig-RemainderFractions}). Three equations emerged describing CO$_2$ remainders:  A. C$_{RL}$ low (eq. \ref{eq:NetCO2RemainderL}), B. C$_{RC}$ central fit to harmonic mean (eq. \ref{eq:NetCO2RemainderC}) , C. C$_{RH}$ high (eq.  \ref{eq:NetCO2RemainderH}). Harmonic mean was used because these are ratio values. 
These three CO$_2$ remainder equations provide the low, central and high remainder fractions shown in figure \ref{Fig-RemainderFractions}.  

\label{eqs:NetCO2Remainders}

 We note that resolution is a known limitation of current climate models. Current models' cell size is 50-100 km, and 1 km cells are expected in $\approx$ 2025. Insufficient granularity is known to result in poor modelling (viz. factor of 2 underestimation of rain \cite{Li2022RaindropPhysics2XprecipitationIncrease}). The 2022 high resolution proto-earth collision model illustrates the effect of higher resolution in another area of science. This proto-earth collision provides a first model of Moon formation \cite{Kegerreis_2022_ImmedOriginMoon}. Low resolution models do not even hint at what this high resolution model shows.

\begin{flalign}
\begin{aligned}
\label{eq:NetCO2RemainderL}
\mathbf{A.) \, Low \, remainder\, CO_2 \qquad} \\
C_{RL}(y) =& 
\frac{(\frac{4}{5{\mathrm e}}+\frac{1}{{\mathrm e}^{ 0.02285 y}})}
     {(\frac{4}{5{\mathrm e}}+1)} \times e^{-0.00155 y + (\frac{y}{(y + 10)} \times ln(1 - 0.2))} \\
     & +( \frac{y}{(y + 10)} \times 0.08)
\end{aligned}
\end{flalign}

\begin{flalign}
\begin{aligned}
\label{eq:NetCO2RemainderC}
\mathbf{B.) \, Central \, remainder\, CO_2  \quad} \\
C_{RC}(y) =& \frac{(\frac{6}{{\mathrm 7e}}+\frac{1}{{\mathrm e}^{ 0.008215 y}})}{(\frac{6}{{\mathrm 7e}}+1)} \times e^{-0.0008 y + (\frac{y}{(y + 10)} \times ln(1 - 0.2))} \\
&+ (\frac{y}{(y + 10)} \times 0.113) \\
 \end{aligned}
\end{flalign}

\begin{flalign}
\begin{aligned}
\label{eq:NetCO2RemainderH}
\mathbf{C.) \, High \, remainder \, CO_2 \quad} \\
C_{RH}(y) =& \frac{(\frac{4}{3{\mathrm e}}+\frac{1}{{\mathrm e}^{ 0.004815 y}})}{(\frac{4}{3{\mathrm e}}+1)} \times e^{-0.0008 y +(\frac{y}{(y+10)} \times ln(1-0.2))} \\
& +(\frac{y}{(y+10)} \times 0.21) \\
\text{Where: } y  =&  \text{year  from  starting  year  }\\
\end{aligned}
\end{flalign}

\subsubsection{Energy forcing to temperature: EEI $\rightarrow$ EEI$_T$}
\begin{figure}[hbt!]
\centering\includegraphics[width=4.5in]{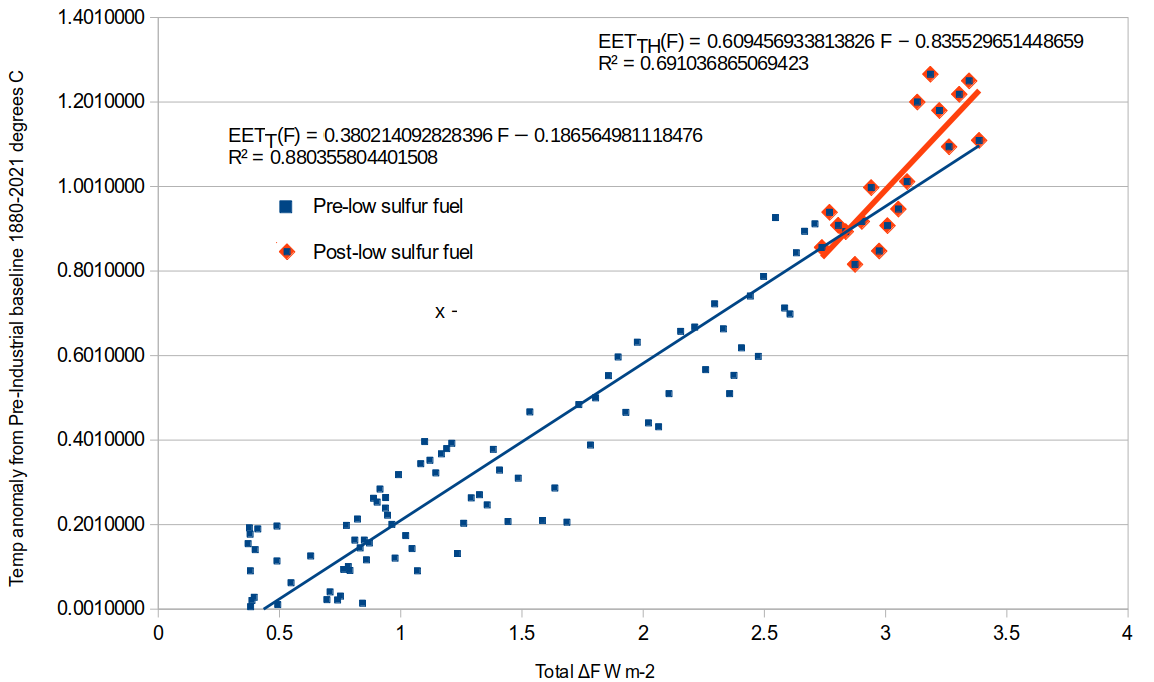}
\caption{\textbf{Fitted equations for EEI to EEI$_T$} Time period 1880 CE to 2021 CE. Based on Hansen 2023 \cite[p. 33]{Hansen2023WarmingInPipeline} supported by Manshausen, et al, and Diamond, et al \cite{Manshausen2022Invisibleshiptracks,Diamond2023CloudSulfurRegulation} }
\label{Fig_Supp_dF_to_T_sans-WW2}
\end{figure}

The core of climate modelling is that GHG concentrations drives energy absorption across the boundary with space, the Earth Energy Imbalance (EEI) (fig. \ref{Fig_Climate_Drivers_diagram}). For OPTiMEM's EEI we use the full set of equations of Byrne et al.\cite[p. 157, Table 2]{Byrne2014RadiativeForcingEquations}, including adjustments for spectrum overlap. The resulting energy forcing is then converted to temperature by our own fitted equations (fig. \ref{Fig_Supp_dF_to_T_sans-WW2}, eq. \ref{eq:T0_F}). These forcing to temperature linear equations are differentiated by condition-change regarding sulfur emissions from shipping \cite{Manshausen2022Invisibleshiptracks,Diamond2023CloudSulfurRegulation,Hansen2023WarmingInPipeline}\footnote{Opposing views on sulfate aerosols attribute the heating to El Ni\~{n}o \cite{Raghuraman2024WarmingSpikeElNinoSouthernOscillation}, or statistical variation \cite{Watson-Parris2024SO2ConfoundedByInternalVariability,Samset2024_2023TempsReflectSSTVariability}. As these scenarios are intended to account for uncertainty, we retain the sulfate aerosol drop scenario.}. 

\begin{flalign}
  \begin{aligned}  
\label{eq:T0_F}
 \text{ Pre-low sulfur fuel o} & \text{il} \qquad \qquad \text{ See fig. \ref{Fig_Supp_dF_to_T_sans-WW2}}\\
EEI_T(F) =&  -0.18656496801124243078 + 0.38021409032645513676 F  \\
 \text{Post-low sulfur fuel } & \text{oil}   \\ 
 EEI_{TH}(F) =& -0.835529651448654 + 0.609456933813825 F   \\
\end{aligned}
\end{flalign}
\qquad \qquad \qquad \qquad \qquad \qquad \qquad \qquad \qquad \text{Where: } \textit{F}  =  \text{ Forcing flux in W} m$^{-2}$

\subsubsection{CH$_4$ emissions estimation}
\label{lb:CH4_EmissionsEstimation}
Anthropogenic emissions of CH$_4$ are from multiple sources as shown in table \ref{tab:IEAMtCH4}, with a total of 62.12\% of CH$_4$ assigned as population driven. The same population equations are used for CH$_4$ population linked estimates as for N$_2$O population linked estimates. See eq's. \ref{eq:GWPProj1}. 

\begin{table}[!hbt]
    \centering
    \caption{IEA sourced reference estimates 2021}
    \begin{tabular}{llll}
\textbf{Category} & \textbf{Mt CH$_4$} & \textbf{\% of total} & \textbf{Assigned relationship} \\
Agriculture  & 141.4           & 39.62\%              & Population CH$_4$         \\
Energy       & 135.2           & 37.88\%              & CO$_2$                 \\
Waste        & 73              & 20.45\%              & Population CH$_4$        \\
{ Other}  & { 7.3}       & { 2.05\%}         & Population CH$_4$ 
    \end{tabular}
    \label{tab:IEAMtCH4}
\end{table}

Using 2021 as the reference point for CH$_4$, two fitted equations with high R$^2$ values were developed relative to carbon (GtC) for the energy category (37.88\%) and Agriculture-Waste-Other (62.12\%). Both equations are power law equations, which means they have lower than linear long-term rate curves. 

\begin{align}
    \begin{split} 
\label{eq:CH4fromCandPop}
 \text{Carbon extraction} & \text{ linked CH$_4$   R$^2$=0.933} \qquad \qquad \text{ }\\
EnergyCH4ppb(C) =&\,  369.2592 \cdot C^{0.29056}  \\
\end{split}
 \begin{split}
 \text{ Population linked} & \text{ CH$_4$   R$^2$=0.952 } \\ 
 AgCH4ppb(p) =& 530.3234 \cdot p^{0.3948}  \\
\end{split}
\end{align}
\qquad \qquad \qquad \qquad \text{Where: } C  =  \text{ gigatonnes of carbon extracted (GtC)}  \qquad p = population (See: eq's. \ref{eq:GWPProj1})

\subsubsection{Permafrost thaw carbon release, CO$_2$ and CH$_4$ }
\label{PermafrostCH4}
Climate economics has cited Shakhova, et al's 2010 proposed release of 50 GtCH$_4$ by 2100 \cite{Shakhova2010MethanEastSiberianShelf}, but this has been superseded by Schuur, et al's 2022 review \cite{Schuur2022PermafrostFeedbacksWarming}. We did not attempt to model outlier permafrost/methane scenarios from Schuur. We use the RCP4.5-RCP8.5 IPCC related figures found in Schuur, et al's table 1. The low scenario is incorporated into the current baseline. The high figures are the permafrost thaw high end scenario. 

Schuur's review treats outlier scenarios, and the current 17 craters in east Siberia certainly appeared abruptly, but they are expected to emit for many years to come, and individually, they represent relatively small increments for each crater \cite{Zolkos2021CH3Craters,Bogoyavlensky2021PermanentCH3emission}.

We can only report that Antarctic methane is a concern of unknown size \cite{Seabrook2025AntarcticSeepEmergenceAndDiscoveryInTheShallowCoastalEnvironment}. 

In OPTiMEM, permafrost carbon inventory is estimated at 1500 Pg. (1 petagram = 1 gigatonne). Of this, maximum yearly carbon release rate of CO$_2$ is 2.97 Pg, and for CH$_4$ is 0.100 Pg. This corresponds to peak emission of 10.88 Pg of CO$_2$ and 0.134 Pg of CH$_4$. Each year's release decrements the total permafrost carbon. 

\begin{figure}[!ht]
\centering\includegraphics[width=6.25in]{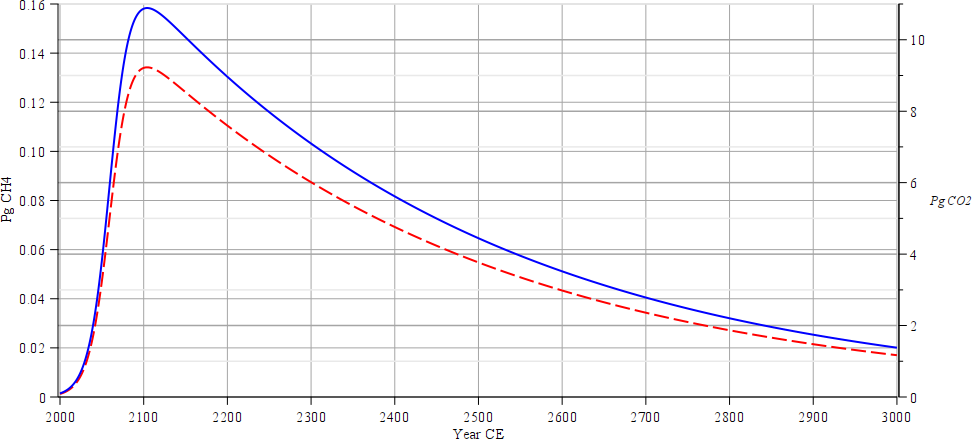}
\caption{\textbf{Petagrams (identical to GtC) of carbon emitted from thawing permafrost by year.} Red dashed curve (left scale) is CH$_4$ release. Blue solid curve (right scale) CO$_2$ emissions. CO$_2$e not shown. Produced from equations (\ref{eq:Schuur} built on Schuur et al. \cite{Schuur2022PermafrostFeedbacksWarming}. Note that this graph is informational, and could only be directly meaningful when expressed within the total model system. This curve does not include the Antarctic, but the Antarctic is a concern \cite{Seabrook2025AntarcticSeepEmergenceAndDiscoveryInTheShallowCoastalEnvironment}.}
\label{Fig_Supp_PermafrostCurves}
\end{figure} 

The permafrost thaw scenarios call eq. \ref{eq:Schuur} with parameters. The high curve delivers $\approx$134 GtC by 2100, primarily as CO$_2$, with roughly 4.4 GtC, $\approx$3.4\% of the total, as CH$_4$. The low scenario is incorporated into baseline model behaviour. The high scenario is the permafrost scenario. 

\begin{align}
    \begin{split} \label{eq:Schuur}
 \text{CO$_2$ \& CH$_4$ permafrost } & \text{emissions high \& low} \qquad \qquad \text{ See fig. \ref{Fig_Supp_PermafrostCurves}}\\
PermafrostC_Ldt&(t, yCO2, yCH4) \\
 yCO_2 =& \, \frac{L_{CO_2}}{1+e^{-k(t-m)}} \cdot CO_2toC \qquad \qquad
 yCH_4 = \, \frac{L_{CH_4}}{1+e^{-k(t-m)}} \cdot CH_4toC \\
\text{Where: } L_{CO_2} =& \, C_{Inv} \cdot I_{CO2}Lim \qquad \qquad \qquad \quad \;\,
L_{CH_4} = \, C_{Inv} \cdot I_{CH_4}Lim \\
C_{Inv} =& \text{1500, decremented by each iteration's C, prior to multiplying by the applicable ratio.} \\
CO_2toC =& \frac{44.009}{12.011} \qquad \qquad \qquad
 \qquad \qquad \, CH_4toC = \, \frac{16.0422}{12.011} \\
t =& \, year \qquad \qquad \qquad \qquad \quad m = \, \text{midpoint = }2060 \\
\text{High release: } I_{CO2}Lim =& \, 0.002259527 \qquad \qquad \qquad \quad \; \;
I_{CH_4}Lim = \, 0.000076581 \\
\text{Low release: } I_{CO2}Lim =& \, 0.000307235 \qquad \qquad \qquad \quad \; \;
I_{CH_4}Lim = \, 0.000014406 \\
\end{split}
\end{align}

 The total CO$_2$ release for the high permafrost thaw scenario is $\approx$1290 Gt CO$_2$ over 980 years. The total methane release for the high scenario is $\approx$44 Gt CH$_4$ over 980 years which is mostly converted to CO$_2$. 

\subsubsection{DICE model scenario}
\label{DICE_Model_scenario}
The DICE family of models climate economics scenario of primarily fossil fuel based exponential growth to 2300 is taken from DICE, as DICE was recently used for a 2021 publication \cite{dietz2021tipping}. DICE assumes a temperature increase up to $+17.68 \degree$C \cite[p. 45]{dietz2021supplement}, implying CO$_2$ levels of 750 to as much as 1600 ppm \cite{Keery2018EoceneCO2Sensitivity} based on geological history. However, using GHG equations to achieve $17.68 \degree$C with CO$_2$ alone requires CO$_2$ in the 20,000 ppm range, and there is no evidence of CO$_2$ remotely close to this level during the PETM. Other gas(es) and/or climate feedback mechanisms were present in the PETM, which warrants great caution regarding overconfidence in climate models showing low temperature responses---including the current one. Methane release was likely a part of the PETM \cite{Frieling2016,Frieling2019,Inglis2020,Inglis2020_Supp}. However, unknown mechanisms may also be involved, perhaps even astronomical influence altering our planet's orbit by a passing star \cite{Kaib2024PassingStars}. Exactly how the PETM peak temperature of 13.5 $\pm 2.6\degree$C occurred is a question that global civilisation does not want to solve by experience---caution is advised.

\begin{figure}[!ht]
\centering\includegraphics[width=6.25in]{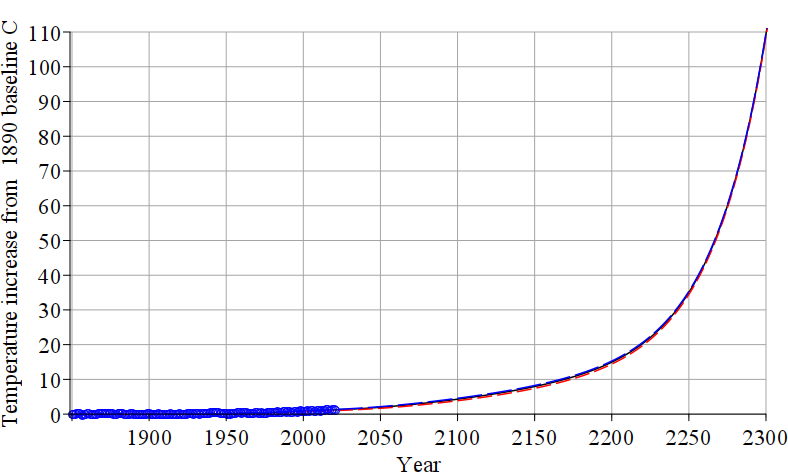}
\caption{\textbf{ Implicit DICE assumptions scenario results: temperature increase $\degree$C.} Blue circles, historical temperature anomaly records. This shows that by the end of the DICE term at 2300, it is quite impossible that civilisation, or most life, would survive. The mean global sea surface temperature (SST) in 2021 was $\approx$16$\degree$C. With current peak SST of 36$\degree$C, some equatorial seas should boil before 2275. In 2300, steam is a primary form of water. This scenario shows that the assumptions made founded on a carbon based economy are not possible.}
\label{Fig_Supp_DICE_Unhinged_Scenario_18.1}
\end{figure}

What the computations of this DICE model scenario show (fig. \ref{Fig_Supp_DICE_Unhinged_Scenario_18.1}) is that current economic models are completely disconnected from climate science. Since 1980, the 41, 31, 21, and 11 year growth rate of carbon extraction has been 1.50, 1.54, 1.83, and 1.19 percent respectively. If we presume a conservative consumption growth rate of 1.4\% per year to 2300, from 2021, this is a multiple of $\approx$50 on today's consumption of $\approx$ 10 GtC/yr, which yields 500 GtC that would need to be burned in 2300 alone---not a credible quantity. 

While figure \ref{Fig_Supp_DICE_Unhinged_Scenario_18.1} is a nice example of exponential growth, it is hard to conceive that a carbon based economy could grow as projected in the DICE/Fund family of IAMs absent an extraordinary carbon source. For instance, harvesting of hydrocarbons from Saturn's moon, Titan, could provide some of the required carbon \cite{Lorenz2008TitanOrganicInventory}. Such an implausible undertaking to wreck Earth's climate does not seem likely. And, the proposition that life as we know it could continue on a world with the seas and lakes boiling is preposterous. However, the shape of this curve does appear to be of the quadratic form used in DICE, just one that accelerates far faster. 

For this DICE climate economics scenario, we assume, as DICE does, continuous growth and maintaining the same primarily carbon energy mix. This DICE scenario attains a global mean temperature increase of 18$\degree$C circa 2205 CE, which temperature is past the limit defined by Nordhaus. However, the CO$_2$ equations did not break in this scenario, so we show the scenario out to the DICE year limit of 2300 CE of $\approx$110 $\degree$C. We do not consider this scenario to be credible, as fossil fuels should peak in the 21st Century, and the PETM was 13 $\pm 2.6 \degree $C \cite[p. 13289]{Burke2018Plio} warmer than our pre-industrial period. Consequently, we did not pursue it further.

\subsection{\emph{e}fold ($\tau$, tau) lifetime of greenhouse gases and half-life}

The equations used for modelling CH$_4$, N$_2$O, and Fgas are the best available descriptors of gas behaviour available from atmospheric science. As also mentioned in the Social Cost of GHG section (\S \ref{SC-GHG}) the true rate varies dependent on a number of factors, including concentration, availability of reactants such as hydroxyl, UV light, and atmospheric mixing by altitude, which changes with atmospheric chemistry and concentration of gases \cite{Orkin2020AtmosLifeHalogenHydrocarbons,Totterdill2016NF3andCFC115}. We do not have a set of climate model results we can cite for these gases as with CO$_2$. Thus, the use of $e^{-\frac{t}{\tau}}$ results should be reasonable except for possible departures due to new influences such as Anderson's rise in atmospheric halogens from the ocean \cite{Anderson2017ChlorineBromineCatalysis,Anderson2018CouplingCatalysis}, or the wild-card impact of volcanism which can cool or warm, and may inject water into the stratosphere with complex effects \cite{StauntonSykes2021HalogensRadiativeForcing,Millan2022HungaTongaHydrationStratosphere}. Put qualitatively, there is some guesswork involved in picking a single $e$fold $\tau$ value, as there is uncertainty that cannot be expressed in a single number.

\begin{figure}[!ht]
\centering\includegraphics[width=3.25in]{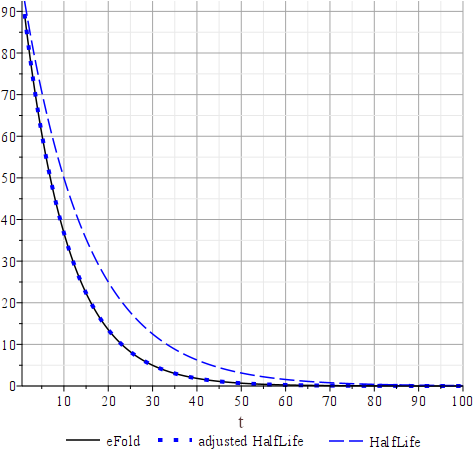}
\caption{\textbf{ $e$fold compared to half-life for eq. \ref{eq:eFold}.} Black curve is $e$fold [$\tau$ = 10]. Dashed blue curve is half-life [$t_{\frac{1}{2}}$ = 10]. Blue dots superimposed on black curve is adjusted half-life. [$t_{\frac{1}{2}}$ = $\tau$ $\cdot ln(2)$ = 10 $\cdot$ ln(2)]}
\label{Fig_Supp_eFold}
\end{figure}

The half-life of gas in the atmosphere is not quite the same as the $e$fold chemical lifetime (or $\tau$). For those more familiar with  half-life ($t_{\frac{1}{2}}$), $e$fold $\tau$ (tau) (eq. \ref{eq:eFold}) is the time required for the concentration of a gas to decrease by $\frac{1}{e}$ ($e^{-1}$) of its original concentration due to a chemical reaction or other sink. To convert $\tau$ to half-life, multiply a $\tau$ value by ln(2) as shown in figure \ref{Fig_Supp_eFold}.
\begin{align}
\begin{split} 
 \text{Half-} & \text{life} \\
 N(t) =& \, N_0 \cdot (\frac{1}{2})^{\frac{t}{t_{\frac{1}{2}}}} \\
\text{Where: } t =& \text{ time} \\
 t_{\frac{1}{2}} =& \text{ half life} \\
 N_0 =& \text{ initial gas quantity} \\
\end{split}
\begin{split}
\label{eq:eFold}
 \text{$e$fold lifetime} & \\
 N(t) =& \, N_0 \cdot e^{-\frac{t}{\tau}}\\
\text{Where: } t =& \text{ time. } \\
 \tau =& \text{ $e$fold atmospheric lifetime. ($\tau$)} \\
 N_0 =& \text{ initial gas quantity} \\
\end{split}
\end{align}

$\qquad \qquad \text{ See fig \ref{Fig_Supp_eFold} with sample parameters.} \qquad \text{Sample parms: }t  = \text{ 1 to 100; }  t_{\frac{1}{2}} \text{ = 10; } t_{\frac{1}{e}} \text{ = 10; } c \text{ = 100} $\\ \\
To align half-life requires conversion, multiplying by the natural logarithm of the $e$fold value for a gas shown with a simple example in figure \ref{Fig_Supp_eFold}. 

\subsubsection{CO$_2$ is the only GHG not modelled using \emph{e}fold $\tau$ (tau)} 
CO$_2$ is a special case in our set of 4 GHGs, having a complex, and deeply modelled characterisation that should be more accurate than the $e$fold model. For this reason, the equations:  A. C$_{RL}$ (eq. \ref{eq:NetCO2RemainderL}), B. C$_{RC}$ (eq. \ref{eq:NetCO2RemainderC}), C. C$_{RH}$ high (eq.  \ref{eq:NetCO2RemainderH}) of figure \ref{Fig-RemainderFractions} fitted to Archer are used \cite{Archer2009AtmosphericLifetimeOfCO2}. A significant fraction, 15\% to 20\% of CO$_2$, stays in the atmosphere "forever", or more precisely, as long as it takes for ocean acidification to be neutralized by reactions with CaCO3 in sediments and in coral reefs. So CO$_2$ can last thousands of years of residence in the atmosphere. CO$_2$ has a very long final time scale which is presented well by Archer, in 'The ultimate cost of carbon' \cite{Archer2020UltimateCostofCarbon}. However, assumptions may be incorrect to some degree, leading to higher long-term  CO$_2$ per Zondervan, et al. \cite{Zondervan2023RockCarbonCO2releaseOffsetsSilicateWeatheringSink}.

The ocean acts as a buffer for CO$_2$, that can both absorb and release CO$_2$. If vapour pressure of dissolved CO$_2$ is higher than that of the atmosphere, the ocean should release CO$_2$. This should come into play in anthropogenic scenarios where CO$_2$ declines on time scales of a few hundred or thousand years. Those scenarios should be taken into account in our fitted equations for the time span shown.

\subsubsection{CH$_4$ $e$-fold}
\label{CH4_efold}
 EPA in 2010 had a CH$_4$ chemical life-time of 9.6 to 12 years and noted that lower hydroxyls may increase atmospheric life \cite{EPA2010CH4_N2O}. EPA in 2023, lists 12 years for CH$_4$ \cite{EPA2023CH4_N2O}. Our internal discussion believes that there may be too much weighting to the high end of range. 
 
 Figure 1 of Holmes 2013 $\tau$ varies from $\approx$8.5 to $\approx$10 years \cite{Holmes2013FutureCH4_OH}. Thus, figure 1 would suggest CH$_4$ midrange was closer to 9.25 years. Holmes cites both Prather 2012 $\tau$ = 11.2 $\pm$ 1.3 y (9.9 to 12.5 y) \cite{Prather2012GHGreactive} and Naik 2013 $\tau$ = 9.7 $\pm$ 1.5 y (8.2 to 11.2) \cite{Naik2013hydroxylCH4lifetime}. Based on Naik and Holmes midpoints, we split the difference, obtaining 9.475 y.

We use CH$_4$ $\tau$ of $\approx$9.5 y. (Equivalent $t_{\frac{1}{2}}$ $\approx$6.6 y.) 

\subsubsection{N$_2$O $e$-fold}
\label{N2O_efold}
In 2010 EPA had an N$_2$O chemical lifetime of 114 years \cite{EPA2010CH4_N2O}. EPA in 2023 recommends 121 years \cite{EPA2023CH4_N2O}. Prather 2015 $\tau$ = 116 $\pm$ 9 y (107 to 125)\cite{Prather2015MeasureModelN2O}. We also performed our own cross-check of $\tau$ for N$_2$O. 

We applied a basic mass mass balance computation using NOAA N$_2$O data \cite{NOAA2022AGGIClimateTable}: 
\begin{align}
\label{eq:N2Omass} 
\begin{split} 
 \text{Mass } & \text{balance equation} \\
 d\frac{[N_2O]}{dt} =& \, S - \frac{[N_2O]}{\tau} \\
\text{Where: } S =& \text{ net sources: sum of all sources}\\ 
&\text{and sinks, except chemical destruction} \\
 [N_2O] =& \text{ mass fraction in the atmosphere} \\
 \tau =& \text{ $e$fold atmospheric lifetime} \\
\end{split}
\begin{split} 
 \tau\text{ solved } & \text{equation }N_2 \text{ mass} \\
 \tau =& \frac{[N_2]}{(S-d\frac{[N_2]}{dt})} \\
\text{Where: } S =& \text{ net sources: sum of all sources}\\ 
&\text{and sinks, except chemical destruction} \\
 [N_2] =& \text{nitrogen mass fraction in the atmosphere} \\
 \tau =& \text{ $e$fold atmospheric lifetime} \\
\end{split}
\end{align}

 When eq. \ref{eq:N2Omass} is worked out, we only use the N$_2$ part of N$_2$O for mass of its nitrogen. So below, that will be written as N$_2$, but it means the mass of nitrogen in the nitrous oxide. \\
 $d\frac{[N_2]}{dt}$ is the yearly change in N$_2$. Practically speaking, this is averaged over a 10 year period. 
 
To obtain the masses for eq. \ref{eq:N2Omass} we reference NOAA GHG data \cite{NOAA2022AGGIClimateTable}, and use some fundamental values.

Global dry air moles (all molecules) = 1.772E+020 mole. So, 1 ppb mole of N$_2$O = 1.77E+11. \\ 
However, the mole fraction of N$_2$O is not constant by altitude. The mole fraction of N$_2$O drops to zero in the mid-stratosphere, with the mid-point of the drop-off around 24 km altitude The net result is that we need to apply a ratio of 96.48\% to our 1.77E+11 mole of N$_2$ = 1.70769E+11. \\
The molecular weight of the N$_2$ portion = 28.014 $\frac{g}{mole}$. \\
Thus, 1.70769E+11 mole N$_2$ $\cdot$ 28.014 $\frac{g}{mole}$ = 4.79 Teragrams (Tg) of N$_2$ in 1 ppb. 

The mean ppb of N$_2$ from 2007-2016 (the Tian decade) = 324.681 ppb. \\
The mean mass of N$_2$, $\overline{[N_2]}$ = 324.681 ppb $\cdot$ 4.79 $\frac{Tg}{ppb}$ = 1555.0 Tg 

The mean change in N$_2$ ppb from 2007-2016 = 0.913 ppb. \\
The mean mass change, $\overline{d\frac{[N_2]}{dt}}$ is 0.913 ppb $\cdot$ 4.79 $\frac{Tg}{ppb}$ = 4.373 Tg. 

The value of $\overline{S}$ is obtained from Tian \cite[Table 1: Bottom-up total source]{Tian2020N2OSources_sinks}. $\overline{S}$ = 17.0 

Substitute these values into eq. \ref{eq:N2Omass} to obtain $\tau$:\\
$ \tau \, = \, \frac{1555.0 \, Tg}{17.0 \,Tg - 4.373 \,Tg} \, = \, 123.1$ y.

We performed further computations (not shown) on previous and current decades using the min, mean, and max values in \emph{Tian's Table 1: Bottom-up total source}. This spanned a range of $\tau$ from 115.4 y to 123.1 y, with a rough mean of 119.2 y. So $\tau$ appears to be increasing a bit from the 1980-2010 period. A small increase is the opposite direction from what a chemist would expect from an increase in concentration of N$_2$O. However, the 1980--2000 period has sparser data with more smoothing, and those are most of the low. This could explain the small an increase in $\tau$ for Tian's 2007--2016 decade from the mean. We make the assumption that Tian, et al chose 2007--2016 because it had the best data. All of which is to say that what $\tau$ truly is, has some uncertainty. On balance, we decided to use our 123.1 value after considerable deliberations. 

We use N$_2$O $\tau$ $\approx$ 123.1 y. (Equivalent $t_{\frac{1}{2}}$ $\approx$  85.3 y.) 

\subsubsection{Fluorinated and chlorofluorocarbon gases (Fgas) $e$-fold}
\label{Fgas_efold}
These gases are primarily fluorine compounds. They have various names, with Other Trace Gases (OTG) being common. For this model three reported gases are the basis for our estimate: SF$_6$, NF$_3$, and a basket of chlorofluorocarbon (CFC) gases, the most common type. Published chemical lives: SF$_6$ 1278 y \cite{Kovacs2017LifetimeOfSF6} , NF$_3$ 509 y \cite{Totterdill2016NF3andCFC115}, CFC 492 y\cite{Totterdill2016NF3andCFC115}. 

The most recent SF$_6$ $\tau$ is from Kouznetzov 2020, ranging from 600 to 2900 years. The most recent paper utilising an SF$_6$ $\tau$ used an estimate of Kovacs \cite[p. 1176]{Loeffel2022SF6impactOnAoA}. This Kovacs best estimate of $\tau$ is 1278 y (1120 to 1475 y) \cite{Kovacs2017LifetimeOfSF6}, while Ray's best estimate published the same year is 850 y (580 to 1400 y) \cite{Ray2107QuantificationSF6lifetime}. These are both much lower than the previously accepted estimate of 3200 years, which was down from 25,000 years \cite{Ko1993SF6SourcesSinks}. We chose the best estimate of of Kovacs to be conservative, because Kovacs' and Ray's estimates terminated at $\approx$1400 y, the Kovacs low range was so far above Ray's best estimate, and Kouznetzov's most recent high end was 2900 y. Thus, SF$_6$ has the greatest degree of uncertainty among these gases.

Deciding on a value for a basket of the long-lived flourocarbon GHGs, a qualitative factor we took into account is that if these gases stop being emitted, the life span of the set rises continuously toward the asymptote of the longest lived gas. This number we chose could be replaced in the future by a more detailed model featuring each halogenated hydrocarbon, and its $\tau$ if it was available. Additionally, ozone is part of the breakdown of halogenated hydrocarbons, and natural halogens are predicted to deplete ozone \cite{Anderson2018CouplingCatalysis,Anderson2017ChlorineBromineCatalysis}. Thus it is plausible that half-lives of these chemicals could change in complex ways \cite{Orkin2020AtmosLifeHalogenHydrocarbons}. Future atmospheres could, potentially, have conditions to raise the atmospheric life span ($\tau$) of all of the halogenated hydrocarbon gases. 

We use the unweighted mean $\tau$ of these three gases, $\approx$760 y. (Equivalent $t_{\frac{1}{2}}$ $\approx$ 527 y.) 

\subsubsection{Interactions between greenhouse gases cause an N$_2$O anomaly that is actually correct.}
\label{SC-GHG_N2O_anomaly}
In the SC-GHG tables \ref{SC-GHG_tables}, damages for N$_2$O are seen to be apparently switched around. The high CO$_2$ scenario is a bit lower than the low CO$_2$ scenario, and the central scenario is in between. This is correct behaviour due to the adjustment equations we implemented from Byrne \& Goldblatt \cite[pp. 19, 157, Table 2]{Byrne2014RadiativeForcingEquations}. These adjustment equations account for the light spectrum overlap of CO$_2$ and CH$_4$ with N$_2$O such that CO$_2$ and CH$_4$ absorption takes precedence over N$_2$O. Thus, as CO$_2$ rises, it takes more GHG effect away from N$_2$O. The effect is minor overall, but enough to be visible in the tables. It is not a bug in the model.

\subsection{Heat model }
 In an over-simplified temperature driven system with stable temperature, ocean heat content reaches a limit if a temperature is maintained, and in that over-simplified model, creates a classic logistic curve at equilibrium.  In OPTiMEM, \emph{real-world climate model drivers are a dynamic system that never reaches equilibrium}. 

 The heart of this model is two iterative loops. The first loop, the carbon-gases loop, starts with GtC, and computes CO$_2$, adds other GHGs, uses those GHGs to compute EEI, and then uses EEI to compute EEI$_T$. The EEI$_T$ is the temperature forcing on the ocean. All the values that are computed at each step are stored in large arrays and written out by the model into .xlsx files. This allows detailed examination and later use of these values. This first iterator runs from year 1 to year 5000. The early values up to roughly 1750-1850 are unlikely to be as reliable those from 1850 to 2021. 
 
The second loop uses the temperature forcing array that the carbon-gases loop produces as input for each year. This loop computes ocean heat using the standard heat equation. It computes values for change in temperature ($\Delta$T), each year's change in ocean heat ($\Delta$Q) and total heat, (Q). The internal values produced during the computations are also written out into .xlsx files. 

These loops are written below as plain language for ease of understanding. The Maple\textsuperscript{\texttrademark} model implements the pseucode. 

\subsubsection{Carbon and gases loop }
\setlength{\parskip}{0pt}
For each year from 1 to 5000 

\qquad Compute driving carbon (GtC) using eq. \ref{eq:1500GtC}

\qquad Compute CO$_2$ from GtC. Multiply GtC x 3.664141204 

\qquad Obtain CH$_4$. 

\qquad \qquad In baseline scenarios, 881.441 ppb + (2.386 x CO$_2$). 

\qquad \qquad \quad Equation fitted for 1990-2021 Ratio of CH$_4$ ppb to CO$_2$ ppm.

\qquad \qquad In permafrost CH$_4$ scenarios use eq. \ref{eq:Schuur}

\qquad \qquad Propagate CH$_4$ $\tau$ to year 5000

\qquad \qquad and oxidize CH$_4$ to CO$_2$

\qquad Propagate CO$_2$ remainder to year 5000 using eq's. \ref{eqs:NetCO2Remainders}, generating 3 CO$_2$ levels, low, central \& high. 

\qquad Obtain N$_2$O using eq.'s \ref{eq:TianEqN2OtoPopGWP}, calling eq.'s \ref{eq:PopProj1} \& \ref{eq:GWPProj1}   

\qquad Obtain F-gases. 7.182 + (3.352 x CO$_2$).  Equation fitted for 1990-2021 Ratio of F-gases ppt to CO$_2$ ppm. 

\qquad Compute EEI for CO$_2$, CH$_4$, N$_2$O, to obtain total EEI. See Byrne \& Goldblatt, 2014 \cite[p. 157, Table 2]{Byrne2014RadiativeForcingEquations}

\qquad Compute EEI$_T$ using eq. \ref{eq:T0_F}. This is the forcing temperature increase.

end loop    
\setlength{\parskip}{6pt}

\begin{figure}[!ht]
\centering\includegraphics[width=5.5in]{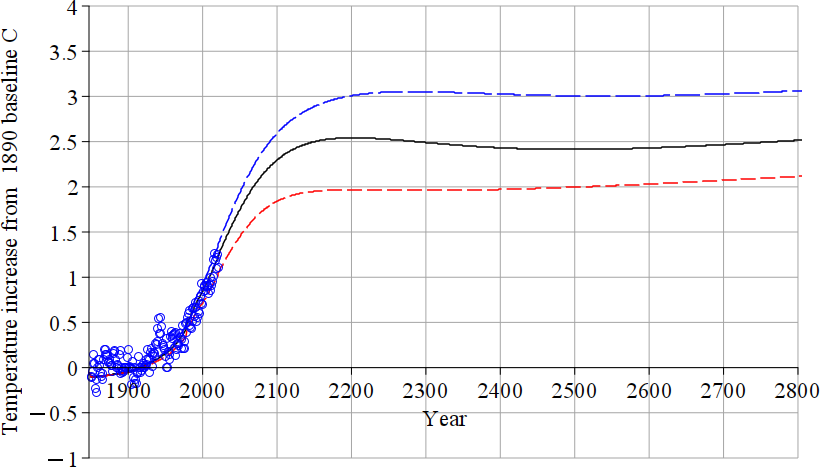}
\caption{\textbf{Temperature increase $\degree$C resulting from baseline carbon scenario.} (Baseline\_PopNorm) Blue circles, historical temperature anomaly records. Upper blue long dash and lower red dash curves are highest and lowest remainder CO$_2$ from CO$_2$ remainder functions \ref{eqs:NetCO2Remainders}. Black curve is the result of Archer density plot harmonic mean, our central value. }
\label{Fig_Supp_TempIncrease_Scenario1_V1}
\end{figure} 

\subsubsection{Ocean heat loop }

The EEI$_T$ array created by the Carbon-gases loop is the input for this Ocean-heat loop. 

\setlength{\parskip}{0pt} 

For each year from 1890 to 5000 

\qquad Subtract cumulative ocean temperature increase from EEI$_T$ (Forcing temperature) to get dT. 

\qquad \quad dT is this year's temperature increase forcing.

\qquad Compute 1 year of ocean heating ($\Delta$Q) using dT (eq. \ref{eq:DeltaQ}) to get $\Delta$Q for the year. 

\qquad Add $\Delta$Q for year to Q total. 

\qquad Compute $\Delta$T ocean temperature increase for this 1 year of ocean heating using eq. \ref{eq:DeltaT}. 

\qquad Add $\Delta$T to cumulative ocean temperature increase. 

end loop
\setlength{\parskip}{6pt}

\subsubsection{Dimensional analysis of heat model and functions}
This section is in $C$ (centigrade) rather than $K$ (Kelvin). The interval is the same. \\

\textbf{Dimensions} \\ J =  Joules; m =  metres; s = seconds; $\degree$C = degrees C; kg = kilograms   \\ 
\textbf{Constants} \\
$k$ = 0.6; $A$ = 3.619E14; $Cs$ = 4200; $M$ = 3688 ·1026 · $A$;  

The bracket [] convention is used to represent dimensions of the equations. \\
Ex. $k[\frac{J}{m \cdot \degree C \cdot s}]$ means that $k$ is measured in Joules per meter-degree-seconds.  \\

\textbf{$\Delta$Q (Heat) in relation to time.}  $\Delta$Q($\Delta$t, $\Delta$FT) 
\begin{align}
    \begin{split} \label{eq:DeltaQ}
 \Delta Q(\Delta t, \Delta FT)[J]  &= k[\frac{J}{m \cdot \text{$\degree$}C \cdot s}] \cdot A[m^2]\ \cdot \frac{\Delta FT[\text{$\degree$}C] }{\Delta x[m]} \cdot \Delta t[s]  \\
 \text{Dimensions } &\Rightarrow J  = \frac{J}{m \cdot \text{$\degree$}C \cdot s} \cdot m^2  \cdot \frac{\text{$\degree$}C }{m} \cdot s 
  \Rightarrow J  = J \frac{m^2}{m \cdot m} \cdot \frac{\text{$\degree$}C \cdot s }{\text{$\degree$}C \cdot s} \\
\text{Where: } \Delta Q &= \text{ heat change for time interval } \qquad
   k = \text{ thermal conductivity }\\
  A &= \text{ area } \qquad
 \Delta FT = \text{ forcing temperature difference }  \\
  \Delta x &= \text{ thickness of warming sections }  \qquad  
  \Delta t = \text{ time interval }  \\ 
  \end{split}
\end{align}

\textbf{$\Delta$T by $\Delta$Q.} $\Delta$T($\Delta$Q, Cs, M) 
\begin{align}
    \begin{split} \label{eq:DeltaT}
 \Delta T(\Delta Q, Cs, M)[\text{$\degree$}C] &= \frac{\Delta Q[J]}{Cs[\frac{J}{kg \cdot [\text{$\degree$}C]}]\cdot M[kg]}   \\
 \text{Dimensions } &\Rightarrow \text{$\degree$}C  = \frac{J \cdot kg \cdot \text{$\degree$}C}{J \cdot \cdot kg}  
 \\
\text{Where: } \Delta T &= \text{ temperature change} \qquad
\Delta Q = \text{ heat change } \\
   Cs &= \text{ heat capacity } \qquad 
 M = \text{ mass } \\
  \end{split}
\end{align}

\subsubsection{Heat model results}
Because the initial run of this model had already been calibrated somewhat in development using a temperature stability model (the over-simplified model mentioned above), the first results fell surprisingly close to recorded data. The curve was, nevertheless, fitted using an $R^2$ minima algorithm over two parameters, initial ocean heat content, and the $\Delta x$ (thickness of warming sections) (eq. \ref{eq:DeltaT}).

 \subsubsection{Assignment of heat fractions from CH$_4$}
For use by the heat conjecture to estimate weather damages, CH$_4$ presents a special problem. Part of CH$_4$ is linked to CO$_2$, and part of it is tied to population. Both are anthropogenic, but it is important that only heat from CH$_4$ that was released because of carbon extraction, or from primarily CO$_2$ based heat should be assigned to CO$_2$. This is implemented by subtracting the population linked CH$_4$ from the total CH$_4$ in a particular year. This leaves energy linked CH$_4$ and permafrost release of CH$_4$, which fraction of total CH$_4$ is multiplied by the total heat due to CH$_4$ and added to the heat from CO$_2$ alone. 

There are arguments pro and con for only including carbon extraction linked CH$_4$ in CO$_2$ total. We leaned toward permafrost CH$_4$ release being due to the majority effect of CO$_2$. And, the CO$_2$ released from permafrost, plus the CH$_4$ converted to CO$_2$ was already assigned to CO$_2$. 

It is only for computation of weather damages that this is done. 

\subsection{Primary scenario results}
\label{lb:CMIP}
\begin{figure}[!ht]
\centering\includegraphics[width=5.5in]{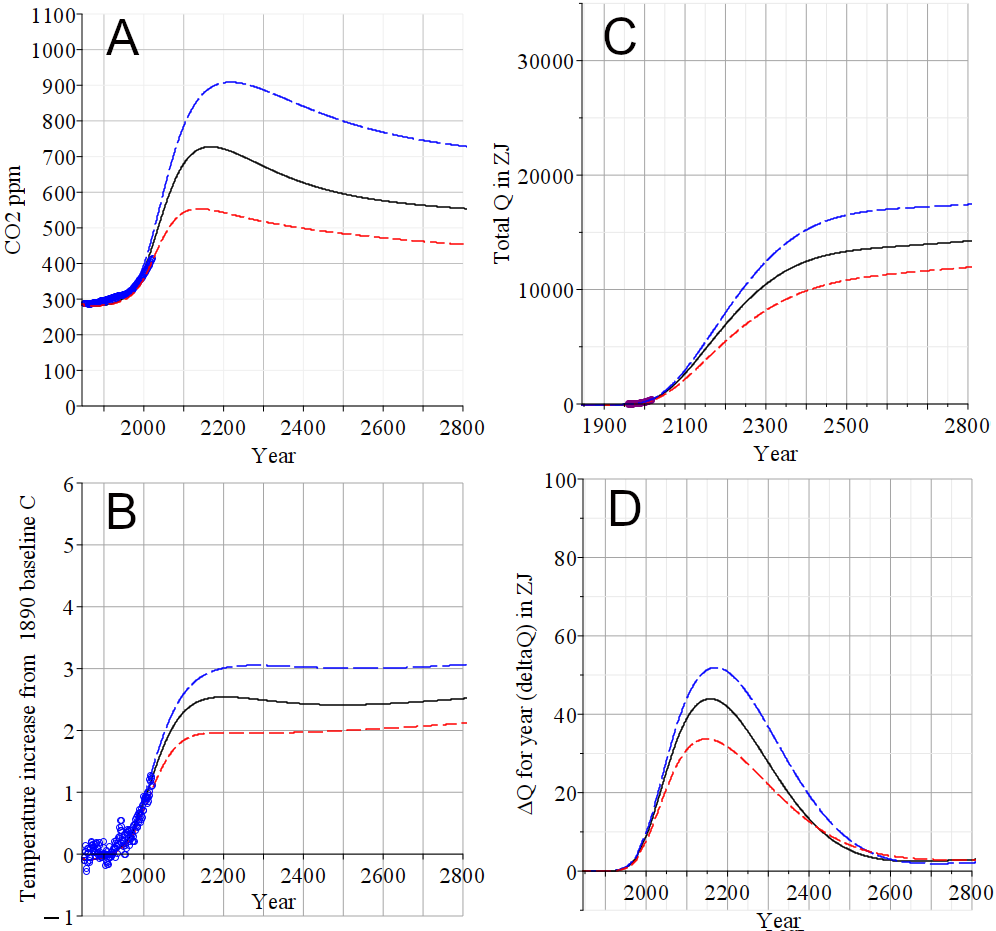}

    \caption{\textbf{Baseline normal population scenario results.} Upper blue long dash and lower red short dash curves are high and low bounds. Black solid curve is central. (\textbf{A}) CO$_2$ ppm; (\textbf{B}) $\Delta$T $\degree$C; (\textbf{C}) 'Q' Total ocean heat content increase (OHC) in ZJ; (\textbf{D}) '$\Delta$Q' (dQ) change in OHC for year. (\textbf{A}, \textbf{B}, \& \textbf{C}, also show fit to existing datasets up to 2021. See figure \ref{Fig_Supp_Scenario_tree} for scenarios. }
    \label{Fig_Supp_Baseline_PopNorm_19.5}
\end{figure}

\begin{figure}[!ht]
\centering\includegraphics[width=5.5in]{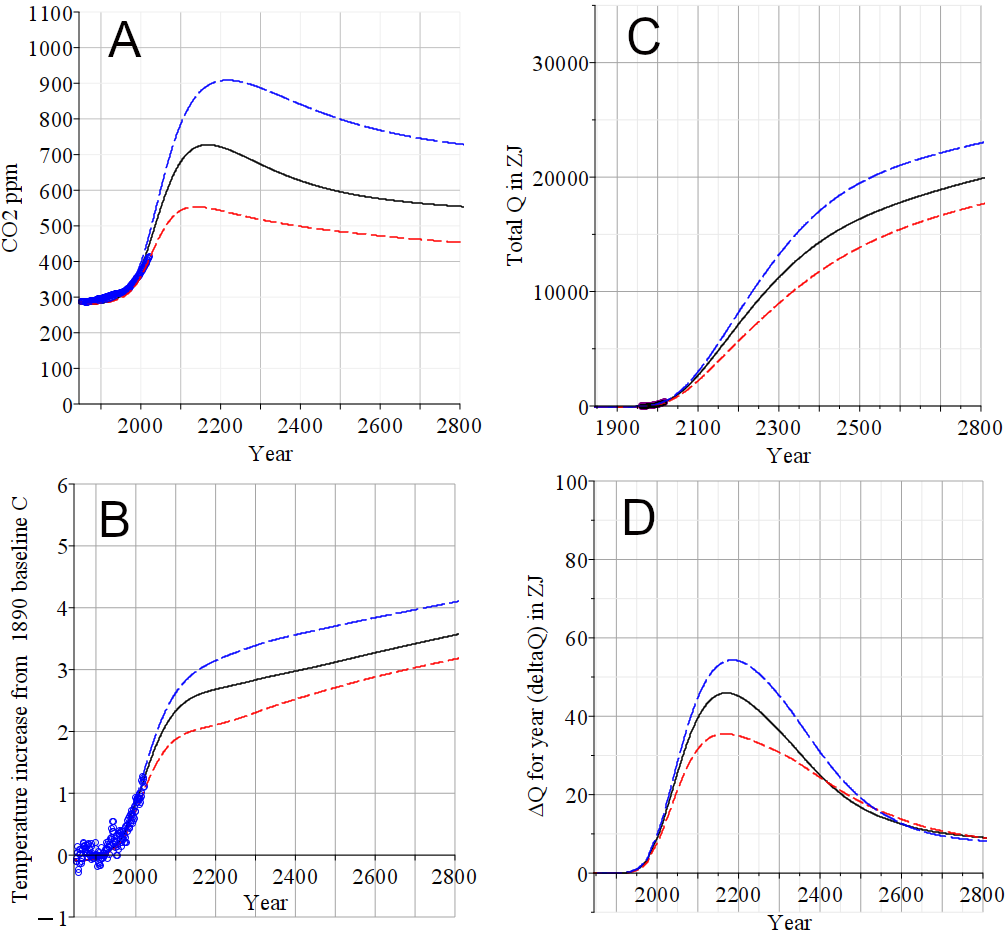}

    \caption{\textbf{Baseline high population scenario results.} Upper long dash blue and lower red dash curves are high and low bounds. Black solid curves are central. (\textbf{A}) CO$_2$ ppm; (\textbf{B}) $\Delta$T $\degree$C; (\textbf{C}) 'Q' Total ocean heat content (OHC) in ZJ; (\textbf{D}) '$\Delta$Q' (dQ) change in OHC for year. (\textbf{A}, \textbf{B}, \& \textbf{C}, also show fit to existing datasets up to 2021. See figure \ref{Fig_Supp_Scenario_tree} for scenarios. }
    \label{Fig_Supp_Baseline_PopHigh_19.5}
\end{figure}

\begin{figure}[!ht]
\centering\includegraphics[width=5.5in]{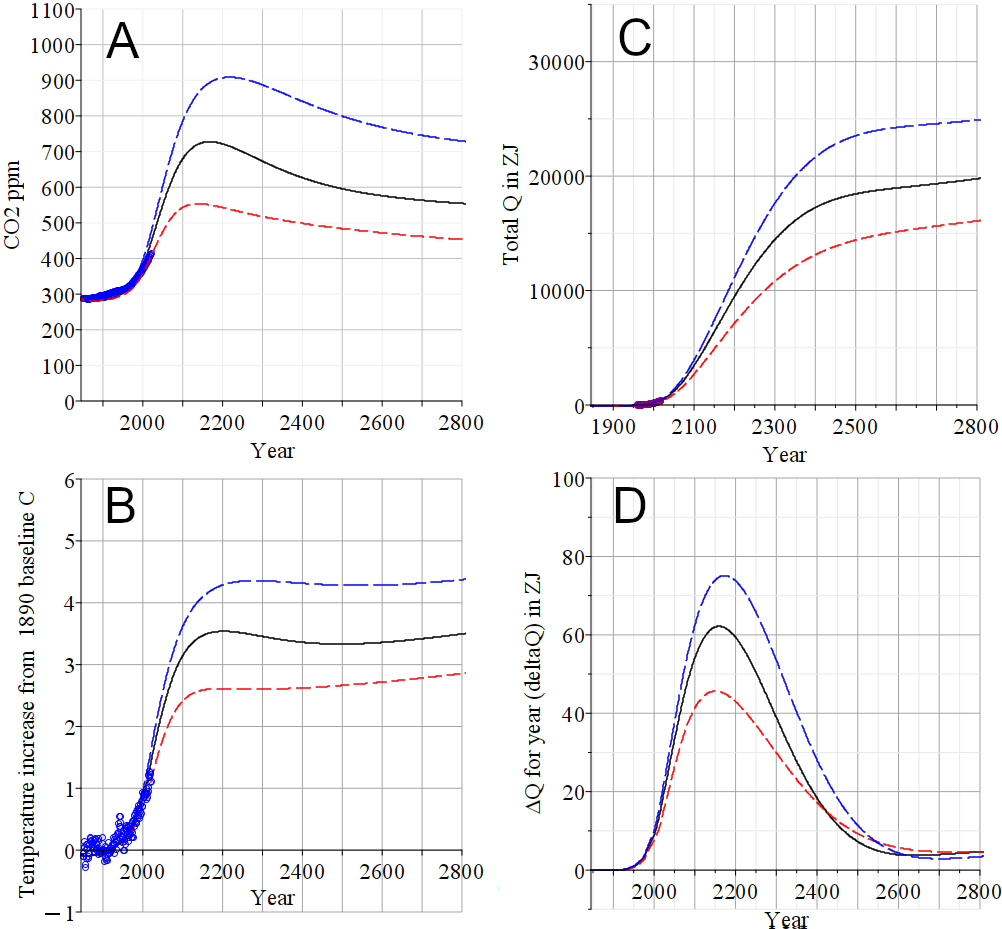}

    \caption{\textbf{Baseline normal population S-aerosol scenario results.} Upper blue long dash and lower red dash curves are high and low bounds. Black solid curves are central. (\textbf{A}) CO$_2$ ppm; (\textbf{B}) $\Delta$T $\degree$C; (\textbf{C}) 'Q' Total ocean heat content (OHC) in ZJ; (\textbf{D}) '$\Delta$Q' (dQ) change in OHC for year. (\textbf{A}, \textbf{B}, \& \textbf{C}, also show fit to existing datasets up to 2021. See figure \ref{Fig_Supp_Scenario_tree} for scenarios. }
    \label{Fig_Supp_Baseline_PopNorm_S_aerosol_19.5}
\end{figure}

\begin{figure}[!ht]
\centering\includegraphics[width=5.5in]{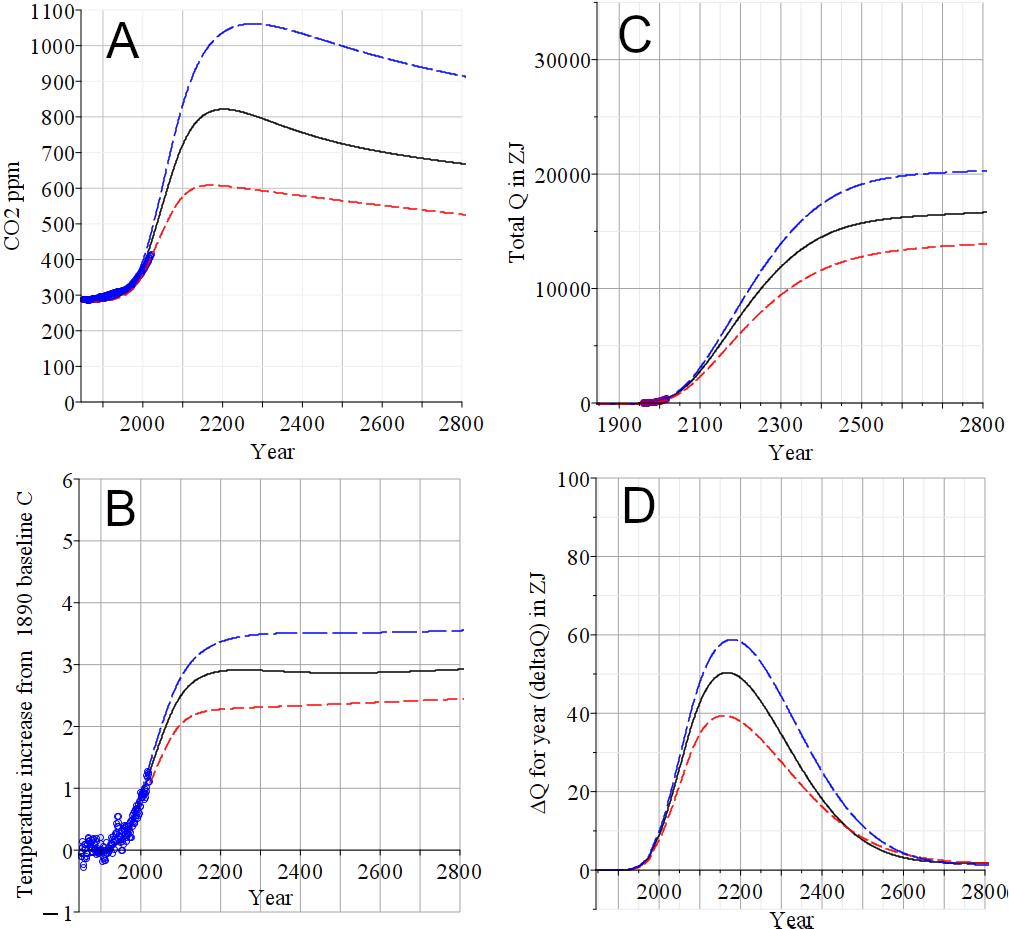}

    \caption{\textbf{Permafrost normal population scenario results.} Upper blue long dash and lower red dash curves are high and low bounds. Black solid curves are central. (\textbf{A}) CO$_2$ ppm; (\textbf{B}) $\Delta$T $\degree$C; (\textbf{C}) 'Q' Total ocean heat content (OHC) in ZJ; (\textbf{D}) '$\Delta$Q' (dQ) change in OHC for year. (\textbf{A}, \textbf{B}, \& \textbf{C}, also show fit to existing datasets up to 2021. See figure \ref{Fig_Supp_Scenario_tree} for scenarios. }
    \label{Fig_Supp_PermaFrost_PopNorm_19.5}
\end{figure}

\begin{figure}[!ht]
\centering\includegraphics[width=5.5in]{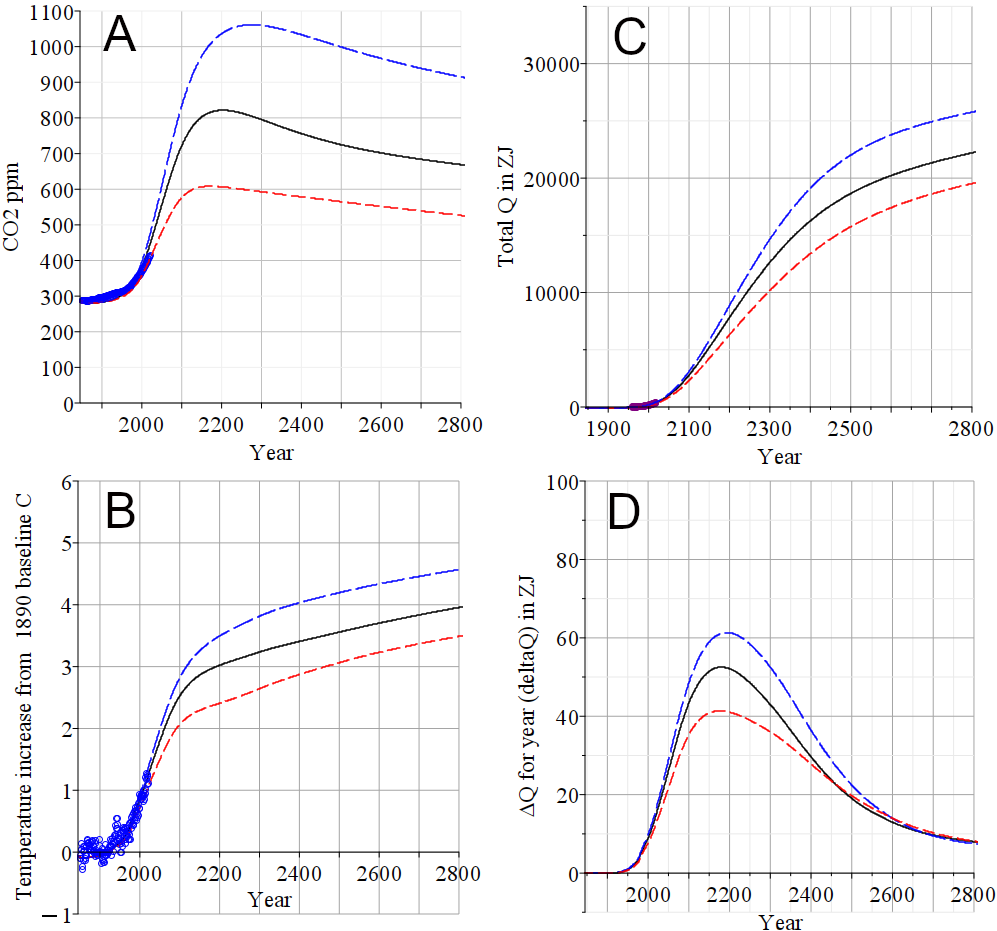}

    \caption{\textbf{Permafrost high population scenario results.} Upper blue long dash and lower red dash curves are high and low bounds. Black solid curves are central. (\textbf{A}) CO$_2$ ppm; (\textbf{B}) $\Delta$T $\degree$C; (\textbf{C}) 'Q' Total ocean heat content (OHC) in ZJ; (\textbf{D}) '$\Delta$Q' (dQ) change in OHC for year. (\textbf{A}, \textbf{B}, \& \textbf{C}, also show fit to existing datasets up to 2021. See figure \ref{Fig_Supp_Scenario_tree} for scenarios. }
    \label{Fig_Supp_PermaFrost_PopHigh_19.5}
\end{figure}

\begin{figure}[!ht]
\centering\includegraphics[width=5.5in]{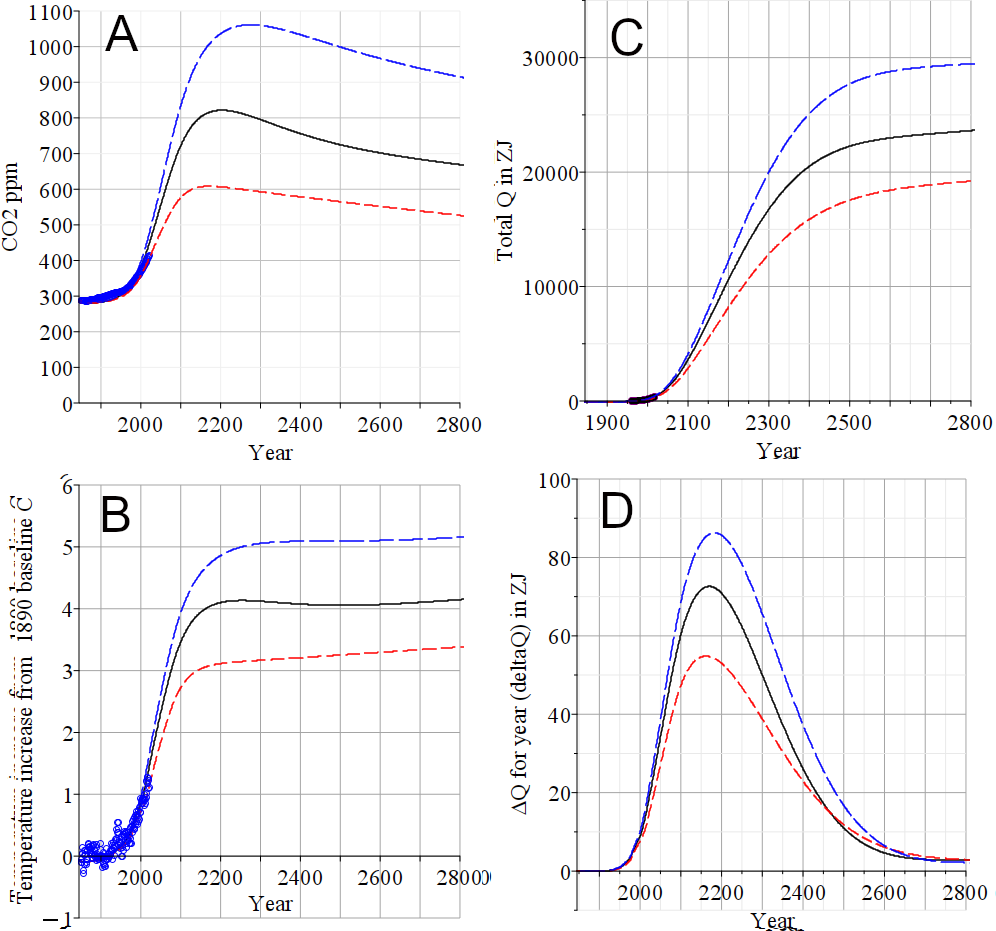}

    \caption{\textbf{Permafrost normal population S-aerosol scenario results.} Upper blue long dash and lower red dash curves are high and low bounds. Black solid curves are central. (\textbf{A}) CO$_2$ ppm; (\textbf{B}) $\Delta$T $\degree$C; (\textbf{C}) 'Q' Total ocean heat content (OHC) in ZJ; (\textbf{D}) '$\Delta$Q' (dQ) change in OHC for year. (\textbf{A}, \textbf{B}, \& \textbf{C}, also show fit to existing datasets up to 2021. See figure \ref{Fig_Supp_Scenario_tree} for scenarios. }
    \label{Fig_Supp_PermaFrost_PopNorm_S-aerosol_19.5}
\end{figure}

Central and low values for all scenarios fall within the central range SSP245 of CMIP6 temperature \cite{CMIP62023ClimateProjections,CMIP62026CopernicusDataTutorials} (Panel B of figures \ref{Fig_Supp_Baseline_PopNorm_19.5} through \ref{Fig_Supp_PermaFrost_PopNorm_S-aerosol_19.5}).  Highest temperature increases for normal population rise to 4$\degree$C and may surpass it (fig's. \ref{Fig_Supp_Baseline_PopHigh_19.5}, \ref{Fig_Supp_Baseline_PopNorm_S_aerosol_19.5}, \ref{Fig_Supp_PermaFrost_PopHigh_19.5}, \ref{Fig_Supp_PermaFrost_PopNorm_S-aerosol_19.5}) ). These high range scenarios land in the lower end of SSP585, exceeding +4$\degree$C and reaching +5$\degree$C. Note that over time, CMIP projections have been a bit lower than empirical data \cite{Carvalho2022CMIP3_CMIP5_CMIP6_recent_future_climate_projections}. 

In figures \ref{Fig_Supp_Baseline_PopNorm_19.5} through \ref{Fig_Supp_PermaFrost_PopNorm_S-aerosol_19.5} the time relationship can be seen between the driving CO$_2$ greenhouse gas (A panels), which slightly leads temperature (B panels), and the buildup of heat energy in the ocean (C panels), which lags CO$_2$ and temperature by 325-400 years. As also discussed in figure \ref{Fig_Supp_dQvQ_Shakhova-Hansen} the change in heat per year appears as either dQ or $\Delta$Q.  The zero crossing of $\Delta$Q should signal the start of several centuries of release of energy from the ocean, which should appear as more energetic weather \cite{Hansen2016IceMeltSeaLevelRiseAndSuperstorms}. The centuries past 2500 CE could be quite dramatic if GHG abatement is done sufficiently. See figure \ref{Fig_Supp_Baseline_PopNorm_n2O-comparison_T}. 

Graph panels in figures \ref{Fig_Supp_Baseline_PopNorm_19.5} through \ref{Fig_Supp_PermaFrost_PopNorm_S-aerosol_19.5} are identically scaled for ease of comparison.

\section{Heat conjecture}
\label{lb:HeatConjecture}
This modelling system's core problem is really two problems that are interlocked, and those problems are related to  criticism levelled at existing climate economics models \cite{STANTON2009Inside30ClimateModels,WEITZMAN2010,WEITZMAN2012}. The two basic problems are finding a logical basis for quantitative projection, and that discontinuities from peak events make smooth curves unrealistic.  

The ocean heat content is not uniform in the water column, and like other, more complex climate models, most of the increased heat in OPTiMEM is in the upper ocean. It takes far longer to warm the deep ocean than the warming that human activity is likely to produce. The ocean is generally divided into 3 layers, with the first 200 metres being the warmest, and therefore having the greatest heat content. 200-1500 metres is the thermocline region, and the deep ocean from the thermocline down is the coldest. In addition, sea surface temperature (SST) varies a great deal by rough latitude and circulation pattern. In OPTiMEM, the entire ocean is has a single total heat content, however, the dynamics of the heat equation contain diffusion by depth which we do not record.  

\subsection{Forward projection of weather damages problem}
\label{Forwardprojectionofweatherdamagesproblem}
\begin{figure}[!hbt]
\centering\includegraphics[width=5.25in]{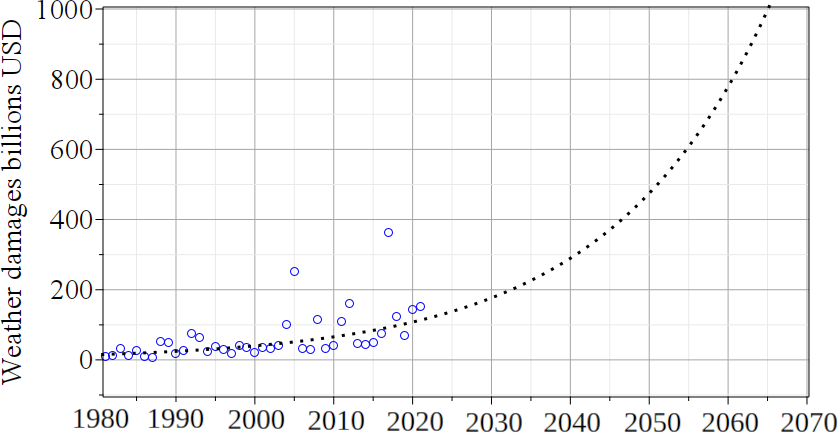}
\caption{\textbf{Projection of exponential fit WDe(y) (eq. \ref{eq:Fite_Dyr1}) to weather damages dataset.} \cite{NOAAbillionDlr} The 0.97 $R^2$ value of this fit was highest of those tested. Real world physics does not exhibit long term exponential behaviour as a rule. We could not justify it. This is discussed in subsection \ref{Forwardprojectionofweatherdamagesproblem} Blue circles, total NOAA weeather damages dataset. Dotted curve, WDe(y).}
\label{Fig_Supp_WD_FitToData_graph}
\end{figure}

The forward projection of weather damages problem is visible in figure \ref{Fig_Supp_WD_FitToData_graph}. It has some similarity to the issues with the DICE model discussed above (fig. \ref{Fig_Supp_DICE_Unhinged_Scenario_18.1}). While narrative examination, geological climate history, and climate models did not support the idea of the low or negligible damages produced by IAMs, could an exponential rise in damages be supported? If so, for how long into the future, and could we construct scenarios that made sense? 

Damages cannot continue to increase exponentially without exponential carbon and heat, and exponential carbon is not reasonable to believe. Nor is exponential heat, which eventually boils the ocean. Such modelling is unhinged from reality---it simply does not apply to planet Earth, and the results of the DICE assumptions model (Supplement: \S6.2.6, fig. 21) bear this out. Thus, we rejected the premise of endless carbon-energy based growth, which means we had to reject a continuing exponential growth model.

The longest time period going forward from 2021 that seemed reasonable to believe may fit an exponential rise in damages is shown by our SC-GHG model's fitted curve (WDe(y) \ref{eq:Fite_Dyr1} (fig. \ref{Fig_Supp_WD_FitToData_graph}) which is perhaps as long as 50 years. That would take us to roughly 2070 CE. This would be a trivially short model projection.

The next curve explored was the logistic equation similar to the curve used in figure \ref{Fig-Expert-Sigmoid}, the type of model that was abandoned. A logistic curve for damages would seem reasonable if temperature is considered in a simple model that goes to a peak equilibrium. However, where should the upper limit of damages be? What should the slope of the linear phase be? The only way to answer this was to turn to some type of climate model. 

\subsection{Heat curve conjecture solution}
\begin{figure}[!ht]
\centering\includegraphics[width=5in]{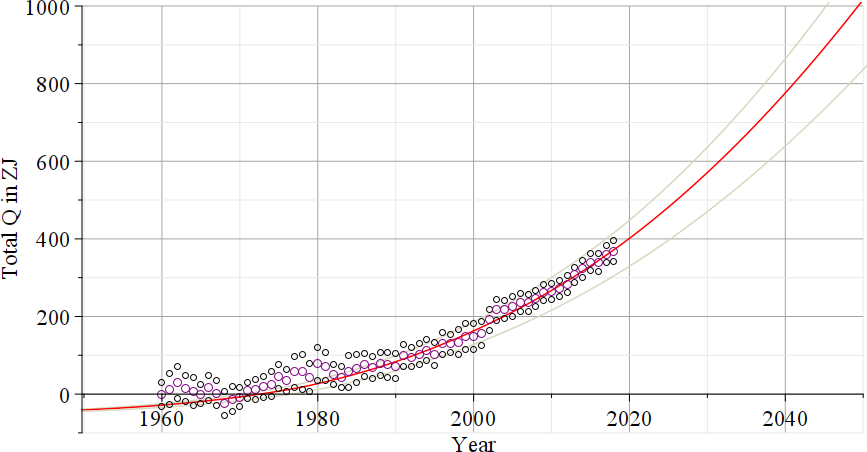}
\caption{\textbf{ Synchrony of heat (Q) model results with ocean heat dataset.} Large circles are central values from von Shuckmann, et al. Small circles are uncertainty bounds. Solid curves are heat model results. Central values fall within the uncertainty bounds, with only small idealisation anomalies from the dataset, which is expected. After 2020 CE, the 3 CO$_2$ variants begin to differentiate, and the high and low track closely with the OHC uncertainty bounds. Note that the earliest technology was the least accurate, with three major data gathering technology shifts, and normalization is imperfect. ZJ = zettajoule = 10$^{21}$ Joules.}
\label{Fig_Supp_OHCMatchToQHeat}
\end{figure} 
\label{Heatcurveconjecturesolution}

\begin{align}
    \begin{split} \label{eq:OHC_Heat_conjecture}
\text{Baseline OHC energy release prior to warming}\\
\overline{Q_B} \cdot W &= \overline{Q_W}  \\
\text{OHC energy release with warming} \\
(\overline{Q_B} + Q_{\uparrow}) \cdot W &= \overline{Q_W} + Q_{\uparrow W} \\
\text{Where:} \qquad \overline{Q_B} &= \text{ mean average OHC baseline before warming}\\
0 < W &< 1, \text{$W$ is fraction of OHC released in a year, and $W$ is small}\\
\overline{Q_W} &= \text{ mean average baseline OHC released each year}\\
Q_{\uparrow} &= \text{OHC increase over }\overline{Q_B} \\
Q_{\uparrow W}  &= \text{ OHC increase over }\overline{Q_B} \text{that is released in a year}\\
  \end{split}
\end{align}

Subject to stochastic variance, a similar fraction of the heat in the upper ocean should be available for release in any given year (eq. \ref{eq:OHC_Heat_conjecture}). Thus, as the amount of heat in the upper ocean rises, a similar fraction of the OHC increase ($Q_{\uparrow}$) should go into producing climate effects, and hence, the increase in damages. Assuming $W$ is constant, the magnitude of the heat energy available for release should rise, on average, synchronously with OHC increase. We assume that weather damages will rise proportionally with the increase in released OHC ($Q_{\uparrow W}$), that is above the baseline OHC energy release ($\overline{Q_W}$). Therefore, this justifies scaling the OHC to fit existing weather damages data. If, at some future time, it is found that the proportion ($W$) changes in response to conditions, then an equation describing how the proportion changes can be substituted for $W$ of equation \ref{eq:OHC_Heat_conjecture}. 

\begin{figure}[!hbt]
\centering\includegraphics[width=6.25in]{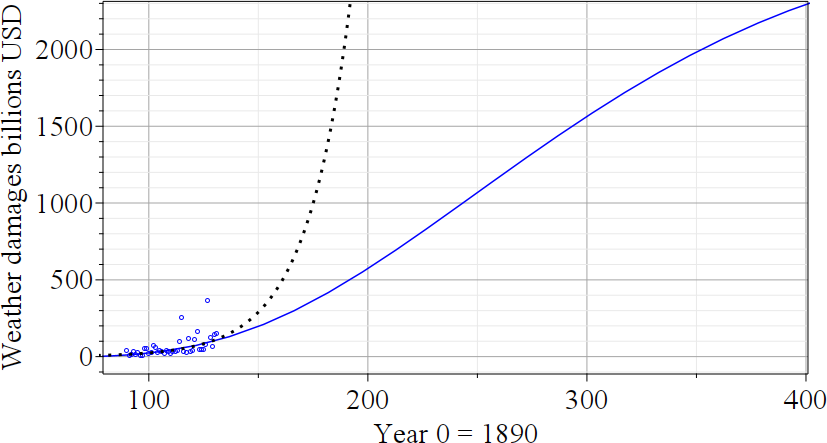}
\caption{\textbf{Fit by scaling of heat curve to WDe(y).} Solid blue is scaled heat curve, for scaling factor 0.2683. Scaling factor is discussed in subsection \ref{Heatcurveconjecturesolution}. Dotted curve is the $WDe(y)$ (eq. \ref{eq:Fite_Dyr1}) exponential weather damages equation fitted to NOAA weather dataset. Blue circles are NOAA weather dataset.}
\label{Fig_Supp_ExptoHeatProjHeatConjecture}
\end{figure}

The OHC curves generated by our OHC climate model validated very well against existing ocean heat data (fig. \ref{Fig_Supp_OHCMatchToQHeat}). The $WDe(y)$ (eq. \ref{eq:Fite_Dyr1}) was fitted to the NOAA weather damages dataset  (fig. \ref{Fig_Supp_ExptoHeatProjHeatConjecture}). Then, the OHC curves were fitted to $WDe(y)$ (eq. \ref{eq:Fite_Dyr1}) over the 42 years of weather damages. This was done using a binary search $R^2$ minima algorithm, which generated a y axis scalar of 0.2752 with $R^2$ = 0.9986. We see by inspection of figure \ref{Fig_Supp_ExptoHeatProjHeatConjecture} the contrast between a continued exponential fit and our more conservative OHC based curve. 

These scaled heat curves were translated to a set of weather damages functions $fWD_m(y)[]$ (eq. \ref{eq:Fit_Dyr2}) (also see eq. \ref{eq:GlobalWeatherDmgs2}. 

\begin{figure}[hbt!]
\centering\includegraphics[width=6.25in]{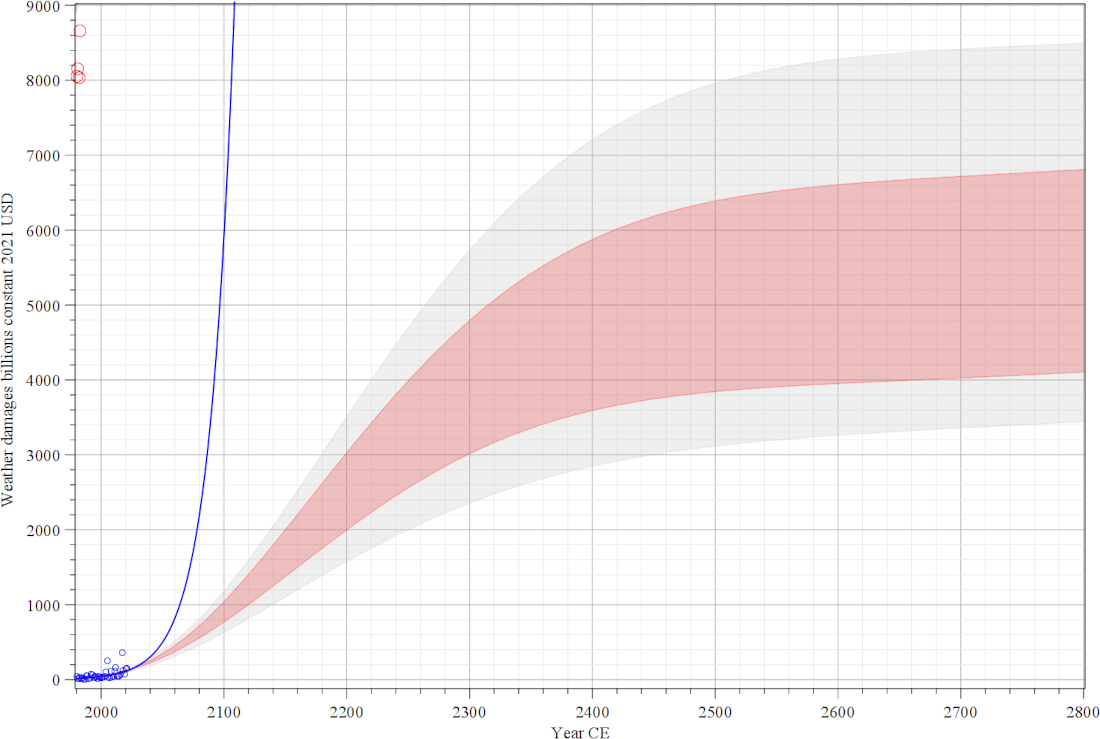}
\caption{\textbf{Heat conjecture weather damages curves estimating uncertainty. This is a slice through Tab A of figure 7 of the main paper.}  Blue datapoints, NOAA weather damages. Blue curve is the trajectory of the $WDe(y)$ (eq. \ref{eq:Fite_Dyr1}) exponential fit equation to NOAA data. Light grey region, shows the outer limits from lowest to highest scenarios. Inner pink region shows the high and low central scenarios. }
\label{Fig_WD_Heat_Conjecture_graph}
\end{figure}

The contrast between the $WDe(y)$ (eq. \ref{eq:Fite_Dyr1}) curve fit and the set of weather damages (WD) curves that result from using the heat conjecture are shown in figure \ref{Fig_WD_Heat_Conjecture_graph}. In this figure the exponential solid blue curve is shown extended to aid visualisation. 

Using these OHC curves, the regions shown in fig. \ref{Fig_WD_Heat_Conjecture_graph} are, taken together, our best estimate of future damages. This method may underestimate average damages due to critical point effects, where a critical point is an extraordinary excursion from these "average" projections. We did not attempt to show uncertainty bounds or standard errors because the range of the Archer-based CO$_2$ remainder equations overwhelm any such estimates. 

It should be kept in mind that each damages curve shown in figure \ref{Fig_WD_Heat_Conjecture_graph} represents a kind of average. For any specific year, damages should fluctuate, and as we see in a detail plot (fig. \ref{Fig_WD_Heat_Conjecture_Risk_Detail_graph}), years actually coincident with any specific WD curve should be rare. 

\subsubsection{Responding to objections to the OHC curve}
Because the OHC has the slowest rise, trailing temperature by centuries, this is raised as an objection by some, because aspects of land climate are thought driven directly and rapidly by temperature. We think that what is implicitly being objected to in the OHC versus temperature ($\Delta$Q vs. $\Delta$T) debate is that OPTiMEM does not include land and cryosphere heat. Particularly the absence of land heat modelling is a valid caveat of OPTiMEM, but including these at this time is not feasible for us---land and ice are high resolution GCM elements that are difficult to model and validate (\ref{lb:Heat}). Against the idea that this land and cryosphere heat content will prove critical, is the literature that shows that the ocean is primary to climate \cite{Raghuraman2024WarmingSpikeElNinoSouthernOscillation,Watson-Parris2024SO2ConfoundedByInternalVariability,Samset2024_2023TempsReflectSSTVariability}. In addition, the period of the enhanced NOAA damages dataset encompasses +0.77$\degree$ C of temperature change. (2021: +0.97$\degree$C, 1980: 0.2$\degree$C \cite{NASA_GSFC2021GlobalTemperature})

Like every model, it will be important to revisit predictions of this one periodically to see how well it does. We think this heat conjecture basis for estimating damages is as sound as any.

\subsection{Risk methods from the heat conjecture and current limitations of our method}
\label{lb:RiskMethods}
It is important that a method be available to estimate risk in any single year, and this allows risk estimation in any arbitrary contiguous set of years. To this end, we used a trailing average concept to estimate standard deviation change over time---a trailing standard deviation. The change in standard deviation allowed fitting of curves to the data produced by the trailing standard deviation.

It is also important to understand when and why our method will fail. We assume that weather damages are the result of release of energy powered by heat.  In the most correct analysis, estimates would be produced by a physics model, such as a refined-detail global climate model, that could be operated in a Bayesian manner. Readers should note that in this paper, OPTiMEM shows rising temperature and heat that we believe to be realistic. Climate models show some suppressive effect on severe weather while heat is going into the earth weather system and OPTiMEM does not contradict this. 

However, if heat change passes zero and becomes negative in the real world, this would mean heat is released from the ocean, and this should create massive storms \cite{Hansen2016IceMeltSeaLevelRiseAndSuperstorms}. Such storms could be much larger than anything that appears while heat is going into the system. \emph{At the point when $\Delta$Q becomes negative, this method will fail.} See panel D of figure \ref{Fig_Supp_Baseline_PopNorm_n2O-comparison_T}. Heat flow is shown as $\Delta$Q as the D panel in figures \ref{Fig_Supp_Baseline_PopNorm_19.5} through \ref{Fig_Supp_PermaFrost_PopNorm_S-aerosol_19.5}.

To estimate risk, a distance dataset $D_+[]$ $(n=18$ of $42)$ is generated. These data are the positive distance subset of each year's total damages distance from $WDe(y)$ (eq. \ref{eq:Fite_Dyr1}). We named our trailing standard deviation dataset $D\sigma[]$. Window size for the trailing standard deviation computation was tested from 3 to 9 sequential elements of $D_+[]$, and the 7 year window yielded the lowest log slope. This lowest log slope was chosen to provide conservative risk estimation. (See \S \ref{EstimationOfsigmafunction}).) To maintain a fixed number of points in the $D_+[]$ dataset, the previous 7 positive distance points are used which produces a more conservative slope than taking the past 7 points and accepting a highly variable number of points below 7. This may be overly conservative.

\begin{figure}[hbt!]
\centering\includegraphics[width=6.25in]{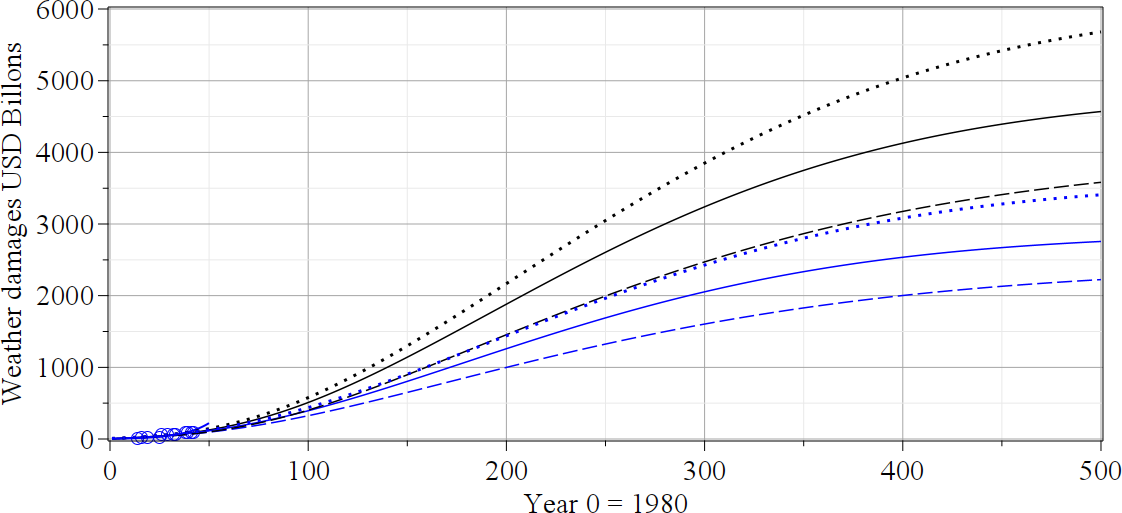}
\caption{\textbf{  1 $\sigma$ risk curves for scenarios} Standard deviation estimation curves based on the trailing standard deviation method discussed below (See section  \ref{EstimationOfsigmafunction}). Dotted curves are Permafrost thaw S-aerosol high (upper, black), and Baseline low (lower, blue) scenarios respectively. Solid curves are Permafrost thaw S-aerosol central (upper, black) and Baseline central (lower, blue) scenarios. Dashed curves are Permafrost thaw low (upper, black) and Baseline low (lower, blue) scenarios. Blue circles are $D\sigma[]$ datapoints for fit. A multiple of these 1 standard deviation curves is added to the appropriate scenario to obtain probability coverage using the Chebyshev equation \ref{eq:Chebyshev1}. These curves will fail when $\Delta$Q heat flow into the ocean turns negative. }
\label{Fig_Supp_1SD_Heat_curves}
\end{figure}

A curve fit to the $D\sigma[]$ dataset produced equation $E\sigma e_M(y)$ (eq. \ref{eq:EsigmaeMYr1}). This $E\sigma e_M(y)$ curve was then used as the target for scaling the heat curves over the 42 years of NOAA data using the binary search $R^2$ minima algorithm used in figure  \ref{Fig_Supp_ExptoHeatProjHeatConjecture}. This binary search generated a $y$ axis scalar of the heat curve of 0.2093 with $R^2$ = 0.99. Those heat curve fits yield figure \ref{Fig_Supp_1SD_Heat_curves}, which shows only the projected 1$\sigma$ curves.

\begin{figure}[hbt!]
\centering\includegraphics[width=6.25in]{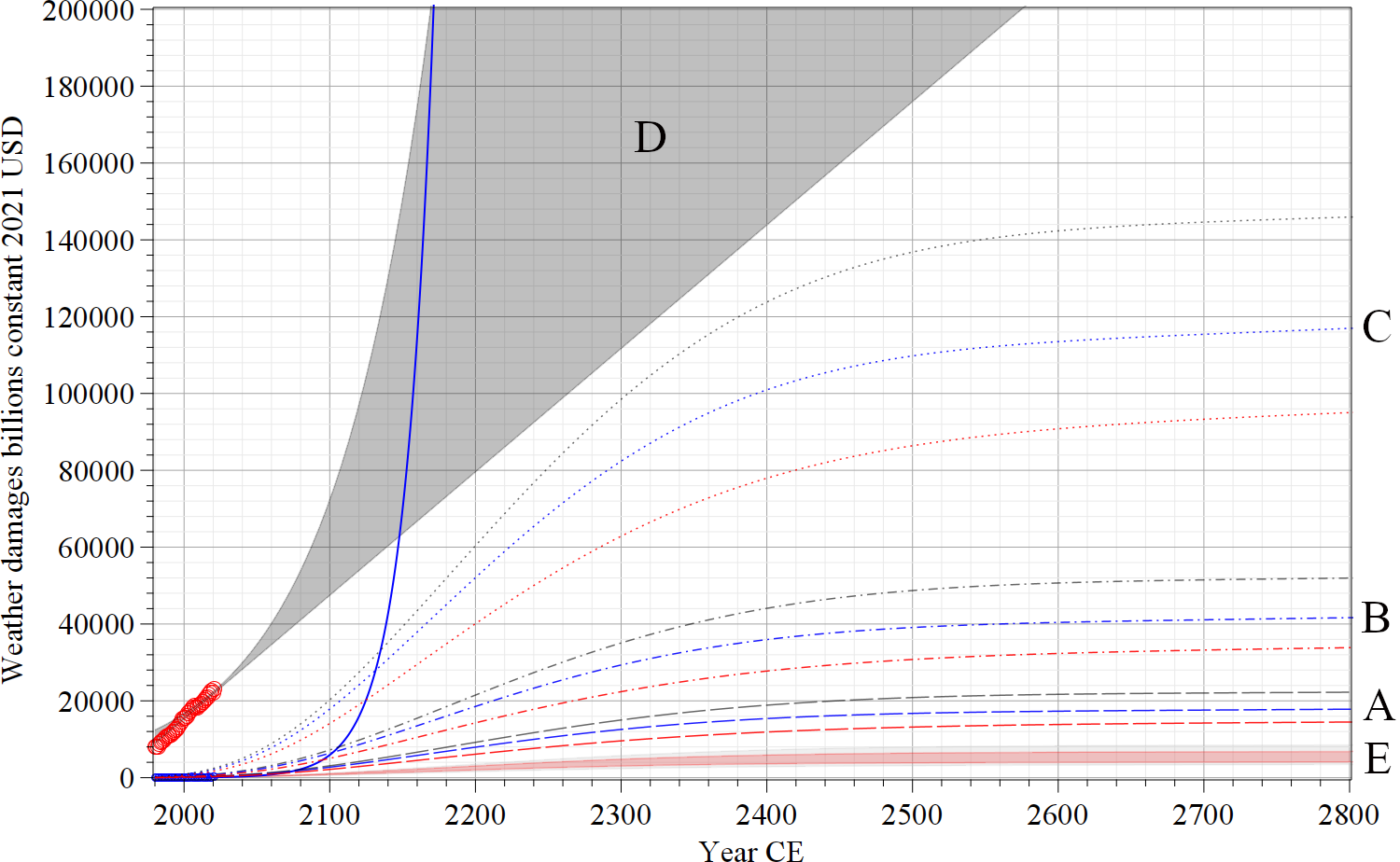}
\caption{\textbf{$\sigma$ risk curves show the risk by year for outlier weather damages in any single year using Permafrost thaw S-aerosol scenarios. This is a slice through Tab A of figure 7 of the main paper.} Dashed curves \textbf{A} (lowest group), are 1:10 years (10\%) risk, for low, central, and high CO$_2$. Solid curves \textbf{B} (middle group) are: 1:100 years (1\% risk), for low, central, and high CO$_2$. Dotted curves \textbf{C} (upper group) are: 1:1000 years (0.1\% risk), for low, central, and high CO$_2$. Gray shaded region \textbf{D} shows US GDP region from a low linear projection to a high 1.48\% per year growth. Shaded region \textbf{E} is the heat conjecture damages curves shown in figure \ref{Fig_WD_Heat_Conjecture_graph} (This is a slice through Tab A of figure 7 of the main paper.). Red circles are datapoints for US GDP. Blue solid circles (lower left) are NOAA weather damages, and blue exponential curve is the least squares curve fit $WDe(y)$ (eq. \ref{eq:Fite_Dyr1} to NOAA weather damages projected well past its validity region. Note that risk curves for slices on negative discount regions will be higher.}
\label{Fig_WD_Heat_Conjecture_Risk_graph}
\end{figure}

Compare figure \ref{Fig_WD_Heat_Conjecture_Risk_graph}  with figure \ref{Fig_WD_Heat_Conjecture_graph}. All  of the shaded regions of figure \ref{Fig_WD_Heat_Conjecture_graph} are compressed into the narrow band at the bottom of figure \ref{Fig_WD_Heat_Conjecture_Risk_graph}. What we see in figure \ref{Fig_WD_Heat_Conjecture_Risk_graph} is that only the 1 in 1000 year weather damages should get near GDP. However, if we look at figure 7 of the main paper, we can see that slices along negative discount rates will produce higher risk curves.

\begin{figure}[hbt!]
\centering\includegraphics[width=6.25in]{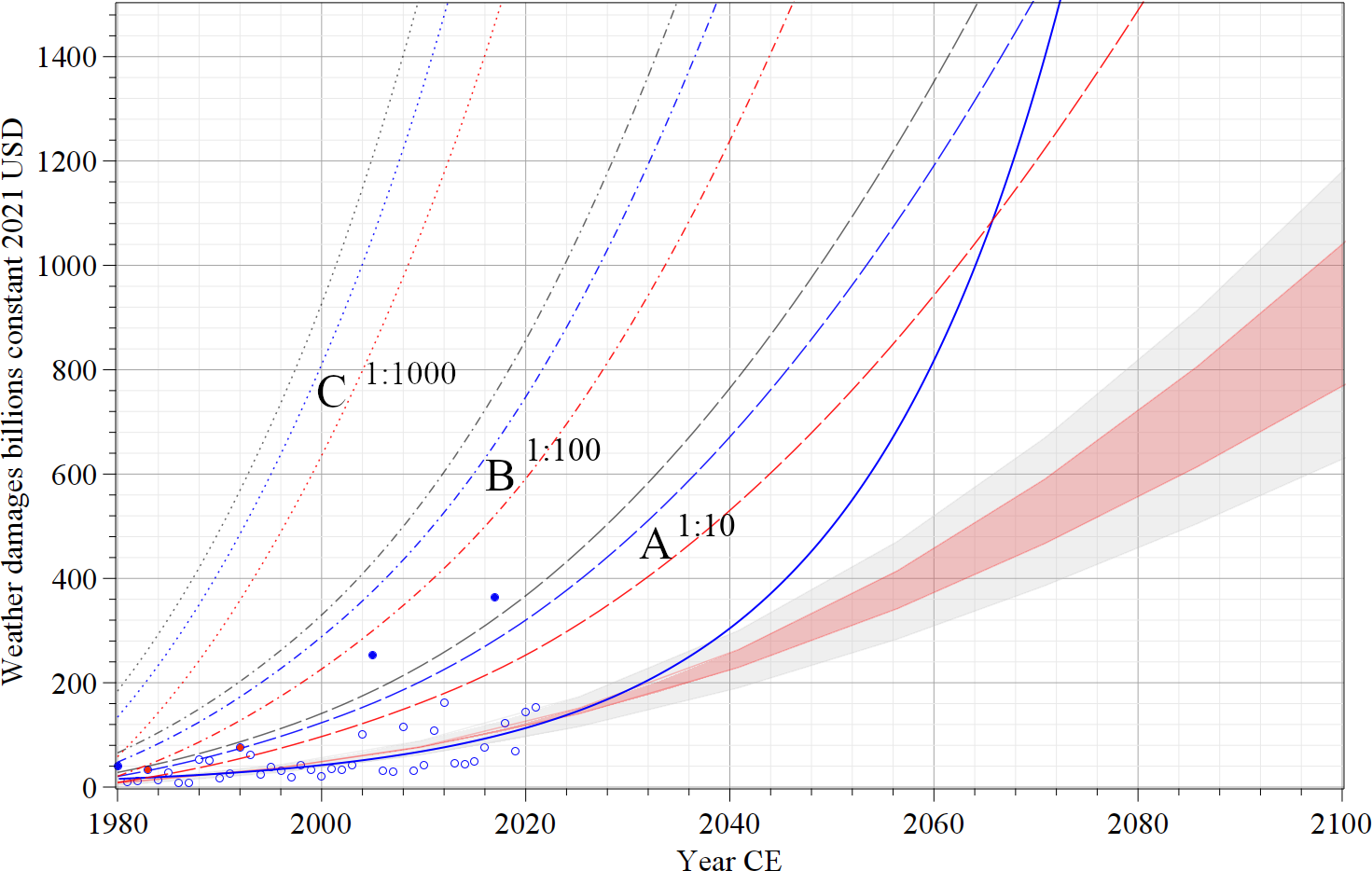}
\caption{\textbf{$\sigma$ risk curves detail based on uncertainty graph. This is the same graph as figure \ref{Fig_WD_Heat_Conjecture_Risk_graph} over a shorter period with risk curves added to it.} Note here the 3 filled blue circles above all three 1:10 year curves for the 42 year period of the NOAA weather damages. These 1:10 year curves, if perfect, should have 4 weather damages years at or above them. There are three 1:10 curves ( group \textbf{A}). Note that the central dashed curve would pick up 2 more weather damages year for a total of 5. This suggests that the 1:10 group are using a correct methodology for determining standard deviation in this case, and that the central and high curves of group \textbf{A} may be closest of the set to the real values. As in figure \ref{Fig_WD_Heat_Conjecture_Risk_graph}, the dashed curves \textbf{A} are 1 in 10 weather damages years; solid curves \textbf{B} are 1 in 100 years; and dotted curves \textbf{C} are 1 in 1000 weather damages years.}
\label{Fig_WD_Heat_Conjecture_Risk_Detail_graph}
\end{figure}

We can see by inspection of the risk detail graph (fig. \ref{Fig_WD_Heat_Conjecture_Risk_Detail_graph}) that there are three years that are above all of the 1:10 weather damages curves, marked as filled circles. In 42 years, one would expect $\approx$ 4 years that meet or or exceed the 1:10 curves. Three years is not perfect, but the central curve of this group A picks up another point. In 42 years of NOAA weather disaster damages, we have not yet had a 1:100 weather year, nor a 1:1000 weather year, which is to be expected.  These weather damages data points vary consistent with chaos theory, for which weather was one of the first applications \cite{Slingo2011UncertaintyChaosLorenz}. 

\subsection{Discussion of qualitative support for risk of outlier damages years.}
IPCC 2022, projects 3 to 39\% extinction of assessed species, "\emph{in terrestrial ecosystems, 3 to 14\% of species assessed will likely face very high risk of extinction at global warming levels of 1.5$\degree$C, increasing up to 3 to 18\% at 2$\degree$C, 3 to 29\% at 3$\degree$C, 3 to 39\% at 4$\degree$C}... \emph{In ocean and coastal ecosystems, risk of biodiversity loss ranges between moderate and very high by 1.5$\degree$C global warming level and is moderate to very high by 2$\degree$C but with more ecosystems at high and very high risk (high confidence), and increases to high to very high across most ocean and coastal ecosystems by 3$\degree$C (medium to high confidence, depending on ecosystem). Very high extinction risk for endemic species in biodiversity hotspots is projected to at least double from 2\% between 1.5$\degree$C and 2$\degree$C global warming levels and to increase at least tenfold if warming rises from 1.5$\degree$C to 3$\degree$C (medium confidence)}." (i.e. doubling = 4\% loss, and 10-fold is 20\% loss) \cite[p. 14]{IPCC_AR6_WG3_Report}.  

Water availability/quality will be reduced, "\emph{At approximately 2$\degree$C global warming, snowmelt water availability for irrigation is projected to decline in some snowmelt dependent river basins by up to 20\%, and global glacier mass loss of 18\% $\pm$ 13\% is projected to diminish water availability for agriculture, hydropower, and human settlements in the mid- to long-term, with these changes projected to double with 4$\degree$C global warming (medium confidence).}" (i.e. double = 40\% decline). "\emph{At global warming of 4$\degree$C, approximately 10\% of the global land area is projected to face increases in both extreme high and low river flows in the same location, with implications for planning for all water use sectors (medium confidence). Challenges for water management will be exacerbated in the near, mid and long term, depending on the magnitude, rate and regional details of future climate change and will be particularly challenging for regions with constrained resources for water management (high confidence)}" \cite[p. 14]{IPCC_AR6_WG3_Report}. 

Food will be significantly affected: "\emph{...food security risks due to climate change will be more severe, leading to malnutrition and micro-nutrient deficiencies, concentrated in Sub-Saharan Africa, South Asia, Central and South America and Small Islands (high confidence). Global warming will progressively weaken soil health and ecosystem services such as pollination, increase pressure from pests and diseases, and reduce marine animal biomass, undermining food productivity in many regions on land and in the ocean (medium confidence)}" \cite[p. 14]{IPCC_AR6_WG3_Report}. 

However, there are problems with the IPCC reports, because both scientific and IPCC report processes contribute, "\emph{the culture of climate science to “err on the side of least drama”, to not  be alarmists, which can be compounded by the consensus processes of the IPCC. Complex risk assessments, while more realistic, are also more difficult to do.}" \cite{Kemp2022ClimateEndgame}. While noting that the criticism of the conservative culture of science applies to the current work, the global temperatures OPTiMEM shows have not been seen for 6 million years or more \cite{Burke2018Plio}, and are in line with IPCC estimates. 

Various authors have less sanguine assessment of  the 2 to 4$\degree$C warming range, warning of catastrophic risks. "\emph{The current carbon pulse is occurring at an unprecedented geological speed and, by the end of the century, may surpass thresholds that triggered previous mass extinctions.}" "\emph{UN Secretary-General Antonio Guterres called climate change an “existential threat.” Academic studies have warned that warming above 5$\degree$C is likely to be “beyond catastrophic”, and above 6$\degree$C constitutes “an indisputable global catastrophe”.}" "\emph{There is ample evidence that climate change could become catastrophic. We could enter such “endgames” at even modest levels of warming.}" \cite{Kemp2022ClimateEndgame}.

To this discussion of risk and uncertainty, we add the grave concern that the rosy projections that are supposed to save civilisation have serious problems. The two regions, Germany and California, that have the most renewables installed have among the highest energy prices, low reliability, and little or no decrease in CO$_2$ over more than a decade. This is due to the intermittent supply problem of renewables \cite{Notton2018IntermittentRenewables,Allison2023InadequacyOfWind} that requires 100\% backup generation capacity to supply power during crucial times of the day and year. To this add the EROEI problem of windmills and solar panels that have $\approx$20 year life spans before replacement, and it appears that renewables are dependent on dispatchable electricity to exist  \cite{Tverberg2021FossilFuelProblem}. Discussion of renewables tends to elide basic facts about the fundamental grid operations requirement to exactly match supply to demand, as well as the economics of electricity production dating back to Insull. It is concerning that the most vocal scientist-advocate for renewables was thoroughly debunked in 2017 \cite{Clack2017MZJrebuttal}, and yet no change of course has occurred.

\section{Social Cost of GHG: SC-CO$_2$, SC-CH$_4$, SC-N$_2$O, SC-Fgas }
\label{SC-GHG}
\begin{figure}[hbt!]
\centering\includegraphics[width=4.25in]{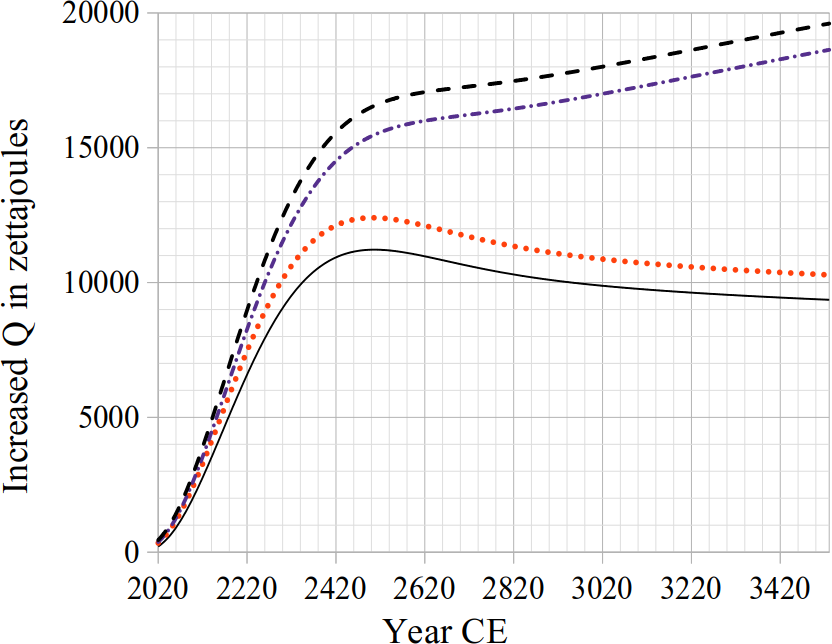}
\caption{\textbf{Heat contribution estimate by GHG category for the next 1500 years. This is a sample graph for 'baseline, normal population, high CO$_2$ remainder' scenario.} Solid black curve is CO$_2$. Red dotted curve is CH$_4$. Blue dot-dash curve is N$_2$O. Upper blue dashed curve is F-gas. In this graph, the y-axis distance from bottom CO$_2$ curve to 0 is CO$_2$ raw contribution. The y-axis distance from CH$_4$ to CO$_2$ is CH$_4$ raw contribution. The y-axis distance from N$_2$O to CH$_4$ is N$_2$O contribution. And y-axis distance from F-gas to N$_2$O is F-gas contribution. See figure \ref{Fig_T_by_gas_contribution}}
\label{Fig_Supp_Q-by-GHG_for-SCC-calc-v1}
\end{figure}

The social cost of carbon for this model is based on empirical datasets. We cannot assume all warming is a result of CO$_2$, because other gases make significant contributions, and those contributions will increase, as shown in sample graph \ref{Fig_Supp_Q-by-GHG_for-SCC-calc-v1}. The cost per tonne of each GHG is computed, starting with assuming that the relative change in heat generated by each GHG category is proportional to the weather damages. The point of assigning a cost to carbon is to spur change in order to cure the problem. That cost should be as accurate as possible, and anthropogenic GHG contributions of other gases than CO$_2$ should be properly accounted for, where they are unrelated to CO$_2$. 

\subsection{Simple English social cost algorithm description}
The social cost of a GHG is the amount of damages attributed to that GHG, divided by the tonnes of the GHG in the atmosphere. The cost per tonne in each year, over the span of years chosen for, is summed, applying a remainder $e$fold adjustment for all GHG except CO$_2$. For CO$_2$ the equations based on Archer's survey of climate models are used for adjustment instead of $e$-fold, which provides 3 different estimates for CO$_2$. After determining the remainder cost for a year, the discount rate is applied for that year. The discount rate is (1+d)$^n$ where $d$ is discount rate, and $n$ is years in the future. These resulting costs for years are summed together to obtain the social cost of the GHG (SC-GHG). The resulting SC-GHG is a global number, not specific to any one nation. 

\subsection{The general equations for SC-GHG:} 

\begin{align}\label{eq:GeneralSC-GHG}
\begin{split}
 SC_{GHG} =& \frac{GWD(fWD_m, y)  \cdot WDFraction(GHG) } {Tonnes
 _{GHG}} \\
\text{Where: } GWD()=& \text{ Global Weather Damages eq. \ref{eq:GlobalWeatherDmgs2}} \\ 
  y =& \text{ year } \\
 fWD_m  =& \text{ (eq. \ref{eq:Fit_Dyr2})  the appropriate function describing weather damage for this GHG}  \\
 WDFraction(GHG)=& \text{ The fraction of total weather damages attributable to this GHG or GHG fraction.} \\
 Tonnes_{ghg}=& \text{ Tonnes of GHG after emission, for year}
\end{split}
\end{align} 

\begin{align}\label{eq:GlobalWeatherDmgs1}
\begin{split}
 GWD(fWD_m, y) =& \frac{fWD_m(y)  \{eq. \ref{eq:Fit_Dyr2} \} } {U_{mGWP}} \times 10^9 \\
\text{Where: } y =& \text{ year } \\
 U_{mGWP}  =& \text{ USA  mean  share  of  gross  world  product } = 0.2573  \\
\end{split}
\end{align} 

\begin{flalign}
  \begin{aligned} \label{eq:Fit_Dyr2}  
fWD_m(y) =& \text{ weather damages scenario function } \\
\text{Where: } &\text{ } fWD_m \text{ may be 1 of the set of 32 WD functions $fWD_m(y)[]$ } \\\
& \text{where each WD function is a scaled heat Maple interpolation object.}
\end{aligned}
\end{flalign}

USA share of global GDP has dropped from 39\% to 25\% at  $\approx$ -0.23\%yr$^{-1}$ from 1960--2019, mostly due to growth of China \cite{FRED2021INTDSRUSM193N,FRED2021NYGDPMKTPCDWLD}. China's growth  is slowing, so the rate of USA GDP share loss should flatten. 

\subsection{CO$_2$ damages}
CO$_2$ damages include CO$_2$ plus the effect of CH$_4$ emissions due to energy extraction, plus the effect of CH$_4$ due to permafrost melting. The concept here is to try to assign to CO$_2$ effects that are caused by carbon consumption.  Essentially, everything not assigned to the Ag sector (which is Waste, Ag, and Other) is CO$_2$ related. To this, is added permafrost permafrost melt damages. This is only done at time of damages calculation.

\subsection{CH$_4$ damages}
\label{CH4_damages}
For this model, CH$_4$ damages are limited to the anthropogenic CH$_4$ for which humans can reasonably make abatement efforts, separate from energy extraction, and permafrost melt. This relationship is shown in table \ref{tab:IEAMtCH4} for 2021, which is in \S\ref{lb:CH4_EmissionsEstimation}. This is modelled using eq. \ref{eq:CH4fromCandPop} 'Population linked CH$_4$', where population ($p$) is obtained from eq. \ref{eq:PopProj1} \& \ref{eq:PopProj2}.  

In this model, when CH$_4$ is converted to CO$_2$, that CO$_2$ is added to CO$_2$ for OHC computation. This conversion is based on gas $e$fold $\tau$ per section \S\ref{CH4_efold}. Remember that first the climate model runs, and then the damages estimation makes use of scaled curves based on the climate OHC. 

We note that radical ozone depletion from interaction between warming, CH$_4$, N$_2$O, \cite{Portman2012StratoOzonDepletion} and natural halogens \cite{Anderson2017ChlorineBromineCatalysis,Anderson2018CouplingCatalysis} should create currently unexplored damages from ultraviolet light over North America, and possibly lengthen half-lives of CH$_4$, N$_2$O, and halogenated hydrocarbons (F-gas) molecules. Without a clear model for reference, no attempt was made to model this possible effect. 

\subsection{N$_2$O damages}
\label{N2O_damages}
N$_2$O damages are in one sense straightforward, as anthropogenic N$_2$O addition overwhelms natural N$_2$O increase from rising temperature. Currently, there is no modelling we know of that proposes natural N$_2$O increase will become a significant component, but deep time is suggestive this could be plausible \cite{ROBERSON2011_N2OandCH4Proterozoic,Junium2018EoceneN2O}.  We assume zero heating effect from other nitrogen gases, as current available literature indicates this is negligible, which may not be entirely correct. 

We also assume a fixed $e$fold $\tau$ N$_2$O in this model per section \S\ref{N2O_efold}, however, the $\tau$ value is unlikely to be completely correct, for two reasons. First, the rate of destruction of N$_2$O should rise with concentration. Second, N$_2$O depletes ozone, and higher concentrations should deplete ozone faster, thus, as ozone level drops the $\tau$ of N$_2$O should increase. Ozone depletes in complex ways, interacting with CO$_2$, CH$_4$, and N$_2$O  \cite{Portman2012StratoOzonDepletion, Prather2023ObservedDecreasingN2Olifetime}, and ozone also depletes from natural halogens \cite{Anderson2017ChlorineBromineCatalysis,Anderson2018CouplingCatalysis}.

In tables, the damages for N$_2$O are seen to be switched, with the low being higher than central and high being lower than central, although the low and high are fairly close. This is correct behaviour due to the adjustment equations we implemented from Byrne \& Goldblatt \cite[p. 157, Table 2]{Byrne2014RadiativeForcingEquations}. These adjustment equations account for light spectrum overlap of CO$_2$ and CH$_4$ with N$_2$O.

Anderson expects 6 orders of magnitude ($10^6$) increase in depletion of ozone over North America in summer due to increased ClO$^+$ (chlorine monoxide ion) from warming. How fast this halogen induced ozone depletion will progress is not clear. However, as halogen ozone depletion rises, this may slow the rate of depletion of N$_2$O, although UV and competitive hydroxyls from ClO$^+$ could make this complex.  Without a clear numeric model with a dataset to compare to, we could not base the rate of N$_2$O breakdown on relative to ozone, because ozone breakdown relative to ClO$^+$ ion, and halogenated hydrocarbons relative to ozone is not happening yet. This would be a significant modelling effort by itself \cite{Anderson2018CouplingCatalysis,Anderson2017ChlorineBromineCatalysis}. We decided not to attempt to create such a model in what is a complex area. Qualitatively, it appears probable that N$_2$O breakdown will rise with concentration, then slow down because ozone is depleted. Thus the $e$fold time of N$_2$O may increase. The impact of Anderson's prediction is not included in this model's results, and it should be quite significant. 

\subsection{Fgas (halogenated hydrocarbon) damages}
\label{Fgas_damages}
Fluorinated hydrocarbons and chlorocarbons are very strong and long-lived GHGs. We assume an $e$fold $\tau$ per section \S\ref{Fgas_efold} of 527 years.   

\subsection{Feasibility of separating damages from anthropogenic GHG emissions from natural variability}
\label{AnthroGHGdifferentiation}
There are multiple ways to interpret this idea that damages might be separated. In this model, we do not have the capacity to differentiate, we simply provide the cost going forward. We did try diligently to attempt to work out separating anthropogenic climate damages, because on its face, at first it sounded reasonable. After examining the literature and trying to come up with a method, it was abandoned as unfeasible and not as logical as it appeared at first glance. The literature that exists is mostly oriented towards proving that some specific weather event’s damages are definitely caused by anthropogenic climate change due to carbon, starting with the assumption they are not. But teasing out the other way, determining what is not caused, from the starting point that it is caused, is another story.  Certain basic premises, such as the presence of large populations and habitations in the way of disaster events not necessarily being related to anthropogenic carbon, are at best highly questionable. 

\subsubsection{World Weather Attribution project}
\label{WorldWeatherAttribution}
To get at this question of attribution properly requires the techniques being developed by World Weather Attribution \cite{WorldWeatherAttribution2024}. They have been working on this problem for a while and have published on their methods \cite{Haustein2017ARealTimeGlobalWarmingIndex_WeatherAttribution,Philip2020AprotocolForProbabilisticExtremeEventAttributionAnalyses,vanOldenborgh2021PathwaysAndPitfallsInExtremeEventAttribution,vanOldenborgh2022AttributingandProjectingHeatwavesIsHardWeCanDoBetter,Friederike2022HowtoProvideUsefulAttributionStatementsLessonsLearnedfromOperationalizingEventAttributioninEurope,ZACHARIAH2023Attributionofthe2015DroughtInMarathwadaIndiaFromAMultivariatePerspective}. At this time, these methods are in development and have been used for some events. The point of the World Weather Attribution project is to be able to say to the public, yes, this event was due to climate change to X extent.  The Climate Change Impacts and Risk Analysis (CIRA) project, has published on methods for analysis of future climate impacts. The CIRA publication list is extensive, comprising: general climate impact risk modelling (4), and specific models in the health (18), infrastructure (14), electricity (4), water resources (8), agriculture (2), and ecosystem sectors (9) \cite{CIRA2024CIRAjournalPubs}. These are examples of work that could be folded into a giant project to instrument multiple GCMs with valid determinations of ranges of economic damages based on geography.  Every type of damages determination will need to be represented, and each damages algorithm will require considerable work. However, it will be some time before a Bayesian approach can be feasible with high resolution GCM models. Using 2024 supercomputing clusters, these models push the limits of current day computing capacity. We suggest using multiple GCMs instead, to generate a dataset such as Archer was able to work with \cite{Archer2009AtmosphericLifetimeOfCO2}. 

\subsubsection{Do GHGs have natural variation and does this make separating damages possible}
If the idea is that GHGs themselves have natural variation, natural CO$_2$ is very stable. A distant second place to fossil fuel production is CO$_2$ from wildfire. 

\begin{quote}
"Globally, CO$_2$ emissions from fossil fuels and land-use change averaged 9 billion metric tons of carbon (GtC) per year since 2000, whereas fire CO$_2$ emissions were approximately 2 GtC per year (5). Eighty percent of the carbon released by fires is taken up by vegetation in subsequent growing seasons; the remaining 20\% remains in the atmosphere much longer and contributes to the build-up of atmospheric CO$_2$. Fire emissions from global forest ecosystems have been increasing since 2000" Zheng, et al. 2023 \cite{Zheng2023RecorodBorealCO2}.
\end{quote}

For this model, records of CO$_2$ production are fitted to actual data and having them line up was a basic step (fig. \ref{Fig_Supp_C&CO2_BaselineScenario1}. CO$_2$ records are quite long-term, and include inputs from natural sources. To add enough CO$_2$ to make serious changes to the CO$_2$ modelling would appear to require some catastrophic events. Such a natural CO$_2$ source could be volcanic, and would require a huge volcanic system to matter much. Such sources do not not appear in OPTiMEM, and fall into tipping points and exogenous "black swan" events. 

There is no known natural source of CH$_4$ that is not linked to increased temperatures, but not all of CH$_4$ sources are in OPTiMEM because there is insufficient data. CH$_4$ escaping from oil, coal and gas extraction are greater are much greater than natural, at least for now. CH$_4$ is one of the potential tipping point concerns. For instance, it is not settled what the impact of methane clathrates from the deep ocean might be, and part of methane seep increase could be from possible deep CH$_4$ sources that may have been capped by methane clathrate in the seabed for thousands or millions of years (\S\ref{table_tipping}. We do not include anthropogenic increases in CH$_4$ from the tropics due to temperature because we don’t have a source for numbers to model it, but such would also be anthropogenic. We include permafrost melt, which is anthropogenic. (\S\ref{PermafrostCH4}). 

N$_2$O has natural and directly human sources. We treat all the increase as anthropogenic based on Tian \cite{Tian2020N2OSources_sinks}, because Tian accounts for the net effect. The only thing that matters for modelling is net increase in N$_2$O, so we treat it as a nearly “black box” increase system that is discussed in \S\ref{Nitrogen_model}.  It is possible that N$_2$O production increased significantly during the PETM from biological sources \cite{ROBERSON2011_N2OandCH4Proterozoic}. This would fall into unknown potential tipping points.

There may be effects on Fgas, N$_2$O and CH$_4$ due to warming induced increase in halogens in the atmosphere breaking them down more rapidly and destroying ozone over the northern hemisphere \cite{Anderson2017ChlorineBromineCatalysis,Anderson2018CouplingCatalysis}. However we did not attempt to model this effect. This kind modelling is suitable for a full-blown GCM. 

In any case, these sources of CO$_2$ and CH$_4$ are anthropogenic, including from increased wildfire. 

\subsubsection{Can natural variability of weather allow separation of anthropogenic from natural?}

Sea level rise acceleration began roughly 1800 \cite{Jevrejeva2008RecentGlobalSeaLevelAccelerationStartedOver200YearsAgo}, with 6 cm of rise in the 19th century, and 19 cm in the 20th century, indicating that anthropogenic warming was in process then. Thus, finding a baseline for a non-warming human world is quite problematic.  Weather is a chaotic dynamics system that never can come to equilibrium. When the heat content of the atmosphere, land, cryosphere and ocean increases, the whole system has higher energy, and variation persists because of the type of system it is. The proper way to make use of the natural variation of the system is to use it to generate risk curves. We do that. See \S\ref{lb:RiskMethods}  (main paper \S\emph{2(c)} ). 

All of the net increase in GHGs is anthropogenic in OPTiMEM (Supplement \S\ref{lb:ClimateModelandScenarios}. Therefore, this should mean that essentially all of the increase in temperature is anthropogenic, and virtually all of the increase in OHC is likewise anthropogenic. That should mean that the OHC projection curves should be showing damages that are anthropogenic (fig. \ref{Fig_WD_Heat_Conjecture_graph}). These damages curves curves do not have variability baked into them, they have uncertainty. Variability is what gives access to risk, as seen in figure \ref{Fig_WD_Heat_Conjecture_Risk_Detail_graph}. Data such as sea level rise, CO$_2$, temperature measurements, Atlantic storm records, and carbon consumption, go back to 1880-1890 CE. The total NOAA weather damages dataset starts in 1980. 

The real world weather damages starting point is 1880-1890, which is the start of the industrial revolution. Sea level  rise has been occurring continuously since that time \cite{Church2011SeaLevelRiseLate19th_Early21stCentury, NASA_GSFC2021GlobalMeanSeaLevelTrend}. Sea level rise starting in 1880 means that weather and damages to the economy should increase as well starting from that point. So what baseline should be chosen? 

In figure \ref{Fig_WD_Heat_Conjecture_graph} one can get a sense for how much  of these increased damages are going to be above and beyond. Any baseline chosen should essentially disappear, dwarfed by what is coming. 

\subsection{Use of Willingness To Pay (WTP) in current SCC models}

The social cost of carbon (SCC) has been defined as \emph{"...how much we are willing to pay today ... to reduce emissions by a unit."} \cite[p. 11]{Stern2021TheEconomicsofImmenseRiskUrgentActionandRadicalChange}, and alternatively as the economic cost \cite{Nordhaus2017RevisitingTheSCC}. The willingness to pay definition used rests on the idea that our willingness to pay can be substituted for the actual cost. We can use the DICE assumptions model results of figure \ref{Fig_Supp_DICE_Unhinged_Scenario_18.1} to show that this idea is unlikely to be true. Boiling the oceans cannot be easy to pay for, and civilisation will be destroyed long before the ocean boils. 

 The WTP concept is borrowed from established practice of small scale environmental evaluations \cite[p. 44]{dietz2021supplement}. For example, when evaluating a football stadium to be built on wilderness reserve land, if willingness to pay to preserve the wilderness land needed for the stadium is less than the present value (PV) of an income stream generated by the stadium, then WTP indicates to build the stadium. Use of WTP in this way is defensible when the impact of wilderness loss is so low that nearly zero can be treated as zero. We accept this kind of method all the time. In engineering, for instance, ignoring very small measurement differences as equal to zero difference appears in the concept of tolerance for parts.  

This logic used for small scale projects is what the DICE model, citing Manne \cite{manne1995merge,manne2005merge}, accepts that what a government is willing to pay for environmental matters on a large scale is the true willingness to pay. This logic precisely ignores the small damage from each economic activity undertaken.  

This concept of WTP as applied to SCC at a global scale makes two questionable assumptions. First, it conflates human willingness to pay with actual damages that human society will experience in the future from failure to pay. This conflation is due to the descriptivist method \cite[p. 47]{Stern2007EconomicsofClimateChangeSternReview} implicitly used by DICE.  Second, it implicitly includes the assumption that error near zero is set to zero. However, when expanding  WTP to a continent or the globe, neither of these assumptions work. This is illustrated with an example based on the wilderness damages case. We chose wilderness damages because what willingness to pay fails to account for is all of the globe that is not developed and/or categorised into human benefits. Thus, wilderness of the globe is as precise an analogue to that which is not developed as we can come up with. 

The USA protects $\approx$21 million hectares of wilderness \cite{DIETZ2015WorldsLArgestWilderness50years}, and global protection is $\approx$424 million hectares \cite{Eidsvik1989InternationalStatusWilderness}. A stadium occupies $\approx$20 hectares. 20 hectares is 953 parts per billion (ppb) of the USA's wilderness and 47 ppb of global wilderness without applying an environmental damages per hectare factor to the stadium's 20 hectares. This level of error is in the nano-scale range. It is defensible to accept that nearly zero can be ignored for 20 hectares (See: \S\ref{WTP_StadiumVGlobe})

However, when dealing with the entire planet, there is no separable number of hectares that can be set aside to use as a divisor to the damages. So we have the planetary hectares both above and below in this fraction, and they cancel to unity. We also need to introduce an environmental damages factor $E_D$. The denominator of this fraction is 1-$E_D$. This value will be too high to ignore as it will not be a billionths fraction approximating zero (eq. \ref{eq:WTP_for_globe}).

Thus, we see that improper use of  WTP applied to the globe neglects environmental damages. Because this is the logic used to justify accepting that the 2\% that government paid for environmental matters in 1995 is the true cost of environmental damages that will occur, that makes this use of WTP invalid. 

Seeking empirical confirmation, damages from one cause that is climate change linked, fire, in one nation, (Australia 2019/2020) reached 17 million ha in a 12 month period. This is 4\% of all global wilderness, which is a level that cannot be ignored. Thus, we had to rethink the concept of willingness to pay when applied to climate. 

\section{Qualitative literature support for this model}
\label{QualitativeSupport}
 Each IPCC report has increased impacts, \emph{"all five [ecosystems, extreme weather, global impacts, economic and ecological impact and abrupt transitions] are now at 'high' or 'very high' for 2$\degree$C to 3$\degree$C of warming"} \cite[p. 5]{Kemp2022ClimateEndgame}. Regarding higher temperatures, \emph{Equilibrium climate sensitivity (ECS) is 'heavy tailed'---higher probability of very high values than of very low values. ... 18\% probability ECS >4.5$\degree$C.} \cite[p. 2]{Kemp2022ClimateEndgame}. And \emph{"... above 5$\degree$C is likely to be 'beyond catastrophic' and 6$\degree$C an 'indisputable global catastrophe'. ...global temperature rise of 6$\degree$C 'imperils even the survival of humans as a species'"} \cite[pp. 3, 5]{Kemp2022ClimateEndgame}. These concerns are echoed by others \cite{RIPPLE2023RiskyFeedbackClimate}. This literature is in line with OPTiMEM's projections (Supplement: \S6.5). 

\subsection{Effects on production} 
The 2022 UNDRR report warns, \emph{"...within the concept of planetary boundaries show a dangerous tendency for the world to move towards a global collapse"} \cite[pp. 51-52]{UNDRR2022OurWorldAtRisk}. UNDRR speaks of human systems, and impact of the \emph{"climate emergency"} on those systems \cite[pp.  ix, xiii, 41, 202, ]{UNDRR2022OurWorldAtRisk}. This report introduces \emph{...life years lost ... the time required to produce economic development and social progress. It provides a way of measuring setbacks to social and economic development across countries and regions} \cite[p. 34]{UNDRR2022OurWorldAtRisk}. 

Burke identifies a non-linear negative effect of temperature on production, saying \emph{"...overall economic productivity is non-linear in temperature for all countries, with productivity peaking at an annual average temperature of 13 $\degree$C and declining strongly at higher temperatures. The relationship is globally generaliseable, unchanged since 1960, and apparent for agricultural and non-agricultural activity in both rich and poor countries."}  \cite{Burke2015GlobalNonlinearEffectOfTemp}. However, if we take into account that $\approx \frac{1}{3}$ of global population should be outside of the human temperature niche by 2100 with 2.7$\degree$C of warming \cite{Lenton2023QuantifyingHumanCostOfGlobalWarming} , it is obvious that the impact could be catastrophic.

The 2022 IPCC report sets 2025 as the year for compliance with AR6 recommendations to hold warming to 1.5$\degree$C maximum temperature\cite[SPM-21]{IPCC_AR6_WG3_Report}. Based on our current modelling experience, we do not believe those targets would hold the  world to this 1.5$\degree$C target, nor can they be met.  

\subsection{Support for extreme weather}
 Ocean current studies  show ocean current circulation speed increase 15\% $\pm$ 12\% per decade \cite[p. 2 \& fig. 1]{Hu2020DeepAccelerationOfGlobalOceanCirculationOverPast2Decades}.  Modelling supports this, and indicates that the acceleration should affect 77\% of the global ocean \cite{Peng2022SurfaceWarmingInducedGlobalAccelerationOfUpperOceanCurrents}, due to upper layer heating making the most active current layer shallower, while volume is largely static.  Modelling also indicates that increasing wind should accelerate warming of the ocean by $\approx$17\% even in context of the AMOC \cite{McMonigal2023WindAccelerateWarming}. The effect of this should be faster heating and greater weather effects for the global ocean. 

 \subsection{Forest and Wildfire}
 \label{lb:ForestAndWildfire}
 Since 2016, the Amazon and global forests have continued to lose ground, with human activity (mostly fire) a major driver \cite{NASA2022UptickAmazonFire,Gatti2021Amazonia_as_a_carbon_source,GlobalForest2022}. The Amazon is losing resilience, and may be near an irreversible transition with major climate effects (not included in OPTiMEM) \cite{Boulton2022Amazonresilience}. The net effect of forestation should be less than simplistic calculations, as albedo and greenhouse gases offset up to $1/3$ of CO$_2$ sequestration \cite{Weber2024FeedbacksOffsetOnethirdForestationCO2removal}. 
 
 2021 was a record year for boreal fire CO$_2$, contributing 23\% of global fire CO$_2$, and \emph{"rapidly increasing forest fire emissions in extratropic"} fire areas suggesting boreal fires could prove problematic climate feedbacks \cite{Zheng2023RecorodBorealCO2,Jones2024GlobalRiseInForestFire_ClimateChangeInExtratropics,Balch2024Fastest_most_destructive_fires2001to2020}.  U.S. western wildfire risk is increasingly dire with 17 of the 20 largest California wildfires by acreage, and 18 of the 20 most destructive wildfires by number of buildings destroyed, occurring since the year 2000.
In four of the last five years (2017, 2018, 2020 and 2021) the total, direct wildfire costs across the Western U.S. have been 5 to 10 times more costly than most every inflation-adjusted wildfire year on record (1980-2021). Australia's 2019-2020 wildfire losses, $\approx$4\% of global wilderness, have already been mentioned. Sub-alpine rocky mountain wildfire is at a 2000 year peak \cite{Higuerae2021ForestsBurnGT2000yr}. Higher night-time temperature increases vapor pressure deficit (VPD) causing 5 additional flammable nights globally 1979-2020. Night wildfire increased intensity by 7.2\% 2003-2020 \cite{Balch2022Warming-night-time-fir}.  Climate related tree disease \cite{Juroszek202030yrReviewClimateCrop,REED2022Beechdisease,Ewing2020BeechDisease} exacerbates fire susceptibility and forest loss. 
 
 \subsection{Agriculture and flora impacts}
A 2021 OECD report discusses the grave threat to the global food system represented by the loss of the AMOC combined with increased global average temperature \cite[p. 152]{OECD2021ManagingClimateRisksDamages}. This report estimates, "\emph{approximately half of the remaining suitable land is lost for wheat and maize. ...gains in suitable area for rice cultivation are dwarfed by losses in suitable area from wheat and maize. ...[AMOC] collapse combined with climate change would have a catastrophic impact.}"

Heat is an agricultural concern  \cite{Parker2020FoodSecurityClimate,MariemPlants2021CO2-starch-nutrition-stress}, with Parker et al, saying, \emph{"Current extreme heat adaptation practices may not be sustainable under future climatic conditions or regulatory constraints."} Karki, et al, using data up to 2016 found, \emph{"...farmers worldwide have been experiencing changes in climate mainly regarding rising temperature, unpredictable and reduced rainfall. A majority have witnessed reduced agricultural production."} \cite{Karki2020Farmersclimate}. "Flash droughts"---rapid, subseasonal drought of 20-45 days---have intensified \cite{Yuan2023FlashDroughtsClimateChange}. Flash drought is harder to predict and mitigate. Thus, not seeing climate model's predicted rising humidity in arid regions is alarming \cite{Simpson2024ObservedHumidityInDryRegionsContradictClimateModels}. We must consider the possibility that climate models are biased toward more desirable outcomes.

\subsection{Cryosphere ice melt}
\label{lb:CryosphereIceMelt}
Climate modelling has used heat diffusion to model ice melt. However, the geological record shows grounded ice retreating as fast as 610 meters in a day \cite{Batchelor2023RapidIcdRetreatHundredsMetres}. Current measurements in Antarctica show ungrounded ice retreating in a range from 1.3-1.7 km y$^{-1}$ \cite{Milillo2022RapidGlacierRetreat}.  New observations of ice melt show striations and highly variable melt should break up Thwaite's shelf much faster, suggesting 0.65-3.2 metres of sea level rise could occur in a decade (i.e. 2032 CE) \cite{Pettit2021ThwaitesEasternCollapse, Fox2022TheComingCollapseSciAm}. Long-term examination generally supports this \cite{Domgaard2025HalfCenturyOfDynamicInstabilityFollowingTheOcean-drivenBreak-upOfWordieIceShelf}. Ice melt in Greenland is also melting faster than expected for similar reasons \cite{Schulz2022ImprovedMeltRate}. In addition, coastal urban areas are subsiding, often as fast as current rate of sea level rise \cite{Ohenhen2024DisappearingCitiesOnUScoasts}. 

\section{Ocean heat, datasets and relationships}
\subsection{General notes for climate economists}
\begin{itemize}
    \item  We agree with Klotsbach (\S\ref{lb:Klotzbach}) that it looks like minimum sea level pressure (MSLP) is probably a better measure to use to gauge severity of storms relative to wind (fig. \ref{Fig_Supp_MillibarstoWind}). Practically speaking, this may improve translating GCM storms into damages.
    \item  We agree with Knutson \cite{Knutson2022WarmingAndHurricanes} that evidence suggests little increase in number of severe storms (fig. \ref{Fig_Supp_AllStorms_1ZJ}B, but large increase in high total storm energy (fig. \ref{Fig_Supp_AllStorms_1ZJ}A) due to warming, a great deal of it due to rain (fig. \ref{Fig_Supp_Rain_Wind_Ratio_PowerLaw}).  This makes basic sense in the physics of ocean heating, because with higher atmospheric temperatures paralleling sea surface temperature (SST), there should be little change in resistance to establishing the necessary heat pipe to the cold upper atmosphere. 
    \item  For near-term climate change, we are still a few years away from being able to judge whether there there is an increase in hurricane activity due to absolute Atlantic SST or because the Atlantic has warmed more than other basins (the relative SST hypothesis) as discussed by Vecchi, et al, \cite[p. 688, Fig 1]{Vecchi2008WhitherHurricaneActivity} \cite{Swanson2008NonlocalityAtlCycloneIntensity}.   
    \item  The ratio between rain energy and storm energy in our analysis is very high. Our analysis confirmed, and goes beyond Landsea's 400:1 ratio \cite{Landsea2018Energy400to1} finding a power law relation with 400:1 at the low end and a high end above 10,000:1 using equations from Emanuel \cite{Emanuel1986AnAirSeaInteractionTheoryforTropicalCyclonesPartISteadyStateMaintenance} (fig. \ref{Fig_Supp_Rain_Wind_Ratio_PowerLaw}). This indicates that relatively small increases in wind energy should be associated with quite large changes in rain.    
\end{itemize}
\subsection{Storm energy}
This section examining fundamentals of storms was done to validate and clarify our understanding of weather dynamics and how it can be expected to respond. 

\subsubsection{Fundamentals of storms}
Storms are driven by heat across a temperature differential. Like any other heat engine, the size of the temperature differential determines the efficiency of the engine. In the case of hurricanes, the stratosphere/tropopause upper end of the heat pipe is $\approx$ -60 $\degree$C. The lower end of the heat pipe is the sea surface temperature (SST), which empirically requires $\approx 26 \degree$C or more  to form a hurricane \cite[p 591]{Emanuel1986AnAirSeaInteractionTheoryforTropicalCyclonesPartISteadyStateMaintenance}. A hurricane will have a heat pipe differential on the order of $85 \degree$C or more. 

To establish the heat pipe into the upper atmosphere, enough calm must prevail to allow evaporating water to form a consistent updraft. As water vapor rises, the air cools, and water condenses out of the rising air. The energy of condensation is the same as the energy of vaporization, but opposite sign, so the heat energy is donated to the cooler upper air. That heat from the surface is mostly radiated out into space. The condensed cool water can only spread out sideways, where it forms clouds. As the heat pipe strengthens, the updraft creates winds toward itself. As the storm enlarges, coriolis effect turns this into a spiral. Higher wind speed increases the rate of evaporation from the ocean, which feeds more rising warm air and water vapor into the heat pipe. When the storm moves over land, it loses energy because there is usually less heat energy and water vapor on land to feed it. However, water is not absolutely required, just heat, and some rare storms are able to maintain quite a high level of energy moving over land given the right conditions. 

\subsubsection{Limitations on storm data}

The most basic limitation is that satellite data only exists since 1979-1980, and full storm track data since $\approx$2004. North Atlantic (NA) storm data gathering formally began in 1851 \cite{Klotzbach2022EmailDatasetAnalysis}. The North Atlantic dataset is, thus, the largest, which makes it the primary data examined for the climate/weather relationship. This dataset is also of special significance for the United States, which is targeted by these storms.

The rainfall estimation equation we used to understand storm relationships to energy (eq. \ref{eq:EmanuelRain}) is less sophisticated than the wind energy equation is, and suffers from lack of detailed empirical data, but this is the generally accepted equation. Its foundation is the average rainfall per unit area estimated for average hurricanes. It would be helpful to conduct studies of precipitation from the eyewall out to the periphery of hurricanes for a large number of hurricanes. This could allow development of equations that can estimate better for simpler equation based models like ours, and confirm how wind energy relates to rain energy. There may be a way to gather this data from satellites. 

\subsubsection{Storm energy}
\label{lb:StormEnergy}
Method: We obtained the accumulated cyclone energy (ACE) dataset \cite{Knapp2018_IBTrACSdatasetV4} and the hurricane HURDAT2 dataset \cite{Landsea2013AtlanticHurricaneDatabaseUncertaintyandPresentationofaNewDatabaseFormat,hurdat2-1851-2021-041922}. 

The Atlantic storm wind energy was calculated (eq. \ref{eq:EmanuelWind}) using Emanuel \cite[p 107]{Emanuel1999PowerOfHurricane} \cite[p 595, eq. 48]{Emanuel1986AnAirSeaInteractionTheoryforTropicalCyclonesPartISteadyStateMaintenance} and the HURDAT2 dataset from 2004-2021. The equation was validated with Emanuel's example, and that of Landsea \cite{Landsea2018Energy400to1}.  Joules were computed for each 6 hour segment by multiplying by 21,600 seconds, and summed for each storm.

\begin{align}
    \begin{split} \label{eq:EmanuelWind}
 E_{Wind}(r_o, M_{s}) &=   1000 \pi \int_{0}^{1.852*r_o} \rho C_d (x^{-\beta} \frac{(1852*M_{S})}{60})^3 x \; dx\\
 \text{Where: } E_{Wind} &= \text{ wind energy in Joules} \\
 r_o &=\text{ Outer storm radius} \\\qquad M_{S} \text{ (Max sustained wind velocity)} &= \; \frac{\text{metre}}{\text{second}} \\ 
 \rho \text{ (air density)} &= 1.15 \;\frac{kg}{m^3} \\ \qquad C_d \text{ (drag)} &= 2\cdot10^{-3}  \\
 \beta \text{ (velocity exponent)}  &= 0.6 
    \end{split}
\end{align}

 Rainfall was computed using eq. \ref{eq:EmanuelRain}. 
 
 \begin{align}
    \begin{split} \label{eq:EmanuelRain}
 E_{Rain}(r_o) &=   \pi \cdot \Delta R \cdot (1852 \cdot r_o)^2 \cdot \Delta H \cdot \\
 \text{Where: } E_{Rain} &= \text{ rain energy in Joules} \\ r_o =\text{ outer storm radius} \\
 \Delta R \text{ (rainfall in 6 hour period)} &=  \frac{0.015 \text{ m}}{4} \qquad \\
 \Delta H \text{ (heat of condensation) }  &= 2.257 \cdot 10^6 \;\frac{J}{kg} \\
    \end{split}
\end{align}

\begin{figure}[hbt!]
\centering\includegraphics[width=6.25in]{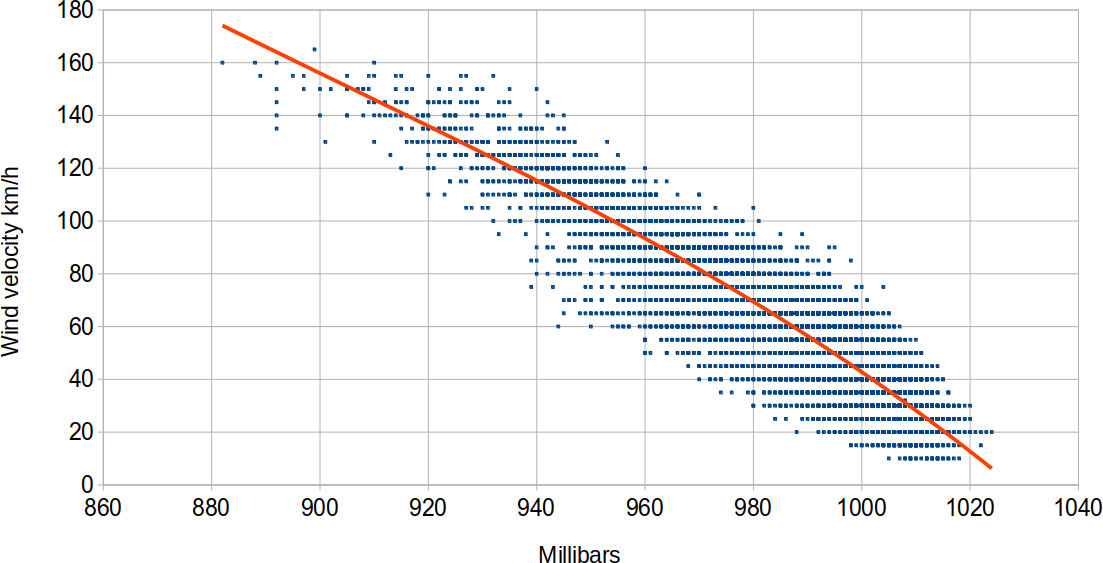}
\caption{\textbf{Millibars to wind velocity.} See eq. \ref{eq:MillibarToWind}. This relationship has a fairly good R$^2$ value of 0.87. So the wind velocity normalization equation is less idealized than eq. \ref{eq:MillibarToWind}. The stratification visible in the data is due to records rounding to the nearest 5 kph. This chart supports Klotzbach's advocacy for using minimum sea level pressure (MSLP) as a better predictor of hurricane damage. MSLP is also simpler to measure.}
\label{Fig_Supp_MillibarstoWind}
\end{figure}   

\begin{figure}[hbt!]
\centering\includegraphics[width=6.25in]{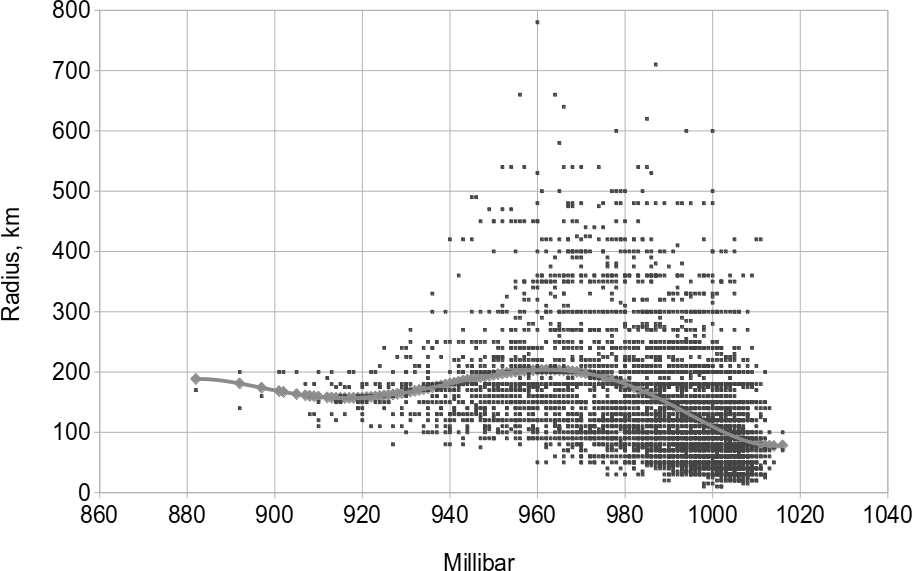}
\caption{\textbf{Millibars to storm radius.} See eq. \ref{eq:MillibarToWind}. For this relationship, there is a high degree of variability, so our computed result in fig. \ref{Fig_Supp_Norm-ZJ-to-ACS}, where ZJ is compared to ACE is a bit idealised. It is useful within the limits of this equation fit.}
\label{Fig_Supp_MillibarstoStormRadius}
\end{figure}   

\begin{align}
    \begin{split} \label{eq:MillibarToWind}
 V_w(m) =& -1.44996275125275\text{ E-05} \, m^3 + 0.0392465858562361 \, m^2 \\ 
&- 36.4069995901403 \, m + 11702.8069337238 \\
R^2 =& \, 0.87  
  \qquad \text{ fig. \ref{Fig_Supp_MillibarstoWind}}\\
 \text{Where: } V_w =&  \text{ wind velocity in km h$^{-1}$ }   \\
  m =& \text{ millibars, for the range 880 to 1016}
  \end{split}
\end{align} 

\begin{align}
    \begin{split} \label{eq:MillibarToRadius}
 R_S(m) =& 1.404524 E-7 m^5 - 6.627979 E-4 m^-4 + 1.2499610361 m^3 \\ & -   1177.5754451 m^2 + 554193.57911 x - 104233593.76 \\
R^2 =& \, 0.20 
  \qquad \text{ fig. \ref{Fig_Supp_MillibarstoStormRadius}}\\
 \text{Where: } R_S =& \text{ Storm radius km }   \\
 m =& \text{ millibars, for the range 880 to 1016}
    \end{split}
\end{align}

\begin{figure}[hbt!]
\centering\includegraphics[width=6.25in]{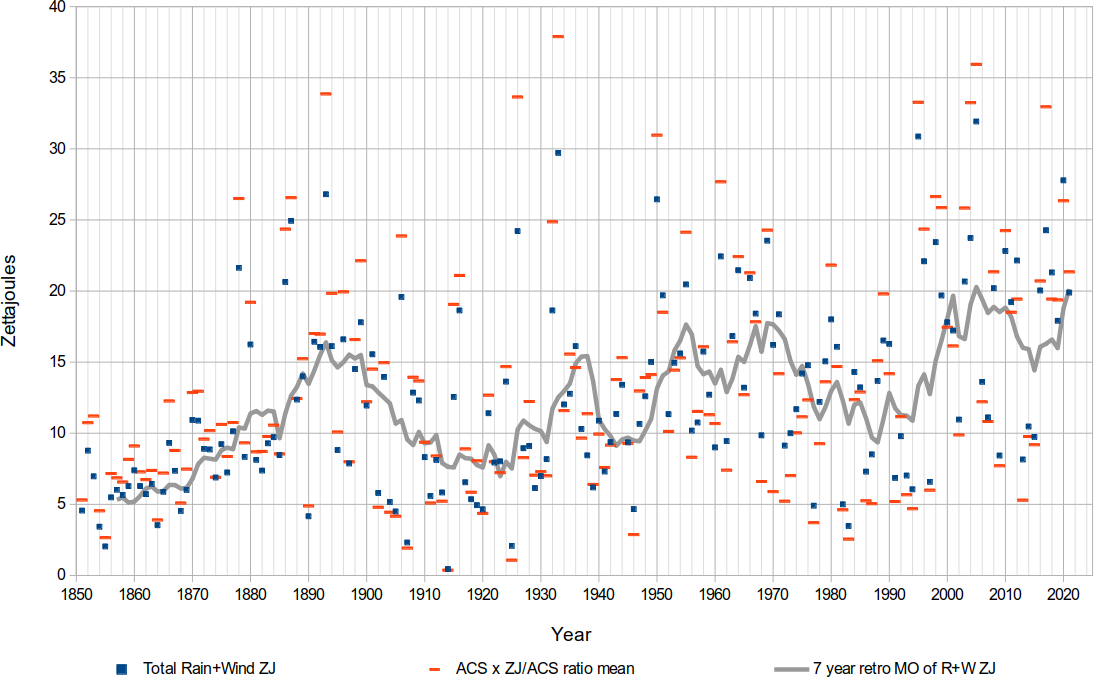}
\caption{\textbf{Normalized storm energy(NSE) year totals compared to scaled ACE for year.   $(NSE/ \overline{ACE})$} Bars: ACE, squares: total rain + wind storm energy. Grey curve: 7 year trailing average of NSE. The North Atlantic multi-decadal oscillation (AMO) is visible in the grey curve. We attempted to use Fourier and other methods to normalize the AMO, and found literature on this before us going back 75 years \cite{ELLIS2016ModulationIceAges_viaPrecessionAndDust-albedoFeedbacks,Loder-Garrett1978_18.6yearCycle,Hachey1948TrendsCyclesSSTCanadianAtlantic}.}
\label{Fig_Supp_Norm-ZJ-to-ACS}
\end{figure}

The HURDAT2 dataset's relationship of millibars to wind velocity and storm radius from 2004 to 2021 was fitted for the purpose of normalizing storm energy between the modern era and the early period. While this does not compensate for missing storms, this makes it possible to estimate storm energy for 1.5 centuries wherein storm radius was not present. Values for each storm entry from 1851 to 2021 were computed, and converted to zettajoules. Normalization is visible in fig. \ref{Fig_Supp_Norm-ZJ-to-ACS} as our computed normalized storm energy (NSE) tends to be closer to the trailing average than the ACE dataset. Scaled ACE is otherwise nicely comparable. 

\subsubsection{Storm energy has a 400:1 or greater ratio of rain energy to wind energy}

\begin{figure}[hbt!]
\centering\includegraphics[width=6.25in]{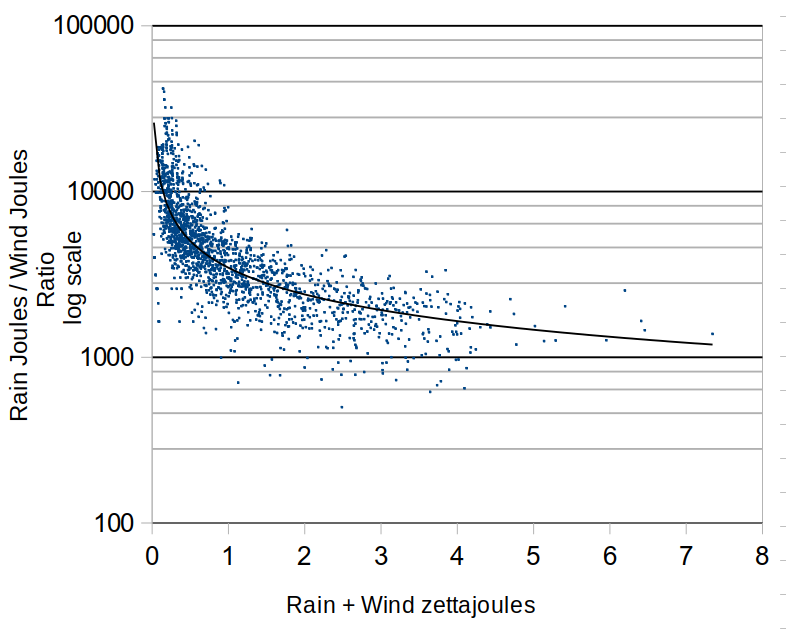}
\caption{\textbf{Normalized storm energy(NSE) year totals compared to ratio of rain energy / wind energy.} Curve: Inverse power law relation between storm energy and rain:wind energy ratio.  The ratio of rain energy declines as wind energy rises, however, we can see that an increase in wind energy is correlated with lower fraction as rainfall. The variance of rain energy / wind energy is quite significant. A 400:1 ratio of rain energy to wind energy is the low end. One zettajoule = 239,234 megatons of TNT. }
\label{Fig_Supp_Rain_Wind_Ratio_PowerLaw}
\end{figure}

Our computations of storm energy of the HURDAT2 dataset generally confirm Landsea's estimate of of a large rain energy to wind energy \cite[400:1]{Landsea2018Energy400to1} ratio.  Our figures correlated well with ACE energy \ref{Fig_Supp_Norm-ZJ-to-ACS}, which suggests they are quite credible. The ratio of rain to wind exhibits an inverse power law (fig. \ref{Fig_Supp_Rain_Wind_Ratio_PowerLaw} eq. \ref{eq:Rain_Wind_Ratio_PowerLaw}. 

The significance of figure \ref{Fig_Supp_Rain_Wind_Ratio_PowerLaw} is that it explains how a climate system can deliver hugely higher energy to land with lower levels of wind. The amount of energy as precipitation is orders of magnitude larger than the amount of energy as wind.

Because the equations (eq. \ref{eq:EmanuelRain}, \ref{eq:EmanuelWind} are approximations, and the rain equation is not terribly sophisticated, what is presented here might be over-determined. We consider the results strongly indicative that the ratio of rain energy to wind energy is very high. The primary message is that This suggests that climate change may deliver far more extra energy as excessive rain than wind. Nevertheless, high  winds should deliver significant rain. The ratio of 400:1 appears to be a lower limit, appearing once in our computed dataset. 

 \begin{align}
    \begin{split} \label{eq:Rain_Wind_Ratio_PowerLaw}
 P(x) &= 3470.42834313298 \cdot x^{-0.535313567104114} \\
R^2 &= 0.574   \qquad \text{ fig. \ref{Fig_Supp_Rain_Wind_Ratio_PowerLaw}}\\
 \text{Where: } P(x) &= \text{ Fitted equation. Inverse power law of } \frac{E_{rain}}{E_{rain} + E_{Wind}}   \\
  x &= \text{total storm energy, rain + wind}
    \end{split}
\end{align}

\subsubsection{Storm energy $\geq$ 1 ZJ as a filter for storm numbers.}
\begin{figure}[hbt!]
\centering\includegraphics[width=6.25in]{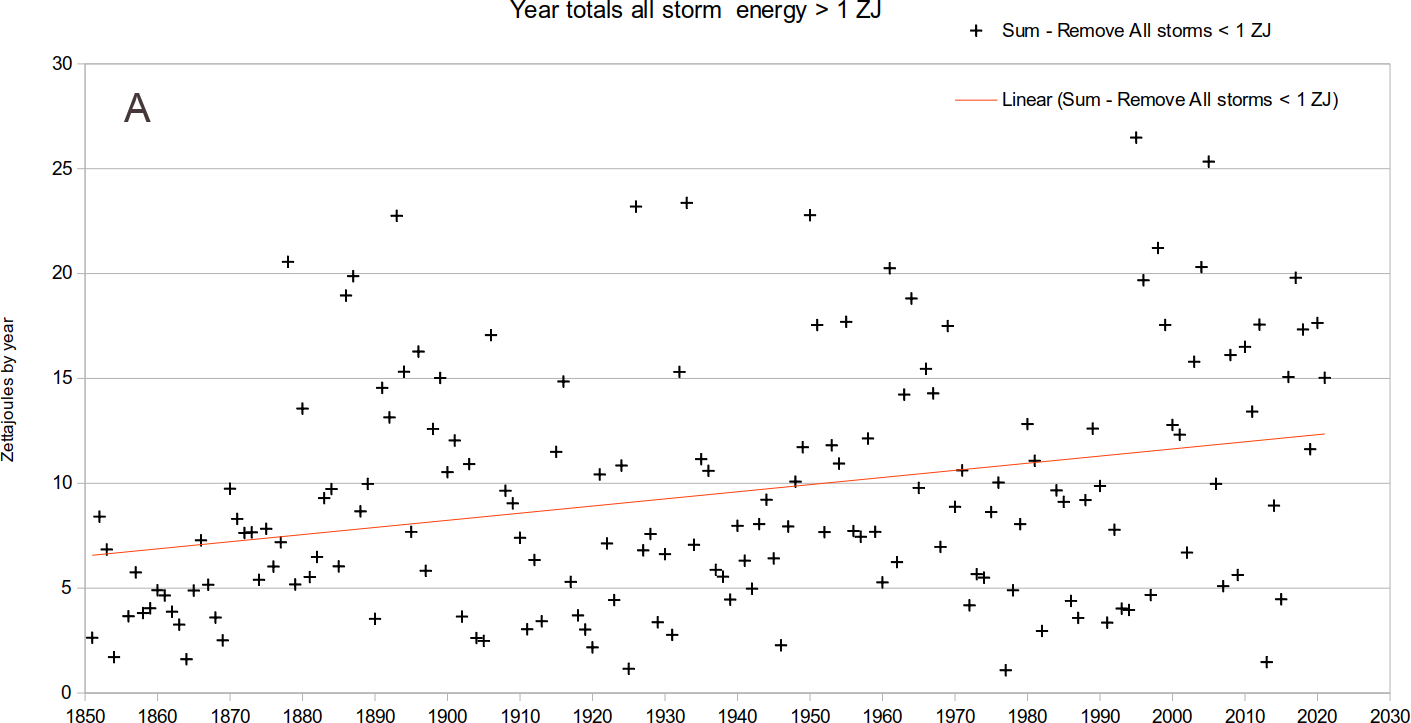}
\centering\includegraphics[width=6.25in]{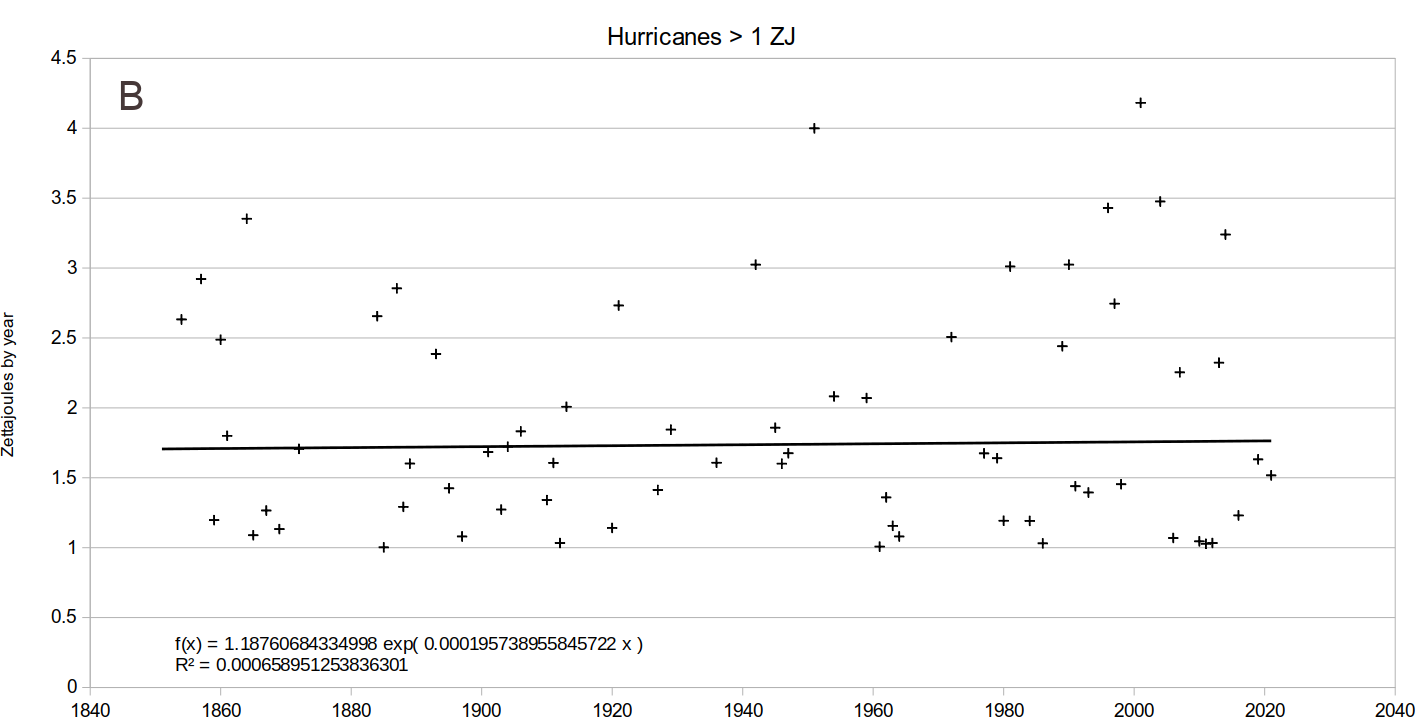}
\caption{\textbf{Normalized storm energy(NSE) year totals > 1 zettajoule.} \textbf{A:} All storms with energy > 1 ZJ. \textbf{B:} All hurricanes with energy > 1 ZJ. There may be a possible slow increase in the outlier large storm energy years. }
\label{Fig_Supp_AllStorms_1ZJ}
\end{figure}

We created two primary views of yearly storm data. In figure \ref{Fig_Supp_AllStorms_1ZJ}A we assigned a cutoff of 1 zettajoule of storm energy as the filter for large storm energy. For scale, hurricane Katrina is rated at 1.57 ZJ in this normalised dataset. With a 1 ZJ filter, we see that many storms not rated as hurricanes pass this filter due to lasting longer, and delivering large amounts of precipitation. 

When the filter is changed to only look at hurricanes $\geq$ 1 ZJ are shown in figure \ref{Fig_Supp_AllStorms_1ZJ}B. For hurricanes $\geq$ the 1 ZJ cutoff, counts and trend is nearly flat, generally agreeing with others \cite{Knutson2022WarmingAndHurricanes}. This may be interpreted to suggest that longer, lower velocity wind ralongerinstorms may be a concern for climate change. 

\subsubsection{Improving weather damages prediction: minimum sea level pressure (MSLP)}
\label{lb:Klotzbach}
The Saffir–Simpson Hurricane Wind Scale (SSHWS) uses maximum sustained winds at 10 meters unobstructed. Klotzbach makes a strong case for minimum sea level pressure (MSLP) being a better predictor/correlate of hurricane damage than the SSHWS on the basis that: A. Maximum wind is difficult to measure, and is not reliably available before 1979; B. The SSHWS scale leaves out rainfall, implicitly assuming it to be a fixed relationship to wind; and C. Storm radius has an $\approx$ 20\% margin of uncertainty at landfall \cite{Klotzbach2020SurfacePressureaMoreSkillfulPredictorofNormalizedHurricaneDamagethanMaximumSustainedWind}. Recent decades have seen the failure of SSHWS to predict damages. MSLP has been recorded for US landfalling hurricanes since 1851, and is much easier to reliably measure.

\subsection{Heat}
\label{lb:Heat}
\begin{figure}[!hbt]
\centering\includegraphics[width=5.25in]{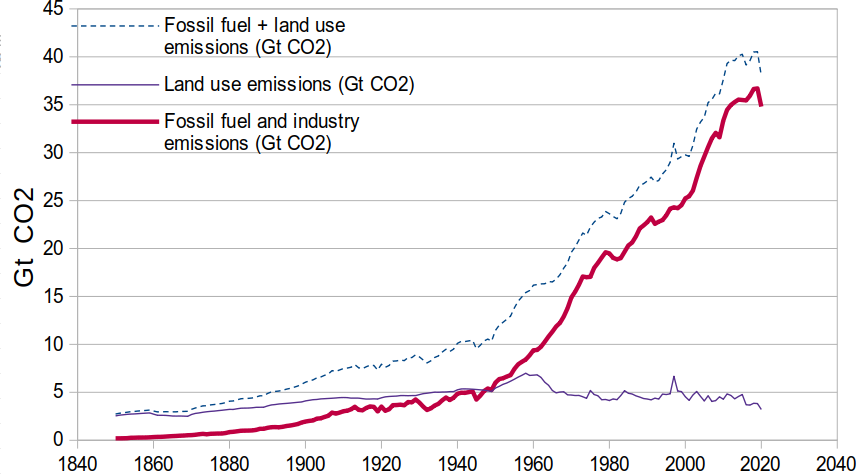}
\caption{\textbf{CO$_2$ emission 1850-2021} \cite{Ritchie2020ourworldinDataCO2} }
\label{Fig_Supp_Gt_CO21850-2021}
\end{figure}

Heat retention is caused by greenhouse gas (GHG) primarily CO$_2$ (fig. \ref{Fig_Supp_Gt_CO21850-2021}). This is the driver for increase in global temperature, and hence heat content. 

\begin{figure}[!hbt]
\centering\includegraphics[width=5.25in]{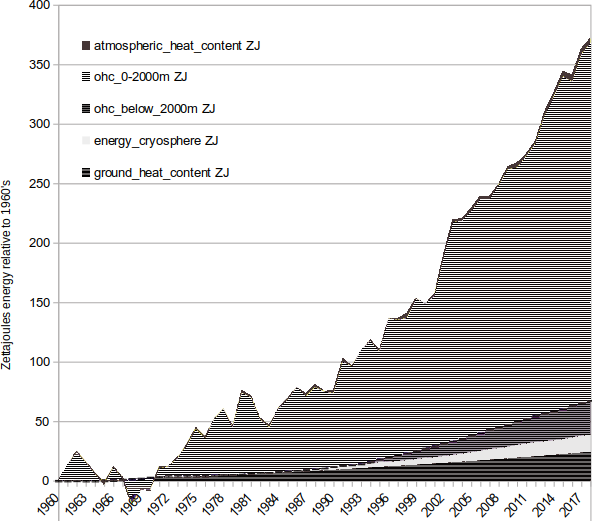}
\caption{\textbf{Figure similar to von Schuckman, et al. from published dataset} \cite{vonSchuckmann2020GCOSEHIEXPv2,vonSchuckmann2020HeatstoredEarthSystem} Here OHC total (below and above 2000 meters) is $\approx$88\% of heat content of the earth's surface. }
\label{Supp_Von_Schuckmann_Full}
\end{figure}

The primary heat buffer and reservoir is the global ocean with 88\% of global heat buffering. Land, ice/permafrost (cryosphere), and atmosphere play relatively minor but still significant roles, as shown in figure \ref{Supp_Von_Schuckmann_Full}. Land and cryosphere heat content modelling is more difficult. What literature exists for land required local computations and data on composition of the land surface and subsurface. Cryosphere modelling was documented to be incorrect \cite{Pettit2021ThwaitesEasternCollapse,Batchelor2023RapidIcdRetreatHundredsMetres,Milillo2022RapidGlacierRetreat}. Both land and ice heat modelling are suitable for a high resolution GCM. Accounting for 88.0\% of earth-surface heat content is sufficient for these purposes. 

\subsection{Ocean heat energy}
The ocean is the primary heat buffer of the globe. Consequently, we modelled the effect of a net total of 18 scenarios on ocean heat as described in \S\ref{lb:ClimateModelandScenarios}. Similar, but much smaller curves should be seen for ground and cryosphere, however, these were not modelled. 

\subsubsection{Ocean heat content is the key to climate and potential damages.}
By these computations, we expect to hit a possible peak global heat content circa 2395 CE, however, this could keep climbing slowly. This heat energy is what gives power to accomplish the physical work of weather phenomena. Climate can be thought of as the overall set of parameters governing the energy of the globe. That energy creates weather. 

In figure \ref{Fig_Supp_OHC_ZJ-v1}A a primary message is that even though CO$_2$ may return to $\approx$400 ppm in our central scenario circa 3300 CE (1,279 years from now), the increased heat will remain far longer, and this drives weather damages. Today's increased total heat energy (Q) since the 1960's is $\approx$360 ZJ. This study only goes out a bit less than 3,000 years to 5000 CE, though we do not show results for SC-GHG beyond a 1,500 year span. However, there should be plenty of excess heat left in the year 10000 CE. 

\begin{figure}[!ht]
\centering\includegraphics[width=6.25in]{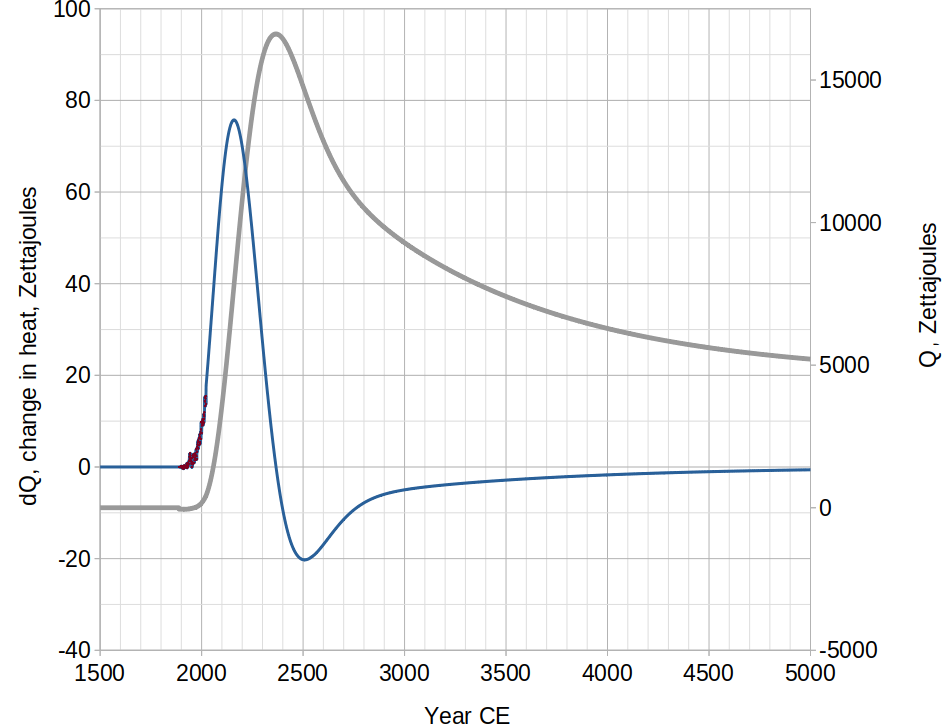}
\caption{\textbf{dQ (change in heat) versus total Q (ocean heat content above 1970's heat content). This figure is for the scenario of figure \ref{Fig_Supp_Baseline_PopNorm_n2O-comparison_T} that does not include updated nitrogen modelling. Thus, panel D can show the effect of net-zero and beyond. It is incorrect to think that after net-zero is achieved climate effects will necessarily become less.} \textbf{Left scale, dQ}: blue curve is \textbf{dQ} (also called \textbf{$\Delta$Q}) model output. \textbf{Black dots} are \textbf{dQ from dataset} for each year. \textbf{Right scale, Q}: thicker grey curve, total OHC, \textbf{Q}.  Both curves are from a Permafrost thaw S-aerosol scenario. The dQ curve shows average projected rate of heat inflow/outflow (dQ) of energy from the ocean. The negative (-dQ) energy, means release of energy from the ocean. Ocean heat release (-dQ or -$\Delta$Q) should maximize $\approx$2500 CE, at -20 ZJ for the year. This maximum outflow of -20 ZJ is $\approx$ 125 years after the peak of ocean heat content which occurs $\approx$2360 CE. \emph{Note that this ocean heat peak (which is also OHC net zero) trails carbon net zero by around 3 centuries, and that following this, energy release should deepen to more than -10 ZJ per year for several centuries.}  Zettajoule = 10$^{21}$ Joules. For comparison, a one megaton nuclear warhead is $\approx 4.18 \times 10^{15}$ Joules \cite{AtomicArchive2020}. So 1 Zettajoule $\approx$ 239,234 one-megaton warheads.}
\label{Fig_Supp_dQvQ_Shakhova-Hansen}
\end{figure} 

Increasing energy going into the ocean on the way up (fig. \ref{Fig_Supp_dQvQ_Shakhova-Hansen} ) should somewhat suppress energy coming out of the ocean that drives weather events relative to heat content. Keep in mind that very small fractions of total ocean heat content represent many megatons TNT of energy. On the downside of these curves, when the net energy in any specific year has more energy leaving the ocean than is going in, then the suppressive effect seen in climate models is no longer present. As the amount of energy leaving the oceans as heat rises to 10 and then 20 Zettajoules in one year, there should be a significant increase in severe weather. This suggests capability of potentially driving extreme storms such as Hansen has discussed \cite{Hansen2016IceMeltSeaLevelRiseAndSuperstorms}. 

OPTiMEM as shown in figure \ref{Fig_Supp_dQvQ_Shakhova-Hansen} does not take into account the effect of rapid ice melt of Greenland and Antarctica into the oceans \cite{Pettit2021ThwaitesEasternCollapse, Fox2022TheComingCollapseSciAm,Batchelor2023RapidIcdRetreatHundredsMetres,Milillo2022RapidGlacierRetreat,Hansen2023WarmingInPipeline,Abram2025EmergingEvidenceOfAbruptChangesInTheAntarcticEnvironment}. This melt will cool the upper layer of the Atlantic, as well as parts of the southern oceans for centuries, along with changing circulation \cite{boers2021amoc,jackson2015amoc,Rahmstorf2015amoc}. Hansen claims the effect of ice melt will be faster heating overall. We cannot model these effects. 

\begin{figure}[!ht]
\centering\includegraphics[width=5.5in]{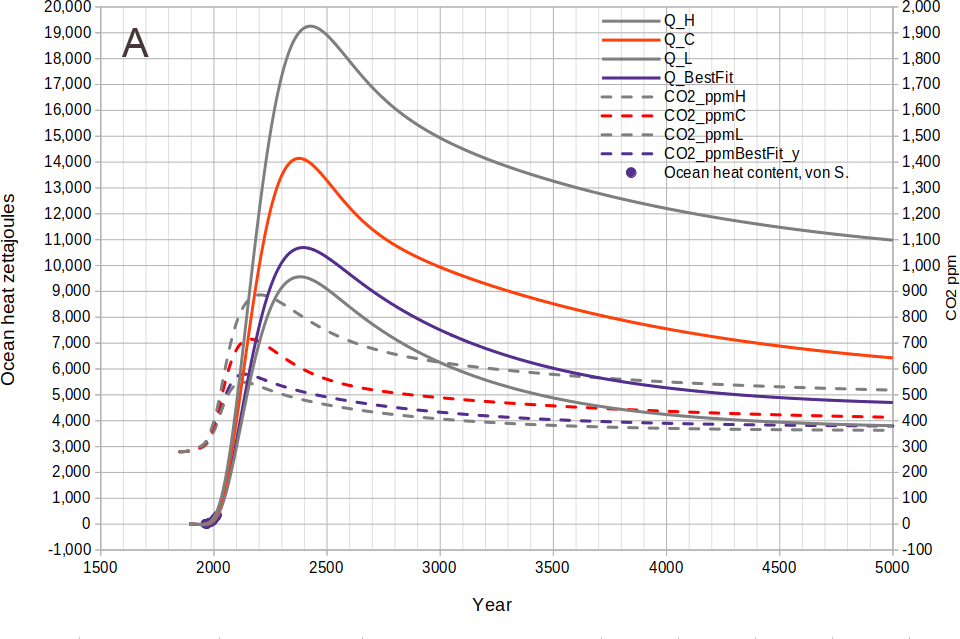}
\centering\includegraphics[width=5.5in]{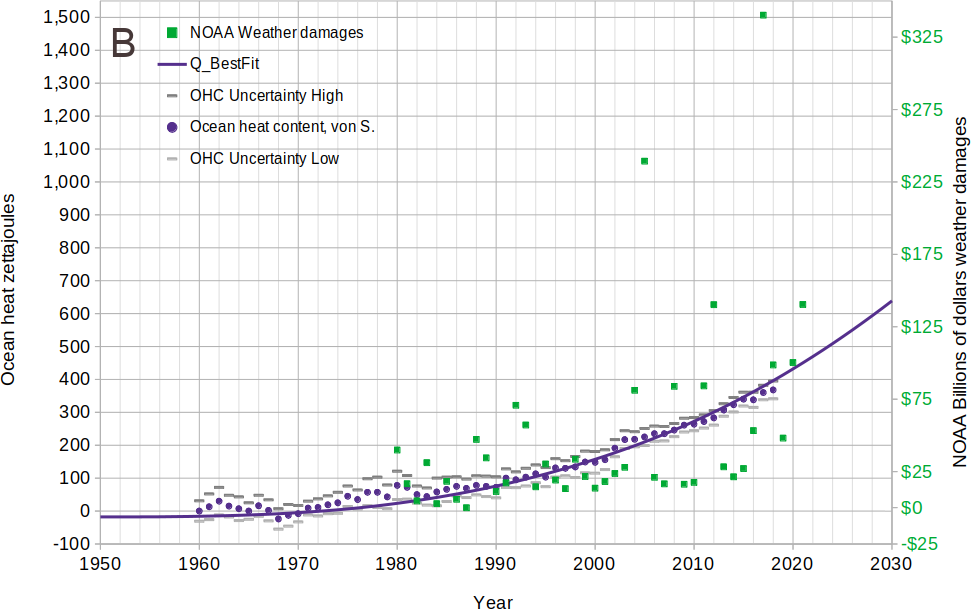}
\caption{\textbf{Upper graph A: Ocean heat content and CO$_2$ to 5000 CE. } Here we see contrasted the level of CO$_2$ above pre-industrial (dashed curves) driving the global heat engine, and the amount of heat expected to accumulate in the world ocean. Visible on the lower left as blue circles is the ocean heat content accumulated since the 1960's. Note the lag between peak CO$_2$, which is net zero carbon, and peak ocean heat, which is net zero energy absorption by the planet. \textbf{Lower graph B: Detail showing ocean heat content compared to scaled NOAA weather damages.} The OHC data points and uncertainty are shown from von Shuckmann \cite{vonSchuckmann2020GCOSEHIEXPv2}. Our heat curve generated from the scenario driving equations fit von Shuckmann's data very well.  Square points are NOAA weather damages by year scaled to the heat (Q) curve. Zettajoule = 10$^{21}$ Joules.}
\label{Fig_Supp_OHC_ZJ-v1}
\end{figure} 

\clearpage 

\begin{figure}[!hbt]
\centering\includegraphics[width=3.5in]{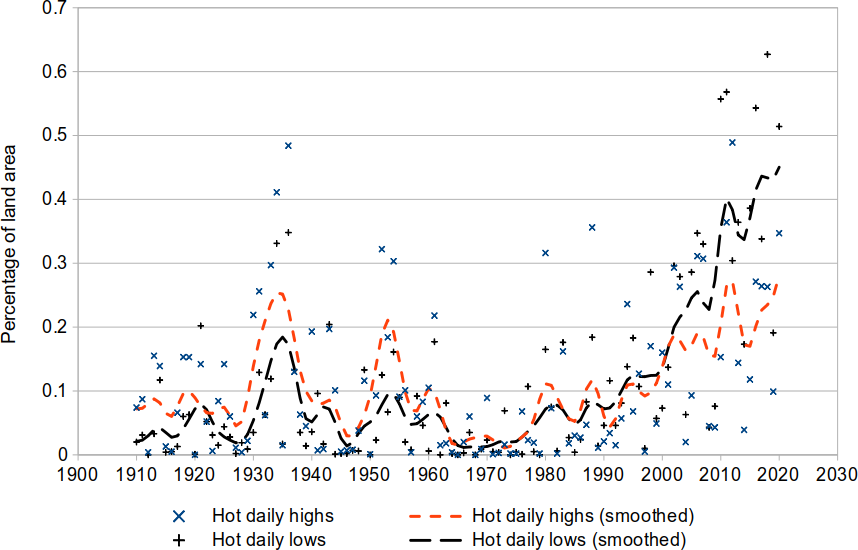}
\caption{\textbf{Area of the Contiguous 48 States with Unusually Hot Summer Temperatures, 1910-2020} \cite{EPA2022GlobalMeanTempPrecip}.  }
\label{Fig_Supp_Heat_Land_USA}
\end{figure}

Heat absorption by land masses is probably $\approx$7.3\% of heat absorbed by the oceans \cite{vonSchuckmann2020HeatstoredEarthSystem}. However, we did not attempt to make use of this  because land heat modelling is much more complex and there is no dataset to validate against. US land temperature hot lows have increased faster than hot highs. This evidences retention of heat on land (fig. \ref{Fig_Supp_Heat_Land_USA}) increasing at night.  River systems are also seeing increasing frequency, duration, and intensity of heatwaves from 1996 to 2021 \cite{Tassone2022HeatwavesStreamsRiversUSA}.  

\begin{figure}[!hbt]
\centering\includegraphics[width=5.5in]{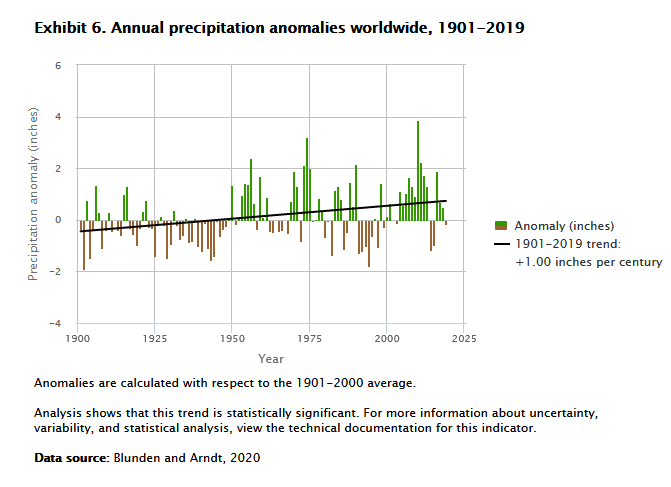}
\caption{\textbf{Figure reproduced from EPA report, Exhibit 6} \cite{EPA2022GlobalMeanTempPrecip}  }
\label{EPA_Global_temp-precip_exhibit-6}
\end{figure}

Precipitation is a secondary indicator of heat, and a degree of movement of heat into the upper atmosphere where it gives up its energy of vaporisation as energy of condensation. Precipitation anomalies (fig. \ref{EPA_Global_temp-precip_exhibit-6}) have increased worldwide. Similarly, even where rainfall has not seen a large increase on hourly or daily scales, precipitation has increased in short bursts, increasing the danger of flash floods \cite{Ayat2022Subhourlyrainintensity}. 

\begin{figure}[!hbt]
\centering\includegraphics[width=6.25in]{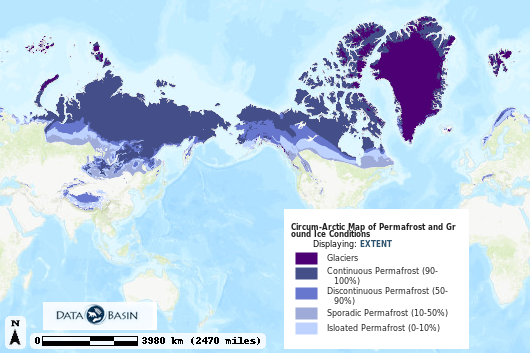}
\caption{\textbf{Arctic permafrost map 2001} \cite{Brown2001CircumArcticICeMap}  }
\label{Fig_Supp_Arctic_Permafrost_Map}
\end{figure}

Land heat absorption in the arctic has the buffer of the cryosphere. Permafrost depth varies up to 610 meters over a large land area (fig. \ref{Fig_Supp_Arctic_Permafrost_Map}). Permafrost held carbon has the capacity to significantly increase the amount of GHG \cite{Schuur2022PermafrostFeedbacksWarming}. Early methane release has begun to a minor extent in the form of 17 methane clathrate explosion craters of Siberia \cite{Zolkos2021CH3Craters}, \cite{Bogoyavlensky2021PermanentCH3emission}, and CH$_4$ and CO$_2$ release from northern permafrost in general \cite{Schuur2022PermafrostFeedbacksWarming}\cite{Anderson2017ChlorineBromineCatalysis}. However, this area of study is moving quickly and current rates of release may need to be revised upward \cite{Morgado2024OsmosisDrivesExplosionsAndMethaneReleaseInSiberianPermafrost}.

\subsection{World war 2 temperature anomaly}
\begin{figure}[!ht]
\centering\includegraphics[width=5.5in]{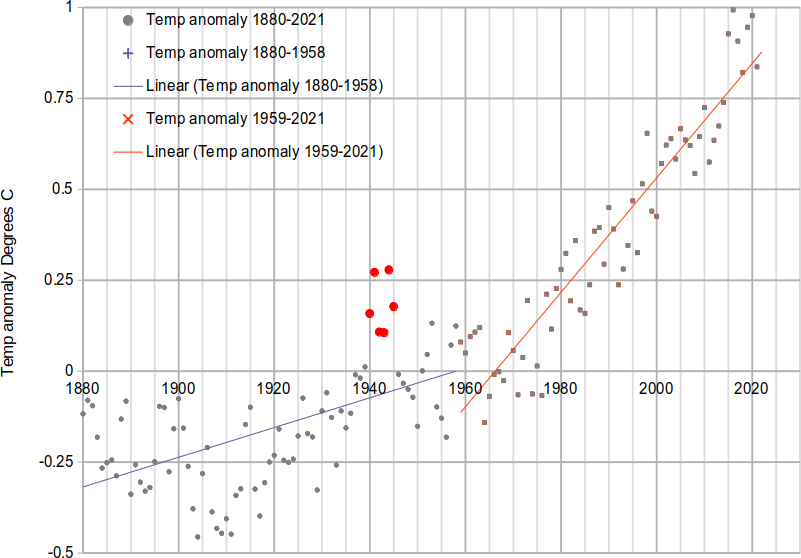}
\caption{\textbf{WW2 temperature anomaly. } \cite{NCEI2022GlobalTempAnomalies}  }
\label{Fig_Supp_NCEI_Temp_Anomaly_WW2}
\end{figure}

NCEI \cite{NCEI2022GlobalTempAnomalies} and other sources reproduce a dataset that contains a temperature anomaly series during WW2 that we decided we could not accept. This anomaly is visible by inspection of figure \ref{Fig_Supp_NCEI_Temp_Anomaly_WW2} as the 1940-1945 red dots. The problem for these 6 years is that there is no reasonable explanation in physics for such an outlier cluster. There are no CO$_2$ spikes recorded commensurate with these values (fig. \ref{Fig_Supp_Gt_CO21850-2021}). CO$_2$ levels should have been extraordinary to achieve this cluster's mean, with an emissions spike on the order of 3X recorded emissions. 

It is possible that sampling bias could be at work. The oceans comprise $\approx$71\% of the planet's surface. Ocean surface temperatures were measured from ships, and WW2 caused a drop in commercial shipping, perhaps causing sampling bias.

Perhaps a drop in sulfate aerosols \cite{Manshausen2022Invisibleshiptracks,Diamond2023CloudSulfurRegulation,Hansen2023WarmingInPipeline} due to WW2's suppression of commercial shipping could be an explanation. We do see a cluster of elevated temperature due to that cause from 2016-2020. However, the great depression starting in 1930 had already cut maritime shipping, and there is no visible effect from this, and naval warships may well have compensated or increased net sulfate aerosols. Additionally, the drop in sulfate aerosols since 2016 should be relatively higher because much more ocean shipping occurs in the present period since 1980 than prior to WW2. Global maritime shipping in the present decade is nearly 11 billion tonnes yr$^{-1}$, and in 1955, 0.8 billion \cite[p. 75]{UNCTAD2022HandbookStatsDevStatus} \cite[p. 3]{UNCTAD1968ReviewMaritimeTrans}. To be consistent, such a WW2 cluster should roughly average on the order of 7\% of the jump we see starting in 2016 CE. However, no such relationship is seen. The average increase of the WW2 cluster is $\approx$0.22 degrees, and the average increase of the 2016-2020 cluster (ignoring 2018) is $\approx$0.17 degrees. Thus, ascribing the WW2 temperature anomaly to a precipitous drop in sulfate aerosol emissions does not seem reasonable. 

There may be a straightforward explanation for this war years phenomenon. The values that are different are winter values, and those begin December of 1939. It is likely that these temperature values were war propaganda for consumption by troops and civilians. We believe the nations making war in cold climates in what was primarily a northern hemisphere war, may have done the same thing, both Allied and Axis. The fact that this begins precisely at the start of the war and ends exactly at the close is suggestive of a propaganda explanation. 

In any case, we do not think that these values can be trusted, even if our explanations are rejected. Consequently, we removed the years 1940-1945 from our temperature dataset for curve fitting purposes, albeit with mild effect. 

We also do not believe it is likely that the larger variance of mean global temperature recorded prior to 1950 is entirely real. The tightening up of values post-1960 is quite clear and ascribed to improved measurement and density of data collection.  Thus the wide spread of mean temperature values before WW2 is most likely the result of measurement error and sparseness of data. 

\clearpage 

\section{Equations and functions, full precision}
\subsection{CO$_2$ parts per million}
\begin{flalign}
\begin{aligned} \label{eq:CO2Yr}
CO2(y) &= fCO2_n(y) \\
\text{Where: } fCO2_n(y) &= \text{ one of 18 CO$_2$ scenarios from OPTiMEM. }\\
y &= \text{ year} 
\end{aligned}  
\end{flalign}
Here, each scenario produces 3 CO$_2$ level curves, one for each  CO$_2$ remainder fraction, low, central and high. 

See: eq.'s \ref{eq:NetCO2RemainderL}, \ref{eq:NetCO2RemainderC}, \& \ref{eq:NetCO2RemainderH}, in \S\ref{eqs:NetCO2Remainders}

\subsection{Weather damages SC-GHG.}
\label{DevelopmentofWeatherDamagesbyYear}

\begin{flalign}
  \begin{aligned} \label{eq:Fit_WDbyScenario}  
f\_\_\_WD_m(y) & \text{ also }fWD_m(y) | = \text{ weather damages scenarios } \\
\text{Where: } & f\_\_\_WD_m \text{ may be }fCO_2WD_m, \; fCH_4WD_m, \; fN_2OWD_m, \; fFgasWD_m\\
& \text{$_m$ = 1 to 18.  For each gas there are 18 potential weather damages scenario functions.} 
\end{aligned}
\end{flalign}

\subsection{\emph{e}Fold atmospheric lifetime}
\begin{flalign}
  \begin{aligned} \label{eq:GasTau}  
Efold(\tau, t) =& \text{ $e$fold atmospheric lifetime in years. See: eq. \ref{eq:eFold} } \\
\text{Where: }\tau = \; &\tau CH_4 = 9.5, \; \tau N_2O = 123.1, \;\tau Fgas = 760 \\
t \; = &\text{span of years}
\end{aligned}
\end{flalign}

\subsection{Global weather damages}
\begin{align}\label{eq:GlobalWeatherDmgs2}
\begin{split}
 GWD(fWD_m, y) =& \frac{fWD_m(y)  \{eq. \ref{eq:Fit_WDbyScenario} \} } {U_{mGWP}} \times 10^9 \\
\text{Where: } y =& \text{ year } \\
 U_{mGWP}  =& \text{USA  mean  share  of  gross  world  product } = 0.2573  \\
\end{split}
\end{align} 
USA share of global GDP has dropped from 39\% to 25\% at  $\approx$ -0.23\%yr$^{-1}$ from 1960--2019, mostly due to growth of China \cite{FRED2021INTDSRUSM193N,FRED2021NYGDPMKTPCDWLD}. China's growth  is slowing, so the rate of USA GDP share loss should flatten. 

\subsection{Social cost of CO$_2$ (SC-CO$_2$}
\begin{flalign}
\begin{aligned}
\label{eq:SCCO2} 
&SCCO2(fCO2_n, C_{Rx}, fCO_2WD_m, y, m, d) \quad  =  \\
& \qquad \qquad \sum_{i = 0}^{m-1} \frac{GWD(fCO_2WD_m, y+i ) \times C_{R_X}(y+i)}{(fCO2_n(y+i)-CO2_{Pre}) \times tCO2_{ppm}} \times \frac{1 }{d^i}   \\
& \text{Where: } \; \; fCO_2WD_m  = \text{Function: Weather damages by scenario attributable to CO$_2$. eq's. } \ref{eq:Fit_WDbyScenario} \\
& \qquad \qquad y  = \text{calendar year } \geq 1890  
 \qquad   m  =  PV_{FL} \text{ limit  of  years to sum }\\
& \qquad \qquad  d  = \text{ discount rate} 
 \qquad CO2_{Pre} = 280 \text{, Preindustrial CO$_2$ ppm} \\
& \qquad \qquad tCO2_{ppm} = 7.797 \text{E9, tonnes of CO$_2$ per ppm of CO$_2$} \\
&\qquad \qquad  C_{R_X}(y)  \mapsto C_{RL}(y) eq. \ref{eq:NetCO2RemainderL} \text{ or } C_{RC}(y) eq.\ref{eq:NetCO2RemainderC} \text{ or }  C_{RH}(y) eq.\ref{eq:NetCO2RemainderH} 
\end{aligned}
\end{flalign}

\subsection{Social cost of CH$_4$ (SC-CH$_4$}
\begin{flalign}
\begin{aligned}
\label{eq:SCCH4}  
&CH4_r(CH_4ppbM, CO2ppmM, n, y) = \\
&\qquad \qquad  \text{Uses eq.'s \ref{eq:CH4fromCandPop},  \ref{eq:PopProj1}  \&  \ref{eq:PopProj2}    See: \S  \ref{CH4_damages}}\\
& \text{Where:  CH4$_r$ is the ratio of non-energy related CH$_4$ to total CH$_4$ } \\
& \qquad \qquad CH_4ppbM  =  \text{array of } CH_4 \text{ppb} \qquad CO2ppmM = \text{array of } CO_2 \text{ppm}  \\ 
& \qquad \qquad n  =  \text{ iteration } \qquad y  = \text{calendar year } \geq 1950  \\ 
 \\
&SCCH4(fCH4_n, \tau CH_4, fCH_4WD_m, y, m, d) \quad  =  \\
& \qquad \qquad \sum_{i = 0}^{m-1} \frac{GWD(fCH_4WD_m, y+i ) \times Efold(\tau CH_4,y+i)}{ (CH4_r(i,y+i) \times (fCH4_n(y+i)-fCH4_n(1950)))\times tCH4_{ppb}} \times \frac{1 }{d^i}   \\
& \text{Where: } \; \; fCH_4WD_m  = \text{Function: Weather damages by scenario attributable to CH$_4$. eq's. } \ref{eq:Fit_WDbyScenario} \\
& \qquad \qquad y  = \text{calendar year } \geq 1950  
 \qquad   m  =  PV_{FL} \text{ limit  of  years to sum }\\
& \qquad \qquad  d  = \text{ discount rate} \\
& \qquad \qquad tCH4_{ppb} = 2.842 \text{E6, tonnes of CH$_4$ per ppb of CH$_4$} \\
\end{aligned}
\end{flalign}

\subsection{Social cost of N$_2$O (SC-N$_2$O)}
\begin{flalign}
\begin{aligned}
\label{eq:SCN2O} 
&SCN2O(fN2O_n, \tau N_2O, fN_2OWD_m, y, m, d) \quad  =  \\
& \qquad \qquad \sum_{i = 0}^{m-1} \frac{GWD(fN_2OWD_m, y+i ) \times Efold(\tau N_2O,y+i)}{ (fN2O_n(y+i)-fN2O_n(1950))\times tN2O_{ppb}} \times \frac{1 }{d^i}   \\
& \text{Where: } \; \; fN_2OWD_m  = \text{ Function: Weather damages by scenario attributable to N$_2$O. eq's. } \ref{eq:Fit_WDbyScenario} \\
& \qquad \qquad y  = \text{calendar year } \geq 1950  
 \qquad   m  =  PV_{FL} \text{ limit  of  years to sum }\\
& \qquad \qquad  d  = \text{ discount rate} \\
& \qquad \qquad tN2O_{ppb} = 7.797 \text{E6, tonnes of N$_2$O per ppb of N$_2$O} \\\\
\end{aligned}
\end{flalign}

\subsection{Social cost of fluorocarbon gases (SC-Fgas)}
\begin{flalign}
\begin{aligned}
\label{eq:SCFgas} 
&SCFgas(fFgas_n, \tau Fgas, fFgasWD_m, y, m, d) \quad  =  \\
& \qquad \qquad \sum_{i = 0}^{m-1} \frac{GWD(fFgasWD_m, y+i ) \times Efold(\tau N_2O,y+i)}{ (fFgas_n(y+i)-fN2O_n(1950))\times tN2O_{ppb}} \times \frac{1 }{d^i}   \\
& \text{Where: } \; \; fFgasWD_m  = \text{Function: Weather damages by scenario attributed to Fgas. eq's. } \ref{eq:Fit_WDbyScenario} \\
& \qquad \qquad y  = \text{calendar year } \geq 1950  
 \qquad   m  =  PV_{FL} \text{ limit  of  years to sum }\\
& \qquad \qquad  d  = \text{ discount rate} \\
& \qquad \qquad tFgas_{ppt} = 2.076 \text{E4, tonnes of Fgas per ppb of Fgas} \\
\end{aligned}
\end{flalign}

\subsection{Risk by year, 10\%, 1\%, and 0.1\% }
\label{RiskbyYear}
To create a risk model for extreme weather damages, we create a dataset of the positive distances from eq. \ref{eq:Fit_WDbyScenario} to estimate $\sigma$ as discussed in: Estimation of $\sigma$ curve risk, see: \ref{lb:RiskMethods}. The result is eq. \ref{eq:EsigmaYr1}, which is multiplied by the Chebyshev risk (eq. \ref{eq:Chebyshev1}) factors $r=1\%$ and $r=0.1\%$ and added to eq. \ref{eq:Fit_WDbyScenario} to produce risk curves. See also \S\ref{EstimationOfsigmafunction} and \ref{Examination_Slope_Change}.

\begin{align}
\begin{split} \label{eq:EsigmaeMYr1}
E\sigma e_M(y) &= {5.07812825162672}\, e^{(0.0771705785402397 \, y)} \qquad \text{High precision construction equation}\\
\text{Where: } y\, &=\text{ year } \\ R^2 &= \, 0.9935
\end{split}  
\end{align} 

\begin{flalign}
\begin{aligned} \label{eq:EsigmaYr1}
E\sigma(f_Q,y) &= x\sigma  \times f_Q(y) \\
\text{Where: } f_Q &= \text{ one of 18 ocean heat conjecture interpolation objects from OPTiMEM. }\\
y &= \text{ year} \\
x\sigma &= \text{ is the scaled fit constant} 
\end{aligned}  
\end{flalign}

\begin{flalign}
  \begin{aligned}  
\label{eq:EsigmaRisk} 
 &(2.23606797750 \times E\sigma(y) ) + fWD_m(y) \text{ for } r=10\% \text{ (1:10)}  \\ 
&(7.07106781187 \times E\sigma(y) ) + fWD_m(y) \text{ for } r=1\% \text{ (1:100)} \\
 &(22.3606797750 \times E\sigma(y) ) + fWD_m(y) \text{ for } r=0.1\% \text{ (1:1,000)} \\ 
&Where: \quad  r= risk \quad r\text{ term is $k$ result of Chebyshev eq. \ref{eq:Chebyshev1}} \\  
& E\sigma(y) = \text{eq. \ref{eq:EsigmaYr1}}  \qquad fWD(y) = \text{eq. \ref{eq:Fit_WDbyScenario} } \\
\end{aligned}
\end{flalign}

\subsection{Use of Chebyshev's theorem (Bienaymé-Chebyshev inequality)}
This theorem states that no more than $\frac{1}{k^2}$ of a distribution's values can be k or more standard deviations away from the mean. Alternatively, at least $1 - \frac{1}{k^2}$ of the distribution's values are less than k standard deviations away from the mean. 

 We use this theorem (Eq. \ref{eq:Chebyshev1}) to estimate risk year events because the NOAA extreme weather data may not be a normal distribution.  Since the applicable distribution is single tailed, $2r$ is substituted for $p$ because $p$ is the double-tailed risk value. 
 
 An arbitrarily chosen risk ($r$) for a year can be selected and solved for ${k}$ using eq. \ref{eq:Chebyshev1}. 
 
Chebyshev’s theorem: 

\begin{align}
\begin{split} \label{eq:Chebyshev1}
2r &= \frac{1}{k^2} \Leftrightarrow{}k=\sqrt{\frac{1}{2r}}  \\   
\text{Where: } r &= \text{ single tailed probability of observations } \geq k \text{ standard deviations } \\
k &= \text{ number of standard deviations} \\
\end{split}
\end{align} 

Here $k$ is the  multiple of $\sigma$ from the mean and $r$ is the single-tailed probability that no more than this value of the distribution falls outside the distance from the mean. In  our usage, the mean is represented by an equation such as $WDe(y)$ (eq.\ref{eq:Fite_Dyr1}), or an OHC curve fitted to it.  
We can then multiply an equation fitted to the $\sigma$ data by $k$ and add this to the baseline fitted equation, to give an estimate for $\sigma$, for a given year or CO$_2$ ppm. In this particular instance, that will be considered a conservative estimate for use by economists. With various values of $k$ generated by setting a desired value of $p$, an arbitrarily chosen probability (i.e risk level) can be estimated. 

\subsection{WDe(y) correlation to data}
\begin{flalign}
  \begin{aligned} \label{eq:Fite_Dyr1}  
WDe(y) =& \num{0.183221172676639} e^{(0.0494335193084322 y)} \qquad \text{High precision construction equation}\\ 
\text{Where: } y=& \text{ year }  \qquad 
R^2= 0.97 
\end{aligned}
\end{flalign}

The first statistic is the $R^2$, for $WDe(y)$ (eq. \ref{eq:Fite_Dyr1}), 0.97. 

A second method of examining WDe(y) is to take a conceptually Bayesian type of approach and see how the slope changes for similar equations, $WDe^*(y)$ starting from a small number of points, adding one point iteratively until the full dataset is reached. This is done in \S \ref{Examination_Slope_Change} and figure \ref{fig_Fit_CurvesYear}, starting with 10 points in the weather damages dataset, up to the full 42 points. In doing so, by inspection, and various measures, the slopes computed begin with wide variation, and settle down to a quite small slope variation from one to the next, summarized in figure \ref{fig_LogSlopes}. This iterated equation set is named $WDe^*(y)$[], which is a set of 32 equations. 

This method was used to generate a standard error for our equation slope. That standard error was not used, because the equation fitted to weather damages is utilised as a target to fit the corresponding segment of our heat equations over the time period of the weather damages dataset, and standard error is overwhelmed by the Archer based equations uncertainty (eq's. \ref{eqs:NetCO2Remainders}).

\subsection{Estimation of weather damages high side standard deviation ($\sigma$) by a trailing standard deviation function }
\label{EstimationOfsigmafunction}
\begin{figure}[hbt!]
\centering\includegraphics[width=5.25in]{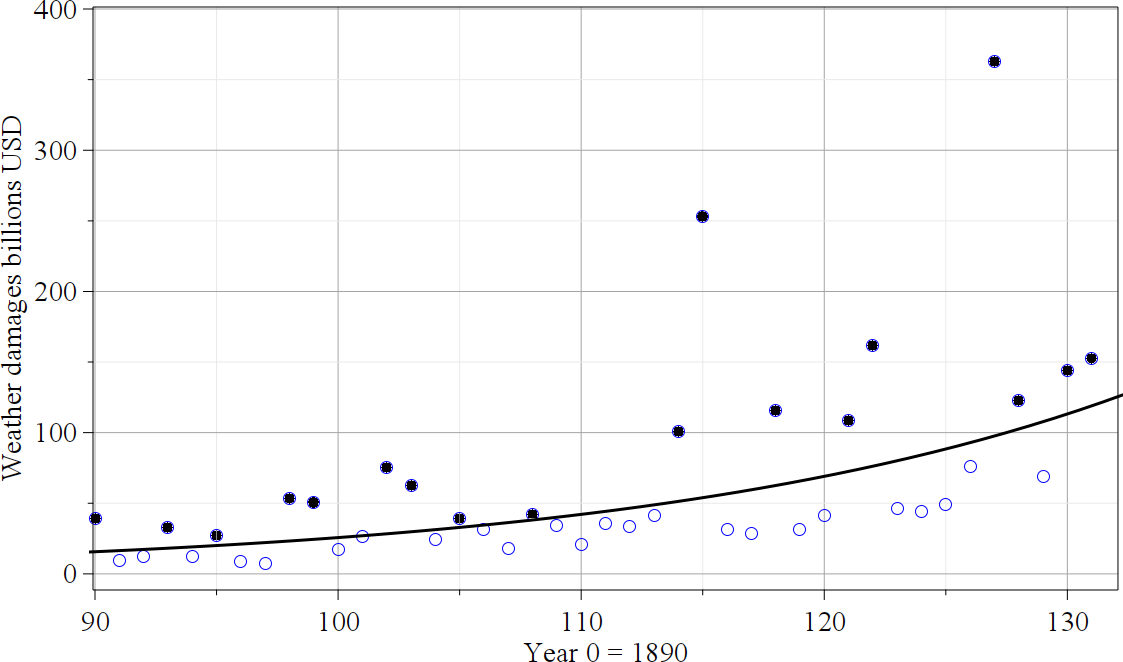}
\caption{\textbf{$\sigma$ curve risk D+[] dataset. } Filled black datapoints are NOAA weather damages above $WDe(y)$ (eq. \ref{eq:Fite_Dyr1}), designated D+[]. The distance of these points from WDe(y) are the $y$ values used to calculate the standard deviation dataset. Black curve WDe(y) is exponential fit equation to NOAA data.}
\label{Fig_Supp_D}
\end{figure}

The concept of this method is to find a way to deal with the problem of what the standard deviation for a dataset like the NOAA billion dollar weather damages \cite{NOAAbillionDlr} should be. The simplistic way to find the standard deviation would be to take the whole dataset shown in figure \ref{Fig_Supp_D}. This would produce a single value. However, a little thought makes it clear that in a situation such as climate change, the standard deviation should be expected to increase. Consequently, if one were to use the simplistic method and project the standard deviation  forward over 300 or 500 years, the result is obviously guaranteed to be wrong. We note that the current 20 year period is in a time when the North Atlantic long term oscillation should be on its way down \cite{Hansen2023WarmingInPipeline}, but we do not see this in the data. At present there is no usable method for normalising these data relative to the North Atlantic oscillation, although we attempted this (not shown), and others have tried \cite{Loder-Garrett1978_18.6yearCycle,Hachey1948TrendsCyclesSSTCanadianAtlantic,ELLIS2016ModulationIceAges_viaPrecessionAndDust-albedoFeedbacks}. 

Inspection of figure \ref{Fig_Supp_D} shows that below the fitted curve, datapoints are far more tightly clustered than above it. Here, it is the high side above the curve that we care about, because what matters most is peak event years. These peak event years will be candidates for catastrophes. 

Consequently, an algorithm modelled on a trailing average within the datapoints above the  curve was developed. This algorithm groups $n$ datapoints, incremented by year, and takes the standard deviation of each group. Values of $n$ from 3 to 9 were tested, which is discussed in the next section (see \S \ref{Examination_Slope_Change} \emph{Examination of change in slope...} ). A value of $n$ = 7 was chosen as the lowest log slope, on the basis that it is the conservative choice.  This $n$ = 7 window generated an equation with the lowest integral value over the region of interest.

\begin{figure}[hbt!]
\centering\includegraphics[width=5.25in]{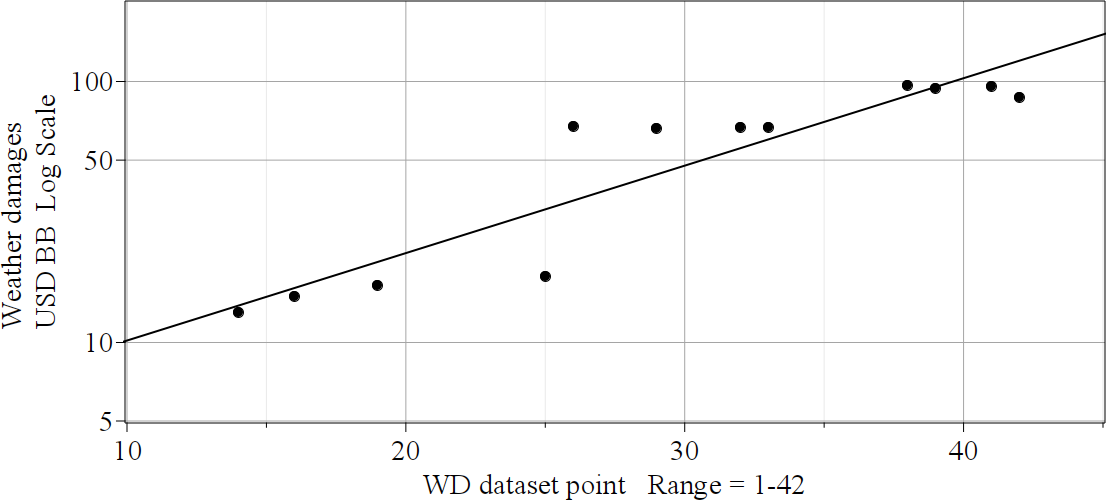}
\caption{\textbf{Trailing standard deviation $E\sigma e_M(y)$ risk curve.} Filled black datapoints are $D\sigma[]$ 7 year trailing standard deviations. Black curve is $E\sigma e_M(y) = \num{5.07812825162672} e^{(0.0771705785402397 y)}$ (eq. \ref{eq:EsigmaeMYr1}) $R^2$ = 0.84.  The first trailing 7 is at 14.}
\label{Fig_Supp_DSigma}
\end{figure}

Thus, $D_+[]$ $(n=18$ of $42)$ contains positive distance of each year's total damages from $WDe(y)$ (eq. \ref{eq:Fite_Dyr1}). Grouping $7$ elements of $D_+[]$ at a time and taking their standard deviation produces the 12 points of $D\sigma[]$.  A curve fit to the $D\sigma[]$ dataset produced equation $E\sigma e_M(y)$ \ref{eq:EsigmaeMYr1}, is shown in figure \ref{Fig_Supp_DSigma}.

\subsection{Examination of change in slope of $WDe^*(y)[]$ varying the included years from n = 10 to 42} 
\label{Examination_Slope_Change} 
This method looks at progression of the fit of variant equations, which can be judged by slope, change in standard deviation of the slope progressions, and standard error for the standard deviation variability. 
\begin{figure}[!ht]
    \centering
    {\includegraphics[width=5.25in]{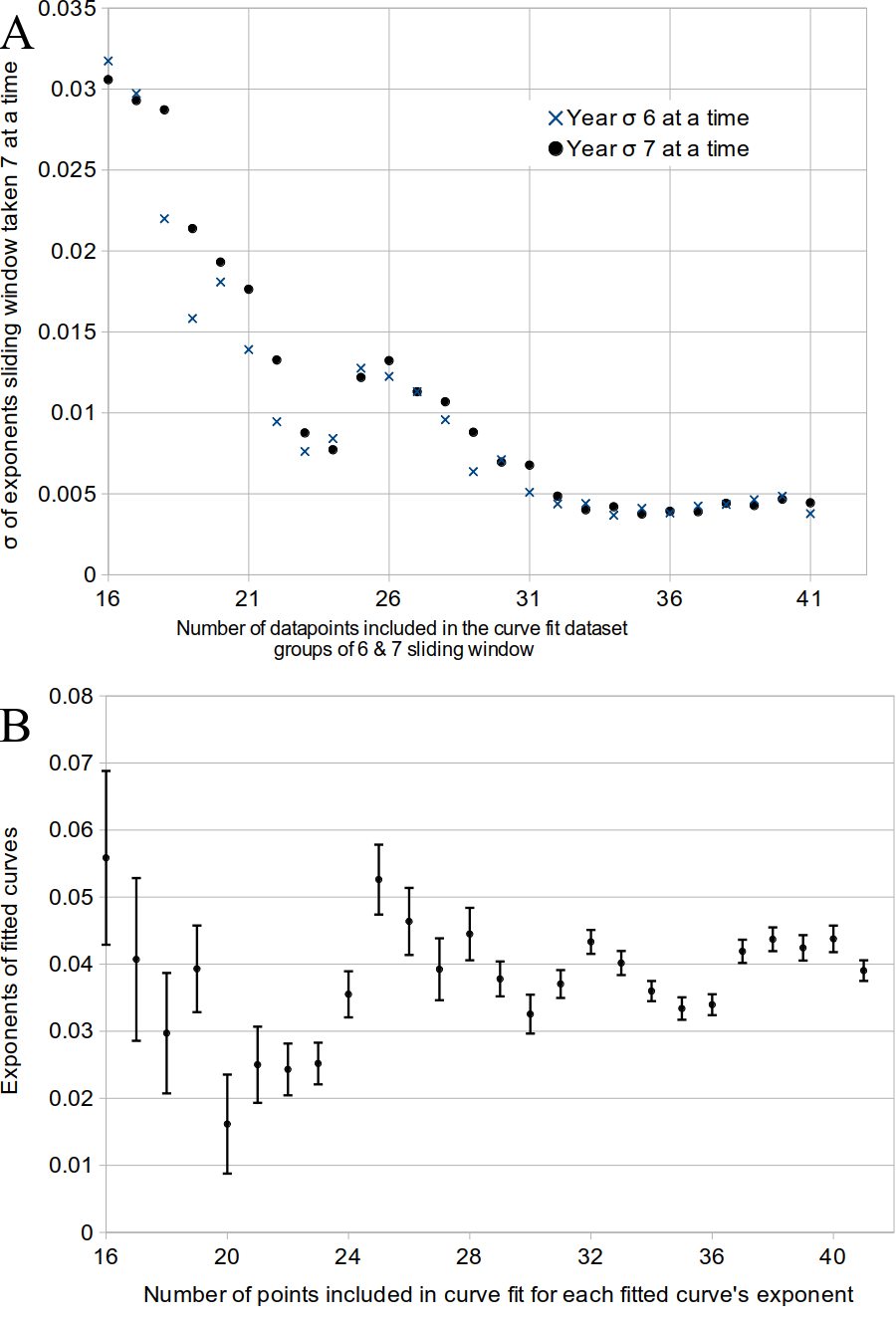} }
    \caption{\textbf{Log slopes summary} Here is shown log slopes, with standard error generated by calculating standard deviation on the slopes for sets of $n$ in a sliding window, low to high. (Left to right). This graph is shows that the exponents for curve fits to NOAA data converge.  Standard error is calculated from the trailing average of a group of 7 exponents. (i.e. the curve fit with 16 points has a standard error calculated from the exponents for curves fitted to 10, 11, 12, 13, 14, 15, and 16 datapoints.) See figure  \ref{fig_Fit_CurvesYear}. }
    \label{fig_LogSlopes}
\end{figure}

\begin{figure}[!ht]
    \centering
    \includegraphics[width=6.5 in]{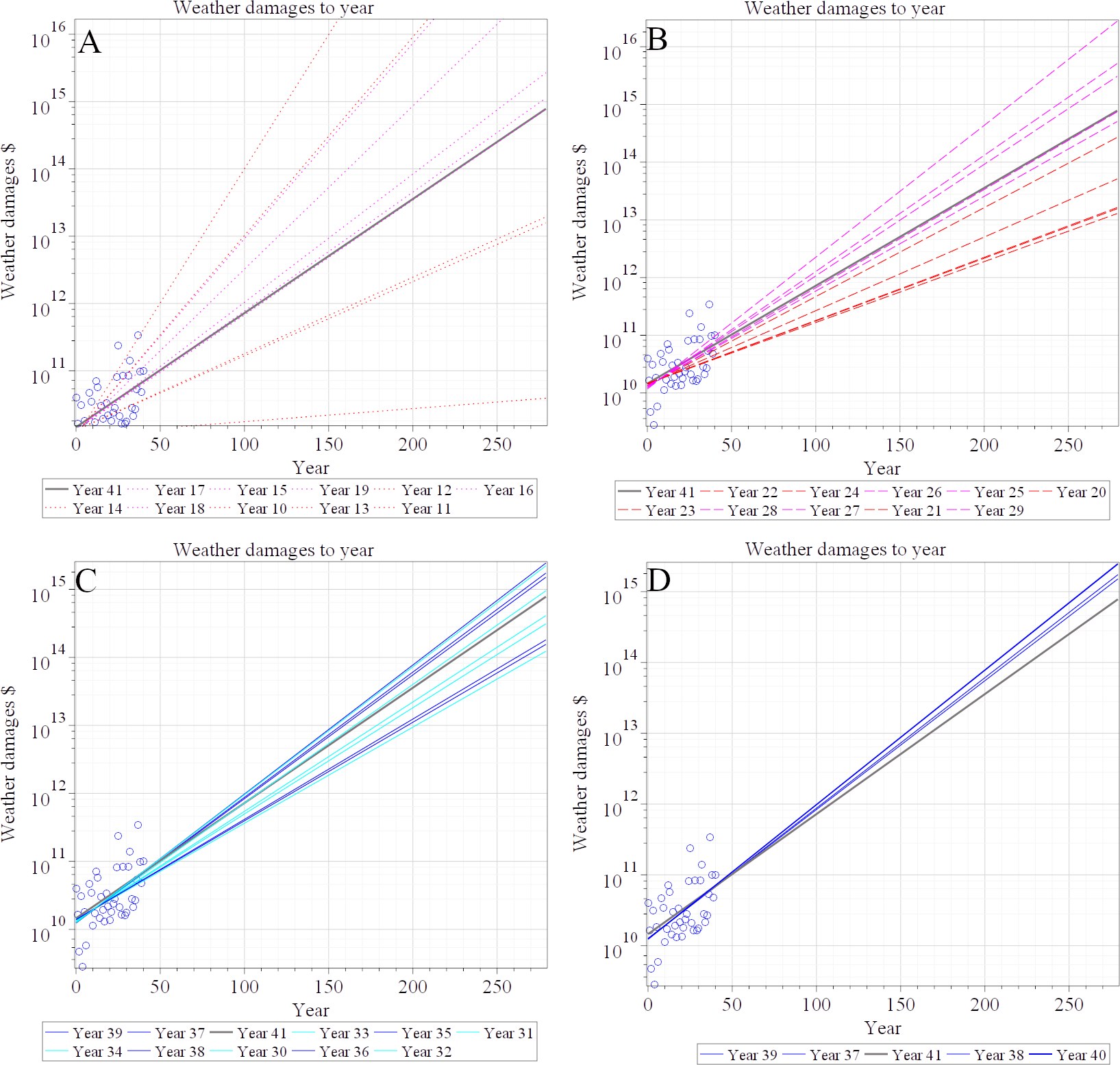}
 \caption{\textbf{$WDe^*(y)[]$ log slopes summary.} data points. \textbf{A} 10-19 points; \textbf{B} 20-29 points; \textbf{C} 30-39 points; \textbf{D} 37-41 points.  Compare to figure \ref{fig_LogSlopes} which shows log slopes starting at 16 points. (The trailing standard deviation requires this.) The variation in the log slopes narrow as expected. }
     \label{fig_Fit_CurvesYear}
\end{figure}

The fits behave as expected. Figures \ref{fig_LogSlopes} and \ref{fig_Fit_CurvesYear} show progression of the log slopes for the fitted curves of the form: 

$ E(x) = C \cdot e^{(M \cdot x)} \qquad
\text{Where: } C \text{ and }  M \text{ are  fit\ constants}$ 

Inspection of the sequence of curve fits to NOAA data from $n=15$ to $n=42$ shows that they rapidly converge and the variation in log slope drops. Figure \ref{fig_LogSlopes} summarises the slopes seen in figure \ref{fig_Fit_CurvesYear}

These provide, by inspection and formally, that these curve fits are reasonable descriptors of the NOAA dataset. 

\clearpage
\subsection{Stadium willingness to pay (WTP) vs global climate WTP}
\label{WTP_StadiumVGlobe}
\subsubsection{WTP for a stadium}
In the case of a single stadium, ignoring environmental damages is defensible taken one by one.  This 20 hectares of wilderness has some arbitrarily assigned monetary valuation. The value of the stadium when it is built, is assigned some PV. If the public is not willing to pay the difference between the PV of the stadium and the valuation of the wilderness, then the stadium should be built. And as we see above, there is no environmental penalty because we have decided to ignore trivial environmental impact as an approximation of zero. \\

\begin{flalign} \label{eq:WTP_for_stadium}
\begin{aligned}
S_{D} &=  \frac{S_{ha} \times E_{D}}{W_{ha}-S_{ha}} \\ 
\text{Where: } S_{D} &= \text{ presumed environmental damages for  one  stadium} \\
S_{ha} &= \text{ stadium  hectares; } W_{ha} = \text{ wilderness  reserve  hectares }\\
E_D &= \; \text{ Environmental damages factor per hectare }0\geq E_D \leq 1 \\
\text{Assume: } S_{D}\, &<\, 10^{-6} \approx \;0  \\
\text{Set: } S_{ha} &=  20 \text{ ha; }  
\text{ USA: } W_{ha} = 20,993,174 \text{ ha; } 
\text{ Global: } W_{ha} = 423,774,398 \text{ ha} \\
E_D &= 0.4 \\ \\
\text{USA: } S_{D}  &=  \frac{20 \text{ ha} \times 0.4}{20,993,174 \text{ ha} - 20 \text{ ha}} \, = \, 3.81 \times 10^{-7} \, \approx 0 \\
\text{Global: } S_{D} \, &= \, \frac{20 \text{ ha}}{423,774,398 \text{ ha} - 20 \text{ ha}}  =  1.88 \times 10^{-8}  \, \approx 0 \\
\end{aligned}
\end{flalign}

\subsubsection{WTP for globe}

Now let us examine damages that apply not just to a stadium, but to all non-wilderness land, all wilderness land, and the  sea. This is justified, because climate is a global issue with impacts everywhere. 
\begin{flalign} \label{eq:WTP_for_globe}
\begin{aligned}
G_{D} &=  \frac{G_{ha} \times E_D}{G_{ha} \times (1-E_D)} = \frac{E_D}{(1-E_D)} \\ 
\text{Where: } G_{D} &= \text{ presumed environmental damages for the globe} \\
G_{ha} &= \text{ global  hectares; } \\
E_D &= \; \text{ Environmental damages factor per hectare }0\geq E_D \leq 1 \\
\text{Assume: } G_{D}\, &<\, 10^{-6} \approx \;0  \\
\text{Set: } G_{ha} &=  51.01 \times 10^9 \text{ ha;  } E_D = 0.05 \\ \\
 G_{D} \, &= \, \frac{0.4}{1-0.05}  =  0.053  \, \neq 0 \\
\end{aligned}
\end{flalign}
Here we can see that WTP cannot ignore global damages because such  damages are not a fraction approximating zero. So, we cannot assume that willingness to pay as commonly used is acceptable. Global  damages cannot be ignored in favour of arbitrarily chosen valuation as was implicitly done for DICE.  

\subsection{Sigmoid/logistic curve used in heuristic survey model} The sigmoid (logistic) function has uses that range from binding models in chemistry and epidemiology's susceptible/recovered curves, to agricultural yield relative to soil nutrients, and curves fitted to ELISA antibody binding results. This type of function should fit the expected behaviour of an economy subject to increasing external shocks until collapse. Parameters for figure \ref{Fig-Expert-Sigmoid} are: $\Delta$ = 99.8, $\delta$ = 0, $m$ = 4.5, $I_{50}$ = 3.65, $\theta$ = 0 to 12.
 
\label{SigmoidExpertEquationText} 
\begin{align}
\begin{split} \label{eq:SigmoidExpertEquation}
D &= \frac{\Delta-\delta}{1+(\frac{\theta}{I_{50}})^{-m}} +\delta \\   
\text{Where: } D &= \text{economic damages }\\
\Delta  &= \text{ maximum percentage damages } \\
\delta &= \text{ minimum percentage  damages }\\
\theta \, &= \text{ temperature }\degree C \text{ above pre-industrial}  \\
I_{50}  &= \text{ 50\% intercept} \\
m \, &= \, slope
\end{split}
\end{align}

\section{Social cost of Greenhouse Gases (SC-GHG) full-size graphics  }
\label{Sect:SC-GHGFullsize}
\begin{figure}[hbt!]
\centering\includegraphics[width=6.25in]{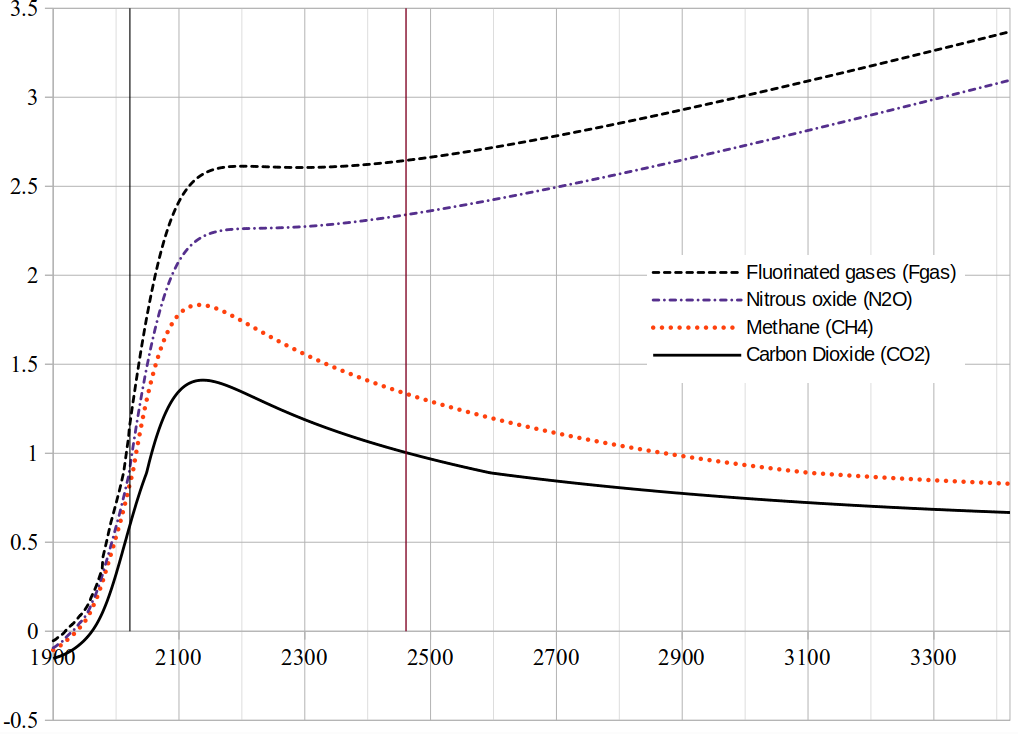}
\caption{\textbf{ Baseline Normal population S-aerosol scenario---Example of gas warming temperature contributions (EEI$_T$).} Black vertical bar at 2022 CE is the start of model generated data. Bar at 2461 CE is the approximate year when this model scenario transitions to warming from N$_2$O $\geq$ warming from CO$_2$. Fluorinated gases (upper black dashed curve) include chlorinated fluorocarbons (CFCs) and other similar gases. These are sometimes called Other Trace Gases (OTGs). A subset of fluorinated gases are in the Montreal Protocol Trace Gases (MPTGs).  N$_2$O (blue dash-dot curve), CH$_4$ (red dotted curve), and CO$_2$ (black solid curve). The upper fluorinated gases curve is the total warming effect of all gases with interference adjustments accounted for. Each curve adds its EEI$_T$ to the curve just below it. The kink in fluorinated gases after 1978 is part of F-gas data. Compare with heat figure \ref{Fig_Supp_Q-by-GHG_for-SCC-calc-v1} to see how much more slowly heat will equalise.}
\label{Fig_T_by_gas_contribution}
\end{figure}

Currently, there are two primary gases that are ascribed a social cost: carbon dioxide/CO$_2$ (SCC/SC-CO$_2$), and nitrous oxide/N$_2$O (SC-N$_2$O). Here we break out the relative contributions by GHG to ocean heat content. In figure \ref{Fig_T_by_gas_contribution}, the relative contributions for greenhouse gases are stacked on top of each other.

Ocean heat lags temperature change by a couple of centuries in complex ways, so one cannot just take the $\Delta$T curves of figure \ref{Fig_T_by_gas_contribution}, look up the year, and allocate proportionally. For the heat conjecture, the model needs to create a heat curve for each GHG fraction as shown in figure \ref{Fig_Supp_Q-by-GHG_for-SCC-calc-v1}.  By comparing these two graphs, we can see that while T influence driving heat equalizes between CO$_2$ and N$_2$O (in this projection) in approximately 2461 CE, heat quantity resulting does not equalise within the 1500 year span presented for this model. 

\clearpage

\subsection{SC-CO$_2$ full-size isosurface for 2025} 
\begin{figure}[!ht]
\centering\includegraphics[width=6.5in]{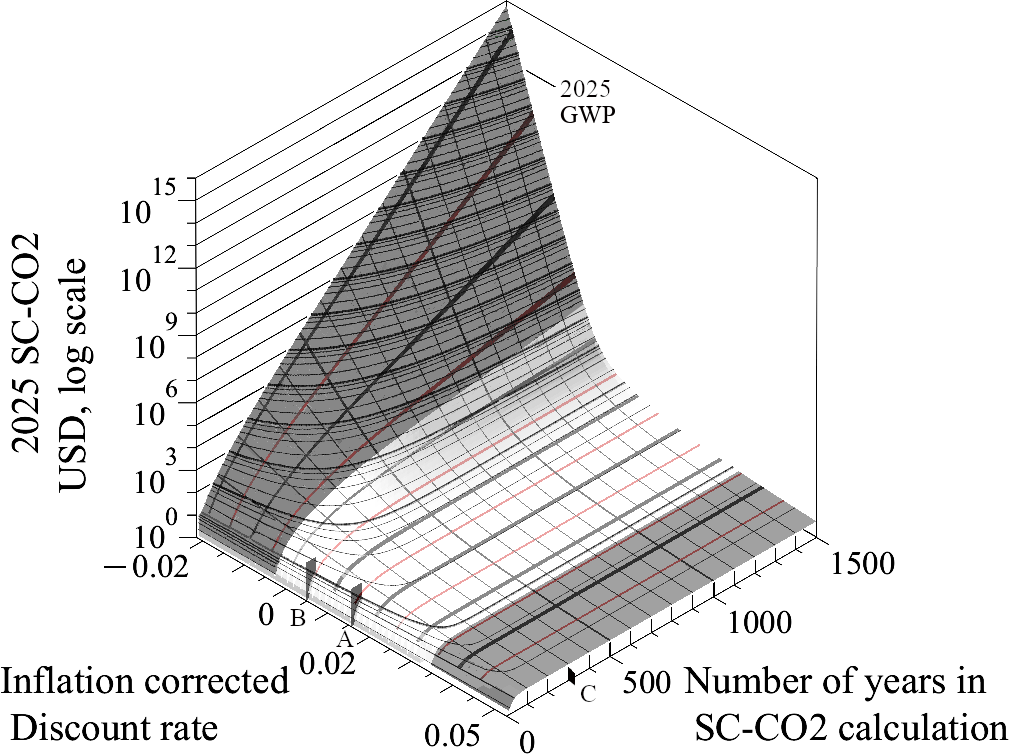}
\caption{\textbf{Social cost (SC) in  real-dollars of emitting 1 tonne of each GHG in 2025.} The x and y axes are, respectively, discount rate and the number of years included in the computation. Discount rate: -2$\sigma$ to +2$\sigma$. Baseline low CO$_2$ (low), and S-aerosol high CO$_2$ with permafrost melt scenario (high)---together provide the thickness of the isosurface. The projected 2025 gross world product (GWP) is presented for reference where applicable. Discount scale marker tabs: \textbf{Tab A} is bond mean discount rate $d_{MT30yr}$ = 1.57\%; \textbf{Tab B} is $d_{rDICE}$ = 0.483\% discount rate.  Range of discount rates is centred on \textbf{Tab A}'s $d_{MT30yr}$ discount rate, with a $\pm$ 2$\sigma$ range, where standard deviation is derived from analysis of 20 and 30 year treasury bill rates. (Supplement: \S3.4) This 2$\sigma$ range is chosen so that it encompasses the DICE model standard discount rate range of 5.1\%. It is unlikely to be realistic to presume a fixed discount rate, so the real world should be expected to take a wandering path relative to this isosurface---keeping in mind that each year has phase space graphs like this. Discount rates are in real dollars. \textbf{Tab C} is 300 year nominal term of our climate bond notes, and the approximate time span of the original DICE model. Time span up to 1500 years is provided. The \textbf{minor Z log scale contours} are 2, 4, 6, and 8 times the previous major contour.  }
\label{Fig_SCCO2_isosurface_fullsize}
\end{figure}  
\clearpage

\subsection{SC-CH$_4$ full-size isosurface for 2025}
\begin{figure}[!ht]
\centering\includegraphics[width=6.5in]{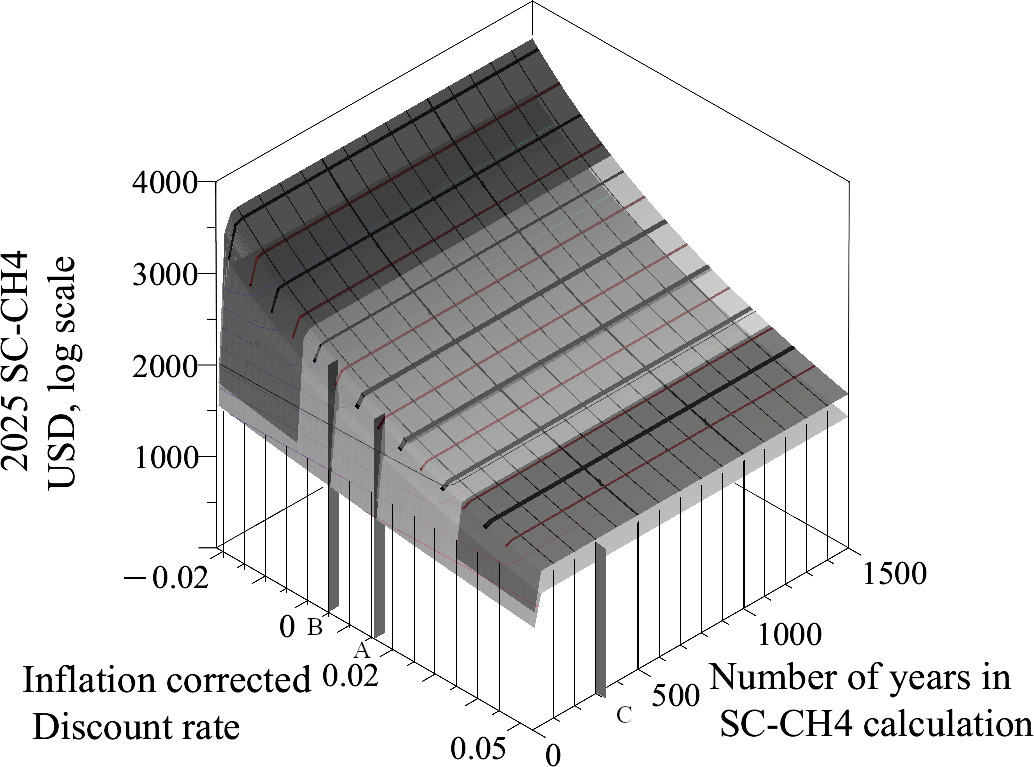}
\caption{\textbf{Social cost (SC) of emitting 1 tonne of CH$_4$ (SC-CH$_4$) in 2025, by discount rate and number of years included in the computation. Discount rate: -2$\sigma$ to +2$\sigma$. Baseline low scenario, and S-aerosol high CO$_2$ scenario---together provide the thickness of the isosurface.}  The visible isosurface is S-aerosol with high CO$_2$ level and permafrost melt. Discount scale marker tabs: \textbf{Tab A} is bond mean discount rate $d_{MT30yr}$ = 1.57\%; \textbf{Tab B} is $d_{rDICE}$ = 0.435\% discount rate (see \S\ref{Risk_and_future_losses_streams}).  Range of discount rates is centred on \textbf{Tab A}'s $d_{MT30yr}$ discount rate, with  $\pm$ 2$\sigma$ range, where standard deviation is derived from analysis of 20 and 30 year treasury bill rates (see \S \ref{CarbonBondInterestRate}). This 2$\sigma$ range is chosen so that it encompasses the DICE model standard maximum discount rate range of 5.1\% \cite{nordhaussztorc2016Rdice}. It is unlikely to be realistic to presume a fixed discount rate, so the real world should be expected to take a wandering path relative to this isosurface. Number of years in computation scale \textbf{Tab C} is 300 year nominal term of our climate bond notes, and the time span of the original DICE model. Time span up to 1500 years is provided. The \textbf{minor Z scale contours} are 2, 4, 6, and 8 times the previous major contour. }
\label{Fig_SCCH4_isosurface_fullsize}
\end{figure}  
\clearpage

\subsection{SC-N$_2$O full-size isosurface for 2025}
\begin{figure}[!ht]
\centering\includegraphics[width=6.5in]{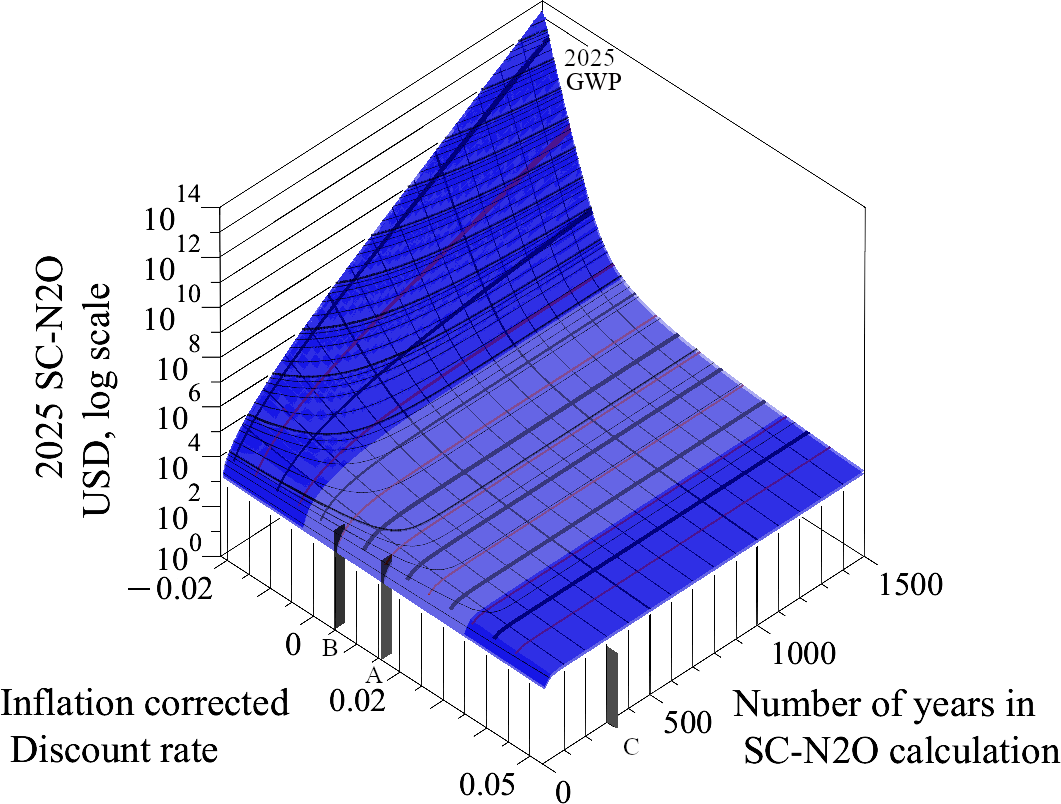}
\caption{\textbf{Social cost (SC) of emitting 1 tonne of N$_2$O (SC-N$_2$O) in 2025, by discount rate and number of years included in the computation. Discount rate: -2$\sigma$ to +2$\sigma$. Baseline low scenario, and S-aerosol high CO$_2$ scenario---together provide the thickness of the isosurface.}  The visible isosurface is S-aerosol with high CO$_2$ level and permafrost melt. Discount scale \textbf{Tab A} is bond mean discount rate $d_{MT30yr}$ = 1.57\%; \textbf{Tab B} is $d_{rDICE}$ = 0.435\% discount rate (see \S\ref{Risk_and_future_losses_streams}).  Range of discount rates is centred on \textbf{Tab A}'s $d_{MT30yr}$ discount rate, with  $\pm$ 2$\sigma$ range, where standard deviation is derived from analysis of 20 and 30 year treasury bill rates (see \S \ref{CarbonBondInterestRate}). This 2$\sigma$ range is chosen so that it encompasses the DICE model standard discount rate range of 5.1\% \cite{nordhaussztorc2016Rdice}. It is unlikely to be realistic to presume a fixed discount rate, so the real world should be expected to take a wandering path relative to this isosurface. Number of years in computation scale \textbf{Tab C} is 300 year nominal term of our climate bond notes, and the time span of the original DICE model. Time span up to 1500 years is provided. The \textbf{minor Z log scale contours} are 2, 4, 6, and 8 times the previous major contour. }
\label{Fig_SCN2O_isosurface_fullsize}
\end{figure}  
\clearpage

\subsection{SC-Fgas fullsize isosurface for 2025}
\begin{figure}[!ht]
\centering\includegraphics[width=6.5in]{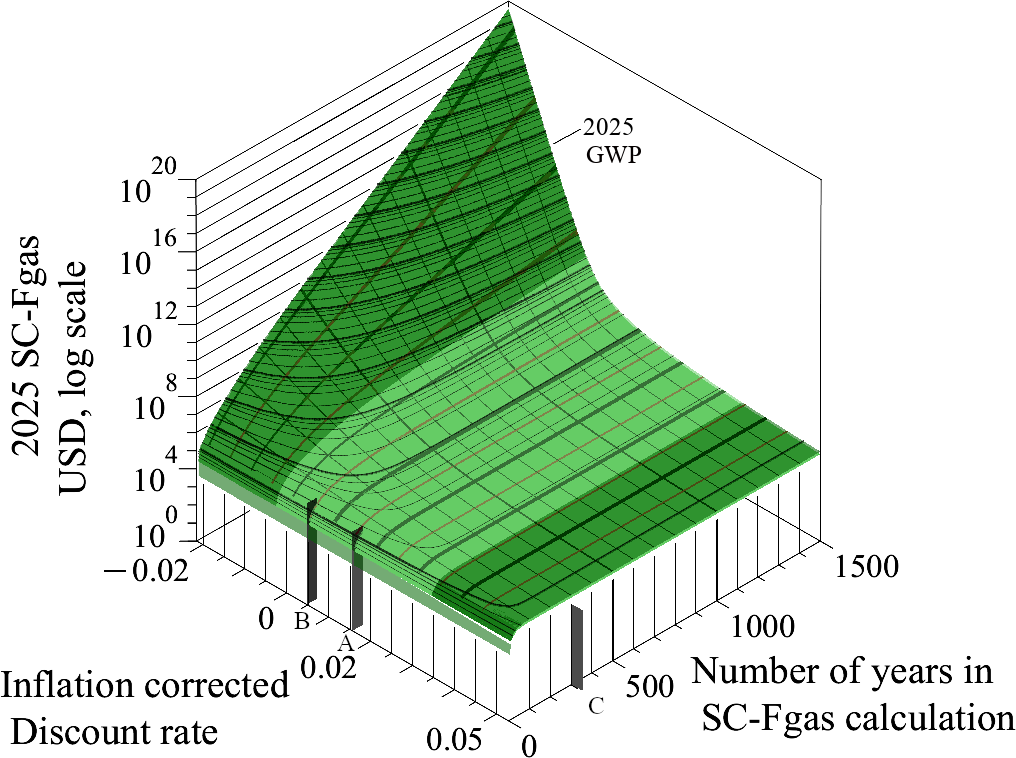}
\caption{\textbf{Social cost (SC) of emitting 1 tonne of flourinated gases (SC-Fgas) in 2025, by discount rate and number of years included in the computation. Discount rate: -2$\sigma$ to +2$\sigma$. Baseline low scenario, and S-aerosol high CO$_2$ scenario---together provide the thickness of the isosurface.}  The visible isosurface is S-aerosol with high CO$_2$ level and permafrost melt. Discount scale \textbf{Tab A} is bond mean discount rate $d_{MT30yr}$ = 1.57\%; \textbf{Tab B} is $d_{rDICE}$ = 0.435\% discount rate (see \S\ref{Risk_and_future_losses_streams}).  Range of discount rates is centred on \textbf{Tab A}'s $d_{MT30yr}$ discount rate, with  $\pm$ 2$\sigma$ range, where standard deviation is derived from analysis of 20 and 30 year treasury bill rates (see \S \ref{CarbonBondInterestRate}). This 2$\sigma$ range is chosen so that it encompasses the DICE model standard discount rate range of 5.1\%. It is unlikely to be realistic to presume a fixed discount rate, so the real world should be expected to take a wandering path relative to this isosurface. Number of years in computation scale \textbf{Tab C} is 300 year nominal term of our climate bond notes, and the time span of the original DICE model. Time span up to 1500 years is provided. The \textbf{minor Z log scale contours} are 2, 4, 6, and 8 times the previous major contour. }
\label{Fig_SCFgas_isosurface_fullsize}
\end{figure}

\clearpage
\section{Summary SC-GHG tables} \label{SC-GHG_tables}

These summary SC-GHG tables only go forward for 95 years incremented 5 years at a time beginning 2020. Keep in mind that these tables assume 1 tonne of carbon is emitted in the year shown in the table, with a term of 300, 500, or 1500 years going forward.  

In the spreadsheet version of these SC-GHG tables it can be noticed that the other GHGs than CO$_2$ also rise due to CO$_2$. This is because as CO$_2$ rises, it moves the other GHG's effects into a higher temperature region, which means slightly higher damages. However, in the 2 decimal place scientific notation used in these tables, this effect is too small to be seen. 

Discount rates shown in these tables are limited to the range from -2$\sigma$  (-0.33\%) to the nominal carbon bond central discount rate (1.57\%). A wider range is visible in the main paper's \S8, fig. \emph{8}. Note that in the 2$\sigma$ range, the lowest discount is -2.2\% and the highest is 5.1\%. Negative discount rates act as positive interest rates.

Thus, there is both risk and uncertainty relevant to these tables that are beyond the ability of this modelling effort to fully address. It remains for the reader to pick the term for the computation that they think is most appropriate, and to understand that the tables and spreadsheets are a limited view of what is visible in figures \ref{Fig_SCCO2_isosurface_fullsize}-\ref{Fig_SCFgas_isosurface_fullsize}. However, the justification for cutting off the span of years for a cost of any GHG is primarily that discount rates cause the cost in later years to go to zero. Also, even prior to 2076, we believe our figures may well be lower than the real world.  We believe this because, as we have stressed, tipping point effects are not fully represented in this strict empirical method, and the link provided by the ocean  heat conjecture with our weather damages dataset means that OPTiMEM is necessarily partial. The question is, to what degree is this model partial?  

\subsection{Using summary SC-GHG tables for damages computations}

We believe that the \textbf{Permafrost thaw}, and \textbf{Permafrost thaw S-aerosol} scenarios in the discount range from 0.0\% to dMT30yr's 1.57\% discount rate have a higher probability---at least we hope so. The negative discount rate range of results will not be sustainable. However, we can drift into the negative discount range quite easily with bad policy. The effects of that are likely to be signalled by devastating consequences that may be unrecoverable for centuries. The primary bad policies are two. First is the use of low (or even negative) EROEI energy sources \cite{Notton2018IntermittentRenewables,Allison2023InadequacyOfWind,Tverberg2021FossilFuelProblem,Clack2017MZJrebuttal}, as this will hamstring societies that do so from being able to deal with damages by building and rebuilding fast enough to compensate. The second bad policy is austerity that lowers the money supply when it needs to be growing.

\subsection{-2.15\% (-2$\sigma$) discount rate SC-GHG tables}

\begin{sidewaystable}[]
   \caption{-2.15\% (-2$\sigma$) discount rate   SC-GHG computations for \textbf{Baseline} scenario: starting the years listed, for each GHG, 300 and 500 year terms. Real dollars. Here $\sigma$ is the standard deviation of 30 year US Treasury bills that are the basis of $dMT30yr$. L=Low, C=Central, H=High. For N$_2$O anomaly explanation see \ref{SC-GHG_N2O_anomaly}} 
\label{PVs_300_yearSCC_B2.15}
\small
\begin{tabular}{crrr|rrr|rrr|rrr}
300 yr   & -2.15\%    & -2.15\%    & -2.15\%    & -2.15\%    & -2.15\%    & -2.15\%    & -2.15\%    & -2.15\%    & -2.15\%    & -2.15\%    & -2.15\%    & -2.15\%    \\
Starting & -2$\sigma$ & -2$\sigma$ & -2$\sigma$ & -2$\sigma$ & -2$\sigma$ & -2$\sigma$ & -2$\sigma$ & -2$\sigma$ & -2$\sigma$ & -2$\sigma$ & -2$\sigma$ & -2$\sigma$ \\
year CE  & CO2 L      & CO2 C      & CO2 H      & CH4 L      & CH4 C      & CH4 H      & N2O L \ref{SC-GHG_N2O_anomaly}     & N2O C      & N2O H \ref{SC-GHG_N2O_anomaly}      & Fgas L     & Fgas C     & Fgas H     \\
2025     & 1.71E+04   & 2.43E+04   & 3.34E+04   & 3.00E+03   & 3.00E+03   & 3.00E+03   & 1.71E+06   & 1.69E+06   & 1.66E+06   & 4.29E+08   & 4.29E+08   & 4.29E+08   \\
2030     & 1.74E+04   & 2.49E+04   & 3.40E+04   & 3.22E+03   & 3.22E+03   & 3.22E+03   & 1.74E+06   & 1.71E+06   & 1.69E+06   & 4.36E+08   & 4.36E+08   & 4.36E+08   \\
2035     & 1.77E+04   & 2.54E+04   & 3.46E+04   & 3.45E+03   & 3.45E+03   & 3.45E+03   & 1.76E+06   & 1.74E+06   & 1.71E+06   & 4.43E+08   & 4.43E+08   & 4.43E+08   \\
2040     & 1.80E+04   & 2.58E+04   & 3.53E+04   & 3.69E+03   & 3.69E+03   & 3.69E+03   & 1.79E+06   & 1.76E+06   & 1.74E+06   & 4.50E+08   & 4.50E+08   & 4.50E+08   \\
2045     & 1.83E+04   & 2.63E+04   & 3.59E+04   & 3.93E+03   & 3.93E+03   & 3.93E+03   & 1.81E+06   & 1.79E+06   & 1.76E+06   & 4.57E+08   & 4.57E+08   & 4.57E+08   \\
\end{tabular} 
\end{sidewaystable}

\begin{sidewaystable}[]
     \caption{-2.15\% (-2$\sigma$) discount rate   SC-GHG computations for \textbf{Baseline S-aerosol} scenario: starting the years listed, for each GHG, 300 and 500 year terms. Real dollars. Here $\sigma$ is the standard deviation of 30 year US Treasury bills that are the basis of $dMT30yr$. L=Low, C=Central, H=High. For N$_2$O anomaly explanation see \ref{SC-GHG_N2O_anomaly}} 
\label{PVs_300_yearSCC_BH2.15}
\small
\begin{tabular}{crrr|rrr|rrr|rrr}
300 yr   & -2.15\%    & -2.15\%    & -2.15\%    & -2.15\%    & -2.15\%    & -2.15\%    & -2.15\%    & -2.15\%    & -2.15\%    & -2.15\%    & -2.15\%    & -2.15\%    \\
Starting & -2$\sigma$ & -2$\sigma$ & -2$\sigma$ & -2$\sigma$ & -2$\sigma$ & -2$\sigma$ & -2$\sigma$ & -2$\sigma$ & -2$\sigma$ & -2$\sigma$ & -2$\sigma$ & -2$\sigma$ \\
year CE  & CO2 L      & CO2 C      & CO2 H      & CH4 L      & CH4 C      & CH4 H      & N2O L \ref{SC-GHG_N2O_anomaly}     & N2O C      & N2O H \ref{SC-GHG_N2O_anomaly}      & Fgas L     & Fgas C     & Fgas H     \\
2025     & 1.93E+04   & 3.11E+04   & 4.49E+04   & 3.18E+03   & 3.52E+03   & 3.70E+03   & 2.72E+06   & 2.69E+06   & 2.66E+06   & 6.81E+08   & 6.82E+08   & 6.83E+08   \\
2030     & 1.97E+04   & 3.18E+04   & 4.58E+04   & 3.49E+03   & 3.91E+03   & 4.09E+03   & 2.76E+06   & 2.73E+06   & 2.70E+06   & 6.92E+08   & 6.94E+08   & 6.95E+08   \\
2035     & 2.00E+04   & 3.24E+04   & 4.67E+04   & 3.85E+03   & 4.31E+03   & 4.48E+03   & 2.81E+06   & 2.77E+06   & 2.74E+06   & 7.04E+08   & 7.05E+08   & 7.06E+08   \\
2040     & 2.03E+04   & 3.30E+04   & 4.76E+04   & 4.24E+03   & 4.72E+03   & 4.89E+03   & 2.85E+06   & 2.81E+06   & 2.78E+06   & 7.15E+08   & 7.17E+08   & 7.17E+08   \\
2045     & 2.06E+04   & 3.37E+04   & 4.85E+04   & 4.66E+03   & 5.14E+03   & 5.31E+03   & 2.89E+06   & 2.85E+06   & 2.82E+06   & 7.26E+08   & 7.28E+08   & 7.29E+08   \\
\end{tabular}
\end{sidewaystable}

\begin{sidewaystable}[]
      \caption{-2.15\% (-2$\sigma$) discount rate   SC-GHG computations for \textbf{Permafrost thaw} scenario: starting the years listed, for each GHG, 300 and 500 year terms. Real dollars. Here $\sigma$ is the standard deviation of 30 year US Treasury bills that are the basis of $dMT30yr$. L=Low, C=Central, H=High. For N$_2$O anomaly explanation see \ref{SC-GHG_N2O_anomaly}} 
\label{PVs_300_yearSCC_S2.15}
\small
\begin{tabular}{crrr|rrr|rrr|rrr}
300 yr   & -2.15\%    & -2.15\%    & -2.15\%    & -2.15\%    & -2.15\%    & -2.15\%    & -2.15\%    & -2.15\%    & -2.15\%    & -2.15\%    & -2.15\%    & -2.15\%    \\
Starting & -2$\sigma$ & -2$\sigma$ & -2$\sigma$ & -2$\sigma$ & -2$\sigma$ & -2$\sigma$ & -2$\sigma$ & -2$\sigma$ & -2$\sigma$ & -2$\sigma$ & -2$\sigma$ & -2$\sigma$ \\
year CE  & CO2 L      & CO2 C      & CO2 H      & CH4 L      & CH4 C      & CH4 H      & N2O L \ref{SC-GHG_N2O_anomaly}     & N2O C      & N2O H \ref{SC-GHG_N2O_anomaly}      & Fgas L     & Fgas C     & Fgas H     \\
2025     & 1.53E+04   & 2.13E+04   & 2.92E+04   & 2.93E+03   & 2.93E+03   & 2.93E+03   & 1.69E+06   & 1.66E+06   & 1.64E+06   & 4.03E+08   & 4.03E+08   & 4.03E+08   \\
2030     & 1.56E+04   & 2.17E+04   & 2.97E+04   & 3.13E+03   & 3.13E+03   & 3.13E+03   & 1.71E+06   & 1.69E+06   & 1.66E+06   & 4.10E+08   & 4.10E+08   & 4.10E+08   \\
2035     & 1.59E+04   & 2.21E+04   & 3.02E+04   & 3.32E+03   & 3.32E+03   & 3.32E+03   & 1.74E+06   & 1.71E+06   & 1.69E+06   & 4.16E+08   & 4.16E+08   & 4.16E+08   \\
2040     & 1.61E+04   & 2.25E+04   & 3.08E+04   & 3.53E+03   & 3.53E+03   & 3.53E+03   & 1.76E+06   & 1.73E+06   & 1.71E+06   & 4.23E+08   & 4.23E+08   & 4.23E+08   \\
2045     & 1.64E+04   & 2.29E+04   & 3.13E+04   & 3.73E+03   & 3.73E+03   & 3.73E+03   & 1.79E+06   & 1.76E+06   & 1.73E+06   & 4.29E+08   & 4.29E+08   & 4.29E+08   \\
\end{tabular}
\end{sidewaystable}

\begin{sidewaystable}[]
   \caption{-2.15\% (-2$\sigma$) discount rate   SC-GHG computations for \textbf{Permafrost thaw S-Aerosol} scenario: starting the years listed, for each GHG, 300 and 500 year terms. Real dollars. Here $\sigma$ is the standard deviation of 30 year US Treasury bills that are the basis of $dMT30yr$. L=Low, C=Central, H=High. For N$_2$O anomaly explanation see \ref{SC-GHG_N2O_anomaly}} 
\label{PVs_300_yearSCC_SH2.15}
\small
\begin{tabular}{crrr|rrr|rrr|rrr}
300 yr   & -2.15\%    & -2.15\%    & -2.15\%    & -2.15\%    & -2.15\%    & -2.15\%    & -2.15\%    & -2.15\%    & -2.15\%    & -2.15\%    & -2.15\%    & -2.15\%    \\
Starting & -2$\sigma$ & -2$\sigma$ & -2$\sigma$ & -2$\sigma$ & -2$\sigma$ & -2$\sigma$ & -2$\sigma$ & -2$\sigma$ & -2$\sigma$ & -2$\sigma$ & -2$\sigma$ & -2$\sigma$ \\
year CE  & CO2 L      & CO2 C      & CO2 H      & CH4 L      & CH4 C      & CH4 H      & N2O L \ref{SC-GHG_N2O_anomaly}     & N2O C      & N2O H \ref{SC-GHG_N2O_anomaly}      & Fgas L     & Fgas C     & Fgas H     \\
2025     & 1.84E+04   & 2.81E+04   & 4.01E+04   & 3.11E+03   & 3.44E+03   & 3.62E+03   & 2.68E+06   & 2.65E+06   & 2.61E+06   & 6.40E+08   & 6.42E+08   & 6.42E+08   \\
2030     & 1.88E+04   & 2.87E+04   & 4.09E+04   & 3.40E+03   & 3.80E+03   & 3.97E+03   & 2.73E+06   & 2.69E+06   & 2.65E+06   & 6.51E+08   & 6.53E+08   & 6.53E+08   \\
2035     & 1.91E+04   & 2.92E+04   & 4.16E+04   & 3.73E+03   & 4.16E+03   & 4.32E+03   & 2.77E+06   & 2.73E+06   & 2.69E+06   & 6.62E+08   & 6.64E+08   & 6.64E+08   \\
2040     & 1.94E+04   & 2.98E+04   & 4.24E+04   & 4.09E+03   & 4.52E+03   & 4.68E+03   & 2.81E+06   & 2.77E+06   & 2.73E+06   & 6.73E+08   & 6.74E+08   & 6.75E+08   \\
2045     & 1.97E+04   & 3.03E+04   & 4.31E+04   & 4.47E+03   & 4.89E+03   & 5.05E+03   & 2.85E+06   & 2.81E+06   & 2.77E+06   & 6.84E+08   & 6.85E+08   & 6.85E+08   \\
\end{tabular}   \end{sidewaystable}
\clearpage

\subsection{0.0\% (zero) discount rate discount rate SC-GHG tables}

\begin{sidewaystable}[]
   \caption{0.0\% (zero) discount rate  SC-GHG computations for \textbf{Baseline} scenario: starting the years listed, for each GHG, 300 and 500 year terms. Real dollars. Here $\sigma$ is the standard deviation of 30 year US Treasury bills that are the basis of $dMT30yr$. L=Low, C=Central, H=High. For N$_2$O anomaly explanation see \ref{SC-GHG_N2O_anomaly}} 
\label{PVs_300_yearSCC_B0.0}
\small
\begin{tabular}{crrr|rrr|rrr|rrr}
300 yr   & 0.0\%    & 0.0\%    & 0.0\%    & 0.0\%    & 0.0\%    & 0.0\%    & 0.0\%    & 0.0\%    & 0.0\%    & 0.0\%    & 0.0\%    & 0.0\%    \\
Starting & zero     & zero     & zero     & zero     & zero     & zero     & zero     & zero     & zero     & zero     & zero     & zero     \\
year CE  & CO2 L    & CO2 C    & CO2 H    & CH4 L    & CH4 C    & CH4 H    & N2O L \ref{SC-GHG_N2O_anomaly}     & N2O C      & N2O H \ref{SC-GHG_N2O_anomaly}    & Fgas L   & Fgas C   & Fgas H   \\
2025     & 1.12E+02 & 1.65E+02 & 2.18E+02 & 2.29E+03 & 2.29E+03 & 2.29E+03 & 2.44E+04 & 2.41E+04 & 2.39E+04 & 2.80E+06 & 2.80E+06 & 2.80E+06 \\
2030     & 1.16E+02 & 1.71E+02 & 2.26E+02 & 2.46E+03 & 2.46E+03 & 2.46E+03 & 2.50E+04 & 2.47E+04 & 2.45E+04 & 2.88E+06 & 2.88E+06 & 2.88E+06 \\
2035     & 1.20E+02 & 1.78E+02 & 2.33E+02 & 2.63E+03 & 2.63E+03 & 2.63E+03 & 2.57E+04 & 2.54E+04 & 2.52E+04 & 2.96E+06 & 2.96E+06 & 2.96E+06 \\
2040     & 1.25E+02 & 1.84E+02 & 2.40E+02 & 2.82E+03 & 2.82E+03 & 2.82E+03 & 2.63E+04 & 2.60E+04 & 2.58E+04 & 3.04E+06 & 3.04E+06 & 3.04E+06 \\
2045     & 1.29E+02 & 1.90E+02 & 2.47E+02 & 3.01E+03 & 3.01E+03 & 3.01E+03 & 2.70E+04 & 2.67E+04 & 2.65E+04 & 3.12E+06 & 3.12E+06 & 3.12E+06 \\
\end{tabular} 
\end{sidewaystable}

\begin{sidewaystable}[]
     \caption{0.0\% (zero) discount rate  SC-GHG computations for \textbf{Baseline S-aerosol} scenario: starting the years listed, for each GHG, 300 and 500 year terms. Real dollars. Here $\sigma$ is the standard deviation of 30 year US Treasury bills that are the basis of $dMT30yr$. L=Low, C=Central, H=High. For N$_2$O anomaly explanation see \ref{SC-GHG_N2O_anomaly}} 
\label{PVs_300_yearSCC_BH0.0}
\small
\begin{tabular}{crrr|rrr|rrr|rrr}
300 yr   & 0.0\%    & 0.0\%    & 0.0\%    & 0.0\%    & 0.0\%    & 0.0\%    & 0.0\%    & 0.0\%    & 0.0\%    & 0.0\%    & 0.0\%    & 0.0\%    \\
Starting & zero     & zero     & zero     & zero     & zero     & zero     & zero     & zero     & zero     & zero     & zero     & zero     \\
year CE  & CO2 L    & CO2 C    & CO2 H    & CH4 L    & CH4 C    & CH4 H    & N2O L \ref{SC-GHG_N2O_anomaly}     & N2O C      & N2O H \ref{SC-GHG_N2O_anomaly}    & Fgas L   & Fgas C   & Fgas H   \\
2025     & 1.25E+02 & 2.06E+02 & 2.86E+02 & 2.38E+03 & 2.63E+03 & 2.78E+03 & 3.65E+04 & 3.68E+04 & 3.68E+04 & 4.35E+06 & 4.38E+06 & 4.40E+06 \\
2030     & 1.29E+02 & 2.14E+02 & 2.96E+02 & 2.61E+03 & 2.94E+03 & 3.08E+03 & 3.78E+04 & 3.80E+04 & 3.79E+04 & 4.48E+06 & 4.52E+06 & 4.53E+06 \\
2035     & 1.34E+02 & 2.22E+02 & 3.06E+02 & 2.88E+03 & 3.25E+03 & 3.39E+03 & 3.90E+04 & 3.91E+04 & 3.90E+04 & 4.62E+06 & 4.65E+06 & 4.66E+06 \\
2040     & 1.39E+02 & 2.31E+02 & 3.16E+02 & 3.19E+03 & 3.57E+03 & 3.70E+03 & 4.02E+04 & 4.03E+04 & 4.02E+04 & 4.75E+06 & 4.78E+06 & 4.80E+06 \\
2045     & 1.44E+02 & 2.39E+02 & 3.26E+02 & 3.52E+03 & 3.90E+03 & 4.03E+03 & 4.15E+04 & 4.15E+04 & 4.13E+04 & 4.88E+06 & 4.92E+06 & 4.93E+06 \\
\end{tabular}
\end{sidewaystable}

\begin{sidewaystable}[]
      \caption{0.0\% (zero) discount rate  SC-GHG computations for \textbf{Permafrost thaw} scenario: starting the years listed, for each GHG, 300 and 500 year terms. Real dollars. Here $\sigma$ is the standard deviation of 30 year US Treasury bills that are the basis of $dMT30yr$. L=Low, C=Central, H=High. For N$_2$O anomaly explanation see \ref{SC-GHG_N2O_anomaly}} 
\label{PVs_300_yearSCC_S0.0}
\small
\begin{tabular}{crrr|rrr|rrr|rrr}
300 yr   & 0.0\%    & 0.0\%    & 0.0\%    & 0.0\%    & 0.0\%    & 0.0\%    & 0.0\%    & 0.0\%    & 0.0\%    & 0.0\%    & 0.0\%    & 0.0\%    \\
Starting & zero     & zero     & zero     & zero     & zero     & zero     & zero     & zero     & zero     & zero     & zero     & zero     \\
year CE  & CO2 L    & CO2 C    & CO2 H    & CH4 L    & CH4 C    & CH4 H    & N2O L \ref{SC-GHG_N2O_anomaly}     & N2O C      & N2O H \ref{SC-GHG_N2O_anomaly}    & Fgas L   & Fgas C   & Fgas H   \\
2025     & 1.01E+02 & 1.48E+02 & 1.96E+02 & 2.24E+03 & 2.24E+03 & 2.24E+03 & 2.42E+04 & 2.39E+04 & 2.37E+04 & 2.65E+06 & 2.65E+06 & 2.65E+06 \\
2030     & 1.05E+02 & 1.54E+02 & 2.02E+02 & 2.40E+03 & 2.40E+03 & 2.40E+03 & 2.48E+04 & 2.45E+04 & 2.43E+04 & 2.72E+06 & 2.72E+06 & 2.72E+06 \\
2035     & 1.09E+02 & 1.59E+02 & 2.08E+02 & 2.55E+03 & 2.55E+03 & 2.55E+03 & 2.54E+04 & 2.51E+04 & 2.49E+04 & 2.80E+06 & 2.80E+06 & 2.80E+06 \\
2040     & 1.13E+02 & 1.64E+02 & 2.15E+02 & 2.71E+03 & 2.71E+03 & 2.71E+03 & 2.61E+04 & 2.58E+04 & 2.55E+04 & 2.87E+06 & 2.87E+06 & 2.87E+06 \\
2045     & 1.17E+02 & 1.70E+02 & 2.21E+02 & 2.87E+03 & 2.87E+03 & 2.87E+03 & 2.67E+04 & 2.64E+04 & 2.62E+04 & 2.95E+06 & 2.95E+06 & 2.95E+06 \\
\end{tabular}
\end{sidewaystable}

\begin{sidewaystable}[]
   \caption{0.0\% (zero) discount rate  SC-GHG computations for \textbf{Permafrost thaw S-Aerosol} scenario: starting the years listed, for each GHG, 300 and 500 year terms. Real dollars. Here $\sigma$ is the standard deviation of 30 year US Treasury bills that are the basis of $dMT30yr$. L=Low, C=Central, H=High. For N$_2$O anomaly explanation see \ref{SC-GHG_N2O_anomaly}} 
\label{PVs_300_yearSCC_SH0.0}
\small
\begin{tabular}{crrr|rrr|rrr|rrr}
300 yr   & 0.0\%    & 0.0\%    & 0.0\%    & 0.0\%    & 0.0\%    & 0.0\%    & 0.0\%    & 0.0\%    & 0.0\%    & 0.0\%    & 0.0\%    & 0.0\%    \\
Starting & zero     & zero     & zero     & zero     & zero     & zero     & zero     & zero     & zero     & zero     & zero     & zero     \\
year CE  & CO2 L    & CO2 C    & CO2 H    & CH4 L    & CH4 C    & CH4 H    & N2O L \ref{SC-GHG_N2O_anomaly}     & N2O C      & N2O H \ref{SC-GHG_N2O_anomaly}    & Fgas L   & Fgas C   & Fgas H   \\
2025     & 1.18E+02 & 1.89E+02 & 2.61E+02 & 2.34E+03 & 2.59E+03 & 2.73E+03 & 3.62E+04 & 3.64E+04 & 3.64E+04 & 4.11E+06 & 4.15E+06 & 4.16E+06 \\
2030     & 1.23E+02 & 1.97E+02 & 2.69E+02 & 2.56E+03 & 2.86E+03 & 3.00E+03 & 3.74E+04 & 3.76E+04 & 3.75E+04 & 4.24E+06 & 4.27E+06 & 4.29E+06 \\
2035     & 1.28E+02 & 2.04E+02 & 2.78E+02 & 2.81E+03 & 3.15E+03 & 3.28E+03 & 3.86E+04 & 3.87E+04 & 3.86E+04 & 4.36E+06 & 4.40E+06 & 4.41E+06 \\
2040     & 1.33E+02 & 2.11E+02 & 2.87E+02 & 3.09E+03 & 3.43E+03 & 3.56E+03 & 3.98E+04 & 3.98E+04 & 3.97E+04 & 4.49E+06 & 4.52E+06 & 4.53E+06 \\
2045     & 1.37E+02 & 2.19E+02 & 2.96E+02 & 3.38E+03 & 3.72E+03 & 3.85E+03 & 4.10E+04 & 4.10E+04 & 4.08E+04 & 4.62E+06 & 4.64E+06 & 4.66E+06 \\
\end{tabular}   \end{sidewaystable}
\clearpage

\subsection{0.435\% (drDICE) discount rate SC-GHG tables}

\begin{sidewaystable}[]
   \caption{0.435\% (drDICE) discount rate  SC-GHG computations for \textbf{Baseline} scenario: starting the years listed, for each GHG, 300 and 500 year terms. Real dollars. Here $\sigma$ is the standard deviation of 30 year US Treasury bills that are the basis of $dMT30yr$. L=Low, C=Central, H=High. For N$_2$O anomaly explanation see \ref{SC-GHG_N2O_anomaly}} 
\label{PVs_300_yearSCC_B0.435}
\small
\begin{tabular}{crrr|rrr|rrr|rrr}
0.435\%  & 0.435\%  & 0.435\%  & 0.435\%  & 0.435\%  & 0.435\%  & 0.435\%  & 0.435\%  & 0.435\%  & 0.435\%  & 0.435\%  & 0.435\%  \\
drDICE   & drDICE   & drDICE   & drDICE   & drDICE   & drDICE   & drDICE   & drDICE   & drDICE   & drDICE   & drDICE   & drDICE   \\
CO2 L    & CO2 C    & CO2 H    & CH4 L    & CH4 C    & CH4 H    & N2O L \ref{SC-GHG_N2O_anomaly}     & N2O C      & N2O H \ref{SC-GHG_N2O_anomaly}    & Fgas L   & Fgas C   & Fgas H   \\
4.96E+01 & 7.43E+01 & 9.61E+01 & 2.17E+03 & 2.17E+03 & 2.17E+03 & 1.41E+04 & 1.40E+04 & 1.39E+04 & 1.23E+06 & 1.23E+06 & 1.23E+06 \\
5.22E+01 & 7.77E+01 & 1.00E+02 & 2.33E+03 & 2.33E+03 & 2.33E+03 & 1.45E+04 & 1.44E+04 & 1.43E+04 & 1.27E+06 & 1.27E+06 & 1.27E+06 \\
5.48E+01 & 8.11E+01 & 1.04E+02 & 2.50E+03 & 2.50E+03 & 2.50E+03 & 1.50E+04 & 1.48E+04 & 1.47E+04 & 1.32E+06 & 1.32E+06 & 1.32E+06 \\
5.75E+01 & 8.47E+01 & 1.08E+02 & 2.68E+03 & 2.68E+03 & 2.68E+03 & 1.54E+04 & 1.52E+04 & 1.51E+04 & 1.36E+06 & 1.36E+06 & 1.36E+06 \\
6.01E+01 & 8.82E+01 & 1.12E+02 & 2.86E+03 & 2.86E+03 & 2.86E+03 & 1.58E+04 & 1.57E+04 & 1.55E+04 & 1.40E+06 & 1.40E+06 & 1.40E+06 \\
\end{tabular} 
\end{sidewaystable}

\begin{sidewaystable}[]
     \caption{0.435\% (drDICE) discount rate  SC-GHG computations for \textbf{Baseline S-aerosol} scenario: starting the years listed, for each GHG, 300 and 500 year terms. Real dollars. Here $\sigma$ is the standard deviation of 30 year US Treasury bills that are the basis of $dMT30yr$. L=Low, C=Central, H=High. For N$_2$O anomaly explanation see \ref{SC-GHG_N2O_anomaly}} 
\label{PVs_300_yearSCC_BH0.435}
\small
\begin{tabular}{crrr|rrr|rrr|rrr}
300 yr   & 0.435\%  & 0.435\%  & 0.435\%  & 0.435\%  & 0.435\%  & 0.435\%  & 0.435\%  & 0.435\%  & 0.435\%  & 0.435\%  & 0.435\%  & 0.435\%  \\
Starting & drDICE   & drDICE   & drDICE   & drDICE   & drDICE   & drDICE   & drDICE   & drDICE   & drDICE   & drDICE   & drDICE   & drDICE   \\
year CE  & CO2 L    & CO2 C    & CO2 H    & CH4 L    & CH4 C    & CH4 H    & N2O L \ref{SC-GHG_N2O_anomaly}     & N2O C      & N2O H \ref{SC-GHG_N2O_anomaly}    & Fgas L   & Fgas C   & Fgas H   \\
2025     & 5.48E+01 & 9.10E+01 & 1.24E+02 & 2.25E+03 & 2.49E+03 & 2.63E+03 & 2.06E+04 & 2.09E+04 & 2.10E+04 & 1.88E+06 & 1.90E+06 & 1.91E+06 \\
2030     & 5.76E+01 & 9.54E+01 & 1.29E+02 & 2.47E+03 & 2.78E+03 & 2.92E+03 & 2.14E+04 & 2.17E+04 & 2.17E+04 & 1.95E+06 & 1.97E+06 & 1.98E+06 \\
2035     & 6.05E+01 & 9.99E+01 & 1.34E+02 & 2.73E+03 & 3.08E+03 & 3.21E+03 & 2.22E+04 & 2.25E+04 & 2.25E+04 & 2.02E+06 & 2.04E+06 & 2.05E+06 \\
2040     & 6.35E+01 & 1.04E+02 & 1.40E+02 & 3.02E+03 & 3.38E+03 & 3.51E+03 & 2.30E+04 & 2.32E+04 & 2.32E+04 & 2.09E+06 & 2.12E+06 & 2.13E+06 \\
2045     & 6.66E+01 & 1.09E+02 & 1.45E+02 & 3.33E+03 & 3.69E+03 & 3.82E+03 & 2.39E+04 & 2.40E+04 & 2.40E+04 & 2.17E+06 & 2.19E+06 & 2.20E+06 \\
\end{tabular}
\end{sidewaystable}

\begin{sidewaystable}[]
      \caption{0.435\% (drDICE) discount rate  SC-GHG computations for \textbf{Permafrost thaw} scenario: starting the years listed, for each GHG, 300 and 500 year terms. Real dollars. Here $\sigma$ is the standard deviation of 30 year US Treasury bills that are the basis of $dMT30yr$. L=Low, C=Central, H=High. For N$_2$O anomaly explanation see \ref{SC-GHG_N2O_anomaly}} 
\label{PVs_300_yearSCC_S0.435}
\small
\begin{tabular}{crrr|rrr|rrr|rrr}
300 yr   & 0.435\%  & 0.435\%  & 0.435\%  & 0.435\%  & 0.435\%  & 0.435\%  & 0.435\%  & 0.435\%  & 0.435\%  & 0.435\%  & 0.435\%  & 0.435\%  \\
Starting & drDICE   & drDICE   & drDICE   & drDICE   & drDICE   & drDICE   & drDICE   & drDICE   & drDICE   & drDICE   & drDICE   & drDICE   \\
year CE  & CO2 L    & CO2 C    & CO2 H    & CH4 L    & CH4 C    & CH4 H    & N2O L \ref{SC-GHG_N2O_anomaly}     & N2O C      & N2O H \ref{SC-GHG_N2O_anomaly}    & Fgas L   & Fgas C   & Fgas H   \\
2025     & 4.55E+01 & 6.76E+01 & 8.75E+01 & 2.13E+03 & 2.13E+03 & 2.13E+03 & 1.40E+04 & 1.39E+04 & 1.38E+04 & 1.17E+06 & 1.17E+06 & 1.17E+06 \\
2030     & 4.78E+01 & 7.06E+01 & 9.09E+01 & 2.28E+03 & 2.28E+03 & 2.28E+03 & 1.44E+04 & 1.43E+04 & 1.42E+04 & 1.21E+06 & 1.21E+06 & 1.21E+06 \\
2035     & 5.02E+01 & 7.36E+01 & 9.42E+01 & 2.42E+03 & 2.42E+03 & 2.42E+03 & 1.48E+04 & 1.47E+04 & 1.46E+04 & 1.25E+06 & 1.25E+06 & 1.25E+06 \\
2040     & 5.26E+01 & 7.67E+01 & 9.77E+01 & 2.57E+03 & 2.57E+03 & 2.57E+03 & 1.52E+04 & 1.51E+04 & 1.50E+04 & 1.29E+06 & 1.29E+06 & 1.29E+06 \\
2045     & 5.50E+01 & 7.98E+01 & 1.01E+02 & 2.72E+03 & 2.72E+03 & 2.72E+03 & 1.57E+04 & 1.55E+04 & 1.54E+04 & 1.33E+06 & 1.33E+06 & 1.33E+06 \\
\end{tabular}
\end{sidewaystable}

\begin{sidewaystable}[]
   \caption{0.435\% (drDICE) discount rate  SC-GHG computations for \textbf{Permafrost thaw S-Aerosol} scenario: starting the years listed, for each GHG, 300 and 500 year terms. Real dollars. Here $\sigma$ is the standard deviation of 30 year US Treasury bills that are the basis of $dMT30yr$. L=Low, C=Central, H=High. For N$_2$O anomaly explanation see \ref{SC-GHG_N2O_anomaly}} 
\label{PVs_300_yearSCC_SH0.435}
\small
\begin{tabular}{crrr|rrr|rrr|rrr}
300 yr   & 0.435\%  & 0.435\%  & 0.435\%  & 0.435\%  & 0.435\%  & 0.435\%  & 0.435\%  & 0.435\%  & 0.435\%  & 0.435\%  & 0.435\%  & 0.435\%  \\
Starting & drDICE   & drDICE   & drDICE   & drDICE   & drDICE   & drDICE   & drDICE   & drDICE   & drDICE   & drDICE   & drDICE   & drDICE   \\
year CE  & CO2 L    & CO2 C    & CO2 H    & CH4 L    & CH4 C    & CH4 H    & N2O L \ref{SC-GHG_N2O_anomaly}     & N2O C      & N2O H \ref{SC-GHG_N2O_anomaly}    & Fgas L   & Fgas C   & Fgas H   \\
2025     & 5.22E+01 & 8.43E+01 & 1.14E+02 & 2.22E+03 & 2.45E+03 & 2.58E+03 & 2.04E+04 & 2.07E+04 & 2.08E+04 & 1.79E+06 & 1.81E+06 & 1.82E+06 \\
2030     & 5.49E+01 & 8.83E+01 & 1.19E+02 & 2.42E+03 & 2.71E+03 & 2.85E+03 & 2.12E+04 & 2.15E+04 & 2.15E+04 & 1.85E+06 & 1.87E+06 & 1.88E+06 \\
2035     & 5.77E+01 & 9.24E+01 & 1.23E+02 & 2.66E+03 & 2.98E+03 & 3.11E+03 & 2.20E+04 & 2.23E+04 & 2.23E+04 & 1.92E+06 & 1.94E+06 & 1.95E+06 \\
2040     & 6.05E+01 & 9.65E+01 & 1.28E+02 & 2.92E+03 & 3.26E+03 & 3.38E+03 & 2.28E+04 & 2.30E+04 & 2.30E+04 & 1.99E+06 & 2.01E+06 & 2.02E+06 \\
2045     & 6.34E+01 & 1.01E+02 & 1.33E+02 & 3.20E+03 & 3.53E+03 & 3.65E+03 & 2.36E+04 & 2.38E+04 & 2.37E+04 & 2.06E+06 & 2.07E+06 & 2.08E+06 \\
\end{tabular}   \end{sidewaystable}
\clearpage

\subsection{1.57\% (dMT30yr) discount rate SC-GHG tables}

\begin{sidewaystable}[]
   \caption{1.57\% (dMT30yr) discount rate  SC-GHG computations for \textbf{Baseline} scenario: starting the years listed, for each GHG, 300 and 500 year terms. Real dollars. Here $\sigma$ is the standard deviation of 30 year US Treasury bills that are the basis of $dMT30yr$. L=Low, C=Central, H=High. For N$_2$O anomaly explanation see \ref{SC-GHG_N2O_anomaly}} 
\label{PVs_300_yearSCC_B1.57}
\small
\begin{tabular}{crrr|rrr|rrr|rrr}
300 yr   & 1.57 \%  & 1.57 \%  & 1.57 \%  & 1.57 \%  & 1.57 \%  & 1.57 \%  & 1.57 \%  & 1.57 \%  & 1.57 \%  & 1.57 \%  & 1.57 \%  & 1.57 \%  \\
Starting & dMT30yr  & dMT30yr  & dMT30yr  & dMT30yr  & dMT30yr  & dMT30yr  & dMT30yr  & dMT30yr  & dMT30yr  & dMT30yr  & dMT30yr  & dMT30yr  \\
year CE  & CO2 L    & CO2 C    & CO2 H    & CH4 L    & CH4 C    & CH4 H    & N2O L \ref{SC-GHG_N2O_anomaly}     & N2O C      & N2O H \ref{SC-GHG_N2O_anomaly}    & Fgas L   & Fgas C   & Fgas H   \\
2025     & 1.43E+01 & 2.13E+01 & 2.62E+01 & 1.95E+03 & 1.95E+03 & 1.95E+03 & 6.36E+03 & 6.31E+03 & 6.28E+03 & 3.36E+05 & 3.36E+05 & 3.36E+05 \\
2030     & 1.55E+01 & 2.27E+01 & 2.78E+01 & 2.09E+03 & 2.09E+03 & 2.09E+03 & 6.55E+03 & 6.50E+03 & 6.47E+03 & 3.52E+05 & 3.52E+05 & 3.52E+05 \\
2035     & 1.68E+01 & 2.42E+01 & 2.93E+01 & 2.24E+03 & 2.24E+03 & 2.24E+03 & 6.75E+03 & 6.70E+03 & 6.66E+03 & 3.68E+05 & 3.68E+05 & 3.68E+05 \\
2040     & 1.80E+01 & 2.57E+01 & 3.09E+01 & 2.40E+03 & 2.40E+03 & 2.40E+03 & 6.96E+03 & 6.91E+03 & 6.87E+03 & 3.85E+05 & 3.85E+05 & 3.85E+05 \\
2045     & 1.93E+01 & 2.72E+01 & 3.25E+01 & 2.57E+03 & 2.57E+03 & 2.57E+03 & 7.18E+03 & 7.12E+03 & 7.08E+03 & 4.02E+05 & 4.02E+05 & 4.02E+05 \\
\end{tabular} 
\end{sidewaystable}

\begin{sidewaystable}[]
     \caption{1.57\% (dMT30yr) discount rate  SC-GHG computations for \textbf{Baseline S-aerosol} scenario: starting the years listed, for each GHG, 300 and 500 year terms. Real dollars. Here $\sigma$ is the standard deviation of 30 year US Treasury bills that are the basis of $dMT30yr$. L=Low, C=Central, H=High. For N$_2$O anomaly explanation see \ref{SC-GHG_N2O_anomaly}} 
\label{PVs_300_yearSCC_BH1.57}
\small
\begin{tabular}{crrr|rrr|rrr|rrr}
300 yr   & 1.57 \%  & 1.57 \%  & 1.57 \%  & 1.57 \%  & 1.57 \%  & 1.57 \%  & 1.57 \%  & 1.57 \%  & 1.57 \%  & 1.57 \%  & 1.57 \%  & 1.57 \%  \\
Starting & dMT30yr  & dMT30yr  & dMT30yr  & dMT30yr  & dMT30yr  & dMT30yr  & dMT30yr  & dMT30yr  & dMT30yr  & dMT30yr  & dMT30yr  & dMT30yr  \\
year CE  & CO2 L    & CO2 C    & CO2 H    & CH4 L    & CH4 C    & CH4 H    & N2O L \ref{SC-GHG_N2O_anomaly}     & N2O C      & N2O H \ref{SC-GHG_N2O_anomaly}    & Fgas L   & Fgas C   & Fgas H   \\
2025     & 1.52E+01 & 2.48E+01 & 3.19E+01 & 2.01E+03 & 2.22E+03 & 2.35E+03 & 8.68E+03 & 9.02E+03 & 9.14E+03 & 4.88E+05 & 5.01E+05 & 5.07E+05 \\
2030     & 1.66E+01 & 2.66E+01 & 3.40E+01 & 2.21E+03 & 2.48E+03 & 2.60E+03 & 9.12E+03 & 9.42E+03 & 9.52E+03 & 5.16E+05 & 5.28E+05 & 5.34E+05 \\
2035     & 1.79E+01 & 2.85E+01 & 3.61E+01 & 2.44E+03 & 2.75E+03 & 2.87E+03 & 9.57E+03 & 9.83E+03 & 9.91E+03 & 5.44E+05 & 5.55E+05 & 5.61E+05 \\
2040     & 1.93E+01 & 3.04E+01 & 3.83E+01 & 2.69E+03 & 3.02E+03 & 3.14E+03 & 1.00E+04 & 1.02E+04 & 1.03E+04 & 5.72E+05 & 5.83E+05 & 5.88E+05 \\
2045     & 2.08E+01 & 3.24E+01 & 4.05E+01 & 2.98E+03 & 3.31E+03 & 3.42E+03 & 1.05E+04 & 1.07E+04 & 1.07E+04 & 6.01E+05 & 6.12E+05 & 6.17E+05 \\
\end{tabular}
\end{sidewaystable}

\begin{sidewaystable}[]
      \caption{1.57\% (dMT30yr) discount rate  SC-GHG computations for \textbf{Permafrost thaw} scenario: starting the years listed, for each GHG, 300 and 500 year terms. Real dollars. Here $\sigma$ is the standard deviation of 30 year US Treasury bills that are the basis of $dMT30yr$. L=Low, C=Central, H=High. For N$_2$O anomaly explanation see \ref{SC-GHG_N2O_anomaly}} 
\label{PVs_300_yearSCC_S1.57}
\small
\begin{tabular}{crrr|rrr|rrr|rrr}
300 yr   & 1.57 \%  & 1.57 \%  & 1.57 \%  & 1.57 \%  & 1.57 \%  & 1.57 \%  & 1.57 \%  & 1.57 \%  & 1.57 \%  & 1.57 \%  & 1.57 \%  & 1.57 \%  \\
Starting & dMT30yr  & dMT30yr  & dMT30yr  & dMT30yr  & dMT30yr  & dMT30yr  & dMT30yr  & dMT30yr  & dMT30yr  & dMT30yr  & dMT30yr  & dMT30yr  \\
year CE  & CO2 L    & CO2 C    & CO2 H    & CH4 L    & CH4 C    & CH4 H    & N2O L \ref{SC-GHG_N2O_anomaly}     & N2O C      & N2O H \ref{SC-GHG_N2O_anomaly}    & Fgas L   & Fgas C   & Fgas H   \\
2025     & 1.35E+01 & 2.00E+01 & 2.47E+01 & 1.91E+03 & 1.91E+03 & 1.91E+03 & 6.33E+03 & 6.28E+03 & 6.26E+03 & 3.24E+05 & 3.24E+05 & 3.24E+05 \\
2030     & 1.46E+01 & 2.14E+01 & 2.61E+01 & 2.04E+03 & 2.04E+03 & 2.04E+03 & 6.52E+03 & 6.47E+03 & 6.44E+03 & 3.39E+05 & 3.39E+05 & 3.39E+05 \\
2035     & 1.58E+01 & 2.27E+01 & 2.74E+01 & 2.18E+03 & 2.18E+03 & 2.18E+03 & 6.72E+03 & 6.66E+03 & 6.63E+03 & 3.54E+05 & 3.54E+05 & 3.54E+05 \\
2040     & 1.69E+01 & 2.40E+01 & 2.88E+01 & 2.31E+03 & 2.31E+03 & 2.31E+03 & 6.92E+03 & 6.87E+03 & 6.83E+03 & 3.69E+05 & 3.69E+05 & 3.69E+05 \\
2045     & 1.81E+01 & 2.54E+01 & 3.03E+01 & 2.45E+03 & 2.45E+03 & 2.45E+03 & 7.14E+03 & 7.08E+03 & 7.04E+03 & 3.85E+05 & 3.85E+05 & 3.85E+05 \\
\end{tabular}
\end{sidewaystable}

\begin{sidewaystable}[]
   \caption{1.57\% (dMT30yr) discount rate  SC-GHG computations for \textbf{Permafrost thaw S-Aerosol} scenario: starting the years listed, for each GHG, 300 and 500 year terms. Real dollars. Here $\sigma$ is the standard deviation of 30 year US Treasury bills that are the basis of $dMT30yr$. L=Low, C=Central, H=High. For N$_2$O anomaly explanation see \ref{SC-GHG_N2O_anomaly}} 
\label{PVs_300_yearSCC_SH1.57}
\small
\begin{tabular}{crrr|rrr|rrr|rrr}
300 yr   & 1.57 \%  & 1.57 \%  & 1.57 \%  & 1.57 \%  & 1.57 \%  & 1.57 \%  & 1.57 \%  & 1.57 \%  & 1.57 \%  & 1.57 \%  & 1.57 \%  & 1.57 \%  \\
Starting & dMT30yr  & dMT30yr  & dMT30yr  & dMT30yr  & dMT30yr  & dMT30yr  & dMT30yr  & dMT30yr  & dMT30yr  & dMT30yr  & dMT30yr  & dMT30yr  \\
year CE  & CO2 L    & CO2 C    & CO2 H    & CH4 L    & CH4 C    & CH4 H    & N2O L \ref{SC-GHG_N2O_anomaly}     & N2O C      & N2O H \ref{SC-GHG_N2O_anomaly}    & Fgas L   & Fgas C   & Fgas H   \\
2025     & 1.47E+01 & 2.35E+01 & 3.02E+01 & 1.98E+03 & 2.18E+03 & 2.31E+03 & 8.64E+03 & 8.97E+03 & 9.09E+03 & 4.71E+05 & 4.83E+05 & 4.89E+05 \\
2030     & 1.59E+01 & 2.52E+01 & 3.21E+01 & 2.16E+03 & 2.43E+03 & 2.55E+03 & 9.08E+03 & 9.37E+03 & 9.47E+03 & 4.97E+05 & 5.08E+05 & 5.14E+05 \\
2035     & 1.72E+01 & 2.69E+01 & 3.41E+01 & 2.38E+03 & 2.67E+03 & 2.79E+03 & 9.52E+03 & 9.77E+03 & 9.85E+03 & 5.23E+05 & 5.34E+05 & 5.39E+05 \\
2040     & 1.85E+01 & 2.87E+01 & 3.60E+01 & 2.61E+03 & 2.92E+03 & 3.03E+03 & 9.96E+03 & 1.02E+04 & 1.02E+04 & 5.49E+05 & 5.60E+05 & 5.65E+05 \\
2045     & 1.99E+01 & 3.05E+01 & 3.80E+01 & 2.87E+03 & 3.17E+03 & 3.27E+03 & 1.04E+04 & 1.06E+04 & 1.06E+04 & 5.76E+05 & 5.87E+05 & 5.91E+05 \\
\end{tabular}   \end{sidewaystable}
\clearpage

\subsection{1500 year SC-GHG tables}
\begin{sidewaystable}[]
   \caption{1500 year term SC-GHG computations for \textbf{Baseline} scenario: forward 1500 years, starting the year listed, for four discount rates. Real dollars. Here $\sigma$ is the standard deviation of 30 year US Treasury bills that are the basis of $dMT30yr$. L=Low, C=Central, H=High. For N$_2$O anomaly explanation see \ref{SC-GHG_N2O_anomaly}} 
\label{PVs_1500_yearSCC_B1}
\small
\begin{tabular}{crrr|rrr|rrr|rrr}
1500 yr  & -2.15\%   & -2.15\%   & -2.15\%   & -2.15\%   & -2.15\%   & -2.15\%   & -2.15\%   & -2.15\%   & -2.15\%   & -2.15\%   & -2.15\%   & -2.15\%   \\
Starting & -2$sigma$ & -2$sigma$ & -2$sigma$ & -2$sigma$ & -2$sigma$ & -2$sigma$ & -2$sigma$ & -2$sigma$ & -2$sigma$ & -2$sigma$ & -2$sigma$ & -2$sigma$ \\
year CE  & CO2 L     & CO2 C     & CO2 H     & CH4 L     & CH4 C     & CH4 H     & N2O L     & N2O C     & N2O H     & Fgas L    & Fgas C    & Fgas H    \\
2025     & 2.60E+15  & 4.28E+15  & 6.35E+15  & 3.00E+03  & 3.00E+03  & 3.00E+03  & 2.77E+13  & 2.74E+13  & 2.70E+13  & 2.70E+19  & 2.70E+19  & 2.70E+19  \\
2030     & 2.60E+15  & 4.28E+15  & 6.35E+15  & 3.22E+03  & 3.22E+03  & 3.22E+03  & 2.77E+13  & 2.74E+13  & 2.70E+13  & 2.70E+19  & 2.70E+19  & 2.70E+19  \\
2035     & 2.60E+15  & 4.28E+15  & 6.35E+15  & 3.45E+03  & 3.45E+03  & 3.45E+03  & 2.77E+13  & 2.74E+13  & 2.70E+13  & 2.70E+19  & 2.70E+19  & 2.70E+19  \\
2040     & 2.60E+15  & 4.28E+15  & 6.35E+15  & 3.69E+03  & 3.69E+03  & 3.69E+03  & 2.77E+13  & 2.73E+13  & 2.70E+13  & 2.70E+19  & 2.70E+19  & 2.70E+19  \\
2045     & 2.60E+15  & 4.28E+15  & 6.35E+15  & 3.93E+03  & 3.93E+03  & 3.93E+03  & 2.76E+13  & 2.73E+13  & 2.69E+13  & 2.70E+19  & 2.70E+19  & 2.70E+19  \\
\end{tabular}
\end{sidewaystable}

\begin{sidewaystable}[]
   \caption{1500 year term SC-GHG computations for \textbf{Baseline} scenario: forward 1500 years, starting the year listed, for four discount rates. Real dollars. Here $\sigma$ is the standard deviation of 30 year US Treasury bills that are the basis of $dMT30yr$. L=Low, C=Central, H=High. For N$_2$O anomaly explanation see \ref{SC-GHG_N2O_anomaly}} 
\label{PVs_1500_yearSCC_B2}
\small
\begin{tabular}{crrr|rrr|rrr|rrr}
1500 yr  & 0.435\%  & 0.435\%  & 0.435\%  & 0.435\%  & 0.435\%  & 0.435\%  & 0.435\%  & 0.435\%  & 0.435\%  & 0.435\%  & 0.435\%  & 0.435\%  \\
Starting & drDICE   & drDICE   & drDICE   & drDICE   & drDICE   & drDICE   & drDICE   & drDICE   & drDICE   & drDICE   & drDICE   & drDICE   \\
year CE  & CO2 L    & CO2 C    & CO2 H    & CH4 L    & CH4 C    & CH4 H    & N2O L \ref{SC-GHG_N2O_anomaly}     & N2O C      & N2O H \ref{SC-GHG_N2O_anomaly}    & Fgas L   & Fgas C   & Fgas H   \\
2025     & 8.02E+01 & 1.18E+02 & 1.57E+02 & 2.17E+03 & 2.17E+03 & 2.17E+03 & 1.50E+04 & 1.48E+04 & 1.47E+04 & 1.91E+06 & 1.91E+06 & 1.91E+06 \\
2030     & 8.30E+01 & 1.21E+02 & 1.62E+02 & 2.33E+03 & 2.33E+03 & 2.33E+03 & 1.54E+04 & 1.52E+04 & 1.51E+04 & 1.95E+06 & 1.95E+06 & 1.95E+06 \\
2035     & 8.57E+01 & 1.25E+02 & 1.66E+02 & 2.50E+03 & 2.50E+03 & 2.50E+03 & 1.58E+04 & 1.57E+04 & 1.55E+04 & 2.00E+06 & 2.00E+06 & 2.00E+06 \\
2040     & 8.85E+01 & 1.29E+02 & 1.70E+02 & 2.68E+03 & 2.68E+03 & 2.68E+03 & 1.62E+04 & 1.61E+04 & 1.60E+04 & 2.05E+06 & 2.05E+06 & 2.05E+06 \\
2045     & 9.13E+01 & 1.33E+02 & 1.74E+02 & 2.86E+03 & 2.86E+03 & 2.86E+03 & 1.67E+04 & 1.65E+04 & 1.64E+04 & 2.09E+06 & 2.09E+06 & 2.09E+06 \\
\end{tabular}
\end{sidewaystable}

\begin{sidewaystable}[]
   \caption{1500 year term SC-GHG computations for \textbf{Baseline S-aerosol} scenario:  forward 1500 years, starting the year listed, for four discount rates. Real dollars. Here $\sigma$ is the standard deviation of 30 year US Treasury bills that are the basis of $dMT30yr$. L=Low, C=Central, H=High. For N$_2$O anomaly explanation see \ref{SC-GHG_N2O_anomaly}} 
\label{PVs_1500_yearSCC_BH1}
\small
\begin{tabular}{crrr|rrr|rrr|rrr}
1500 yr  & -2.15\%    & -2.15\%    & -2.15\%    & -2.15\%    & -2.15\%    & -2.15\%    & -2.15\%    & -2.15\%    & -2.15\%    & -2.15\%    & -2.15\%    & -2.15\%    \\
Starting & -2$\sigma$ & -2$\sigma$ & -2$\sigma$ & -2$\sigma$ & -2$\sigma$ & -2$\sigma$ & -2$\sigma$ & -2$\sigma$ & -2$\sigma$ & -2$\sigma$ & -2$\sigma$ & -2$\sigma$ \\
year CE  & CO2 L      & CO2 C      & CO2 H      & CH4 L      & CH4 C      & CH4 H      & N2O L \ref{SC-GHG_N2O_anomaly}     & N2O C      & N2O H \ref{SC-GHG_N2O_anomaly}      & Fgas L     & Fgas C     & Fgas H     \\
2025     & 2.60E+15   & 4.78E+15   & 8.11E+15   & 3.18E+03   & 3.52E+03   & 3.70E+03   & 4.37E+13   & 4.40E+13   & 4.33E+13   & 4.33E+19   & 4.33E+19   & 4.33E+19   \\
2030     & 2.60E+15   & 4.78E+15   & 8.12E+15   & 3.49E+03   & 3.91E+03   & 4.09E+03   & 4.36E+13   & 4.39E+13   & 4.33E+13   & 4.33E+19   & 4.33E+19   & 4.33E+19   \\
2035     & 2.60E+15   & 4.78E+15   & 8.12E+15   & 3.85E+03   & 4.31E+03   & 4.48E+03   & 4.36E+13   & 4.39E+13   & 4.33E+13   & 4.33E+19   & 4.33E+19   & 4.33E+19   \\
2040     & 2.60E+15   & 4.78E+15   & 8.12E+15   & 4.24E+03   & 4.72E+03   & 4.89E+03   & 4.35E+13   & 4.38E+13   & 4.32E+13   & 4.33E+19   & 4.33E+19   & 4.33E+19   \\
2045     & 2.60E+15   & 4.77E+15   & 8.12E+15   & 4.66E+03   & 5.14E+03   & 5.31E+03   & 4.35E+13   & 4.38E+13   & 4.32E+13   & 4.33E+19   & 4.33E+19   & 4.33E+19   \\
\end{tabular}   \end{sidewaystable}

\begin{sidewaystable}[]
   \caption{1500 year term SC-GHG computations for \textbf{Baseline S-aerosol} scenario:  forward 1500 years, starting the year listed, for four discount rates. Real dollars. Here $\sigma$ is the standard deviation of 30 year US Treasury bills that are the basis of $dMT30yr$. L=Low, C=Central, H=High. For N$_2$O anomaly explanation see \ref{SC-GHG_N2O_anomaly}} 
\label{PVs_1500_yearSCC_BH2}
\small
\begin{tabular}{crrr|rrr|rrr|rrr}
1500 yr  & 0.435\%  & 0.435\%  & 0.435\%  & 0.435\%  & 0.435\%  & 0.435\%  & 0.435\%  & 0.435\%  & 0.435\%  & 0.435\%  & 0.435\%  & 0.435\%  \\
Starting & drDICE   & drDICE   & drDICE   & drDICE   & drDICE   & drDICE   & drDICE   & drDICE   & drDICE   & drDICE   & drDICE   & drDICE   \\
year CE  & CO2 L    & CO2 C    & CO2 H    & CH4 L    & CH4 C    & CH4 H    & N2O L \ref{SC-GHG_N2O_anomaly}     & N2O C      & N2O H \ref{SC-GHG_N2O_anomaly}    & Fgas L   & Fgas C   & Fgas H   \\
2025     & 8.75E+01 & 1.45E+02 & 2.06E+02 & 2.25E+03 & 2.49E+03 & 2.63E+03 & 2.19E+04 & 2.23E+04 & 2.23E+04 & 2.97E+06 & 2.99E+06 & 3.00E+06 \\
2030     & 9.05E+01 & 1.50E+02 & 2.11E+02 & 2.47E+03 & 2.78E+03 & 2.92E+03 & 2.28E+04 & 2.30E+04 & 2.31E+04 & 3.04E+06 & 3.06E+06 & 3.08E+06 \\
2035     & 9.35E+01 & 1.54E+02 & 2.17E+02 & 2.73E+03 & 3.08E+03 & 3.21E+03 & 2.36E+04 & 2.38E+04 & 2.38E+04 & 3.12E+06 & 3.14E+06 & 3.15E+06 \\
2040     & 9.66E+01 & 1.59E+02 & 2.23E+02 & 3.02E+03 & 3.38E+03 & 3.51E+03 & 2.44E+04 & 2.46E+04 & 2.46E+04 & 3.20E+06 & 3.22E+06 & 3.23E+06 \\
2045     & 9.97E+01 & 1.64E+02 & 2.29E+02 & 3.33E+03 & 3.69E+03 & 3.82E+03 & 2.53E+04 & 2.54E+04 & 2.53E+04 & 3.27E+06 & 3.29E+06 & 3.30E+06 \\
\end{tabular}   \end{sidewaystable}

\begin{sidewaystable}[]
   \caption{1500 year term SC-GHG computations for \textbf{Permafrost thaw}  \cite{Schuur2022PermafrostFeedbacksWarming} scenario: forward 1500 years, starting the year listed, for four discount rates. Real dollars. Here $\sigma$ is the standard deviation of 30 year US Treasury bills that are the basis of $dMT30yr$. L=Low, C=Central, H=High. For N$_2$O anomaly explanation see \ref{SC-GHG_N2O_anomaly}} 
\label{PVs_1500_yearSCC_S1}
\small
\begin{tabular}{crrr|rrr|rrr|rrr}
-2.15\%    & -2.15\%    & -2.15\%    & -2.15\%    & -2.15\%    & -2.15\%    & -2.15\%    & -2.15\%    & -2.15\%    & -2.15\%    & -2.15\%    & -2.15\%    \\
-2$\sigma$ & -2$\sigma$ & -2$\sigma$ & -2$\sigma$ & -2$\sigma$ & -2$\sigma$ & -2$\sigma$ & -2$\sigma$ & -2$\sigma$ & -2$\sigma$ & -2$\sigma$ & -2$\sigma$ \\
CO2 L      & CO2 C      & CO2 H      & CH4 L      & CH4 C      & CH4 H      & N2O L \ref{SC-GHG_N2O_anomaly}     & N2O C      & N2O H \ref{SC-GHG_N2O_anomaly}      & Fgas L     & Fgas C     & Fgas H     \\
2.54E+15   & 4.03E+15   & 5.81E+15   & 2.93E+03   & 2.93E+03   & 2.93E+03   & 2.76E+13   & 2.73E+13   & 2.68E+13   & 2.63E+19   & 2.63E+19   & 2.63E+19   \\
2.54E+15   & 4.03E+15   & 5.81E+15   & 3.13E+03   & 3.13E+03   & 3.13E+03   & 2.76E+13   & 2.72E+13   & 2.68E+13   & 2.63E+19   & 2.63E+19   & 2.63E+19   \\
2.54E+15   & 4.03E+15   & 5.81E+15   & 3.32E+03   & 3.32E+03   & 3.32E+03   & 2.76E+13   & 2.72E+13   & 2.68E+13   & 2.63E+19   & 2.63E+19   & 2.63E+19   \\
2.54E+15   & 4.03E+15   & 5.81E+15   & 3.53E+03   & 3.53E+03   & 3.53E+03   & 2.75E+13   & 2.72E+13   & 2.67E+13   & 2.63E+19   & 2.63E+19   & 2.63E+19   \\
2.54E+15   & 4.04E+15   & 5.82E+15   & 3.73E+03   & 3.73E+03   & 3.73E+03   & 2.75E+13   & 2.72E+13   & 2.67E+13   & 2.63E+19   & 2.63E+19   & 2.63E+19   \\
\end{tabular}   \end{sidewaystable}

\begin{sidewaystable}[]
   \caption{1500 year term SC-GHG computations for \textbf{Permafrost thaw}  \cite{Schuur2022PermafrostFeedbacksWarming} scenario: forward 1500 years, starting the year listed, for four discount rates. Real dollars. Here $\sigma$ is the standard deviation of 30 year US Treasury bills that are the basis of $dMT30yr$. L=Low, C=Central, H=High. For N$_2$O anomaly explanation see \ref{SC-GHG_N2O_anomaly}} 
\label{PVs_1500_yearSCC_S2}
\small
\begin{tabular}{crrr|rrr|rrr|rrr}
1500 yr  & 0.435\%  & 0.435\%  & 0.435\%  & 0.435\%  & 0.435\%  & 0.435\%  & 0.435\%  & 0.435\%  & 0.435\%  & 0.435\%  & 0.435\%  & 0.435\%  \\
Starting & drDICE   & drDICE   & drDICE   & drDICE   & drDICE   & drDICE   & drDICE   & drDICE   & drDICE   & drDICE   & drDICE   & drDICE   \\
year CE  & CO2 L    & CO2 C    & CO2 H    & CH4 L    & CH4 C    & CH4 H    & N2O L \ref{SC-GHG_N2O_anomaly}     & N2O C      & N2O H \ref{SC-GHG_N2O_anomaly}    & Fgas L   & Fgas C   & Fgas H   \\
2025     & 7.33E+01 & 1.06E+02 & 1.40E+02 & 2.13E+03 & 2.13E+03 & 2.13E+03 & 1.49E+04 & 1.47E+04 & 1.46E+04 & 1.81E+06 & 1.81E+06 & 1.81E+06 \\
2030     & 7.58E+01 & 1.09E+02 & 1.44E+02 & 2.28E+03 & 2.28E+03 & 2.28E+03 & 1.53E+04 & 1.51E+04 & 1.50E+04 & 1.86E+06 & 1.86E+06 & 1.86E+06 \\
2035     & 7.83E+01 & 1.12E+02 & 1.48E+02 & 2.42E+03 & 2.42E+03 & 2.42E+03 & 1.57E+04 & 1.55E+04 & 1.54E+04 & 1.90E+06 & 1.90E+06 & 1.90E+06 \\
2040     & 8.08E+01 & 1.16E+02 & 1.52E+02 & 2.57E+03 & 2.57E+03 & 2.57E+03 & 1.61E+04 & 1.59E+04 & 1.58E+04 & 1.94E+06 & 1.94E+06 & 1.94E+06 \\
2045     & 8.33E+01 & 1.19E+02 & 1.55E+02 & 2.72E+03 & 2.72E+03 & 2.72E+03 & 1.65E+04 & 1.64E+04 & 1.62E+04 & 1.99E+06 & 1.99E+06 & 1.99E+06 \\
\end{tabular}   \end{sidewaystable}

\begin{sidewaystable}[]
   \caption{1500 year term SC-GHG computations for \textbf{Permafrost thaw S-aerosol} scenario: forward 1500 years, starting the year listed, for four discount rates. Real dollars. Here $\sigma$ is the standard deviation of 30 year US Treasury bills that are the basis of $dMT30yr$. L=Low, C=Central, H=High. For N$_2$O anomaly explanation see \ref{SC-GHG_N2O_anomaly}} 
\label{PVs_1500_yearSCC_SH3}
\small
\begin{tabular}{crrr|rrr|rrr|rrr}
1500 yr  & -2.15\%    & -2.15\%    & -2.15\%    & -2.15\%    & -2.15\%    & -2.15\%    & -2.15\%    & -2.15\%    & -2.15\%    & -2.15\%    & -2.15\%    & -2.15\%    \\
Starting & -2$\sigma$ & -2$\sigma$ & -2$\sigma$ & -2$\sigma$ & -2$\sigma$ & -2$\sigma$ & -2$\sigma$ & -2$\sigma$ & -2$\sigma$ & -2$\sigma$ & -2$\sigma$ & -2$\sigma$ \\
year CE  & CO2 L      & CO2 C      & CO2 H      & CH4 L      & CH4 C      & CH4 H      & N2O L \ref{SC-GHG_N2O_anomaly}     & N2O C      & N2O H \ref{SC-GHG_N2O_anomaly}      & Fgas L     & Fgas C     & Fgas H     \\
2025     & 2.54E+15   & 4.88E+15   & 7.75E+15   & 3.11E+03   & 3.44E+03   & 3.62E+03   & 4.43E+13   & 4.37E+13   & 4.30E+13   & 4.22E+19   & 4.22E+19   & 4.22E+19   \\
2030     & 2.54E+15   & 4.88E+15   & 7.75E+15   & 3.40E+03   & 3.80E+03   & 3.97E+03   & 4.42E+13   & 4.37E+13   & 4.30E+13   & 4.22E+19   & 4.22E+19   & 4.22E+19   \\
2035     & 2.54E+15   & 4.88E+15   & 7.76E+15   & 3.73E+03   & 4.16E+03   & 4.32E+03   & 4.42E+13   & 4.36E+13   & 4.29E+13   & 4.22E+19   & 4.22E+19   & 4.22E+19   \\
2040     & 2.54E+15   & 4.88E+15   & 7.76E+15   & 4.09E+03   & 4.52E+03   & 4.68E+03   & 4.41E+13   & 4.36E+13   & 4.29E+13   & 4.22E+19   & 4.22E+19   & 4.22E+19   \\
2045     & 2.54E+15   & 4.88E+15   & 7.76E+15   & 4.47E+03   & 4.89E+03   & 5.05E+03   & 4.41E+13   & 4.35E+13   & 4.28E+13   & 4.22E+19   & 4.22E+19   & 4.22E+19   \\
\end{tabular}   \end{sidewaystable}

\begin{sidewaystable}[]
   \caption{1500 year term SC-GHG computations for \textbf{Permafrost thaw S-aerosol} scenario: forward 1500 years, starting the year listed, for four discount rates. Real dollars. Here $\sigma$ is the standard deviation of 30 year US Treasury bills that are the basis of $dMT30yr$. L=Low, C=Central, H=High. For N$_2$O anomaly explanation see \ref{SC-GHG_N2O_anomaly}} 
\label{PVs_1500_yearSCC_SH4}
\small
\begin{tabular}{crrr|rrr|rrr|rrr}
1500 yr  & 0.435\%  & 0.435\%  & 0.435\%  & 0.435\%  & 0.435\%  & 0.435\%  & 0.435\%  & 0.435\%  & 0.435\%  & 0.435\%  & 0.435\%  & 0.435\%  \\
Starting & drDICE   & drDICE   & drDICE   & drDICE   & drDICE   & drDICE   & drDICE   & drDICE   & drDICE   & drDICE   & drDICE   & drDICE   \\
year CE  & CO2 L    & CO2 C    & CO2 H    & CH4 L    & CH4 C    & CH4 H    & N2O L \ref{SC-GHG_N2O_anomaly}     & N2O C      & N2O H \ref{SC-GHG_N2O_anomaly}    & Fgas L   & Fgas C   & Fgas H   \\
2025     & 8.51E+01 & 1.34E+02 & 1.87E+02 & 2.22E+03 & 2.45E+03 & 2.58E+03 & 2.18E+04 & 2.21E+04 & 2.21E+04 & 2.82E+06 & 2.84E+06 & 2.85E+06 \\
2030     & 8.79E+01 & 1.39E+02 & 1.92E+02 & 2.42E+03 & 2.71E+03 & 2.85E+03 & 2.26E+04 & 2.28E+04 & 2.29E+04 & 2.89E+06 & 2.91E+06 & 2.92E+06 \\
2035     & 9.08E+01 & 1.43E+02 & 1.97E+02 & 2.66E+03 & 2.98E+03 & 3.11E+03 & 2.34E+04 & 2.36E+04 & 2.36E+04 & 2.96E+06 & 2.98E+06 & 2.99E+06 \\
2040     & 9.38E+01 & 1.47E+02 & 2.03E+02 & 2.92E+03 & 3.26E+03 & 3.38E+03 & 2.42E+04 & 2.44E+04 & 2.43E+04 & 3.03E+06 & 3.05E+06 & 3.06E+06 \\
2045     & 9.68E+01 & 1.52E+02 & 2.08E+02 & 3.20E+03 & 3.53E+03 & 3.65E+03 & 2.50E+04 & 2.51E+04 & 2.51E+04 & 3.11E+06 & 3.13E+06 & 3.14E+06 \\
\end{tabular}   \end{sidewaystable}
\clearpage

\vskip 20pt
 Most computations and graphs in this paper were performed using Maple™ \cite{MapleCite2022-Computations}.


\bibliographystyle{RS} 
\bibliography{references2,Refs-+4C} 

\end{document}